\documentclass[a4paper,notitlepage,oneside,12pt,pdflatex]{book}

\usepackage{multirow}
\usepackage[tight]{subfigure}
\usepackage{url}
\usepackage{float}
\usepackage{ifpdf}
\usepackage{amsmath}	
\usepackage{enumerate}
\usepackage[ruled,linesnumbered]{algorithm2e}
\usepackage{framed}
\usepackage{lscape}
\usepackage[numbers,sort]{natbib}
\usepackage{amsfonts}

\usepackage{newtxtext,newtxmath}
\usepackage[T1]{fontenc}

\newcommand{\titlepagetitle}{Distributed Quantum Computing Utilizing Multiple Codes on Imperfect Hardware}

\usepackage{epigraph}

\usepackage{fancyhdr}
\makeatletter
\renewcommand{\chaptermark}[1]{\markboth{{\@chapapp}\ \thechapter. \MakeUppercase{#1}}{}}

\makeatother
\newcommand{\urlwofont}[1]
{
\url{#1}
}

\usepackage{setspace}

\usepackage{CJKutf8}

\usepackage{cmap}
\usepackage{graphicx,color}
\usepackage{epstopdf}
\usepackage[bookmarks=true,bookmarksnumbered=true,%
bookmarkstype=toc,pdftitle={\titlepagetitle},%
colorlinks=true,%
citecolor=black,%
linkcolor=black,%
urlcolor=black,%
bookmarksopen=true,%
pdfpagemode=UseNone,%
plainpages=false,%
pdfsubject={Academic Year 2016 Dissertation},%
pdfauthor={Shota Nagayama},%
pdfkeywords={Quantum computer architecture}]{hyperref}


\makeatletter
\def\caption{%
   \ifx\@captype\@undefined
     \@latex@error{\noexpand\mycaption outside float}\@ehd
     \expandafter\@gobble
   \else
     \refstepcounter\@captype
     \expandafter\@firstofone
   \fi
   {\@dblarg{\@caption\@captype}}%
}

\long\def\@caption#1[#2]#3{%
  \par
  \addcontentsline{\csname ext@#1\endcsname}{#1}%
    {\protect\numberline{\csname the#1\endcsname}{\ignorespaces #2}}%
  \begingroup
    \@parboxrestore
    \if@minipage
      \@setminipage
    \fi
    \normalsize%
    \def\@tempa{#2}
    \def\@tempb{#3}
    \ifx\@tempa\@tempb%
      \@makecaption{\csname fnum@#1\endcsname}{{\bf \ignorespaces #3}}\par
    \else
      \@makecaption{\csname fnum@#1\endcsname}{{\bf \ignorespaces #2}~\ignorespaces #3}\par
    \fi
  \endgroup}
\makeatother

\begin{document}

\baselineskip=3.9mm

\newcommand{\titlepageyear}{2016}
\newcommand{\titlepagehead}{Doctoral Dissertation  \hspace{1cm} Academic Year \titlepageyear{}}
\newcommand{\titlepagelogo}{fig/pen_1.pdf}
\newcommand{\titlepageauthor}{Shota Nagayama}
\newcommand{\titlepageuniv}{%
  Graduate School of Media and Governance \\
  Keio University \\
  5322 Endo Fujisawa, Kanagawa, JAPAN 2520882}
\newcommand{\titlepagemajor}{Media and Governance}

\thispagestyle{empty}
\pagenumbering{roman}

\begin{rmfamily}
\begin{center}

\begin{large} \titlepagehead{} \end{large}

\vspace{3cm}

\begin{Huge} \titlepagetitle{}\\ \end{Huge}

\vspace{1.8cm}

\begin{center}
\includegraphics[height=4cm]{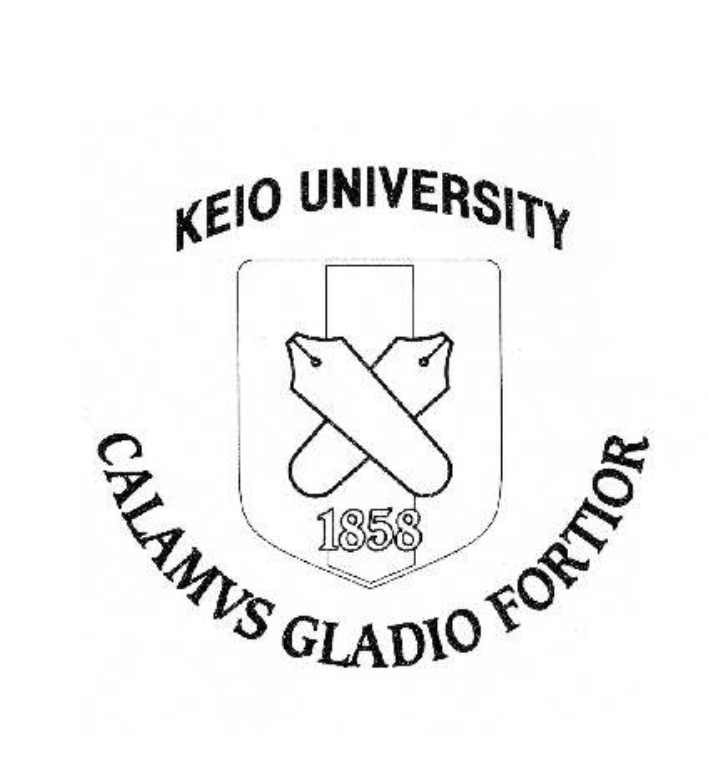}

\end{center}

\vspace{1.8cm}

\begin{LARGE} \titlepageauthor{} \end{LARGE}

\vspace{1cm}

\begin{large} \titlepageuniv{} \end{large}

\vspace{0.5cm}

{\rm\it
A dissertation submitted in partial fulfillment of the requirements\\
for the degree of Doctor of Philosophy
}

\bigskip

Copyright \copyright{} 2016 by \titlepageauthor{}

\bigskip


\vfill

\end{center}
\end{rmfamily}

\cleardoublepage

\baselineskip=1.2\normalbaselineskip
\chapter*{Abstract}
Quantum bits have technological imperfections.
Additionally, the capacity of a component that can be implemented feasibly is limited.
Therefore, distributed quantum computation is required to scale up quantum computers able to solve usefully large problems.

This dissertation presents the design of components of quantum CPUs and of quantum memories taking into account imperfections.
Quantum CPUs employ a quantum error correcting code which has faster logical gates
and quantum memories employ a code which is superior in space resource requirements.
This new quantum computer architecture aimed to realize distributed computation
by connecting quantum computer each of which consists of multiple quantum CPUs and multiple quantum memories.

This dissertation focuses on quantum error correcting codes,
giving a practical, concrete method for tolerating static losses such as faulty devices for the surface code.
To validate this method, I analyzed the resource consumption of cases where faulty devices exist
and quantified the increase of resource consumption by numerical simulation with practical assumptions.
I found that a yield of functional qubits of
90\% is marginally capable of building large-scale systems,
by culling the poorer 50\% of chips during post-fabrication testing.
Yield 80\% is not usable even when discarding 90\% of generated lattices.

For the internal connections between quantum CPU and memory components in a quantum computer
and for connections of quantum computers, this dissertation gives a fault-tolerant method to connect quantum components
that employ heterogeneous quantum error correcting codes.
I have validated this method and quantified the resource consumption of the error management by numerical simulation.
I found that the scheme, which discards any quantum state in which any error is detected,
always achieves an adequate logical error rate regardless of physical error rates
in exchange for increased resource consumption.

Additionally, this dissertation gives a new extension of the surface code suitable for quantum memories.
This code is shown to require fewer physical qubits to encode a logical qubit than conventional codes.
This code achieves the reduction of 50\% physical qubits per a logical qubit.

Collectively, the elements to construct distributed quantum computation by connecting quantum computers are brought together to
propose a distributed quantum computer architecture.

{\bf Keywords: }{Quantum computer architecture, Distributed quantum computation, Quantum error correction, Surface code quantum computation, faulty qubit, quantum error correction code interoperability}\\

\begin{flushleft}
{\bf Thesis Committee:}\\
Supervisor: \\
\hspace{7em}Prof.~Jun Murai, Keio University\\
\smallskip
Co-Adviser:\\
\hspace{7em}Assoc.~Prof.~Rodney Van Meter, Keio University  \\
\hspace{7em}Dr.~Dominic~Horsman, Durham University, U.K. \\
\hspace{7em}Prof.~Hiroyuki Kusumoto, Keio University  \\
\end{flushleft}

\addcontentsline{toc}{chapter}{Abstract}
\chapter*{\begin{CJK}{UTF8}{min}要旨\end{CJK}}
\begin{CJK}{UTF8}{min}
量子ビットの原理的・技術的な不完全性を解決するには，量子計算リソースが必要である．
しかし，一つの部品に実装可能な計算リソース量は限られている．
したがって，量子コンピュータが規模や性能において飛躍的な発展を遂げるためには，
量子コンピュータによる分散処理を実現しなければならない．

本研究では，多数の量子CPUや量子メモリを耐不完全性を考慮して設計・接続し，
分散処理により計算リソースを確保して，
実用的な規模の問題を解ける量子コンピュータアーキテクチャを構築する．
本アーキテクチャでは，計算リソースを節約するために
部品毎の役割を明確化し，各役割に適切な量子エラー訂正符号を採用する．

本研究は以下に述べる三つの成果から構成される．
まず，エラー訂正符号に着目し，未解決の不完全性だった不良量子ビットへの対応手法を開発した．
シミュレーションにより本手法のエラー訂正能力を検証し，消費計算リソースの増加量を明らかにした．
この結果，量子ビットの正常動作率が九割の環境では，生産した計算チップの50\%が実用可能であると判明した．
また，動作率八割では，10\%も実用できない事が分かった．

次に，異なる量子エラー訂正符号を用いる部品間を，耐不完全性を考慮して繋ぐ接続手法を開発した．
本手法は，単一量子コンピュータ内の量子部品の接続に加えて，異なる量子エラー訂正符号を用いる量子デバイス間の相互接続を実現する．
本手法の有効性を検証するため，シミュレーションにより，必要な計算リソース量を明らかにした．
この結果，エラーを訂正するよりも，エラーが発見された通信リソースを廃棄するエラー管理方法が，本接続手法に適している事が分かった．

さらに，メモリ向きの新しい量子エラー訂正符号を開発した．
計算の結果，本符号は，従来のエラー訂正符号の二倍の空間効率で量子メモリを構成出来る事が分かった．

これらの研究により，量子コンピュータの相互接続によるネットワーク型分散コンピュータを構成する要素が整った。
これらを組み合わせ，実現可能な分散量子コンピュータアーキテクチャを提案する．

\textbf{キーワード: }{量子コンピュータアーキテクチャ, 分散量子計算, 量子エラー訂正符号, Surface Code 量子計算, 不良量子ビット, 量子エラー訂正符号変換}
\end{CJK}

\chapter*{Acknowledgment}
Portions of this dissertation are adapted from papers under copyrights by the American Physical Society~\cite{PhysRevA.93.042338,PhysRevA.95.012321} and by the IOP Publishing~\cite{1367-2630-19-2-023050}.

\vspace{1cm}

First, I would like to thank all the people who support, help and encourage my research life.

\vspace{1cm}
I gratefully acknowledge Professor Jun Murai, my thesis supervisor.
He has kept giving insightful comments.
His words always indicate how IT and IT researchers should be.
I gratefully acknowledge Associate Professor Rodney D. Van Meter.
He has kindly and patiently led me from my undergraduate days.
He has given me many opportunities, such as,
to work on research topics I got interested in,
to attend many conferences
and 
to collaborate with great people.
He has always respected my interests.
I learned how to enjoy research and how to make good research from him.
If I had not met him, I might not have become a researcher.
I acknowledge Postdoctoral Research Associate Dominic Horsman from Durham University.
My surface code life could not start without him.
I acknowledge Professor Hiroyuki Kusumoto. 
His advice based on wide knowledge from physics to informatics are always pertinent.
%

I also acknowledge Project Associate Professor Shigeya Suzuki,
for his professional advice, guidance and encouragement.
I acknowledge Dr. Austin Fowler, Dr. Simon Devitt, Dr. Byung-Soo Choi
and Dr. Takahiko Satoh,
for great collaborations. 
I also acknowledge Professor Ken Brown.
My work has not been accomplished without their
professional advice.

I would like to thank professors, associate professors and assistant professors in MAUI project.
I acknowledge Professor Osamu Nakamura.
His comments have always been straight to the points of discussions.
His so severe comments in research meetings were actually very tough to me, and, at the same time roused me so much.
I acknowledge Associate Professor Keisuke Uehara.
He has kindly paid attention to me in the laboratory and kept grasping my situation.
I thank 
Professor Jin Mitsugi,
Professor Keiko Okawa,
Professor Keiji Takeda,
Project Assistant Professor Achmad Husni Thamrin,
Project Assistant Professor Ucu Maksudi
Project Associate Professor Masaaki Sato
and Dr. Kenji Saito
for their kindly and insightful comments and for keeping their eyes on me.

I also acknowledge Professor Yoshihisa Yamamoto and members of the FIRST Quantum Information Processing Project.
I learned many things about quantum information processing in the project.
The project gave me great opportunities.
Connections I made in this project are very important assets for my research life.

I also thank members of AQUA group in the laboratory,
Takaaki Matsuo for I have enjoyed work with him and collaboration in assistance of classes,
Takafumi Oka for his interests in mathematics has reminded me the importance of basic theories,
and Shinnosuke Ozawa for his persistent work for ORF has encouraged me.

I thank 
Project Research Associate Noriatsu Kudo, for his regular and wide advice about the process to get the Ph.D degree and about the laboratory life,
Dr. Takeshi Matsuya, his addiction to research and hobbies has encouraged me,
Project Research Associate Yohei Kuga, I aspire to be widely informed as him,
and
Dr. Shiori Suzuki Sawada, she is close in grade, I have many good memories of considering various things to get the Ph.D degree together and I thank her friendship.
I thank Dr. Masayoshi Mizutani for I often consult his doctoral dissertation to write my doctoral dissertation.

I thank
Katsuhiro Horiba, I have learned many things from his gentle manner and his attitude keeping always a smile,
Yuki Sato, his very serious attitude to research has encouraged me,
Hirotaka Nakajima, for his political power and his way how to get on in the world serve as useful references,
Syuhei Kimura, I have much to learn from his devotion for education,
Tsugumasa Suzuki, for his persistent efforts to establish new educating style tells the importance of continuation,
Masaki Minami, for his wide knowledge about methodology of education is evocative,
Rajoria Nitish, for his research topics involving physics is actually rare in this laboratory and I have felt a kind of fellowship,
and Andrey Ferriyan, for his cheerfulness has been helpful when I was in hard situation.

I also thanks to 
Hiromichi Morimoto, for his well-seen concern,
Iori Mizutani, his hard work has encouraged me that I cannot allow myself to get overtaken by him as a senior student,
Yohei Kezuka, I have some fellowship with him working on encoding,
Masato Miyazawa, for organizing graduate students' meeting,
Takuya Yui, for his serious attitude to everything has encouraged me,
Keiko Shigeta, for her graceful concern,
and Shota Kawamoto for organizing graduate students' meeting.

I also thanks to living-in-laboratory members,
Haruka Nakashima, for her passion for tidying up,
Kohei Suzuki, for his network operations which support my research life and for his nice funny rap,
Kotaro Oki, for his network operations which support my research life,
Mariko Kobayashi, her office politics, organizing skills and her mental stability are outstanding and are good models to learn,
Masahiko Hara, his bland tone organizing skill is a good model to learn,
Ryosuke Abe, for his seriousness even for problems of somebody else has encouraged me and I thank him for Student Assistant in my ICT foundation class,
Shun Kinoshita, his management skills at controlling hates has a lot to learn and Vainglory he introduced me has been good relaxation to write the dissertation,
Yuka Sasaki, for her passion for tidying up,
Yuma Kurogome, for I have been encouraged by his passion to technologies and research,
Yusei Yamanaka, he sometimes reminded me that a man should be selfish to his/her interests,
Yusuke Takahashi, I hope that he be happy,
Akira Syoji, his curiosity to various things has inspired me and I feel that his kind friendship may be now producing the basic mood of the office, 
I like the change of his interest previously aimed to playing games to technologies including that forms the foundation advanced games,
Ryoma Kawaguchi, I'm sorry but I cannot conceive of acknowledgment about you more than about Mr. Yachi in the meaning of the number of lines,
I thank him for the ease to stay/work together backed by his engaging frankness and thank him for Student Assistant in my ICT foundation class,
Sena Kuwabara, for his cooking, hopefully if his snorer were a bit less,
Shuya Osaki, I guess he can be a bit more selfish without getting disliked, actually I think he get more loved (or possibly liked) by representing his humanity,
Ryo Konishi, his knowledge and interests in applications and platforms are somehow new and fresh to me and were good for refresh in writing the dissertation,
Yasunobu Toyoda, his kindness helped me at various times and I hope his kindness granted always to everyone, and I thank him for Student Assistant in my ICT foundation class,
Akio Yasui, actually sometimes I'm confused his name with Yasuo Akita, I believe this time I'm correct, and I thank him for encouraging fresh talks about girls,
Jiu Jie, I would like him to find new hobbies other than the special videos made in Japan,
Kaku Girai, I hope he learn beautiful Japanese somewhere else,
Kazunori Jo, his unique character often relaxed me strongly stressed by the dissertation,
Maiko Wakatsuki, I thank her for taking over laboratory tasks from me,
Reo Kobayashi, he reminds me that I can have a thicker skin,
Yuta Sugafuji, for his passion to entertain people, I would like him to learn technologies with passion, even 10\% of the passion to be Mr. SFC as an entertainer is great enough,
Hirotaka Kumagai, for his interests in AIs has inspired me,
Ryusei Morishima, I hope that he live calm life though I see talents to be Hentai Gentleman in him,
Yuji Kurihara, for his slave comedian character is so funny, I hope him to get employed in white company,
and 
Yusuke Suzuki, his wide techniques has interested me,
finally, I thank all of those above for interesting, delightful and funny laboratory life.

I gratefully acknowledge the secretaries of the Murai Laboratory,
Yasue Watanabe, Yukie Shibuya and Yumiko Usami.
They helped me with various things such as scheduling and complex paper works.

\vspace{1cm}

I also acknowledge the StarBED Project~\cite{Miyachi2012}.
My big simulation could not be accomplished without their support.
I acknowledge JSPS KAKENHI Grant Number 25:4103, DC2, the scholarship by Japan Society for the Promotion of Science.

\vspace{1cm}

I thank Asuka Nakajima for 
her slow-tempo character has relaxed me and for
she sometimes plans travels and tries to
get me like a houseplant out of home.

I thank my friends sharing a dark history,
Yasuhiro Sogawa, for he knows me very much, BTW his happily sulky face in his wedding party was so good,
Yasuyo Onari, for she knows me very much, and every time I see her on TV, I feel happy and get encouraged, 
and Yukiko Otsuka Nakano and Kazuaki Nakano, for they taught me that my friend's child was so cute to me and relaxed me, and for her good life advice.
I believe we can create something but a dark history next time, yet, I love our dark history.

I thank Yudai Yamagishi and Yuta Tokusashi, for their friendship, I enjoyed the Ph.D student life so much with them.
I thank Megumi Nakasato, for her omiyage was good. She knows good snacks which make me happy.

\vspace{1cm}

I thank Bose Corporation, their noise canceling earphone is great and has saved me from kinds of noises,
caused mainly by servers, by air conditioners and sometimes by people.
Noise cancellation is useful of course when I work, and also when I sleep.

I thank Hikaru Utada, her voice has mitigated my stress.
Her songs have had me sleep well even when I'm so stressed and feel hard to sleep, 
with the noise canceling earphone mentioned above.
I like 
\begin{CJK}{UTF8}{min}
荒野の狼
\end{CJK} 
the best in her 6th album, Fant\^ome.

I thank D anime store, for providing BGM to write this dissertation.
I apologize for having occupied network bandwidth though I did not watch animes seriously during writing the dissertation.

I thank Sheryl Nome, she has encouraged my enduring effort.
Her character, serious, with endeavor but a bit unskillful 
attitude to closest people and to herself is so humanly beautiful.

I thank Walkure, for I often listen to them as BGM when writing this dissertation,
though now their songs remind me the stress and the pressure during writing the dissertation.

I thank Mr. Yachi, though I don't know his given name, 
he has sold many necessary things to us in the convenient store behind the campus.
His attitude to work, playing Puzzle and Dragons during his working time,
often gave warnings to me not to be too serious to work.

I thank Kuro, Shiro and gone Senmu, the cats living/lived around Murai laboratory.
They taught me animal therapy is good.

\vspace{1cm}

I thank my family, my father Ryuji, my mother Hiroko, my brother Taiki, my sister Hikari,
my grandfather Masaru, my grandmother Kinuko, gone my grandfather Takenori and gone my grandmother Kyoko
for their support and for keeping their eyes on me.

\vspace{1cm}

Finally, I thank Japanese Robot Animes and their creators, especially Gundams.
My motivation to work on science is to make a wonderful world come true such as theirs,
however, actually the world I wish to realize is not their world 
but the world I have been dreaming from my childhood days to now and I'm going to dream in the future.

\addcontentsline{toc}{chapter}{Acknowledgement}

\baselineskip=5.0mm
\tableofcontents
\addcontentsline{toc}{chapter}{Contents}

\listoffigures
\addcontentsline{toc}{chapter}{List of Figures}

\listoftables
\addcontentsline{toc}{chapter}{List of Tables}

\mainmatter
\renewcommand{\thepage}{\arabic{page}}\setcounter{page}{1}
\renewcommand{\baselinestretch}{1.5}
\baselineskip=1.2\normalbaselineskip

\clearpage
\chapter{Introduction}
\label{chap:intro}
The presence of Information Technology today is ubiquitous.
Every individual and organization has computer devices,
many connected via the Internet.
The Internet, hence the IT world, covers the entire planet today.
Sometimes individuals and organizations communicate one-on-one and
sometimes they communicate as a group on the Internet immediately and easily, even if they are on the opposite sides of the planet.
Thanks to the IT infrastructure, many everyday activities such as commerce are transferred from the physical world to the IT world.
This power accelerates human lives and increases their flexibility.

While the Internet connects everyone and every place,
local networking can also aggregate 
large numbers of individual, powerful computers to create scalable distributed-memory systems capable of solving large problems
~\cite{bluegene}.
However, there still remains problems we cannot solve even with such aggregated computational power.
Feynman initially conceived of quantum computation to execute quantum simulation,
a problem that conventional computers 
find impossible to solve efficiently as the size of the problem grows
~\cite{feynman1982simulating}.

\section{Importance of quantum computation}
Quantum computation is important for two key reasons.
The first is the limit of the improvement of the classical computation's fabrication.
The development of classical CPU technology has obeyed Moore's Law and the number of transistors in a chip
doubled every two years for several decades~\cite{Moore'sLaw}.
However, the semiconductors used in classical computation face fundamental physical limitations as features approach atomic size. For example, they
suffers from leakage currents
caused by quantum tunneling, resulting in heating
since the devices implemented on the semiconductor have thickness of only tens of atoms today.
Classical CPU developers are making efforts to suppress such quantum effects,
however, it is clear that there will be a limit.
The ultimate solution to this problem is computation \emph{utilizing} the quantum effect.

The second is the lure of quantum algorithms~\cite{MontanaroNpj2016}.
There exist many problems which can not be solved by conventional computation based on classical physics.
Quantum computing is a promising technology that may give new computational power to human beings and open the door to a new scientific world.
The archetypal quantum algorithm is Shor's algorithm, found in 1994~\cite{shor:1994factor}.
Shor's algorithm factors large numbers $N$ in $O(log^3 N)$ time, while classical algorithms require superpolynomial time.
The most popular influence of Shor's algorithm is that some of the encryption systems used today will be broken;
the most common public key sharing algorithm depends on the fact that efficient classical algorithms have not been found yet
and hence on the difficulty of factoring large numbers~\cite{Boneh1995}.
The number of states of a quantum system increases exponentially 
with the number of particles hence simulation is not scalable on classical computers
~\cite{Buluta108,e12112268,RevModPhys.86.153}.
Efficient quantum simulation will contribute to chemistry, materials, high-energy physics, superconductivity and nanotechnologies.
Grover's algorithm is for unstructured search~\cite{PhysRevLett.79.325,PROP:PROP493}.
It is applied to problems such as minimum value search including the minimum of an unknown functions,
graph connectivity determination and pattern matching and SAT problems
~\cite{quant-ph/9607014,Durr2004minimum,Ramesh2003103,Montanaro2015}.

There are other useful quantum algorithms.
The Quantum Algorithm Zoo cites 314 papers at the time of writing this dissertation~\cite{qlzoo}.

\section{Problems}
\label{sec:problem}
Though many useful quantum algorithms have been found,
there is still a ways to go to construct a quantum computer which can solve large problems practically.
For example, there is no quantum bit (qubit) implementation which can sustain a state with sufficient fidelity from the start of a long computation to the end.
DiVincenzo summarized the physical conditions necessary for building a practical quantum system in 2000~\cite{quant-ph/0002077},
\begin{enumerate}
 \item A scalable physical system with well characterized qubits
 \item The ability to initialize the state of the qubits to a simple fiducial state, such as $\vert000...\rangle$
 \item Long relevant decoherence times, much longer than the gate operation time
 \item A “universal” set of quantum gates
 \item A qubit-specific measurement capability
 \item The ability to interconvert stationary and flying qubits
 \item The ability faithfully to transmit flying qubits between specified locations.
\end{enumerate}
The first five are for a quantum computer and the last two are for quantum communication.
Obviously some items of those criteria include compromises;
are coherence times much longer than the gate times but much shorter than the computation time actually enough?
Though many teams are working hard to realize any of several physical systems to be the fundamental technologies of quantum computing systems,
the coherence times range from nanoseconds to seconds~\cite{0034-4885-74-10-104401,ladd10:_quantum_computers}.

Accepting compromises on such factors can be justified by the introduction of fault-tolerant quantum computation achieved by special quantum error correcting codes.
Fault-tolerant quantum computation makes the quantum computation tolerant against the very fragile quantum states
by encoding a logical qubit in a large number of physical qubits without the necessity of decoding during the computation
~\cite{quant-ph/9608026,steane:10.1098/rspa.1996.0136,knill96:concat-arxiv,PhysRevA.52.R2493,steane:10.1038/20127,quant-ph/9705052,PhysRevA.71.042322,steane02:ft-qec-overhead,preskill98:_reliab_quant_comput,PhysRevLett.96.020501,QIC.8.0399,
PhysRevA.82.022323,
svore2007noise,1367-2630-11-1-013061}.
In fact, DiVincenzo's first five criteria only guarantee that the fault-tolerance system itself can be run.
The fault-tolerant quantum computation has large overhead both in space and in time,
hence the five criteria do not guarantee the ability to solve problems of practically meaningful sizes.
Processing Shor's algorithm to factor a number described with $N$ bits requires a quantum register with at least $2N+2$ high-quality qubits
~\cite{VanMeter:2005Fast_quantum_modular_exponentiation,shor:1994factor,Fowler:2004:ISA:2011827.2011828,Vedral:1996Quantum_networks_for_elementary_arithmetic_operations,Beckman:1996Efficient_networks_for_quantum_factoring,Pavlidis:2014:FQM:2638682.2638690,Takahashi:2006:QCS:2011665.2011669}.
If the resource requirements
to represent $2N+2$ high-quality qubits
in a fault-tolerant fashion are too high, we cannot even start,
let alone finish, the computation. Therefore, fully scalable quantum computers are required
~\cite{nielsen-chuang:qci,VanMeter:2013:BBQ:2507771.2494568,cirac2000scalable,yao10:_scal_room_temp_arch,chiaverini05:qft-impl,PhysRevB.76.174507}.

Topological quantum computation, especially the surface code
(introduced in Chapter \ref{chap:qcbasic})
extended originally from Kitaev's toric code,
has been developed as a fully scalable fault-tolerant quantum computation mechanism
~\cite{kitaev2003ftq,bravyi1998qcl,Raussendorf20062242,PhysRevLett.98.190504,
raussendorf07:_topol_fault_toler_in_clust,:/content/aip/journal/jmp/43/9/10.1063/1.1499754,PhysRevA.86.032324,1402-4896-2009-T137-014020}.
The surface code qubits are grouped in ``plaquettes'' which
consist of four neighboring
qubits in the lattice. Each plaquette is associated with a stabilizer measurement (quantum parity check).
There are two types of stabilizers -- Z stabilizers and X stabilizers -- enabling the correction of arbitrary errors.
Error syndromes are associated with pairs of sequential stabilizer measurements that differ.

The surface code has two advantages compared to other quantum error correcting codes:
its high feasibility because it requires operations only between nearest-neighboring qubits and
its higher physical state error rate threshold, nearly 1\%,
allowing it to work with a broad range of physical technologies and achieve an
arbitrary logical error rate by using longer code distances
~\cite{Fowler:2009High-threshold_universal_quantum_computation_on_the_surface_code}.
It has been shown that the fault-tolerant quantum computation including the surface code can tolerate several quantum imperfections besides unintended quantum state changes, such as
leakage errors~\cite{Lo-Spiller-Popescu:iqci,Aliferis:2007:FQC:2011706.2011715},
losses~\cite{quant-ph/9705052,
PhysRevLett.97.120501,
PhysRevA.75.010302,
RevModPhys.79.135}
including dynamic losses~\cite{PhysRevA.81.022317,PhysRevLett.102.200501,1409.4880}
and static losses~\cite{1367-2630-19-2-023050}.
A leakage error is the change of a physical state from the computational space to unused space
e.g. if the qubit zero and one states are defined as energy levels of an atom,
then leakage is finding the atom in a third energy level.
Leakage errors can be corrected by building dedicated units on each qubit~\cite{PhysRevB.91.085419}.
Dynamic loss is such as photon generation failure and dynamic loss of other qubit carriers.
Whiteside and Fowler numerically revealed that the threshold of dynamic loss rate is between 0.1\% and 1\% with practical assumptions~\cite{1409.4880}.
Static loss, where qubits are lost from the beginning of the computation to the end, may be tolerated by methods similar
to ones for dynamic loss. However, precise analysis for static loss has not been achieved prior to this dissertation.

Those characteristics make the surface code quantum computation the most promising form of fault-tolerant quantum computation.
However, there are still several problems which prevent us from building the surface code for a practical quantum computer
that should be solved at the computer system level.

\subsection{Imperfections caused in fabrication}
\label{subsec:int:imp}
Static loss is the presence of imperfections 
such as devices incapable of trapping single electrons for use as qubits, incapable of high-fidelity gates, etc.
In some fixed physical systems, fabrication imperfections could result in static losses.
For example, DiVincenzo offered an architecture for superconducting hardware for the surface code~\cite{1402-4896-2009-T137-014020},
in which a superconducting loop which does not show the appropriate quantum effect will be a missing site in the qubit layout.
Likewise, Jones et al. proposed an architecture for scalable quantum computation with self-assembled quantum dots used 
to trap electrons, which are used as qubits~\cite{Jones:2012Layered_Architecture_for_Quantum_Computing}.
There very likely will be defective quantum dots which cannot trap a single electron,
leaving holes in the code.
During initial boot stage, qubits are calibrated;
if qubits cannot be tuned to hold a single quantum, 
or if they cannot be tuned to match their neighbors, 
they can be declared not working.

To tolerate static loss, we have two choices:
design a microarchitecture to work around missing qubits,
or adapt the syndrome collection and processing to tolerate loss.
Van Meter et al. proposed a system in which the microarchitecture can create
the regular 2-D lattice even when some qubits are faulty~\cite{van-meter10:dist_arch_ijqi}.
However, this requires the ability to couple qubits across a distance spanning several qubit sites.
Stace et al. showed that qubit loss is acceptable
when performing the surface code
and that there is a trade-off between the loss rate and the state error rate.
Theoretically, if no errors occur in the qubits during the syndrome extraction process,
the logical state can be repaired after loss of less than $p_{loss}=50\%$ of the qubits.
They introduced the concept of a ``superplaquette'', which 
consists of several plaquettes that surround defective qubits.
They showed that, under the assumption that the superplaquette operators can be measured perfectly,
a threshold error rate existed for qubit loss rates below 50\%.
Barrett et al. showed that dynamic loss in the 3-D topological quantum computation
is acceptable up to $p_{loss}=24.9\%$~\cite{PhysRevLett.105.200502}.
This latter approach, however, cannot be used if a device (used to bond together qubits the 3-D topological lattice)
in the quantum computer is permanently faulty, leading to a column in time of lost qubits.
We measure the six face qubits in a unit cell for syndrome extraction in the 3-D topological computation 
because the six qubits are the output of $\Pi_i X_i \otimes _{q_j\in ngbr(q_i)}Z_j$ where $q_i$ are the face qubits.
A lost qubit merges $\Pi_i X_i \otimes _{q_j\in ngbr(q_i)}Z_j$ of two unit cells~\cite[Chapter 20]{lidar-brun:qec}.
A column in time of lost qubits from the beginning of the computation to the end works as a logical qubit
because we do not have stabilizer
$X_0\otimes _{q_j\in ngbr(q_0)}Z_j$
on the first 2-D surface of the 3D cluster state
where the lost column starts from $q_0$
and
$X_0'\otimes _{q_j\in ngbr(q_0')}Z_j$
on the last 2-D surface,
therefore the merged $\Pi_i X_i \otimes _{q_j\in ngbr(q_i)}Z_j$ of the merged unit cells is not closed.
Hence another solution is required for faulty devices in 3D topological code.
And therefore a practical method and an estimation for tolerance against static loss is required to complete
our toolkit of strategies against the full set of quantum imperfections.

\subsection{Excessive resource consumption}
\label{subsec:int:rr}
The most complete architecture proposed for the surface code requires at least $4.57\times10^7$ physical qubits and $10.81$ hours to solve 1024-bit factorization~\cite{1312.2316},
even with some ideal assumptions in a simplified flat qubit placement design.
In fact, such a design is also practically difficult
because of the size of the chip and because of integration challenges such as control lines~\cite{Hille1500707,1608.06335}.
It is not clear yet whether this number of qubits is feasible.
For architectures such as shown in~\cite{PhysRevB.76.174507}, which fits 84 flux qubits in a 12mm by 5mm area,
if we assume that a chip is limited to 84 qubits,
then a full system could require $5.44\times 10^5$ chips, a fabrication challenge perhaps beyond our technical capabilities.
Resource reduction even by a few percent will increase the feasibility,
and advance the construction of the first quantum computer.
Reduction of the number of required qubits would be a valuable contribution to the construction of a practical quantum computer.

\subsection{Requirement of architectural support for internal/external heterogeneously encoded fault-tolerant quantum communication}
\label{subsec:int:arch}
To build a large quantum computer from small components having limited abilities,
quantum communication and an efficient design of networking to connect such components are required.
The purpose of a quantum connection is to create quantum states known as Bell pairs shared between
two subsystems to enable quantum teleportation~\cite{PhysRevLett.70.1895}
moving data around to support large scale computation.
Bell pairs are discussed in Subsection~\ref{subsec:basic:ent}.

The design of connections between internal components is related to the computer architecture.
Oskin et al. and Copsey et al. assumed the existence of internal connections between different components
in their hierarchical memory architectures,
but did not focus on the creation of those internal connections~\cite{976922,copsey02:_quant-mem-hier}.
Components for the optical interconnection and some simple architectures utilizing them have been proposed
~\cite{PhysRevA.89.022317,ahsan2015designing,kim03:_1100_port_mems}.
Photonic implementations that fulfill DiVincenzo's 6th and 7th criteria can be employed as the flying qubits carried in the internal connections
~\cite{0957-0233-21-1-012002,0953-2048-25-6-063001,:/content/aip/journal/apl/104/8/10.1063/1.4866582}.
The advantage of optical connections is the ability to distribute photons;
qubits can go to wherever fibers are connected as long as the conversion is
realized between a photonic qubit and the computational qubit
~\cite{Inagaki:13,Herbst15:free-space143km}.

The design of the connections of computers is distributed quantum computation
~\cite{RevModPhys.82.665,buhrman03:_dist_qc,buhrman1998quantum,PhysRevA.89.022317,Chien:2015:FOU:2810396.2700248,broadbent2010measurement,crepeau:_secur_multi_party_qc}.
To achieve distributed quantum computation,
we need quantum networks, quantum internetworks and a new quantum computer architecture.
Quantum networks and quantum internetworks consist of quantum nodes
that correspond to classical switching hubs or routers
and deliver quantum information between two arbitrary quantum computers~\cite{kimble08:_quant_internet,Azuma_2015,6246754,munro2011designing,vanmeter2014book,Devitt:arXiv:1410.3224,
Aparicio:2011:PDQ:2089016.2089029,doi:10.1117/12.893272,
Inagaki:13,Herbst15:free-space143km}.
The new quantum computer architecture must support networking
and then must support internal routing among components~\cite{quant-ph/0607065}.

Due to again the vulnerability of quantum states to noise and the lossy photonic channel,
quantum communications must support error correction
for both internal connections of a quantum node and connections between quantum nodes
~\cite{briegel98:_quant_repeater,dur:PhysRevA.59.169,PhysRevA.85.062326,Jiang30102007,PhysRevA.79.032325,PhysRevLett.104.180503,munro2010quantum,1367-2630-15-2-023012}.

\section{Contribution of this dissertation}
The primary contribution of this dissertation is a practical quantum computer architecture which tolerates quantum imperfections
with efficient computational resource requirements and can be built from computational components
that can feasibly built,
solving the problems and requirements stated in Section \ref{sec:problem}.

My contributions to working around static losses and overhead of the estimation are
 \begin{itemize}
  \item A method for constructing the circuits for error correction to give practical adaption to static losses
	  to ``superplaquettes'' proposed but treated ideally in ~\cite{PhysRevA.81.022317,PhysRevLett.102.200501,PhysRevLett.105.200502}.
	  Both statically lost data qubits and ancilla qubits are tolerable.
 \item Analysis of tolerable yield of functional qubits by numerical simulation, showing that yield 90\% is marginally capable of building large-scale systems,
	 by culling the poorer 50\% of chips during post-fabrication testing, assuming randomly distributed nonfunctional qubits.
	 Yield 80\% is not usable even when discarding 90\% of generated lattices.
 \item Correlations between the logical error rate and a dozen characteristics of the lattice,
	 which contribute to guiding the construction of an ensemble of quantum computation chips
	 good enough to compose a large scale fault-tolerant quantum computer
	 ~\cite{RevModPhys.82.665,buhrman03:_dist_qc,buhrman1998quantum,PhysRevA.89.022317,
	 Chien:2015:FOU:2810396.2700248,broadbent2010measurement,crepeau:_secur_multi_party_qc}.
	 The deepest depth of parity check circuits and the biggest number of data qubits owned by a parity check unit
	 have largest and the next largest correlation with the logical error rate of the lattice, respectively.
\end{itemize}

My contributions to new denser packed surface code with the universal set of logical gates are
\begin{itemize}
 \item Denser packing of logical qubits of the surface code, good for quantum memories,
	 with the reduction of 50\% physical qubits per logical qubit.
 \item Analysis based on error chain comparison with conventional surface code qubits.
 \item An universal set of logical gates for more densely packed logical qubits
	 and direct conversion from other surface code qubits.
\end{itemize}

My contributions to fault-tolerant quantum switched backplane supporting networking are
\begin{itemize}
 \item A practical architecture for fault-tolerant quantum internal communication supporting heterogeneously encoding with scalable switching,
	 extended from ~\cite{PhysRevA.89.022317,ahsan2015designing,kim03:_1100_port_mems}.
 \item Numerical analysis of three possible schemes with assumptions based on experimental results.
 \item Demonstration that the scheme in which any error detected 
	 in a resource state
	 gets discarded
	 always achieves about a certain logical error rate
	 regardless of physical error rates in exchange for resource consumption.
\end{itemize}

\section{Contents and Structure}
This dissertation is divided into eight chapters.
This first chapter is the overview of this dissertation.
Chapter \ref{chap:qcbasic} is a brief summary of basics of quantum computation, quantum error correction,
especially the surface code, and the existing quantum computer architectures.
Chapter \ref{chap:defective} describes the mechanism for tolerating static losses.
Chapter \ref{chap:deformation} demonstrates the dense packing of the deformation-based surface code with logical universal gates.
Chapter \ref{chap:hetero} shows the fault-tolerant Bell pair creation schemes for heterogeneously encoded Bell pairs between quantum components connected optically.
Chapter \ref{chap:sdqcarch} describes the quantum computer architecture for scalable distributed quantum computation.
Chapter \ref{chap:eval} shows the performance of the components proposed in this dissertation.
Chapter \ref{chap:conc} is the conclusion.


\clearpage
\chapter{Quantum Computing Systems}
\label{chap:qcbasic}
  The advantages and motivation of quantum computation were covered in Chapter \ref{chap:intro}.
The criteria to achieve quantum computation, DiVincenzo's criteria were also covered in Chapter \ref{chap:intro}.
This Chapter summarizes the background knowledge necessary to read dissertation.
The basic concepts of quantum information processing;
imperfections in qubits and technologies to protect against them;
related quantum computer architecture; quantum networking,
and distributed quantum computer architecture are explained.

\section{Quantum Information Basics}
\label{subsec:qib}
In this Section, basic descriptions, characteristics and operations of
quantum information are summarized.
Superposition, interference, entanglement and unitarity are key characteristics for quantum computation.
Especially universal gate set, entanglement swapping and quantum teleportation
are the key characteristics for this dissertation.

\subsection{Qubit description}
Researchers have been unlocking the secrets of quantum mechanics for about one hundred years,
and determined 
that energy comes in discrete amounts, rather than continuous,
and 
that the measured value of a sufficiently small object will be determined in
a probabilistic fashion and the state of such a matter is
described with a wave amplitude that gives the probability, where the amplitude can be a complex number.
Such a small matter is called a ``quantum''. ``Quantum'' is a collective term
to describe matters which obey quantum mechanics.

When a quantum is measured, its value is determined by the axis of measurement we execute,
depending on the probability amplitudes. This probabilistic process results in
that a quantum state cannot be reproduced from the observed value,
because we cannot infer the original probability amplitudes from the observed value.
Such characteristics are utilized to define a quantum bit, or a qubit.

A qubit utilizes two states of quantum observables.
For example, the energy levels are one type of observable of an electron.
Two levels of the energy levels can be used to describe a qubit,
such as, the ground state $\vert g \rangle$ for $\vert 0 \rangle$
and the first excited state $\vert e \rangle$ for $\vert 1 \rangle$.
Another example is the polarization of a photon.
Horizontal polarization $\vert H \rangle$ and the vertical polarization $\vert V \rangle$ 
can be used for the two states to define a qubit.

For a qubit encoded on such a quantum state,
the wave function of the qubit can be described using the state vector, or
Dirac ket, notation,
\begin{equation}
\vert\Psi \rangle = \cos \frac{\theta}{2} \vert0\rangle + 
e^{i\phi}\sin \frac{\theta}{2} \vert1\rangle
= \begin{pmatrix}\cos \frac{\theta}{2}\\e^{i\phi}\sin \frac{\theta}{2}\end{pmatrix}
= \begin{pmatrix}\alpha\\\beta\end{pmatrix}.
\label{equ:qubit}
\end{equation}
Here, our two basis vectors are
$\vert0\rangle \equiv \begin{pmatrix} 1 \\ 0 \end{pmatrix}$,
$\vert1\rangle \equiv \begin{pmatrix} 0 \\ 1 \end{pmatrix}$.
Note that the basis states are orthogonal, $\langle \phi \vert \psi \rangle = 0$
for any two basis vectors $\vert \phi \rangle$ and $\vert \psi \rangle$.
This quantum state has two degrees of freedom, called the value and the phase,
which are parameterized by $\theta$ and $\phi$.
This equation describes an important characteristic of a qubit:
both basis vectors can exist together in this generalized state.
This characteristic is called the \emph{superposition},
and the coefficients $\cos \frac{\theta}{2}$ and $e^{i\phi} \sin \frac{\theta}{2}$
describe the probability amplitudes of the basis.

\begin{figure}[t]
 \begin{center}
  \includegraphics[width=200pt]{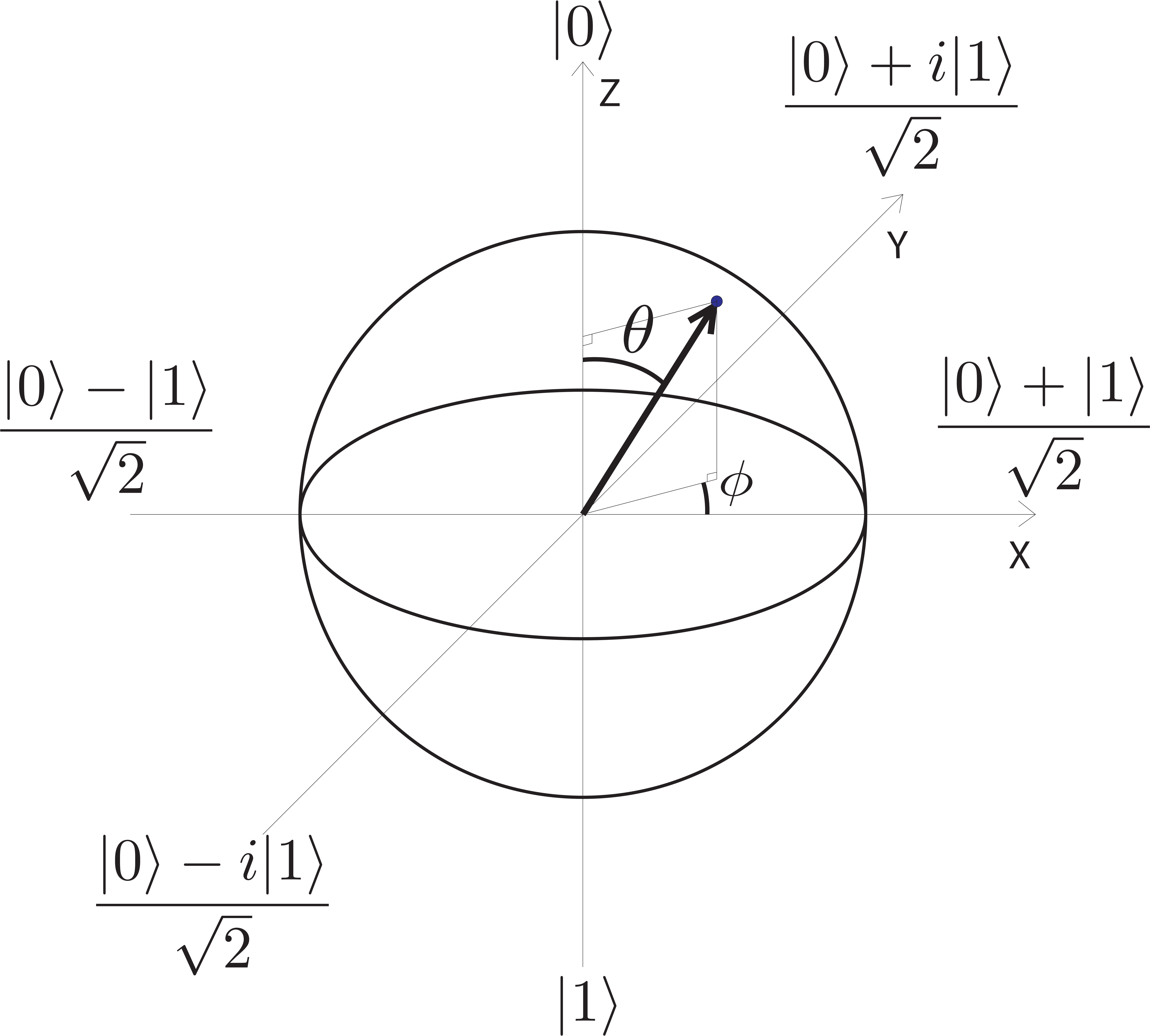}
  \caption[The visualization of the state space of
  single qubit, known as the Bloch sphere.]{The X-Z plane is the real plane and the Y axis
  is the imaginary axis.
  Any two vectors pointing in the opposite directions are the orthogonal states.
  $\theta$ determines the real value and $\phi$ determines the imaginary value.
  }
  \label{fig:bloch}
 \end{center}
 \end{figure}

Figure \ref{fig:bloch} shows the visualization of the state space of a single qubit.
A quantum state is a point on the surface of the sphere, called the Bloch sphere.
In real 3-space, orthogonal states are at right angles, but in the space where the Bloch sphere is described,
orthogonal states show up at opposite points.
Because each point on the surface represents a quantum state, the number of states
a qubit can be in is infinite.
We can choose any axis to measure a qubit,
though by convention we limit our choice to one of the three axes, X, Y or Z,
without loss of generality.
When we measure a qubit along an axis, a value which corresponds to the
maximum or minimum value of the axis in the state space
will be measured probabilistically.
There are three matrices, one corresponding to each axis.
\begin{equation}
X = \begin{pmatrix}0 & 1 \\ 1 & 0\end{pmatrix},
\label{eqx}
\end{equation}
\begin{equation}
Y = \begin{pmatrix}0 & -i \\ i & 0\end{pmatrix},
\label{eqy}
\end{equation}
and
\begin{equation}
Z = \begin{pmatrix}1 & 0 \\ 0 & -1\end{pmatrix}.
\label{eqz}
\end{equation}
The eigenvectors and corresponding eigenvalues of $Z$ are
\begin{equation}
\vert 0 \rangle = \begin{pmatrix} 1 \\ 0 \end{pmatrix}
\end{equation}
for $+1$ eigenvalue and
\begin{equation}
\vert 1 \rangle = \begin{pmatrix} 0 \\ 1 \end{pmatrix}
\end{equation}
for $-1$ eigenvalue.
When we ``measure'' $\vert\Psi\rangle$,
the probability that an eigenvalue is observed 
can be formulated as
\begin{equation}
\langle \Psi \vert  M^\dag M \vert \Psi \rangle 
\end{equation}
where $M$ is $\vert 0 \rangle \langle 0 \vert$ for the $+1$ eigenvalue
and $\vert 1 \rangle \langle 1 \vert$ for the $-1$ eigenvalue.
Therefore,
the probability to observe the $+1$ eigenvalue is 
\begin{equation}
\langle \Psi \vert 0 \rangle \langle 0 \vert \Psi \rangle = \vert\cos\frac{\theta}{2}\vert^2,
\end{equation}
and the probability to observe the $-1$ eigenvalue is
\begin{equation}
 \langle \Psi \vert 1 \rangle \langle 1 \vert \Psi \rangle = \vert e^{i\phi} \sin\frac{\theta}{2}\vert^2
\end{equation}
Each base can be described with other two basis.
$X$ and $Z$ are chosen generally.

Obviously those coefficients have a relationship, $\vert \cos \frac{\theta}{2} \vert ^2 + \vert e^{i\phi} \sin \frac{\theta}{2} \vert ^2 = 1$.
To maintain this relationship, all qubit operations must be unitary.
However, the quantum measurement operation just described is not unitary, sometimes the sum appears not to be 1. In such case, the coefficients are renormalized to make the sum 1.

\subsection{Multiple qubits notation}
To think about the description of multiple qubits,
let's start from the description of two qubits.
Similar to classical computation, two qubits have four orthogonal basis states, $\vert 0_a0_b \rangle$, $\vert 0_a1_b \rangle$, $\vert 1_a0_b \rangle$ and $\vert 1_a1_b \rangle$ in a $2^2$ dimension space
where $a$ is the first qubit and $b$ is the second qubit.
The orthogonal basis vector constraint $\langle \phi \vert \psi \rangle = 0$ still holds.
Therefore the two qubit state is formulated as
\begin{eqnarray}
  \vert\Phi\rangle
  &=& \alpha\vert0_a0_b\rangle + \beta\vert0_a1_b\rangle
  + \gamma\vert1_a0_b\rangle + \delta\vert1_a1_b\rangle\\
  &=&
   \alpha
  \begin{pmatrix}
    1\\
    0\\
    0\\
    0
  \end{pmatrix}
  +
  \beta
  \begin{pmatrix}
    0\\
    1\\
    0\\
    0
  \end{pmatrix}
  +
  \gamma
  \begin{pmatrix}
    0\\
    0\\
    1\\
    0
  \end{pmatrix}
  +
  \delta
  \begin{pmatrix}
    0\\
    0\\
    0\\
    1
  \end{pmatrix}\\
  &=&
  \begin{pmatrix}
    \alpha\\
    \beta\\
    \gamma\\
    \delta
  \end{pmatrix}.
\end{eqnarray}
As with the one qubit description, the absolute value of the square of each coefficient is the probability
of the state being measured. The total probability of the states must be 1.
For example, one possible state is written as
\begin{equation}
  \begin{split}
    \vert\Phi\rangle 
    &=
    \frac{1}{\sqrt{2}}(\vert0_a\rangle + \vert1_a\rangle) \otimes
    \frac{1}{\sqrt{2}}(\vert0_b\rangle + \vert1_b\rangle)\\
    &=
    \frac{1}{2}\vert0_a0_b\rangle + \frac{1}{2}\vert0_a1_b\rangle
    + \frac{1}{2}\vert1_a0_b\rangle + \frac{1}{2}\vert1_a1_b\rangle\\
    &=
    \frac{1}{2}
    \begin{pmatrix}
      1\\
      1\\
      1\\
      1
    \end{pmatrix},
  \end{split}
  \label{notation2qubit}
\end{equation}
with the possible measured values and the corresponding probabilities
are shown in table \ref{tab:2qubits}.
\begin{table}[b]
\begin{center}
\caption{An example of the state and probability table of two qubits.}
\begin{tabular}{|c|c|}
\hline
state & probability\\
\hline
\hline
$\vert 0 0\rangle$ & 25\% \\
\hline
$\vert 0 1\rangle$ & 25\% \\
\hline
$\vert 1 0\rangle$ & 25\% \\
\hline
$\vert 1 1\rangle$ & 25\% \\
\hline
\end{tabular}
\label{tab:2qubits}
\end{center}
\end{table}

Similar to the two qubit case, $n$ qubits have $2^n$ basis from $\vert 000...000\rangle$ to $\vert 111...111\rangle$.
The n-qubit state can be generalized as 
\begin{eqnarray}
 \vert \Psi \rangle =&&
 \alpha_0 \vert 000...000\rangle
  + \alpha_1 \vert 000...001\rangle
  + \alpha_2 \vert 000...010\rangle
  + ...\\
  &&+ \alpha_{n^2-2} \vert 111...110\rangle
  + \alpha_{n^2-1} \vert 111...111\rangle.
\label{eq:nqubits}
\end{eqnarray}
and
\begin{eqnarray}
 \vert \psi \rangle = \Sigma^{2^n-1}_{i=0}\alpha_i\vert i \rangle\\
 \Sigma^{2^n-1}_{i=0}\vert \alpha_i \vert ^2 = 1
\label{eq:nqubitsnom}
\end{eqnarray}
where $\alpha_i \in \mathfrak{C}$.
Equation \ref{eq:nqubitsnom} is the normalization condition.

\subsection{Entanglement}
\label{subsec:basic:ent}
We can discover from the notation of two qubits that 
the probability amplitudes hence the probabilities of the two qubits can be dependent, e.g. in a state such as:
\begin{equation}
  \vert\Phi^+\rangle =
  \frac{1}{\sqrt{2}}\vert0_a0_b\rangle + 0\vert0_a1_b\rangle
  + 0\vert1_a0_b\rangle +   \frac{1}{\sqrt{2}}\vert1_a1_b\rangle,
\end{equation}
Table \ref{tab:entangle} shows the probabilities corresponding to this state.
\begin{table}[b]
\begin{center}
\caption{State and probability table of two qubits in a entangled state, $\vert \Phi^+\rangle$.}
\begin{tabular}{|c|c|}
\hline
state & probability\\
\hline
\hline
$\vert 0 0\rangle$ & 50\% \\
\hline
$\vert 0 1\rangle$ & 0\% \\
\hline
$\vert 1 0\rangle$ & 0\% \\
\hline
$\vert 1 1\rangle$ & 50\% \\
\hline
\end{tabular}
\label{tab:entangle}
\end{center}
\end{table}
In this state, 0 must be observed on the second qubit if 0 is already observed on the first qubit
and 1 must be observed on the second qubit if 1 is already observed on the first qubit.
So is the opposite order of measurements.
This characteristic of qubits is called \emph{entanglement}. Entanglement is one of
the critical effects which distinguish quantum systems from classical ones.
Entangled states cannot be represented as products of single qubit states
such as in equation \ref{notation2qubit}. 
Such inseparablility tells that the states of qubits in the multiple qubits state are dependent.
Compared to the \emph{entangled state}, a multiple qubit state which can be described as a product of single qubit states is called \emph{separable state}.

Maximally entangled two qubits are often called a Bell pair.
There are four types of Bell states:
\begin{eqnarray}
   \vert\Phi^+\rangle = \frac{1}{\sqrt{2}}\vert0_a0_b\rangle + \frac{1}{\sqrt{2}}\vert1_a1_b\rangle \\
   \vert\Phi^-\rangle = \frac{1}{\sqrt{2}}\vert0_a0_b\rangle - \frac{1}{\sqrt{2}}\vert1_a1_b\rangle \\
   \vert\Psi^+\rangle = \frac{1}{\sqrt{2}}\vert0_a1_b\rangle + \frac{1}{\sqrt{2}}\vert1_a0_b\rangle \\
   \vert\Psi^-\rangle = \frac{1}{\sqrt{2}}\vert0_a1_b\rangle - \frac{1}{\sqrt{2}}\vert1_a0_b\rangle.
\end{eqnarray}
These four states can be used as a basis for two-qubit states.

More and more qubits can be connected by entanglement,
as Equation \ref{eq:nqubits} and obeying Equation \ref{eq:nqubitsnom}.

\subsection{No-cloning theorem}
Unfortunately, in contrast to the classical states, quantum states cannot be copied.
To prove the quantum no-cloning theorem, let's assume
an unitary operator $U$ which copies any quantum state to an ancilla qubit $\vert a \rangle$ as
\begin{eqnarray}
 U \vert \psi \rangle \vert a\rangle = \vert \psi \rangle \otimes \vert \psi \rangle
\label{nct1}
\end{eqnarray}
and
\begin{eqnarray}
 U \vert \phi \rangle \vert a\rangle = \vert \phi \rangle \otimes \vert \phi \rangle.
\label{nct2}
\end{eqnarray}
Then if we take the inner product of Equations \ref{nct1} and \ref{nct2},
\begin{eqnarray}
 \langle a \vert \langle \phi \vert U ^ {-1}
 U \vert \psi \rangle \vert a\rangle 
&=& \langle \phi \vert \otimes \langle \phi \vert
 \psi \rangle \otimes \vert \psi \rangle \\
\langle \phi \vert \psi \rangle
&=&
\langle \phi \vert \psi \rangle ^2
\label{nct3}
\end{eqnarray}
Obviously, Equation \ref{nct3} holds only if 
$\langle \phi \vert \psi \rangle = 0$
or
$\langle \phi \vert \psi \rangle = 1$.
Therefore, our quantum ``copy machine'' only works for the cases,
$\vert \phi \rangle \perp \vert \psi \rangle$ or $\vert \phi \rangle \parallel \vert \psi \rangle$.
Hence $U$ can copy specific two states, but cannot copy arbitrary state.
An example of the $U$ and the specific two states are $CNOT$ gate and $\vert 0 \rangle$ and $\vert 1 \rangle$.
After applying $CNOT(c,a)$ between the copied qubit and an ancilla qubit $\vert 0\rangle$,
\begin{eqnarray}
 CNOT(c,a) \vert 0_c \rangle \otimes \vert 0_a \rangle = \vert 0_c \rangle \otimes \vert 0_a \rangle \\
 CNOT(c,a) \vert 1_c \rangle \otimes \vert 0_a \rangle = \vert 1_c \rangle \otimes \vert 1_a \rangle 
\end{eqnarray}
are achieved. Qubit $c$ and qubit $a$ are separable and $\vert 0 \rangle$ and $\vert 1 \rangle$ are copied as desired.
Apparently another arbitrary state $\alpha \vert 0 \rangle + \beta \vert 1 \rangle$ cannot be copied
since an arbitrary two qubit state
\begin{eqnarray}
 CNOT(c,a) (\alpha \vert 0_c \rangle + \beta \vert 1_c \rangle) \vert a \rangle
  = \alpha \vert 0_c0_a \rangle + \beta \vert 1_c1_a \rangle
\end{eqnarray}
other than $\vert 0 \rangle$ and $\vert 1 \rangle$ hence other than $\alpha \ne 0$ and $\beta \ne 0$
is not separable and the two qubits are entangled.

\subsection{Quantum gates}
\label{subsec:qgates}
The quantum gates 
that we use in quantum computation
are described by unitary operations.
For our purposes, we will need one- and two-qubit gates.

\subsubsection{\emph{Single qubit gate}}
The three matrices in Equations \ref{eqx}, \ref{eqy} and \ref{eqz} also act as basic operations
on a single qubit.
\begin{eqnarray}
X\vert \Psi \rangle &=& 
\begin{pmatrix}
0 & 1 \\
1 & 0
\end{pmatrix}
(\alpha \vert 0 \rangle + \beta \vert 1 \rangle)\nonumber\\
&=&\beta \vert 0 \rangle + \alpha \vert 1 \rangle\\
Y\vert \Psi \rangle &=& 
\begin{pmatrix}
0 & -i \\
i & 0
\end{pmatrix}
(\alpha \vert 0 \rangle + \beta \vert 1 \rangle)\nonumber\\
&=&-i\beta \vert 0 \rangle + i\alpha \vert 1 \rangle
\label{pauliybefore} \\
&=&\beta \vert 0 \rangle - \alpha \vert 1 \rangle
\label{pauliyafter}\\
Z\vert \Psi \rangle &=& 
\begin{pmatrix}
1 & 0 \\
0 & -1
\end{pmatrix}
(\alpha \vert 0 \rangle + \beta \vert 1 \rangle)\nonumber\\
&=&\alpha \vert 0 \rangle - \beta \vert 1 \rangle
\end{eqnarray}
In the derivation from Equation \ref{pauliybefore} to Equation \ref{pauliyafter},
the global phase $i$ is removed. 
Since only relative phase between the two basis vector is observable and important,
we can remove global phase which is applied to both basis vector.
By removing the global phase $i$, $Y$ rotation becomes the product of $Z$ rotation and $X$ rotation.
Because 
$X$ exchanges the coefficients of $\vert 0 \rangle$ and $\vert 1 \rangle$ such as 
$\vert0\rangle \rightarrow \vert1\rangle$ and
$\vert1\rangle \rightarrow \vert0\rangle$, this is also called a NOT gate.
$X$, $Y$, $Z$ and $I$ are called Pauli matrices. $I$ is the identity gate,
\begin{equation}
I = \begin{pmatrix} 1 & 0 \\ 0 & 1 \end{pmatrix}.
\end{equation}

An interesting extentions of $X$, $Y$ and $Z$ rotations
are rotations of arbitrary degrees.
\begin{eqnarray}
 R_x(\theta) &=& e^{-\theta X/2} \\
 &=& \cos \frac{\theta}{2}I - i \sin \frac{\theta}{2}X \\
 &=& 
\begin{pmatrix}
 \cos \frac{\theta}{2} & -i \sin \frac{\theta}{2} \\
 -i \sin \frac{\theta}{2} & \cos \frac{\theta}{2} 
\end{pmatrix}\\
 R_y(\theta) &=& e^{-\theta Y/2} \\
 &=& \cos \frac{\theta}{2}I - i \sin \frac{\theta}{2}Y \\
 &=&
\begin{pmatrix}
 \cos \frac{\theta}{2} & - \sin \frac{\theta}{2} \\
 \sin \frac{\theta}{2} & \cos \frac{\theta}{2} 
\end{pmatrix}\\
 R_z(\theta) &=& e^{-\theta Z/2} \\
 &=& \cos \frac{\theta}{2}I - i \sin \frac{\theta}{2}Z \\
 &=&
\begin{pmatrix}
 e^{-i\theta 2} & 0\\
 0 & e^{i\theta /2}
\end{pmatrix}
\end{eqnarray}
Actually, any two of those three arbitrary rotations
can be used to achieve an arbitrary unitary change to one qubit state.

While we can achieve any rotation,
especially $S$ gate and $T$ gate often appear 
in quantum computation.
\begin{equation}
 S = \begin{pmatrix}1 & 0\\0&i\end{pmatrix}
\end{equation}
\begin{equation}
 T = \begin{pmatrix}1 & 0\\0&e^{i\pi / 4}\end{pmatrix}
\end{equation}
$S$ gate is square root of $Z$ gate and
$T$ gate is square root of $S$ gate,
hence
\begin{eqnarray}
 T^4 = S^2 = Z.
\end{eqnarray}

Another basic gate known as the Hadamard gate is
\begin{equation}
H = \frac{1}{\sqrt{2}}\begin{pmatrix}1 & 1\\1&-1\end{pmatrix}.
\end{equation}
The basis vectors in the $X$ basis are
$\vert+\rangle = \frac{1}{\sqrt{2}}\begin{pmatrix}1\\1\end{pmatrix}$ and
$\vert-\rangle = \frac{1}{\sqrt{2}}\begin{pmatrix}1\\-1\end{pmatrix}$.
With this $H$ and the eigenvalues of $Z$ and $X$,
\begin{equation}
H\times(\alpha \vert 0 \rangle + \beta \vert 1 \rangle)=
(\alpha \vert + \rangle + \beta \vert - \rangle)
\end{equation}
is derived.
$H$ swaps the relative relationships of the state with $Z$ axis and with $X$ axis.

\subsubsection{\emph{Two qubit gate}}
The most basic two qubit gate is the controlled-not (CNOT) gate,
\begin{equation}
CNOT = 
\begin{pmatrix}
  1 & 0 & 0 & 0 \\
  0 & 1 & 0 & 0 \\
  0 & 0 & 0 & 1 \\
  0 & 0 & 1 & 0 \\
\end{pmatrix}.
\end{equation}
This gate exchanges the coefficient of $\vert0\rangle$ and $\vert1\rangle$
of the second, target qubit, only if the first, control qubit is $\vert1\rangle$. Table \ref{tab:CNOT}
is the truth table of the CNOT gate.
\begin{table}[b]
  \begin{center}
    \caption[The truth table of the CNOT gate.]
   {Qubit $a$ is the control qubit and qubit $b$ is the target qubit.}
    \label{tab:CNOT}
    \begin{tabular}{c|c||c|c}
      \hline
      $a_{in}$ & $b_{in}$ & $a_{out}$ & $b_{out}$ \\
      \hline
      \hline
      0 & 0 & 0 & 0\\
      \hline
      0 & 1 & 0 & 1\\
      \hline
      1 & 0 & 1 & 1\\
      \hline
      1 & 1 & 1 & 0\\
      \hline
    \end{tabular}
  \end{center}
\end{table}
In this dissertation I use the notation $CNOT(a,b)$
to describe a CNOT gate in which qubit $a$ is the control qubit
and qubit $b$ is the target qubit.

$CZ$ gate and $\sqrt{SWAP}$ gate also often appear,
\begin{eqnarray}
 CZ &=&
\begin{pmatrix}
 1 & 0 & 0 & 0 \\
 0 & 1 & 0 & 0 \\
 0 & 0 & 1 & 0 \\
 0 & 0 & 0 & -1
\end{pmatrix}\\
\sqrt{SWAP} &=&
\begin{pmatrix}
 1 & 0 & 0 & 0 \\
 0 & \frac{1}{2}(1+i) & \frac{1}{2}(1-i) & 0 \\
 0 & \frac{1}{2}(1-i) & \frac{1}{2}(1+i) & 0 \\
 0 & 0 & 0 & 1
\end{pmatrix}.
\end{eqnarray}
$CNOT$, $CZ$ and $\sqrt{SWAP}$ gates are (potentially) entangling gates
and can be converted to one another.
$\sqrt{SWAP}$ can occur in some physical systems, but is not often used at the logical level.

The SWAP gate is a two-qubit gate which also often appears to change the physical location of qubits.
\begin{equation}
SWAP = 
\begin{pmatrix}
  1 & 0 & 0 & 0 \\
  0 & 0 & 1 & 0 \\
  0 & 1 & 0 & 0 \\
  0 & 0 & 0 & 1 \\
\end{pmatrix}
\end{equation}
This operation swaps the state of the first qubit and the second qubit completely.

\subsubsection{Universal gate set}
An universal quantum gate set is a fixed set of quantum gates,
with which the universal quantum computation is enabled.
Solovay-Kitaev decomposition showed that any quantum gates can be 
approximated with a small fixed set of quantum gates,
such as, $H$ gate and $T$ gate
~\cite{PhysRevA.87.052332, dawson05:_solov_kitaev_theorem, 
selinger12:_effic_cliff_T, 
selinger2015efficient, 
jones2012faster}.
With a two-qubit gate for multiple qubit control such as entanglement, let's say $CNOT$ gate, 
an universal quantum gate set is achieved.
Additionally, a fixed set of gates is more suitable to fault-tolerant quantum computation.
Therefore $CNOT$ gate, $H$ gate and $T$ gate is a convenient universal set.

\subsection{Interference}
Quantum computation proceeds by the creation of interference among probability amplitudes of 
basis states in superposition.
One of the simplest example of the interference can be seen with the Hadamard gate.
Let's consider how Hadamard gate affects $\vert \Psi \rangle = \frac{1}{\sqrt{2}}(\vert 0 \rangle + \vert 1 \rangle)$,
\begin{eqnarray}
 H \vert \Psi \rangle &=&
\frac{1}{\sqrt{2}}
\begin{pmatrix}
 1 & 1 \\
 1 & -1
\end{pmatrix}
\frac{1}{\sqrt{2}}(\vert 0 \rangle + \vert 1 \rangle) \\
&=&
\frac{1}{2}
\begin{pmatrix}
 1 & 1 \\
 1 & -1
\end{pmatrix}
 \begin{pmatrix}
  1 \\
  1
 \end{pmatrix}\\
&=&
 \begin{pmatrix}
  1 \\
  0
 \end{pmatrix}\\
&=&
\vert 0 \rangle.
\end{eqnarray}
As above, by the Hadamard gate, the probability amplitudes of $\vert 0 \rangle$ and $\vert 1 \rangle$ interfere.
Constructive interference occurs at the $\vert 0 \rangle$ component in the state vector
and deconstructive interference occurs at the $\vert 1 \rangle$.
Thanks to the existence of the phase component, $e^{i\phi}$ in Equation \ref{equ:qubit},
there will happen more complex interference.
Quantum algorithms are designed to manipulate the terms in the state vector to take advantage of interference and 
leave the probability amplitude of the solution to the problem high when they finish.

\subsection{Quantum teleportation}
Quantum teleportation is a protocol to transfer a quantum state to a remote place,
by consuming a Bell pair~\cite{PhysRevLett.70.1895}.
First we have a qubit to transfer at station $A$,
$\alpha \vert 0_{qA} \rangle + \beta \vert 1_{qA} \rangle$, and
a Bell pair is shared at station $A$ and $B$,
$\frac{1}{\sqrt{2}}(\vert 0_{BellA}0_{BellB} \rangle + \vert 1_{BellA}1_{BellB} \rangle)$.
\begin{eqnarray}
 \vert\phi'\rangle &=& (\alpha \vert 0_{qA} \rangle + \beta \vert 1_{qA} \rangle) \otimes \frac{1}{\sqrt{2}}(\vert 0_{BellA}0_{BellB} \rangle + \vert 1_{BellA}1_{BellB} \rangle) \nonumber \\
 &=& \frac{1}{\sqrt{2}}(\alpha \vert 0_{qA}0_{BellA}0_{BellB} \rangle + \alpha\vert 0_{qA}1_{BellA}1_{BellB} \rangle + \nonumber \\
  & &\beta \vert 1_{qA}0_{BellA}0_{BellB} \rangle + \beta\vert 1_{qA}1_{BellA}1_{BellB} \rangle).
\end{eqnarray}
We begin by applying a $CNOT$ gate between A's two qubits, hence $CNOT(q_{qA},q_{BellA})$,
\begin{eqnarray}
 \vert\phi''\rangle &=& \frac{1}{\sqrt{2}}(\alpha \vert 0_{qA}0_{BellA}0_{BellB} \rangle + \alpha\vert 0_{qA}1_{BellA}1_{BellB} \rangle + \nonumber\\
                    &&\beta \vert 1_{qA}1_{BellA}0_{BellB} \rangle + \beta\vert 1_{qA}0_{BellA}1_{BellB} \rangle)\nonumber\\
 &=& \frac{1}{2}(
  \alpha \vert +_{qA}0_{BellA}0_{BellB} \rangle +
  \alpha \vert -_{qA}0_{BellA}0_{BellB} \rangle + \nonumber\\
 &&\alpha\vert +_{qA}1_{BellA}1_{BellB} \rangle + 
  \alpha\vert -_{qA}1_{BellA}1_{BellB} \rangle +  \nonumber\\
 &&\beta \vert +_{qA}1_{BellA}0_{BellB} \rangle -
  \beta \vert -_{qA}1_{BellA}0_{BellB} \rangle + \nonumber\\
 && \beta\vert +_{qA}0_{BellA}1_{BellB} \rangle - 
  \beta\vert -_{qA}0_{BellA}1_{BellB} \rangle)
  \label{equ:telepo0}
\end{eqnarray}
Here, measuring $X_{qA}$ and $Z_{BellA}$, then we get 
one of the following
state corresponding to the measured values.
\begin{eqnarray}
 &&+,0 \rightarrow \alpha \vert 0_{BellB} \rangle + \beta\vert 1_{BellB} \rangle \nonumber \\
 &&+,1 \rightarrow \alpha \vert 1_{BellB} \rangle + \beta\vert 0_{BellB} \rangle \nonumber \\
 &&-,0 \rightarrow \alpha \vert 0_{BellB} \rangle - \beta\vert 1_{BellB} \rangle \nonumber \\
 &&-,1 \rightarrow \alpha \vert 1_{BellB} \rangle - \beta\vert 0_{BellB} \rangle
\end{eqnarray}
Hence, apply $X_{BellB}$ if we get $1$ from $Z_{BellA}$ and apply $Z_{BellB}$ if we get $-$ from $X_{qA}$, then we deterministically get 
\begin{equation}
\alpha \vert 0_{BellB} \rangle + \beta\vert 1_{BellB} \rangle.
\end{equation}
This is the quantum teleportation.

 \subsection{Entanglement swapping}
\begin{figure}[t]
   \includegraphics[width=15cm]{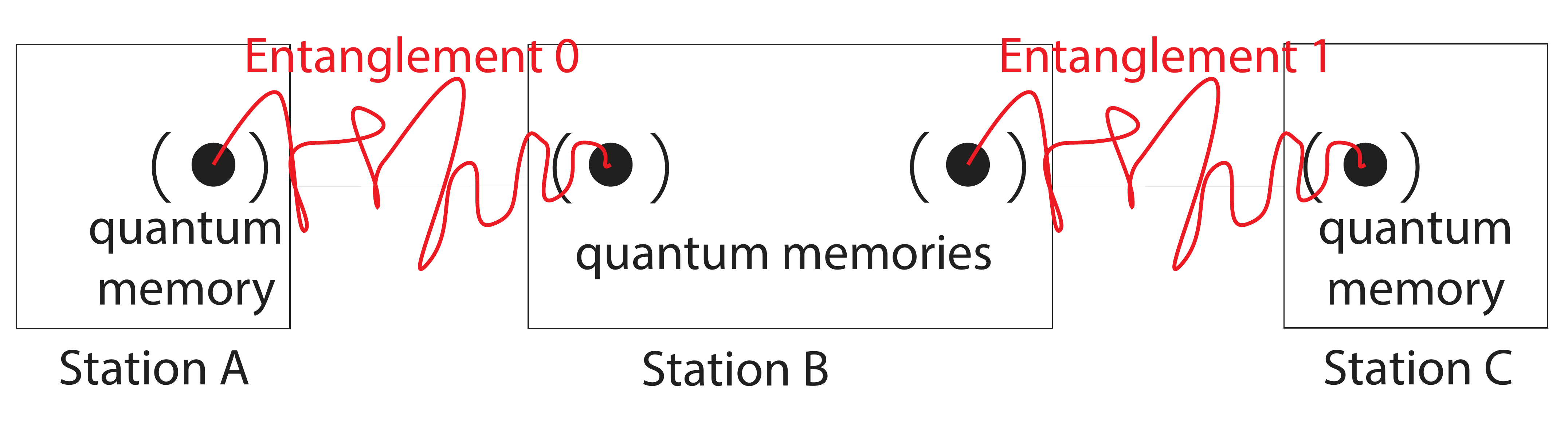}
   \caption[Entanglement Swapping.]{There are two sets of entangled pairs, between Station A and Station B and between Station B and Station C.
   By entanglement swapping, an entangled pair between Station A and Station C remains afterwards.}
   \label{fig:es}
  \end{figure}

  Entanglement swapping connects two Bell pairs into one.
  This operation is often executed when there are linearly connected distant three stations 
  and raw Bell pairs can be directly created only between two neighboring stations,
  as depicted in Figure ~\ref{fig:es}.
  In this dissertation, entanglement swapping is used to create Bell pairs between arbitrary two components.
 \begin{eqnarray}
  \vert \psi_{A0,B0,B1,C1} \rangle
 = \frac{1}{\sqrt{2}}(\vert 0_{A0}0_{B0} \rangle + \vert 1_{A0}1_{B0} \rangle)
  \otimes
  \frac{1}{\sqrt{2}}(\vert 0_{B1}0_{C1} \rangle + \vert 1_{B1}1_{C1} \rangle) \nonumber \\ 
 = \frac{1}{2}(\vert 0_{A0}0_{B0}0_{B1}0_{C1} \rangle + \vert 0_{A0}0_{B0}1_{B1}1_{C1} \rangle +
  \vert 1_{A0}1_{B0}0_{B1}0_{C1} \rangle + \vert 1_{A0}1_{B0}1_{B1}1_{C1} \rangle),
  \label{equ:opcross1}
 \end{eqnarray}
where $A$, $B$ and $C$ indicate stations and $0$ and $1$ indicate two entangled pairs, respectively.
Then the qubits $B0$ and $B1$ are measured in the Bell state basis
by a CNOT gate and two measurement in $X$ basis and $Z$ basis.
Applying CNOT (B0, B1), then get
\begin{eqnarray}
 \vert \psi_{A0,B0,B1,C1} \rangle
 &=& \frac{1}{2}(\vert 0_{A0}0_{B0}0_{B1}0_{C1} \rangle + \vert 0_{A0}0_{B0}1_{B1}1_{C1} \rangle + \nonumber\\
 \vert 1_{A0}1_{B0}1_{B1}0_{C1} \rangle + \vert 1_{A0}1_{B0}1_{B1}0_{C1} \rangle) \nonumber \\
 \\
 &=& \frac{1}{2\sqrt{2}}(
  \vert 0_{A0}+_{B0}0_{B1}0_{C1} \rangle +
  \vert 0_{A0}-_{B0}0_{B1}0_{C1} \rangle + \nonumber \\
&&  \vert 0_{A0}+_{B0}1_{B1}1_{C1} \rangle +
  \vert 0_{A0}-_{B0}1_{B1}1_{C1} \rangle + \nonumber \\
&&  \vert 1_{A0}+_{B0}1_{B1}0_{C1} \rangle -
  \vert 1_{A0}-_{B0}1_{B1}0_{C1} \rangle + \nonumber \\
&&  \vert 1_{A0}+_{B0}0_{B1}1_{C1} \rangle -
  \vert 1_{A0}-_{B0}0_{B1}1_{C1} \rangle
  ).
  \label{equ:opcross2}
\end{eqnarray}
Measuring $B0$ in the $X$ basis and B1 in the $Z$ basis,
we get one of the following results:
\begin{eqnarray}
 &&+,0 \rightarrow \frac{1}{\sqrt{2}}(\vert 0_{A0}0_{C1} + \vert 1_{A0}1_{C1})\rangle \nonumber \\
 &&+,1 \rightarrow \frac{1}{\sqrt{2}}(\vert 0_{A0}1_{C1} + \vert 1_{A0}0_{C1})\rangle \nonumber \\
 &&-,0 \rightarrow \frac{1}{\sqrt{2}}(\vert 0_{A0}0_{C1} - \vert 1_{A0}1_{C1})\rangle \nonumber \\
 &&-,1 \rightarrow \frac{1}{\sqrt{2}}(\vert 0_{A0}1_{C1} - \vert 1_{A0}0_{C1})\rangle.
\end{eqnarray}
By applying $X$ and $Z$ depending on the measured values,
we get
\begin{equation}
 \frac{1}{\sqrt{2}}(\vert 0_{A0}0_{C1} + \vert 1_{A0}1_{C1})\rangle 
\end{equation}
hence a desired Bell pair between station $A$ and $C$ is achieved.

\subsection{Quantum circuit}
\begin{figure}[t]
  \begin{center}
    \includegraphics[width=300pt]{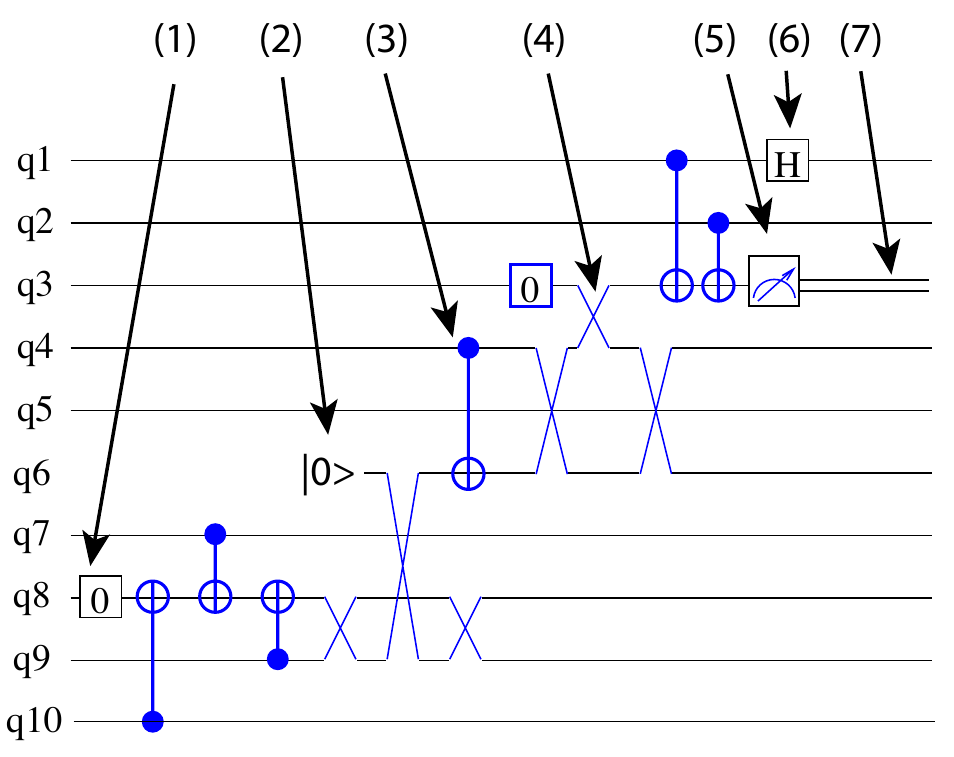}
    \caption[An example of quantum circuits.]
   {q1, q2, ... q10 are qubits.
   Time goes from the left to the right.
   (1) The initialization in $\vert 0 \rangle$ in this dissertation.
   (2) The general description of initialization in $\vert 0 \rangle$.
   (3) The CNOT gate. The dot represents the control qubit and the circle
   represents the target qubit. 
   (4) The SWAP gate. 
   (5) The measurement in $Z$ axis. To describe a measurement in another axis, the axis is represented as a subscript.
   (6) The Hadamard gate. Other single qubit gate is represented such as $X$ in the box instead of $H$ for the $X$ gate.
   (7) Classical information is often represented with double lines.
   }
    \label{fig:circ}
  \end{center}
\end{figure}

Figure \ref{fig:circ} shows a quantum circuit.
Each horizontal line represents a qubit.
Gates on the lines are sequentially executed on corresponding qubits from the left to the right, like the music notation.
Gates on multiple qubits are multiple qubit gates.
Characters and numbers in the most left of this Figure are qubit identifiers, there sometimes are initial states of qubits.
Gates in the same lateral coordinates are executed simultaneously.
The symbols of gates used in this research are described in Figure \ref{fig:circ}.
Gates bounded with a square are single qubit gates.
Gates which crosses several qubits are two or more qubit gates.

\subsection{Single qubit gate supported by an ancilla qubit}
\label{subsec:ancillaonequbit}
For environments where rotations for non-Clifford gates are difficult to implement,
single qubit gates supported by ancilla gates have been developed~\cite{PhysRevA.86.032324}.

\subsubsection{\emph{$S$ gate supported by an ancilla qubit}}
The $S$ gate is a non-Clifford gate, the square root of the $Z$ gate.
The $\vert Y \rangle$ ancilla state where
\begin{equation}
 \vert Y \rangle = \frac{1}{\sqrt{2}}(\vert 0 \rangle + i \vert 1 \rangle)
\end{equation}
is used to execute the $S$ gate.
Figure \ref{fig:Sancilla} shows the circuit for applying an $S$ gate to an arbitrary state $\vert \psi \rangle$.
\begin{figure}[t]
 \begin{center}
\includegraphics[width=10cm]{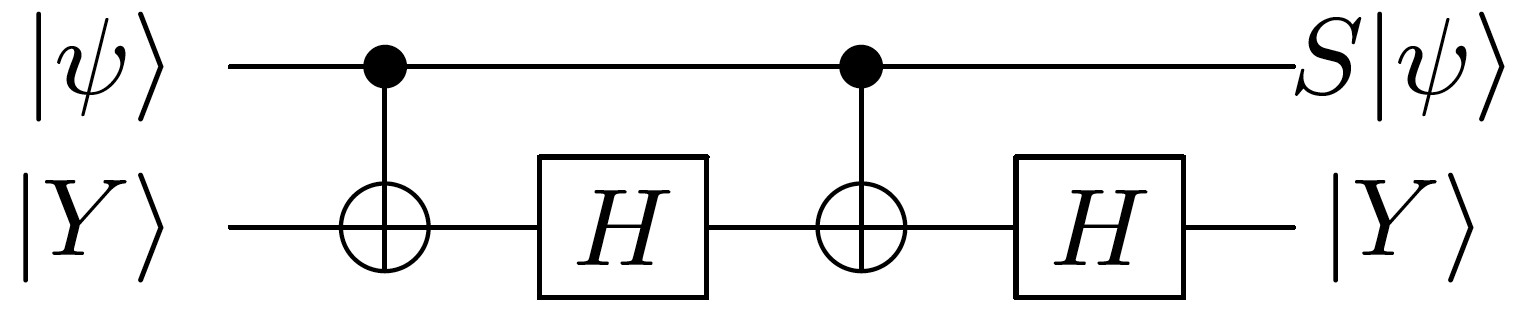}
  \caption[$S$ gate supported by an ancilla qubit.]
  {Eventually $\vert Y \rangle$ is kept and can be reused.}
  \label{fig:Sancilla}
 \end{center}
\end{figure}
The first state is
\begin{eqnarray}
 \vert \Psi \rangle &=& \vert \psi_q \rangle \otimes \vert Y_y \rangle \\
  &=& (\alpha \vert 0_q \rangle + \beta \vert 1_q \rangle) \otimes (\frac{1}{\sqrt{2}}(\vert 0_y \rangle + i \vert 1_y \rangle))  \\
  &=& \frac{1}{\sqrt{2}} (\alpha \vert 0_q 0_y \rangle + i \alpha \vert 0_q 1_y \rangle + \beta \vert 1_q 0_y \rangle + i \beta \vert 1_q 1_y \rangle).
\end{eqnarray}
Apply $CNOT(q,y)$, then get
\begin{eqnarray}
 \vert \Psi ' \rangle &=& 
\frac{1}{\sqrt{2}} (
\alpha \vert 0_q 0_y \rangle 
+ i \alpha \vert 0_q 1_y \rangle
+ i \beta \vert 1_q 0_y \rangle) 
+ \beta \vert 1_q 1_y \rangle .
\end{eqnarray}
Applying $H(y)$,
\begin{eqnarray}
\vert \Psi '' \rangle &=& 
\frac{1}{\sqrt{2}} (
(\alpha + i\alpha) \vert 0_q 0_y \rangle 
+ (\alpha - i \alpha) \vert 0_q 1_y \rangle 
+ (i\beta + \beta) \vert 1_q 0_y \rangle 
+ (i\beta - \beta) \vert 1_q 1_y \rangle). \nonumber \\
\end{eqnarray}
Applying $CNOT(q,y)$, 
\begin{eqnarray}
\vert \Psi ''' \rangle &=& 
\frac{1}{\sqrt{2}} (
(\alpha + i\alpha) \vert 0_q 0_y \rangle 
+ (\alpha - i \alpha) \vert 0_q 1_y \rangle 
+ (i\beta - \beta) \vert 1_q 0_y \rangle 
+ (i\beta + \beta) \vert 1_q 1_y \rangle). \nonumber \\
\end{eqnarray}
Applying $H(y)$,
\begin{eqnarray}
\vert \Psi '''' \rangle &=& 
\frac{1}{\sqrt{2}} (
   \alpha \vert 0_q 0_y \rangle 
+ i\alpha \vert 0_q 1_y \rangle 
+ i\beta \vert 1_q 0_y \rangle 
-  \beta  \vert 1_q 1_y \rangle) \\
&=&
(\alpha \vert 0_q \rangle + i\beta \vert 1_q \rangle) \otimes \frac{1}{\sqrt{2}} (\vert 0_y \rangle + i \vert 1_y \rangle)\\
&=& S \vert \psi_q \rangle \otimes \vert Y_y \rangle.
\end{eqnarray}
Eventually we get $S \vert \psi \rangle$ and $\vert Y \rangle$ as separable states.
This $\vert Y \rangle$ can be reused for other $S$ gates.

\subsubsection{\emph{$T$ gate supported by an ancilla qubit}}
The $T$ gate is a non-Clifford gate, the square root of the $S$ gate.
The $\vert A \rangle$ ancilla state where
\begin{equation}
 \vert A \rangle = \frac{1}{\sqrt{2}}(\vert 0 \rangle + e^{i\pi/4} \vert 1 \rangle)
\end{equation}
is used for the $T$ gate.
\begin{figure}[t]
 \begin{center}
  \includegraphics[width=10cm]{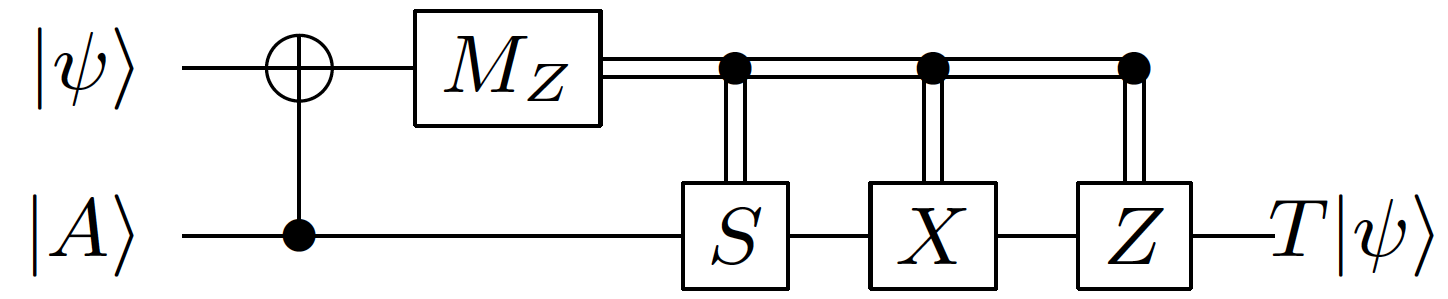}
  \caption[$T$ gate supported by an ancilla qubit.]
  {Eventually $\vert A \rangle$ is consumed and the desired state appears where $\vert A \rangle$ was.}
  \label{fig:Sancilla}
 \end{center}
\end{figure}
The first state is
\begin{eqnarray}
 \vert \Psi \rangle &=& \vert \psi_q \rangle \otimes \vert A_a \rangle \\
 &=& (\alpha \vert 0_q \rangle + \beta \vert 1_q \rangle) 
  \otimes 
  \frac{1}{\sqrt{2}}(\vert 0_a \rangle + e^{i\pi/4} \vert 1_a \rangle) \\
 &=& 
  \frac{1}{\sqrt{2}}(
  \alpha \vert 0_q 0_a \rangle + \alpha e^{i\pi/4} \vert 0_q1_a \rangle
  \beta \vert 1_q 0_a \rangle + \beta e^{i\pi/4} \vert 1_q1_a \rangle
  ).
\end{eqnarray}
Applying $CNOT(a,q)$,
\begin{eqnarray}
 \vert \Psi ' \rangle &=& CNOT(a,q)
  \frac{1}{\sqrt{2}}(
  \alpha \vert 0_q 0_a \rangle + \alpha e^{i\pi/4} \vert 0_q1_a \rangle
  \beta \vert 1_q 0_a \rangle + \beta e^{i\pi/4} \vert 1_q1_a \rangle
  ) \\
 &=&
  \frac{1}{\sqrt{2}}(
  \alpha \vert 0_q 0_a \rangle + \alpha e^{i\pi/4} \vert 1_q1_a \rangle
  \beta \vert 1_q 0_a \rangle + \beta e^{i\pi/4} \vert 0_q1_a \rangle
  ). \\
\end{eqnarray}
By measuring $Z_q$ and finding $+1$ eigenvalue, we get
\begin{eqnarray}
 \vert \Psi_{+1} '' \rangle &=&  
  \alpha \vert 0_a \rangle +  \beta e^{i\pi/4} \vert 1_a \rangle,
\end{eqnarray}
so the desired state.
If we find $-1$ eigenvalue at the $Z_q$ measurement, we get
\begin{eqnarray}
 \vert \Psi_{-1} '' \rangle &=& 
  \beta \vert 0_a \rangle + \alpha e^{i\pi/4} \vert 1_a \rangle
\end{eqnarray}
By applying $Z_aX_aS_a$,
\begin{eqnarray}
 \vert \Psi_{-1} ''' \rangle &=& 
  Z_aX_aS_a (\beta \vert 0_a \rangle + \alpha e^{i\pi/4} \vert 1_a \rangle)\\
  &=& 
  \begin{pmatrix}
   1 & 0\\
   0 & -1
  \end{pmatrix}
  \begin{pmatrix}
   0 & 1 \\
   1 & 0
  \end{pmatrix}
  \begin{pmatrix}
   1 & 0 \\
   0 & i 
  \end{pmatrix}
  (\beta \vert 0_a \rangle + \alpha e^{i\pi/4} \vert 1_a \rangle)\\
 &=& 
  \begin{pmatrix}
   1 & 0\\
   0 & -1
  \end{pmatrix}
  \begin{pmatrix}
   0 & 1 \\
   1 & 0
  \end{pmatrix}
  (\beta \vert 0_a \rangle + \alpha e^{i\pi/2}e^{i\pi/4} \vert 1_a \rangle)\\
 &=& 
  \begin{pmatrix}
   1 & 0\\
   0 & -1
  \end{pmatrix}
  \begin{pmatrix}
   0 & 1 \\
   1 & 0
  \end{pmatrix}
  (\beta \vert 0_a \rangle + \alpha e^{i3\pi/4} \vert 1_a \rangle)\\
 &=& 
  \begin{pmatrix}
   1 & 0\\
   0 & -1
  \end{pmatrix}
  (\alpha e^{i3\pi/4} \vert 0_a \rangle + \beta \vert 1_a \rangle)\\
 &=& 
  \begin{pmatrix}
   1 & 0\\
   0 & -1
  \end{pmatrix}
  (\alpha \vert 0_a \rangle + \beta  e^{-i3\pi/4}  \vert 1_a \rangle)\\
 &=& 
  (\alpha \vert 0_a \rangle + \beta  e^{i\pi} e^{-i3\pi/4}  \vert 1_a \rangle)\\
 &=& 
  (\alpha \vert 0_a \rangle + \beta  e^{i\pi/4} \vert 1_a \rangle).
\end{eqnarray}
Hence, the desired state is achieved.

\subsection{Density matrix}
The density matrix is another notation for describing quantum states.
A single qubit state $\vert \psi \rangle = \alpha \vert 0\rangle + \beta \vert 1 \rangle$ is described 
with density matrix denoted by $\rho$ is
\begin{eqnarray}
\rho = \vert \psi \rangle \langle \psi \vert &=&
\begin{pmatrix}
 \alpha \\
 \beta
\end{pmatrix}
\begin{pmatrix}
 \alpha^* & \beta^*
\end{pmatrix}\\
&=&
\begin{pmatrix}
 \alpha \alpha^* & \alpha \beta^* \\
 \alpha^* \beta & \beta \beta^*
\end{pmatrix}.
\end{eqnarray}

The power of density matrix is 
that it can describe not only quantum probability amplitudes but also classical probability amplitudes of being in different states.
Let's think of a quantum state 
$\vert \psi \rangle = \frac{1}{2}\vert 0\rangle + \frac{\sqrt{3}}{2}\vert 1 \rangle$
in which 
the superposition consists of a probability amplitude corresponding to a 25\%
chance of observing $\vert0 \rangle$, and 
an amplitude corresponding to a 75\% chance of observing $\vert 1\rangle$.
\begin{eqnarray}
\rho_{pure} &=& \vert \psi \rangle \langle \psi \vert \\
&=&
\begin{pmatrix}
 \frac{1}{2}\\
 \frac{\sqrt{3}}{2}
\end{pmatrix}
\begin{pmatrix}
 \frac{1}{2} & \frac{\sqrt{3}}{2}
\end{pmatrix}\\
&=&
\begin{pmatrix}
 \frac{1}{4} & \frac{\sqrt{3}}{4} \\
 \frac{\sqrt{3}}{4} & \frac{3}{4}
\end{pmatrix} \label{equ:dm:quantum}
\end{eqnarray}

Actually the density matrix can be considered as a frequency distribution,
hence we can denote a classically probabilistic state.
For example, a classically probabilistic state
where $\vert0 \rangle$ occupies classically 25\% and $\vert 1\rangle$ occupies classically 75\% is
\begin{eqnarray}
\rho_{mixed1} &=&
\frac{1}{4}\vert 0 \rangle \langle 0 \vert +
\frac{3}{4}\vert 1 \rangle \langle 1 \vert  \\
&=&
\frac{1}{4}
\begin{pmatrix}
 1 \\
 0
\end{pmatrix}
\begin{pmatrix}
 1 & 0
\end{pmatrix}
+
\frac{3}{4}
\begin{pmatrix}
 0 \\
 1
\end{pmatrix}
\begin{pmatrix}
 0 & 1
\end{pmatrix}\\
&=&
\begin{pmatrix}
 \frac{1}{4} & 0 \\
 0 & \frac{3}{4}
\end{pmatrix}. \label{equ:dm:classical}
\end{eqnarray}

Combination of quantum probabilities and classical probabilities is also possible,
such as,
where $\frac{1}{\sqrt{2}}(\vert 0 \rangle + \vert 1 \rangle)$ occupies classically 50\% and $\vert 1\rangle$ occupies classically 50\%.
The density matrix is
\begin{eqnarray}
\rho_{mixed2} &=&
\frac{1}{2}
\begin{pmatrix}
 \frac{1}{\sqrt{2}}\\
 \frac{1}{\sqrt{2}}
\end{pmatrix}
\begin{pmatrix}
 \frac{1}{\sqrt{2}} & \frac{1}{\sqrt{2}}
\end{pmatrix}
+
\frac{1}{2}
\begin{pmatrix}
 0 \\ 
 1
\end{pmatrix}
\begin{pmatrix}
 0 & 1
\end{pmatrix}\\
&=&
\begin{pmatrix}
 \frac{1}{4} & \frac{1}{4} \\
 \frac{1}{4} & \frac{3}{4}
\end{pmatrix}. \label{equ:dm:quantum-classical}
\end{eqnarray}

The diagonal elements of the density matrix describe population of the basis in the state.
Hence $Tr(\rho)=1$ is always satisfied.

For all three of the states in 
Eq. \ref{equ:dm:quantum}, 
Eq. \ref{equ:dm:classical}
and Eq. \ref{equ:dm:quantum-classical},
the state has a 25\% chance of being measured in the $+1$ eigenvalue
and
has a 75\% chance of being measured in the $-1$ eigenvalue.
However, the physical states and the processes of the observation differ.
States consisting of only quantum probabilities may occur during some quantum algorithms.
States consisting of only classical probabilities are just a ``junk'' state from the point of view of quantum computation.
States consisting of combination of quantum and classical probabilities may be happening as a result of some errors which we might be able to correct.
However, it will not generate interference effects so that the algorithm may run properly.
From the density matrix, we can distinguish the existence of classical probabilities by 
calculating $Tr(\rho^2)$.
\begin{eqnarray}
 Tr(\rho_{pure}^2) &=& Tr(
\begin{pmatrix}
 \frac{1}{4} & \frac{\sqrt{3}}{4} \\
 \frac{\sqrt{3}}{4} & \frac{3}{4}
\end{pmatrix}^2
)\\
&=&
Tr(
\begin{pmatrix}
 \frac{1}{4} & \frac{\sqrt{3}}{4} \\
 \frac{\sqrt{3}}{4} & \frac{3}{4}
\end{pmatrix}
)\\
&=&
1
\end{eqnarray}

\begin{eqnarray}
Tr(\rho_{mixed1}^2) &=& Tr
(\begin{pmatrix}
 \frac{1}{4} & 0 \\
 0 & \frac{3}{4}
\end{pmatrix}^2
 ) \\
&=&
Tr(
\begin{pmatrix}
 \frac{1}{16} & 0 \\
 0 & \frac{9}{16}
\end{pmatrix}
)\\
&=& 
\frac{5}{8}
\end{eqnarray}

\begin{eqnarray}
Tr(\rho_{mixed2}^2) &=& Tr(
\begin{pmatrix}
 \frac{1}{4} & \frac{1}{4} \\
 \frac{1}{4} & \frac{3}{4}
\end{pmatrix} ^2
) \\
&=&
Tr(
\begin{pmatrix}
 \frac{1}{8} & \frac{1}{4} \\
 \frac{1}{4} & \frac{5}{8}
\end{pmatrix}
)\\
&=& \frac{3}{4}
\end{eqnarray}

As exemplified above, $Tr(\rho^2) = 1$ indicates that there only are quantum probability amplitudes
hence the qubit is in a ``pure'' state,
and $Tr(\rho^2) < 1$ indicates that the state includes classical probabilities and 
the state is so called ``mixed'' state.
The off-diagonal elements of the density matrix describe quantum coherence of this state;
$\vert 0\rangle \langle 1 \vert$ shows evolution of the state from $\vert 1 \rangle$ to $\vert 0 \rangle$
and
$\vert 1\rangle \langle 0 \vert$ shows evolution of the state from $\vert 0 \rangle$ to $\vert 1 \rangle$.
Classical probabilities do not have this evolution, hence $\rho$ which has classical probabilities 
results in $Tr(\rho^2) < 1$.

The density matrix of a state which is in $\vert \psi_i \rangle$ for classical probability $p_i$ is
\begin{eqnarray}
 \rho \equiv \sum_i p_i \vert \psi_i \rangle \langle \psi_i \vert
\label{dmpure}
\end{eqnarray}
where $\vert \psi_i \rangle$ is the $i$-th possible pure state.
Therefore the pure state is a specific case of the mixed state, only 1 pure state exists classically, 
some $p_j=1$ and all $p_i =0, i\ne j$.

To describe a mixed state with the Bloch sphere, from Eq. \ref{equ:qubit},
\begin{eqnarray}
\vert \psi \rangle &=& \cos \frac{\theta}{2} \vert 0 \rangle + e^{i\phi}\sin \frac{\theta}{2} \vert 1 \rangle \\
&=& \cos \frac{\theta}{2} \vert 0 \rangle + (\cos \phi + i \sin \phi) \sin \frac{\theta}{2} \vert 1 \rangle \\
&=& \cos \frac{\theta}{2} \vert 0 \rangle + (\cos \phi  \sin \frac{\theta}{2} + i \sin \phi \sin \frac{\theta}{2}) \vert 1 \rangle
\end{eqnarray}
is derived and its density matrix is described as
\begin{eqnarray}
 \vert \psi \rangle \langle \psi \vert &=& 
\begin{pmatrix}
\cos (\frac{\theta}{2}) \\ 
\cos \phi  \sin (\frac{\theta}{2}) + i \sin \phi \sin (\frac{\theta}{2})
\end{pmatrix}
\begin{pmatrix}
\cos (\frac{\theta}{2}) &  \cos \phi  \sin (\frac{\theta}{2}) - i \sin \phi \sin (\frac{\theta}{2})
\end{pmatrix} \nonumber \\
\\
&=&\begin{pmatrix}
\cos ^2 (\frac{\theta}{2}) &  \cos (\frac{\theta}{2}) \sin (\frac{\theta}{2}) (\cos \phi  - i \sin \phi) \\ 
(\cos \phi + i \sin \phi) \sin (\frac{\theta}{2})\cos (\frac{\theta}{2}) & \cos^2 \phi \sin^2 (\frac{\theta}{2}) + \sin^2 \phi \sin^2 (\frac{\theta}{2})
\end{pmatrix}\\
&=&\begin{pmatrix}
\cos ^2 (\frac{\theta}{2}) &  \cos (\frac{\theta}{2}) \sin (\frac{\theta}{2}) (\cos \phi  - i \sin \phi) \\ 
(\cos \phi + i \sin \phi) \sin (\frac{\theta}{2})\cos (\frac{\theta}{2}) & \sin^2 (\frac{\theta}{2})
\end{pmatrix}\\
&=&
\frac{1}{2}
\begin{pmatrix}
1+\cos\theta &  \sin\theta(\cos \phi  - i \sin \phi)\\ 
(\cos \phi + i \sin \phi) \sin \theta & 1-\cos\theta
\end{pmatrix}\\
&=&
\frac{1}{2}
\begin{pmatrix}
1+\cos\theta &  \sin\theta\cos \phi  - i \sin\theta\sin \phi\\ 
\cos \phi \sin \theta + i \sin \phi \sin \theta & 1-\cos\theta
\end{pmatrix}\\
&=&
\frac{1}{2}(I + X\sin\theta\cos\phi + Y\sin\theta\sin\phi + Z\cos\theta).
\end{eqnarray}
$(\sin\theta\cos\phi, \sin\theta\sin\phi, \cos\theta)$, the coefficients for $X$, $Y$ and $Z$ respectivelly, tells the coordinate of the state on the Bloch sphere.
For a mixed state, the density matrix is described as 
\begin{eqnarray}
 \rho &=& \sum_i p_i\frac{1}{2}(I + X\sin\theta_i\cos\phi_i + Y\sin\theta_i\sin\phi_i + Z\cos\theta_i) \\
      &=& \frac{I}{2} + \frac{X}{2} \sum_i p_i\sin\theta_i\cos\phi_i 
                      + \frac{Y}{2} \sum_i p_i\sin\theta_i\sin\phi_i 
                      + \frac{Z}{2} \sum_i p_i\cos\theta_i \nonumber \\
\end{eqnarray}
where $\sum_i p_i = 1$ and we can find a vector
\begin{eqnarray}
r = 
 \begin{pmatrix}
\sum_i p_i\sin\theta_i\cos\phi_i \\
\sum_i p_i\sin\theta_i\sin\phi_i \\
\sum_i p_i\cos\theta_i 
 \end{pmatrix}.
\end{eqnarray}
This vector is called the Bloch vector.
The norm of the Bloch vector satisfies $\vert \vert r \vert \vert \le 1$.
$\vert \vert r \vert \vert = 1$ is satisfied only when the state is a pure state.
Hence the vector of a mixed state in the Bloch sphere ends inside the sphere.

To calculate the effects of a quantum gate on a density matrix,
apply the quantum gate from both sides of the density matrix, such as 
\begin{eqnarray}
 X \rho_{mixed2} X 
&=&
\begin{pmatrix}
 0 & 1 \\
 1 & 0
\end{pmatrix}
\begin{pmatrix}
 \frac{1}{4} & \frac{1}{4} \\
 \frac{1}{4} & \frac{3}{4}
\end{pmatrix}
\begin{pmatrix}
 0 & 1 \\
 1 & 0
\end{pmatrix} \\
&=&
\begin{pmatrix}
 \frac{3}{4} & \frac{1}{4} \\
 \frac{1}{4} & \frac{1}{4}
\end{pmatrix}.
\end{eqnarray}

Measurement of a density matrix $\rho$ 
can be described with a measurement operator $M_m$
such as $M_0 = \vert 0 \rangle \langle 0 \vert$ to find $0$.
The probability to observe $m$ can be described with the posterior probability $p(m|i)$ that $m$ is measured after $i$-th pure state of the mixed state is chosen,
\begin{eqnarray}
 p(m|i) &=& \langle \psi_i \vert M_m^\dag M_m \vert \psi_i \rangle \\
        &=& Tr(M_m^\dag M_m \vert \psi_i \rangle \langle \psi_i \vert)\\
 p(m) &=& \sum_i p_i p(m|i) \\
      &=& \sum_i p_i \langle \psi_i \vert M_m^\dag M_m \vert \psi_i \rangle \\
      &=& \sum_i p_i Tr(M_m^\dag M_m \vert \psi_i \rangle \langle \psi_i \vert) \\
      &=& Tr(M_m^\dag M_m \sum_i p_i \vert \psi_i \rangle \langle \psi_i \vert) \\
      &=& Tr(M_m^\dag M_m \rho)
\end{eqnarray}
The preserved density matrix after the measurement is described with
the posterior probability $p(i|m)$ which is the probability that the $i$-th pure state gets chosen after $m$ is observed,
\begin{eqnarray}
 \rho_m  &=& \sum_i p(i|m) \vert \psi_i^m \rangle \langle \psi_i^m \vert\\
         &=& \sum_i p(i|m) \frac{M_m\vert \psi_i \rangle}{\sqrt{\langle \psi_i \vert M_m^\dag M_m \vert \psi_i \rangle}}
                        \frac{\langle \psi_i \vert M_m^\dag}{\sqrt{\langle \psi_i \vert M_m^\dag M_m \vert \psi_i \rangle}} \\
         &=& \sum_i p(i|m) \frac{M_m \vert \psi_i \rangle \langle \psi_i \vert M_m^\dag}{\langle \psi_i \vert M_m^\dag M_m \vert \psi_i \rangle}.
\label{equ:rho_m2}
\end{eqnarray}
Then by the Bayes' theorem $p(i|m) = \frac{p(m|i)}{p_m}$, Equ. \ref{equ:rho_m2} is
\begin{eqnarray}
 \rho_m  &=& \sum_i \frac{p(m|i)}{p_m} \frac{M_m \vert \psi_i \rangle \langle \psi_i \vert M_m^\dag}{\langle \psi_i \vert M_m^\dag M_m \vert \psi_i \rangle}\\
         &=& \sum_i \frac{p(m|i)}{p_m} \frac{M_m \vert \psi_i \rangle \langle \psi_i \vert M_m^\dag}{p(m|i)}\\
         &=& \sum_i \frac{M_m \vert \psi_i \rangle \langle \psi_i \vert M_m^\dag}{p_m}\\
         &=& \frac{M_m \rho M_m^\dag}{p_m}\\
         &=& \frac{M_m \rho M_m^\dag}{Tr(M_m^\dag M_m \rho)}.
\end{eqnarray}

  \section{Nearly perfect quantum computation on imperfect systems}
\label{sec:qc_imperfect}
Mainly, four types of imperfections are addressed in the quantum computation research field today:
state errors, dynamic losses, static losses and leakage errors.
First, state errors received the bulk of researchers' attention and means to tolerate state errors have been proposed~\cite{PhysRevA.52.R2493,nielsen-chuang:qci,quant-ph/9608026,steane:10.1098/rspa.1996.0136}.
Those codes utilize redundancy to correct quantum errors.
Those codes uses gate-based circuits to extract error syndromes,
projecting states to find whether the states have errors or not.
Then after classically analyzing syndromes, error corrections are applied.
Here I first show the basics of the state errors and quantum error correcting codes for state errors, which are later extended to tolerate other imperfections.

Readers interested in quantum error correction are referred to \cite{RevModPhys.87.307,0034-4885-76-7-076001,lidar-brun:qec,quant-ph/9705052}

\subsection{State errors}
The decoherence changes the state of a qubit,
so the probabilities that the bases observed.
Thus, quantum errors should be analog.
State errors have been modeled to several ways~\cite{nielsen-chuang:qci}.
The depolarizing channel is the most thoroughly investigated error model.
In the depolarizing channel,
an qubit may stochastically get depolarized to the completely mixed state $\frac{I}{2}$.
Hence an arbitrary state $\rho$ through a depolarizing channel is changed as
\begin{equation}
 \rho \rightarrow p\frac{I}{2} + (1-p) \rho
  \label{equ:depolarizing0}
\end{equation}
where $p$ is the error rate.
\begin{equation}
 p\frac{I}{2} + (1-p) \rho = (1-\frac{3p}{4})\rho + \frac{1}{4}X\rho X + \frac{1}{4}Z\rho Z + \frac{1}{4}Y\rho Y
  \label{equ:depolarizing1}
\end{equation}
since
\begin{equation}
 \frac{I}{2} = \frac{1}{4}\rho + \frac{1}{4}X\rho X + \frac{1}{4}Z\rho Z + \frac{1}{4}Y\rho Y.
 \label{equ:depolarizing2}
\end{equation}
This formula implies that a depolarizing error can be factored into three types of errors: an $X$ error, a $Z$ error and a $Y$ error.
Quantum error correcting codes are designed to detect those errors.
Meanwhile, decoherence practically occurs in analog fashion. 
Such an analog quantum error can be treated as a discrete $X$ error, $Z$ error or $Y$ error.
A quantum error correcting code ``projects'' the system into either a correct state or a discrete errored state.
Therefore a quantum error correcting code which works with the depolarizing error model can correct any quantum errors.

\subsection{Stabilizer code}
The stabilizer code is the most major quantum error correcting code group which utilizes abelian group~\cite{quant-ph/9705052}.
There are operators $U$ which do not change certain states, such as
\begin{equation}
 \vert \psi \rangle = U \vert \psi \rangle.
\end{equation}
Such operators are called stabilizers of the state.
Products of stabilizers are also stabilizers, hence,
the most simplest set of stabilizers including $I^{\otimes n}$ are called stabilizer generators.
A full set of stabilizer generators,
including the same number of stabilizer generators with the number of qubits,
specifies a qubit state so that
it can be used as a compact notation to describe a quantum state.
The simplest example is that $X$ does not change
$\frac{1}{\sqrt{2}}\begin{pmatrix} 1\\1\end{pmatrix}$:
\begin{equation}
X
\times \frac{1}{\sqrt{2}}\begin{pmatrix} 1\\1\end{pmatrix}
= \begin{pmatrix}0 & 1 \\ 1 & 0\end{pmatrix}
\times \frac{1}{\sqrt{2}}\begin{pmatrix} 1\\1\end{pmatrix}
= \frac{1}{\sqrt{2}}\begin{pmatrix} 1\\1\end{pmatrix}
\end{equation}
However, this simplest example cannot be used for quantum computation.
Only $\frac{1}{\sqrt{2}}\begin{pmatrix} 1\\1\end{pmatrix}$ is allowed by an $X$ stabilizer on a qubit,
hence there is no degree of freedom available to serve as a variable computational qubit.

Generally, stabilizers are used on two or more qubits, such as
\begin{eqnarray}
X\otimes X \times (\alpha(\vert 00 \rangle + \vert 11 \rangle) + \beta (\vert 01 \rangle + \vert 10 \rangle))
&=& X\otimes X \times 
(\alpha
\begin{pmatrix} 
1\\0\\0\\1
\end{pmatrix}
+
\beta
\begin{pmatrix} 
0\\1\\1\\0
\end{pmatrix})\\
&=& \begin{pmatrix}
0 & 0 & 0 & 1 & \\
0 & 0 & 1 & 0 & \\
0 & 1 & 0 & 0 & \\
1 & 0 & 0 & 0 & 
\end{pmatrix}
(\alpha
\begin{pmatrix} 
1\\0\\0\\1
\end{pmatrix}
+
\beta
\begin{pmatrix} 
0\\1\\1\\0
\end{pmatrix})\nonumber \\
\\
&=& \alpha(\vert 00 \rangle + \vert 11 \rangle) + \beta (\vert 01 \rangle + \vert 10 \rangle). \nonumber \\
\end{eqnarray}
where $\alpha(\vert 00 \rangle + \vert 11 \rangle) + \beta (\vert 01 \rangle + \vert 10 \rangle)$ is
used as a logical qubit, $\alpha$ and $\beta$ correspond to 
$\cos \frac{\theta}{2}$ and $e^{i\phi}\sin \frac{\theta}{2}$ in Equation \ref{equ:qubit}.

Larger sets of stabilizer generators are possible.
For example, $X_aX_b$ and $X_bX_c$ are the stabilizer generators
of 
\begin{equation}
\alpha (\vert 0_a0_b0_c \rangle + \vert 1_a1_b0_c \rangle + \vert 0_a1_b1_c \rangle + 1_a0_b1_c \rangle)
+\beta (\vert 1_a1_b1_c \rangle + \vert 0_a0_b1_c \rangle + \vert 1_a0_b0_c \rangle + 0_a1_b0_c \rangle).
\end{equation}
Another example is $ZZ$ on 
\begin{equation}
\alpha \vert 00 \rangle 
+\beta \vert 11 \rangle.
\end{equation}
Those states can be used as a logical qubit, since they have a degree of freedom for $\alpha$ and $\beta$.

In fact, the degree of freedom of the quantum state, the number of qubits and the number of independent stabilizer generators have the relationship,
\begin{equation}
 (\textrm{the\ degree\ of\ freedom}) = 2^{\#\textrm{qubit}-\#(\textrm{stabilizer generator})}.
  \label{equ:sc:freedom}
\end{equation}
We can choose arbitrary numbers of qubits and of stabilizer generators depending on the purpose.

Surprisingly, stabilizers have great power in quantum error correction.
Let's think of $CNOT$ gates on
\begin{equation}
\vert \psi \rangle = \alpha (\vert 0_a0_b\rangle+ \vert 1_a1_b\rangle) + \beta (\vert 0_a1_b\rangle+ \vert 1_a0_b\rangle)
\end{equation}
which is stabilized by $X_aX_b$.
$I_aI_b$ does not change $\vert \psi \rangle$.
Hence, a set of $CNOT$ gates, $CNOT(c,a)$ and $CNOT(c,b)$, does not change $\vert \psi \rangle$
because $0_c$ applies $I_aI_b$ and because $1_c$ applies $X_aX_b$ for an arbitrary qubit $c$.

Let's assume qubit $c$ is in $\frac{1}{\sqrt{2}}(\vert 0 \rangle + \vert 1 \rangle)$, which is the $+1$ eigenvalue of the $X$ axis.
Then, the $CNOT$ gates work as below.
The first state is
\begin{eqnarray}
 \vert \Psi_{c,a,b} \rangle &=& \vert +_c \rangle \otimes \vert \psi_{a,b}\rangle \\
  &=& \frac{1}{\sqrt{2}}(\vert 0_c \rangle + \vert 1_c \rangle) 
\otimes 
(\alpha (\vert 0_a0_b\rangle + \vert 1_a1_b\rangle) + \beta (\vert 0_a1_b\rangle+ \vert 1_a0_b\rangle))\nonumber \\
 \\
 &=& \frac{1}{\sqrt{2}}(
    \alpha \vert 0_c0_a0_b \rangle 
  + \alpha \vert 0_c1_a1_b\rangle
  + \beta  \vert 0_c0_a1_b\rangle
  + \beta  \vert 0_c1_a0_b\rangle \nonumber \\ &&
  + \alpha \vert 1_c0_a0_b \rangle 
  + \alpha \vert 1_c1_a1_b\rangle
  + \beta \vert 1_c0_a1_b\rangle
  + \beta \vert 1_c1_a0_b\rangle). \nonumber \\
\end{eqnarray}
Applying $CNOT(c,a)$,
\begin{eqnarray}
 \vert \Psi_{c,a,b}' \rangle &=& CNOT(c,a) \vert \Psi_{c,a,b} \rangle \\
 &=& \frac{1}{\sqrt{2}}(
    \alpha \vert 0_c0_a0_b \rangle 
  + \alpha \vert 0_c1_a1_b\rangle
  + \beta  \vert 0_c0_a1_b\rangle
  + \beta  \vert 0_c1_a0_b\rangle \nonumber \\ &&
  + \alpha  \vert 1_c1_a0_b \rangle 
  + \alpha  \vert 1_c0_a1_b\rangle
  + \beta \vert 1_c1_a1_b\rangle
  + \beta \vert 1_c0_a0_b\rangle). \nonumber \\
\end{eqnarray}
Applying $CNOT(c,b)$,
\begin{eqnarray}
 \vert \Psi_{c,a,b}'' \rangle &=& CNOT(c,b) \vert \Psi_{c,a,b}' \rangle \\
 &=& \frac{1}{\sqrt{2}}(
    \alpha \vert 0_c0_a0_b \rangle 
  + \alpha \vert 0_c1_a1_b\rangle
  + \beta  \vert 0_c0_a1_b\rangle
  + \beta  \vert 0_c1_a0_b\rangle \nonumber \\ &&
  + \alpha \vert 1_c1_a1_b \rangle 
  + \alpha \vert 1_c0_a0_b\rangle
  + \beta \vert 1_c1_a0_b\rangle
  + \beta \vert 1_c0_a1_b\rangle) \nonumber \\
 \\
&=& \frac{1}{\sqrt{2}}(
\vert 0_c \rangle (\alpha (\vert 0_a0_b\rangle + \vert 1_a1_b\rangle) 
                 + \beta  (\vert 0_a1_b\rangle + \vert 1_a0_b\rangle))\nonumber \\
&&+
\vert 1_c \rangle (\alpha (\vert 0_a0_b\rangle + \vert 1_a1_b\rangle) 
                 + \beta  (\vert 0_a1_b\rangle + \vert 1_a0_b\rangle))\nonumber \\
&=&
\frac{1}{\sqrt{2}}(\vert 0_c + \vert 1_c \rangle)
\otimes 
(\alpha (\vert 0_a0_b\rangle + \vert 1_a1_b\rangle) 
+ \beta  (\vert 0_a1_b\rangle + \vert 1_a0_b\rangle))\nonumber \\
&=& \vert +_c \rangle \otimes \vert \psi_{a,b}\rangle.
\end{eqnarray}

Let's think of a $Z$ error on qubit $a$.
Then, the $CNOT$ gates work as below.
The first state is
\begin{eqnarray}
 Z_a \vert \Psi_{c,a,b} \rangle &=& \vert +_c \rangle \otimes Z_a \vert \psi_{a,b}\rangle \\
  &=& \frac{1}{\sqrt{2}}(\vert 0_c \rangle + \vert 1_c \rangle) 
\otimes 
(\alpha (\vert 0_a0_b\rangle - \vert 1_a1_b\rangle) + \beta (\vert 0_a1_b\rangle- \vert 1_a0_b\rangle))\nonumber \\
 \\
 &=& \frac{1}{\sqrt{2}}(
    \alpha \vert 0_c0_a0_b \rangle 
  - \alpha \vert 0_c1_a1_b\rangle
  + \beta  \vert 0_c0_a1_b\rangle
  - \beta  \vert 0_c1_a0_b\rangle \nonumber \\ &&
  + \alpha \vert 1_c0_a0_b \rangle 
  - \alpha \vert 1_c1_a1_b\rangle
  + \beta \vert 1_c0_a1_b\rangle
  - \beta \vert 1_c1_a0_b\rangle). \nonumber \\
\end{eqnarray}
Applying $CNOT(c,a)$,
\begin{eqnarray}
 \vert \Psi_{c,a,b}' \rangle &=& CNOT(c,a) \vert \Psi_{c,a,b} \rangle \\
 &=& \frac{1}{\sqrt{2}}(
    \alpha \vert 0_c0_a0_b \rangle 
  - \alpha \vert 0_c1_a1_b\rangle
  + \beta  \vert 0_c0_a1_b\rangle
  - \beta  \vert 0_c1_a0_b\rangle \nonumber \\ &&
  + \alpha  \vert 1_c1_a0_b \rangle 
  - \alpha  \vert 1_c0_a1_b\rangle
  + \beta \vert 1_c1_a1_b\rangle
  - \beta \vert 1_c0_a0_b\rangle). \nonumber \\
\end{eqnarray}
Applying $CNOT(c,b)$,
\begin{eqnarray}
 \vert \Psi_{c,a,b}'' \rangle &=& CNOT(c,b) \vert \Psi_{c,a,b}' \rangle \\
 &=& \frac{1}{\sqrt{2}}(
    \alpha \vert 0_c0_a0_b \rangle 
  - \alpha \vert 0_c1_a1_b\rangle
  + \beta  \vert 0_c0_a1_b\rangle
  - \beta  \vert 0_c1_a0_b\rangle \nonumber \\ &&
  + \alpha \vert 1_c1_a1_b \rangle 
  - \alpha \vert 1_c0_a0_b\rangle
  + \beta \vert 1_c1_a0_b\rangle
  - \beta \vert 1_c0_a1_b\rangle) \nonumber \\
 \\
&=& \frac{1}{\sqrt{2}}(
\vert 0_c \rangle (\alpha (\vert 0_a0_b\rangle - \vert 1_a1_b\rangle) 
                 + \beta  (\vert 0_a1_b\rangle - \vert 1_a0_b\rangle))\nonumber \\
&&-
\vert 1_c \rangle (\alpha (\vert 0_a0_b\rangle - \vert 1_a1_b\rangle) 
                 + \beta  (\vert 0_a1_b\rangle - \vert 1_a0_b\rangle))\nonumber \\
&=&
\frac{1}{\sqrt{2}}(\vert 0_c - \vert 1_c \rangle)
\otimes 
(\alpha (\vert 0_a0_b\rangle - \vert 1_a1_b\rangle) 
+ \beta  (\vert 0_a1_b\rangle - \vert 1_a0_b\rangle))\nonumber \\
&=& \vert -_c \rangle \otimes Z_a \vert \psi_{a,b}\rangle.
\end{eqnarray}
Therefore, by measuring $X_c$ after those $CNOT$ gates, we can check the error syndromes of qubits and can detect $Z_a$.
Since in fact we also find $-_c$ when we have a $Z$ error on qubit $b$,
hence the $X_aX_b$ stabilizer is used to \emph{detect} an $Z$ error.
By utilizing multiple stabilizers and by classical processing of error syndromes, we can specify the error location and \emph{correct} the error.

Such $X^{\otimes n}$ stabilizers which detect $Z$ errors are often called $X$ stabilizers.
By symmetry, we can detect an $X$ error by a $Z$ stabilizer.
We can compose a large quantum error correcting code by using more qubits and more stabilizer generators,
which can correct multiple errors.
Additionally, by repeating these operations, we can detect errors on qubit $c$.
Those characteristics of quantum information lead to the stabilizer code.

\subsection{Surface code and the planar code, the simplest form of the surface code}
\label{surface_code}
Surface code quantum computation is regarded as
a promising technique for fault tolerant quantum computation.
Briefly, the surface code originally has two advantages:
\begin{description}
  \item[High feasibility]~\\The surface code requires only nearest neighbor interaction between physical qubits on the lattice.
  \item[High state error threshold 1\%]~\\The surface code is robust against physical quantum state error and it has been shown that the threshold (a physical error rate where the logical error rate rises above the physical error rate) is nearly 1\%.
\end{description}
and later robustness against dynamic loss is shown:
\begin{description}
 \item[Tolerance against loss rate less than 50\% in ideal environment]~\\
		With perfect state error extraction and perfect error/loss correction,
		at state error rate of 0\% loss rate of 50\% is tolerable including both of dynamic loss and of static loss, 
		and at loss rate 0\% state error rate of 12\% is tolerable~\cite{PhysRevA.81.022317,PhysRevLett.102.200501}.
		Between those two points, the relationship between the tolerable state error rate and the tolerable loss rate changes more or less linearly,
		except that at loss rates of $>40\%$ the tolerable state error rates are less than the linear fit.
 \item[Tolerance against dynamic loss below 1\%]~\\
		Even with no state errors, the loss rate has to be less than 1\%~\cite{1409.4880}.
		Otherwise, a surface code qubit of a longer code distance does not have better logical error rate than that of a shorter code distance.
\end{description}

The surface code is a means for encoding logical qubits on a form of entangled 2-D lattice,
consisting of many qubits.
A surface code lattice can be made on a nearest neighbor architecture.
The nearest neighbor architecture uses quantum interaction
only between nearest neighboring qubits.
In a regular lattice, each qubit is entangled with its neighbors,
giving a specific large, entirely entangled state.
This fact makes it potencially possible to fabricate devices using planar photolithography, including
quantum dot, superconducting, and planar ion trap structures.
It gives the quantum processor extensibility by adding
one more row of qubits and control devices along the outside edge of the lattice,
making it one of the most feasible current
proposals for building a scalable quantum computer.

The surface code is characterized by
the method for fixing the physical errors which occur on physical qubits in the lattice,
the logical qubit encoding,
and the means of executing logical gates between logical qubits.
In this subsection the surface code is briefly explained by describing the simplest form of the surface code qubit called planar code. The \textit{defect}-based code is following in the next subsection.

\subsubsection{Stabilizers for error correction}
The lattice is divided into many plaquettes which indicate stabilizer generators of the surface code and 
the state of the lattice is maintained by repeatedly measuring sets of stabilizers.

\begin{figure}[t]
 \begin{center}
  \includegraphics[width=10cm]{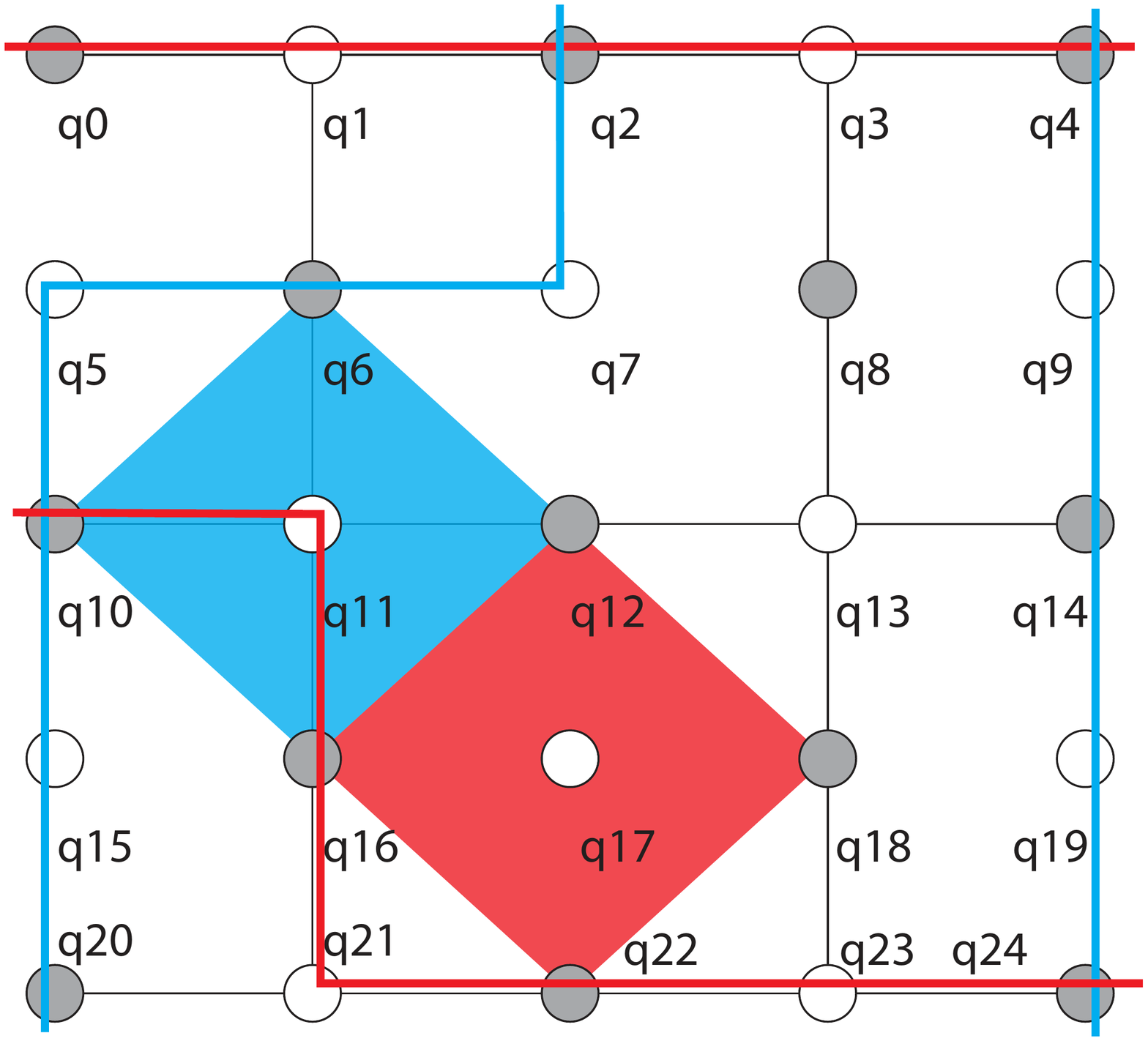}
  \caption[Example of a surface code encoding a single logical qubit,
  describing a Z stabilizer (red diamond), an X stabilizer (blue diamond),
  two instances of the Z operator (red lines) and two instances of the X operator (blue lines).]
  {The gray circles are data qubits, and the white circles are syndrome qubits.
  Qubits q12, q16, q18 and q22 are included in the Z stabilizer and
  qubits q6, q10, q12 and q16 are included in the X stabilizer.
  Other qubits are also involved in other corresponding stabilizers.
  The west and the east boundaries of the lattice are for the Z operator.
  Other possible lines between the west and the east boundaries also have the same Z operator.
  The blue lines have the same X operator.
  The north and the south boundaries of the lattice are for the X operator.
  Other equivalent logical operators are formed by multiplying a line by associated X (Z) stabilizers.
  }
  \label{fig:sc:stabilizers}
 \end{center}
\end{figure}

\begin{figure}[t]
 \begin{center}
  \includegraphics[width=200pt]{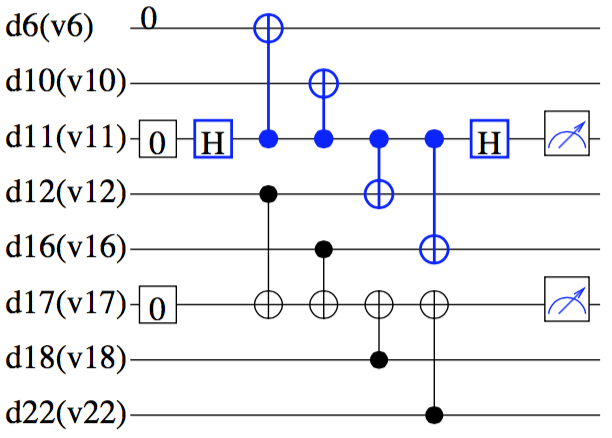}
  \caption[Circuits for each stabilizer in Figure \ref{fig:sc:stabilizers} and in Table \ref{tab:sc:stabilizers}.]
  {The top half is an X stabilizer and
  The bottom half is a Z stabilizer.
  Each face and vertex of the lattice has the same stabilizer circuit.
  All stabilizers are executed concurrently in the absence of faulty components.
  The only gates required for stabilizer operation are
  INIT in Z basis, CNOT, SWAP and H gates and MEASUREMENT in Z basis.
  Boxes containing 0 are the INIT gates.}
  \label{fig:sc:stabilizer-circuits}
 \end{center}
\end{figure}

The surface code corrects errors in each unit and the code space is protected as a whole.
Figure \ref{fig:sc:stabilizers} shows the layout of normal unit stabilizers of a planar code, which is the simplest form of the logical qubit descriptions of the surface code.
The lines in the figure are just a visual guide demarking plaquettes; each syndrome qubit is actually physically coupled to four neighbors.
Table \ref{tab:sc:stabilizers} and Figure \ref{fig:sc:stabilizer-circuits} show the stabilizer representation and the circuits of the stabilizers marked
in Figure \ref{fig:sc:stabilizers}, respectively.
\begin{table}[b]
  \begin{center}
   \caption[Stabilizer representation of the stabilizers colored in Figure \ref{fig:sc:stabilizers}.]
   {The upper line is a Z stabilizer and the lower is an X stabilizer.}
   \label{tab:sc:stabilizers}
   \begin{tabular}[t]{cccccc}
    q6 &q10 &q12 &q16 &q18 &q22\\
    \hline
       &    &Z   &Z   &Z   &Z\\
    X  &X   &X   &X   &    & \\
   \end{tabular}
  \end{center}
\end{table}
The stabilizers measure the parity of the lattice qubits involved.
Normally, in the surface code, qubits are initialized as the parity is even (+1 eigenstate).
When the states of an odd number of qubits that belong to the stabilizer are flipped,
the parity becomes odd (-1 eigenstate).

\subsubsection{Logical operator}
A complete lattice has no degrees of freedom; a lattice with $N$ qubits has
$N$ independent stabilizers, and hence is a fully-specified quantum state.
The surface code defines a logical operator, using the degree of freedom introduced at a set of lattice boundaries in the planar code.

A lattice boundary is a terminal of a logical operator; hence a pair of boundaries introduces a logical operator and two pairs of different boundaries can generate a set of a logical X operator and a logical Z operator so that a single logical qubit is introduced.
The planar code has two pairs of boundaries so that it has a degree of freedom which encodes a logical qubit on a 2-D lattice.
Any path between a pair of boundaries defines the same logical operator.

In the basic planar code, logical two-qubit operations are performed by transversal operations, arranging the planes stacked in the third dimension and coupling corresponding pairs of qubits.
Lattice surgery improved the planar code, adding the capability to execute a logical two-qubit gate more easily than transversal operations~\cite{Horsman:2012lattice_surgery}.
It implements a logical two-qubit gate by connecting and disconnecting two planar code qubits on a 2-D lattice along an edge.

\subsubsection{Logical measurement and redundancy}
To measure a logical qubit, take the parity of the measurement results on the physical qubits composing a logical operator.
Parities of measurements on any path should have the same value, so that the logical measurement has redundancy against measurement failure.
This is shown in Figure \ref{fig:sc:stabilizers}.

\begin{figure}[t]
 \begin{center}
(a)
  \includegraphics[width=135pt]{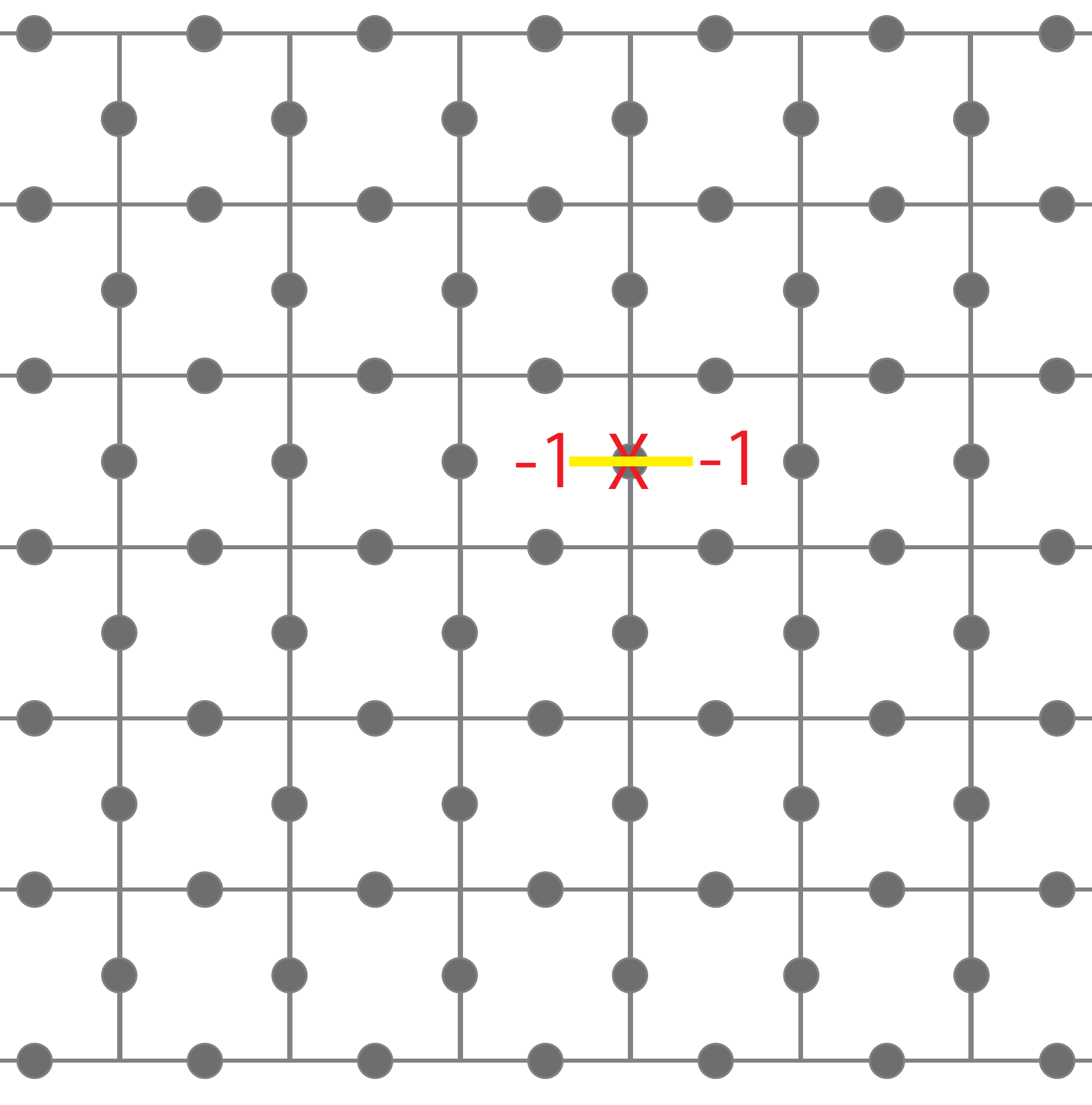}
(b)
  \includegraphics[width=135pt]{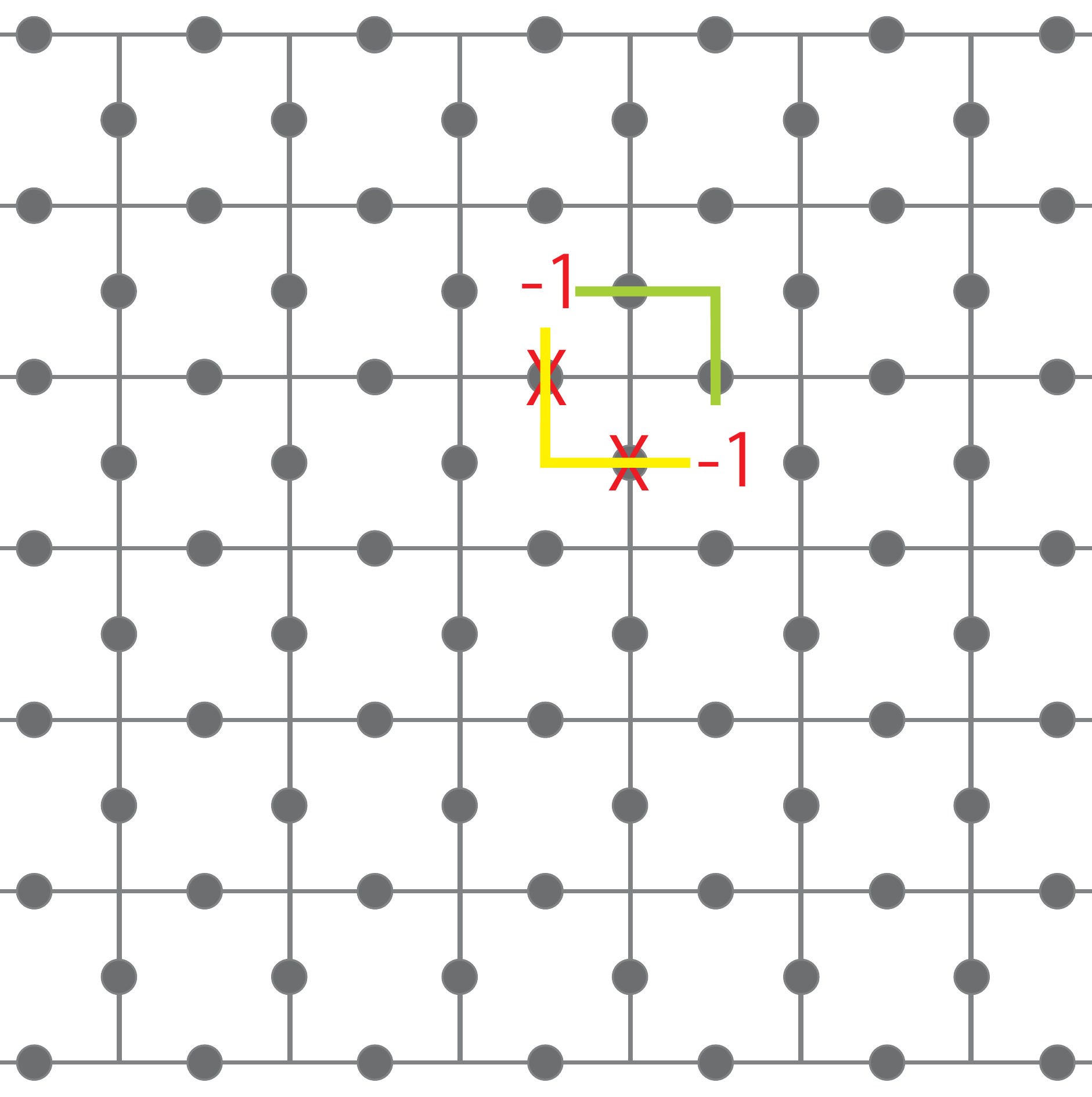}\\
(c)
  \includegraphics[width=135pt]{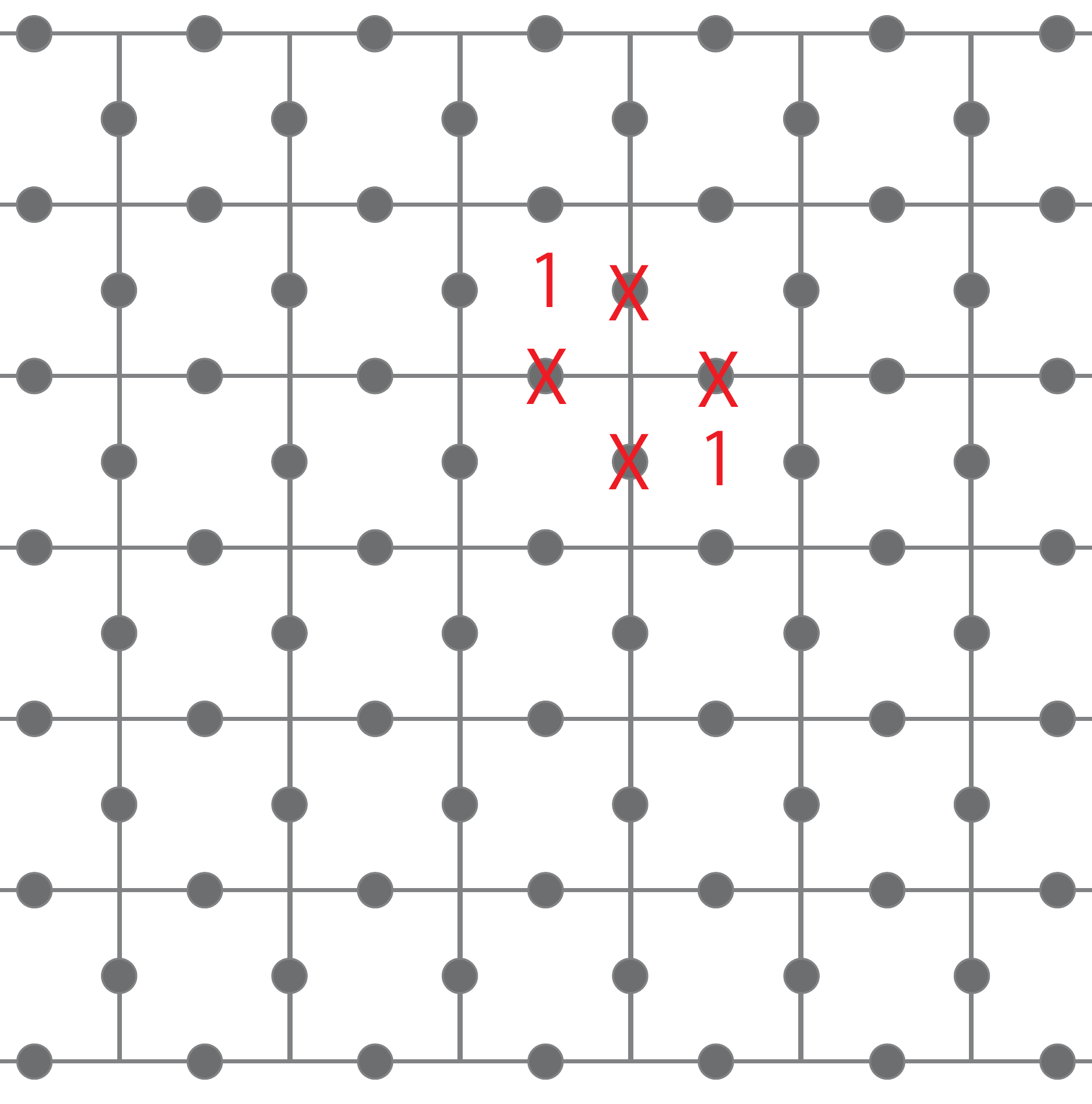}
(d)
  \includegraphics[width=135pt]{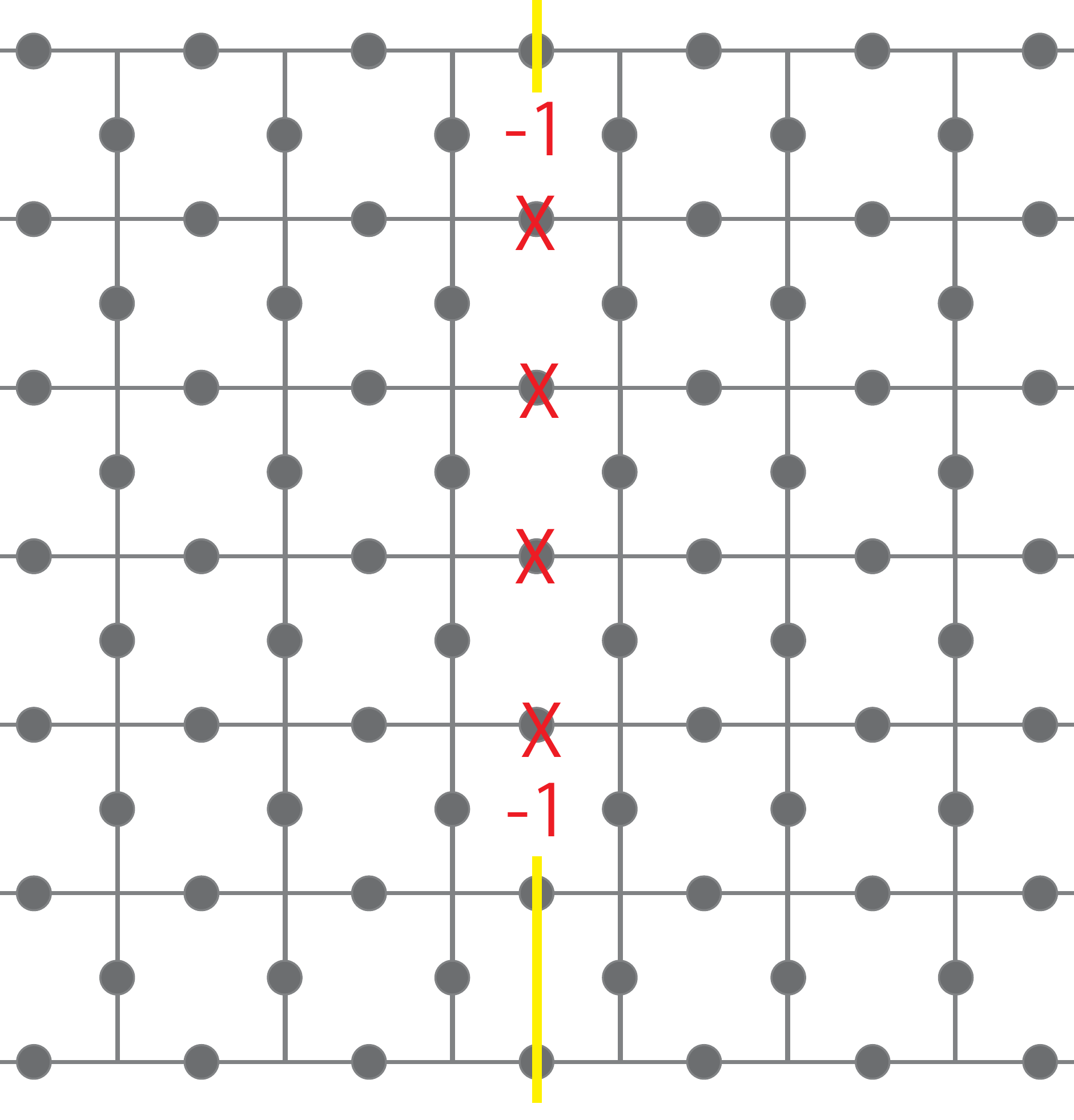}
  \caption[Correctable and uncorrectable errors of the surface code.]
  {Data qubits are described with dots and the lines indicate plaquettes.
  Syndrome qubits are omitted for visibility.
  (a) A correctable single error.
  The red 'X' indicates that an X error occurs on the underlying data qubit.
  The corresponding two Z stabilizers which share the data qubit return -1 eigenvalues.
  It is easy to interpret the error chain from the eigenvalues and the yellow line is the error chain.
  The errored data qubit is under the expected error chain.
  This is correctable.
  (b) A correctable case with two errors.
  We can consider two error chains from the eigenvalues of stabilizer measurement, the yellow one and the green one.
  Either of them is valid.
  Obviously the yellow one is valid, and operating X gates on qubits underlying the green one generates the trivial error cycle described in (c).
  (c) Topologically trivial error cycle.
  This does not affect the logical state of the surface code; this does not affect the states of data qubits on boundaries.
  (d) Example of a mis-correcting error chain.
  Four X errors occur in the center of the lattice.
  The matching algorithm can pair the two -1 plaquettes, for a distance of 4, or pair each -1 with the neighboring boundary of the X operator, for a total distance of 3.
  Because three errors are more probable than four errors, the matching expects that three errors occur and the yellow error chains are expected.
  After applying X gates on the data qubits under the error chains, a logical X operator is generated connecting the two boundaries.
  This is a logical error.
}
  \label{fig:sc:errors}
 \end{center}
\end{figure}

A change in the error syndrome of a stabilizer indicates 
that the stabilizer is the termination of an error chain.
In the normal case, an isolated X (or Z) error, two neighboring
stabilizers will both show -1 eigenstates, and
the error is easily isolated as shown in Figure \ref{fig:sc:errors} (a).
Because two errors on any plaquette cancel and leave the plaquette in the
+1 eigenstate, a series of errors in a neighborhood likely results
in two -1 plaquettes separated by some distance, surrounded by +1 plaquettes.
If an error chain is connected to the boundary
of the lattice, the termination will be hidden.
So, an error chain running between the two boundaries
will be a logical error.

Applying the same flip operation as the original error
it the first method of the error correction, because it 
fixes the states of each stabilizer (Figure \ref{fig:sc:errors} (a)).
We have to identify pairs of error terminations by decoding the detected error information.
This problem can be mapped to the graph theory problem known as ``minimum weight perfect matching'',
a common solution for which is the Blossom V algorithm~\cite{citeulike:4384001}.

Many different possible chains can connect two units with -1 eigenvalues, as depicted in Figure \ref{fig:sc:errors} (b).
Any chain of the two units works as the second method of the error correction and the logical state is corrected.
The exact error chain may not be chosen and then a cycle of errors appears.
Such a trivial error cycle does not affect logical states (Figure \ref{fig:sc:errors} (c)).
Thus, choosing a chain between -1 units is not a problem.
The important problem in the error correction is to pair up the most probable sets of units.
Longer chains of errors occur with lower probability, and the matching algorithm weights such possibilities accordingly.

The distance between the two boundaries for a operator is the code distance of surface code,
shown in Figure \ref{fig:sc:errors} (d).
The larger the code distance, the higher the tolerance against errors.
In the figure, four errors between the two boundaries for the X operator are fatal,
because the matching algorithm fails to pair them properly.
If the two boundaries were farther apart,
more errors would be required to cause the error correction to fail.

\begin{figure}[t]
 \begin{center}
  \includegraphics[width=15cm]{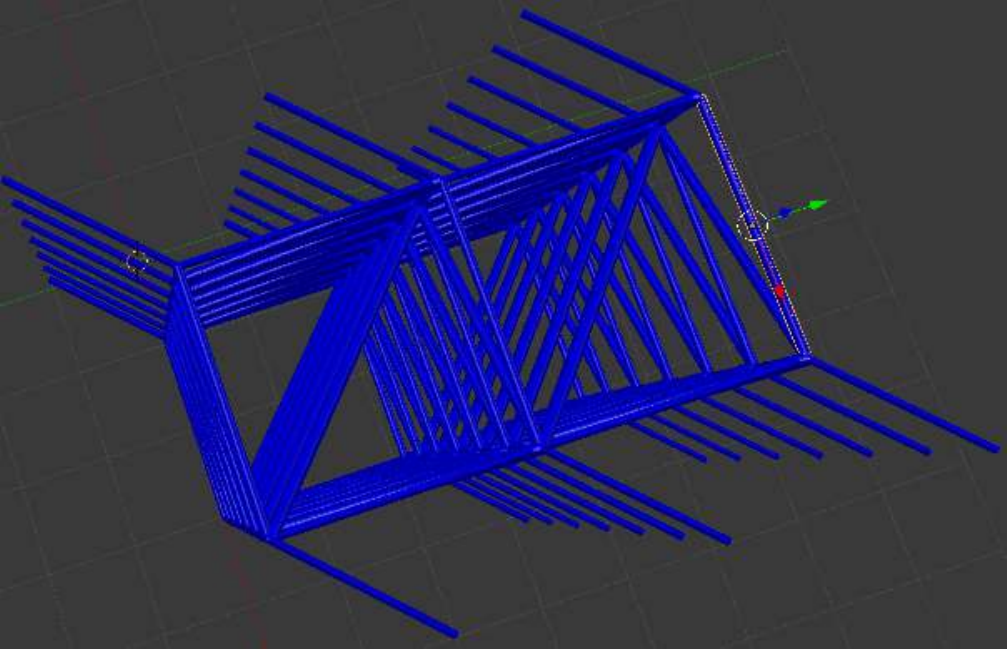}
  \caption[The matching nest for the distance 3 surface code.]
  {A stabilizer measurement corresponds to a vertex and a qubit error corresponds to a edge.
  The ends of the nest correspond to the boundaries of the lattice, hence they do not have measurement values.
  If an error occurs on a data qubit, the stabilizers the data qubit is stabilized
  by will get a different measurement result than the prior round.
  Then the corresponding vertices create and hold nodes for the minimum weight perfect matching.
  Lines for the minimum weight perfect matching are created by Dijkstra's algorithm
  on this nest, searching from a node for other nodes~\cite{Dijkstra1959}.
  The weight of a line is defined by the weights of edges that the search goes through to create the line,
  which corresponds to the possibility of the errors which result in the line.
  }
  \label{fig:sc:normal_nest}
 \end{center}
\end{figure}

A \textit{nest} is used to prepare a network for minimum weight perfect matching.
Figure \ref{fig:sc:normal_nest} depicts a nest of a perfect lattice
of the surface code, output by the Autotune Software created by Fowler et al.~\cite{Fowler:2012autotune}.
Each vertex of the nest corresponds to a stabilizer value
and each edge corresponds to a possible error.
Edges which do not have two vertices are at the boundaries of the lattice.
As time advances, the nest expands along the Z axis, creating new vertices and edges when measuring stabilizers.
A stabilizer which measures a different eigenvalue from the last stabilizer measurement creates and holds a node on the corresponding vertex.
Because an ancilla error (or measurement error) will happen only once, three cycles with an error on the middle cycle would produce the eigenvalue sequence +1, -1, +1. The two transitions will be recorded in the nest as two nodes. A data qubit error results in errors on two neighboring stabilizers, so that an error after the first measurement would give the sequence +1, -1, -1 in two separate places (or only one if the qubit is on a boundary). In this case, the two transitions result in the creation of two horizontal neighbor nodes in the nest. The matching algorithm will match the two vertical nodes of a stabilizer error or the two horizontal nodes of a qubit error. 
Lines for the matching between nodes are created with Dijkstra's algorithm on the nest~\cite{Dijkstra1959}.
The weight of a line is given by the sum of weights of the edges which compose the line.
Minimum weight perfect matching based on those weights selects the most probable physical errors,
therefore it works as error correction.

In deleting the oldest round of error syndromes, there can be an error syndrome which is temporally matched to a syndrome which is not to be deleted. If this error syndrome is deleted, the left pair will be matched to another syndrome, leading to unintended behaviors. To avoid this behavior, Autotune employs a means that the syndrome to be deleted is retained until its pair is deleted.

\subsubsection{Non-Clifford gates}
The surface code does not support $T$ gate and $S$ gate in transversal and fault-tolerant way.

The realizes fault-tolerant $T$ gate by utilizing an ancilla state,
as shown in subsection \ref{subsec:ancillaonequbit}.
The ancilla state required for such $T$ gate is
$\vert A \rangle = \frac{1}{\sqrt{2}}(\vert 0 \rangle + e^{i\pi/4}\vert 1 \rangle)$ state.
The surface code cannot directly create $\vert A \rangle$,
since $T$ gate itself is required to create $\vert A \rangle$, such as $\vert A \rangle = T \vert +\rangle$.
So the surface code employs state injection and
magic state distillation for $\vert A \rangle$ ~\cite{bravyi2005uqc,PhysRevA.86.052329}.
The state injection encodes a state on a physical qubit into an error correcting code. This process is not fault-tolerant, hence, the magic state distillation ``distills'' the desired state with high fidelity.

Additionally,
as shown in subsection \ref{subsec:ancillaonequbit},
ancilla-supported $T$ gate probabilistically requires $S$ gate.
The surface code does not support $S$ gate in transversal and fault-tolerant way,
therefore ancilla-supported $S$ gate is also required.
The ancilla state for $S$ gate is $\vert Y \rangle$ state,
which is created via state injection and magic state distillation, too.

\subsection{The \textit{defect}-based surface code}
The \textit{defect}-based approach uses the same stabilizer circuits, but treats the surface as a whole differently.
A \textit{defect} is a region of the lattice where the physical qubits are measured and not stabilized.
A pair of defects works as a pair of boundaries and it generates a topologically non-trivial loop
so that it holds two degree of freedom: one degree of freedom expressed in the parity of any path of X operators on physical qubits between the two defects
and one degree of freedom in a loop of Z operators on physical
qubits on the boundary circling one of the defects, or any loop of qubits that encircles one of the defects.
Thus, a pair of defects works as a logical qubit.
Any X operator chain between the two defects results in a logical X operation and any loop of Z operators around either defect results in a logical Z operation.
This may be done intentionally to execute the logical gate, but if a chain of X errors or a loop of a chain of Z errors occurs undetected, a logical error to the state is caused.

\begin{figure}[t]
 \begin{center}
  \includegraphics[width=200pt]{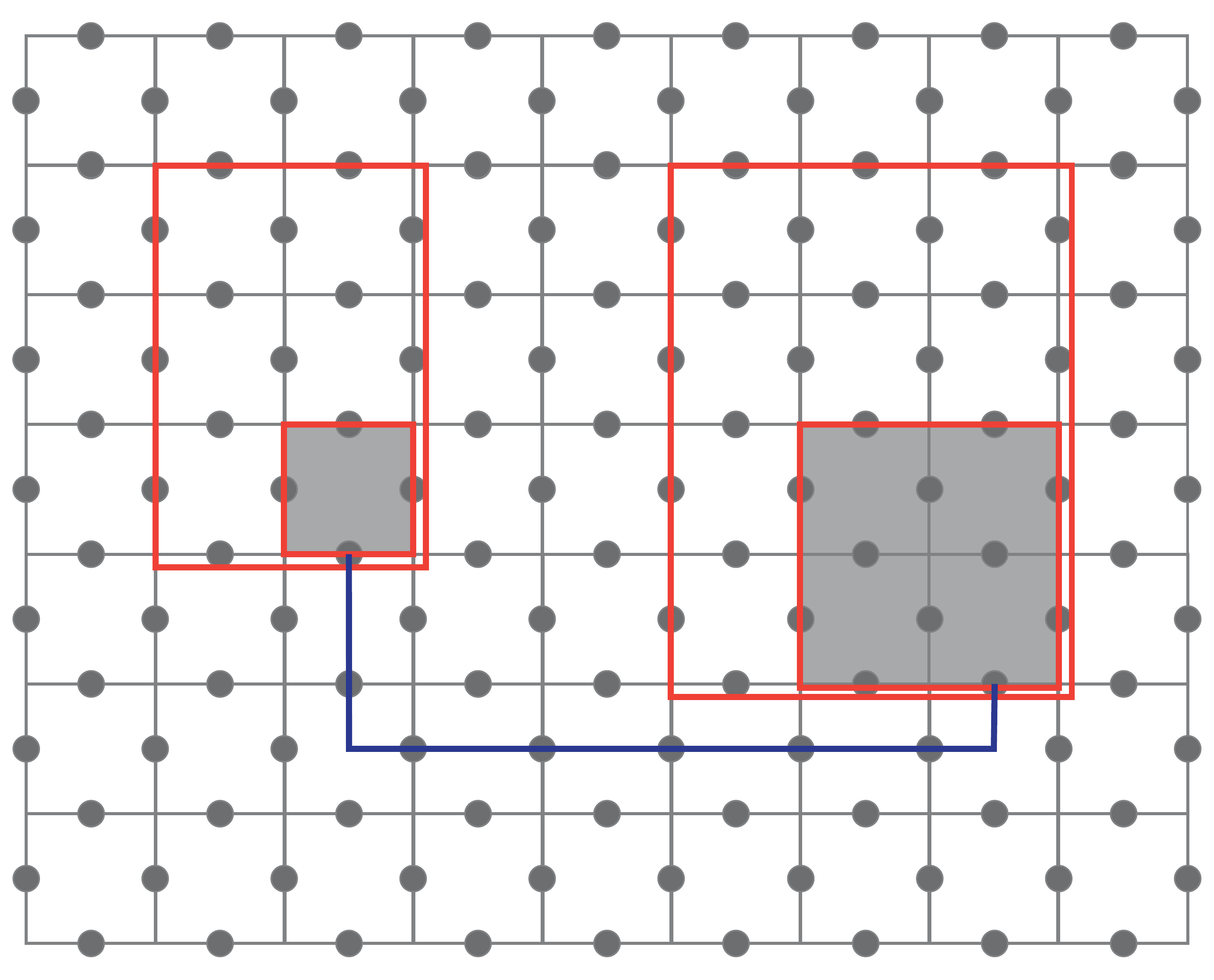}
  \caption[Picture of the defect-based surface code.]
  {A logical qubit is encoded in the lattice:
  there are two defects (gray faces).
  The degrees of freedom on the boundary of the defects
  determine the state of the qubit and 
  the qubits on each four red rectangles has same
  parity depending on the degree of freedom.
  This is the reason why surface code is robust against
  measurement error.
  Physical operation chain on the blue path between
  the two defects results in a logical operation.
  By definition, a logical error is an unintended logical operation.
  Logical errors occur due to a change at each site on a path
  connecting two defects or the boundary of the lattice.
  The concept of ``code distance'' is shown here.
  The longer the distances between the defects and boundary are,
  the more fault-tolerant the computer is.
  This is why surface code is robust against
  memory error and operation error.
  }
  \label{fig:sc:encoding}
 \end{center}
\end{figure}

Parities of measurements on any path or non-trivial loop encircling the defect should have the same value, so that the logical measurement has redundancy against measurement failure.

\begin{figure}[t]
  (a)\includegraphics[width=7cm]{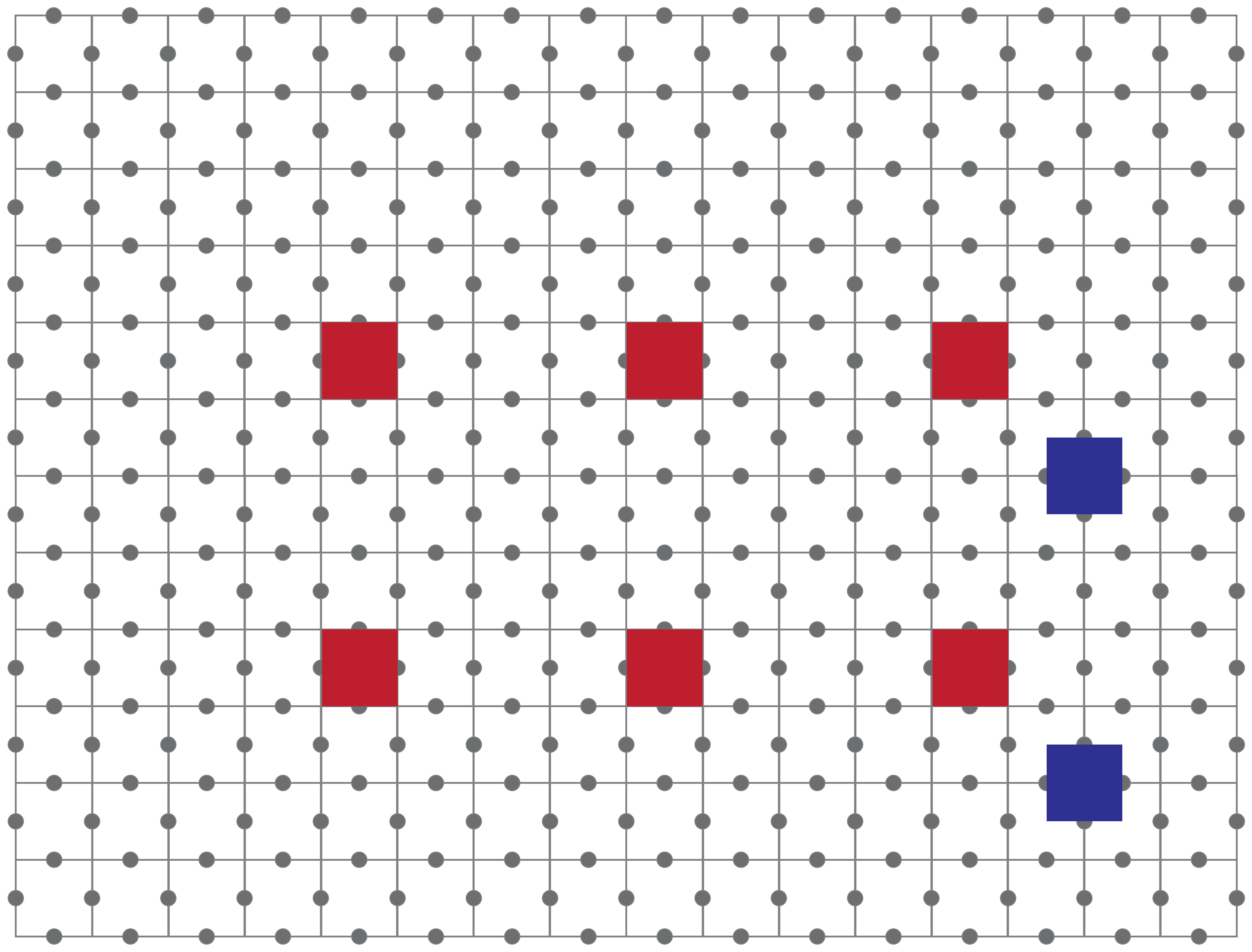}
  (b)\includegraphics[width=7cm]{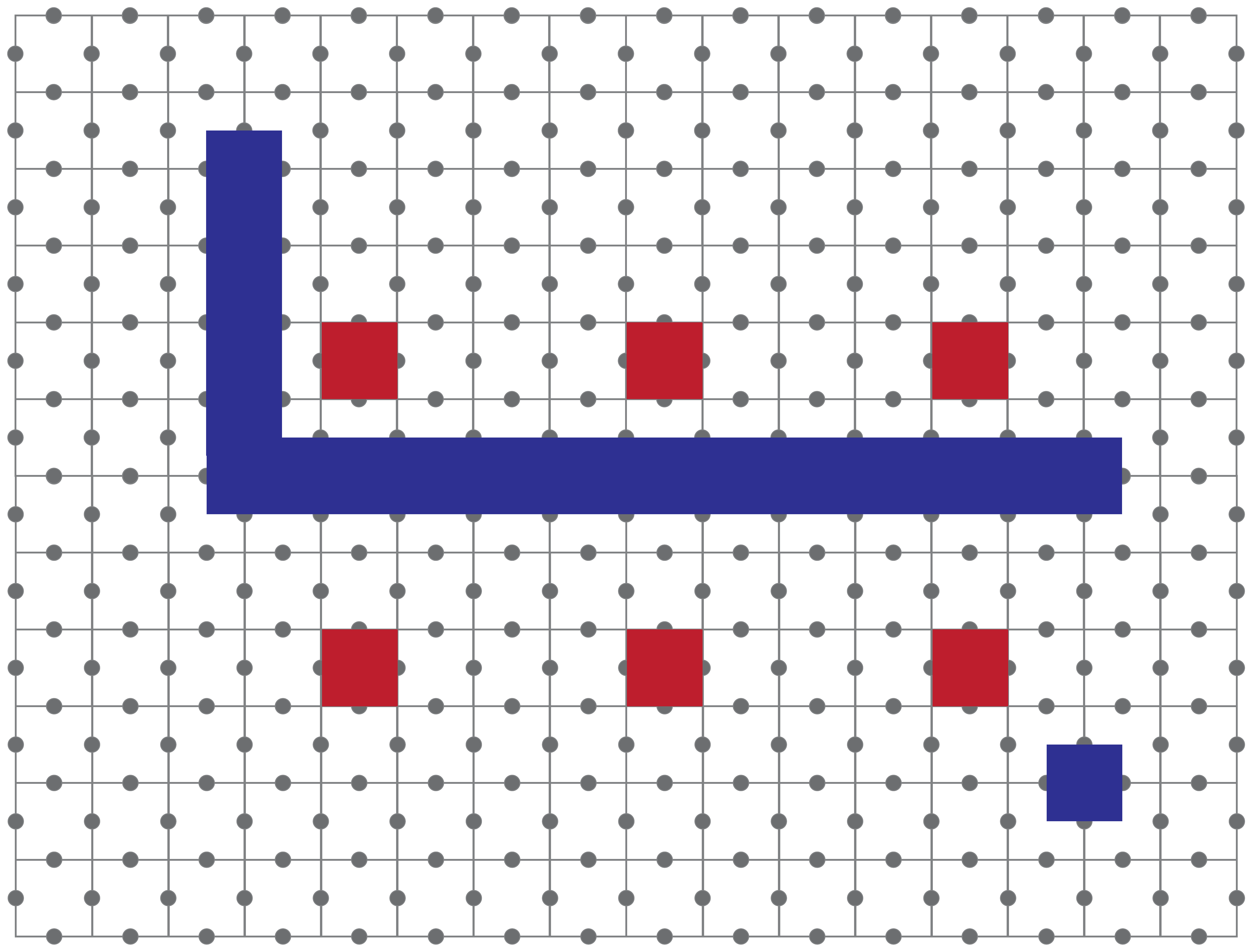}\\
  (c)\includegraphics[width=7cm]{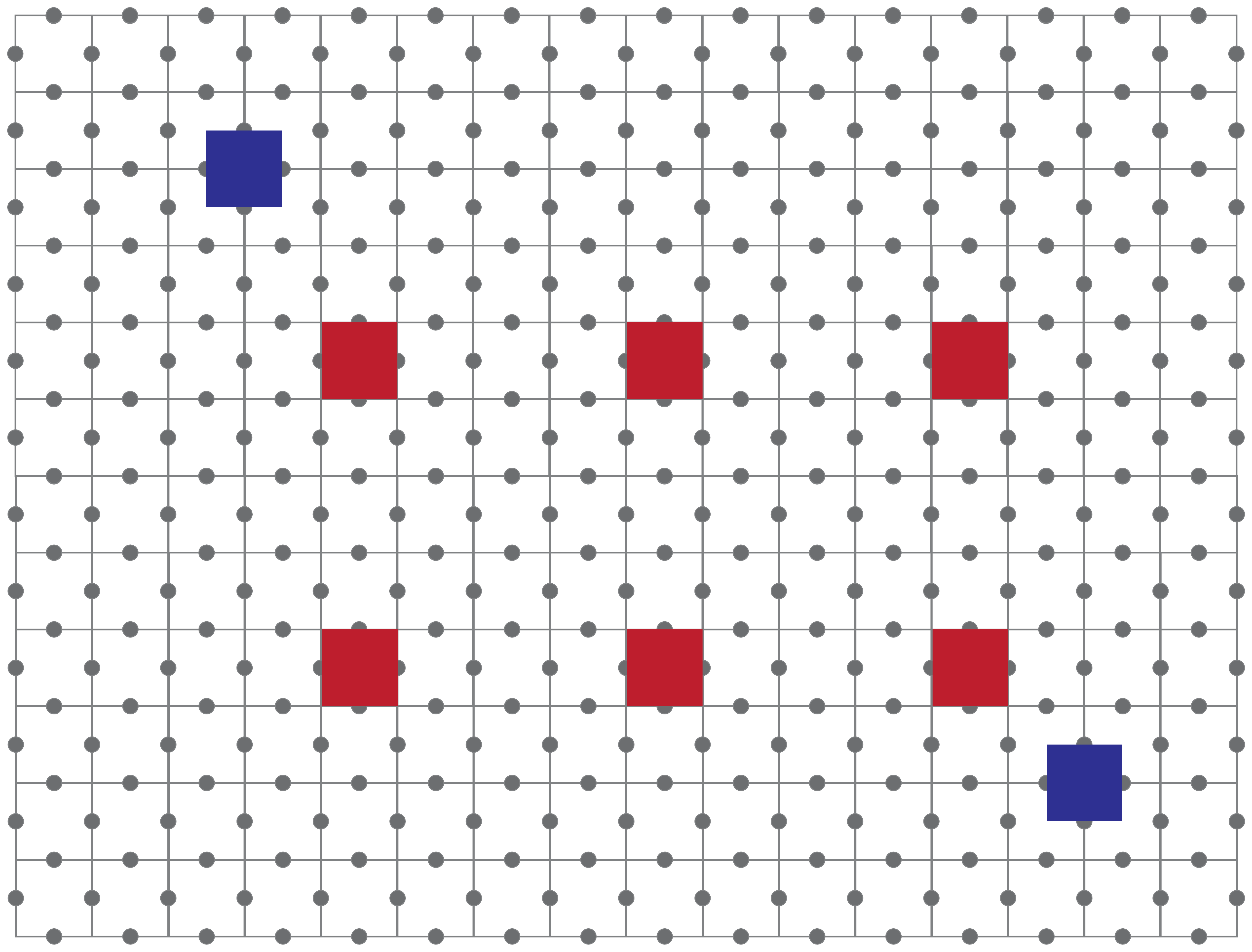}
  (d)\includegraphics[width=7cm]{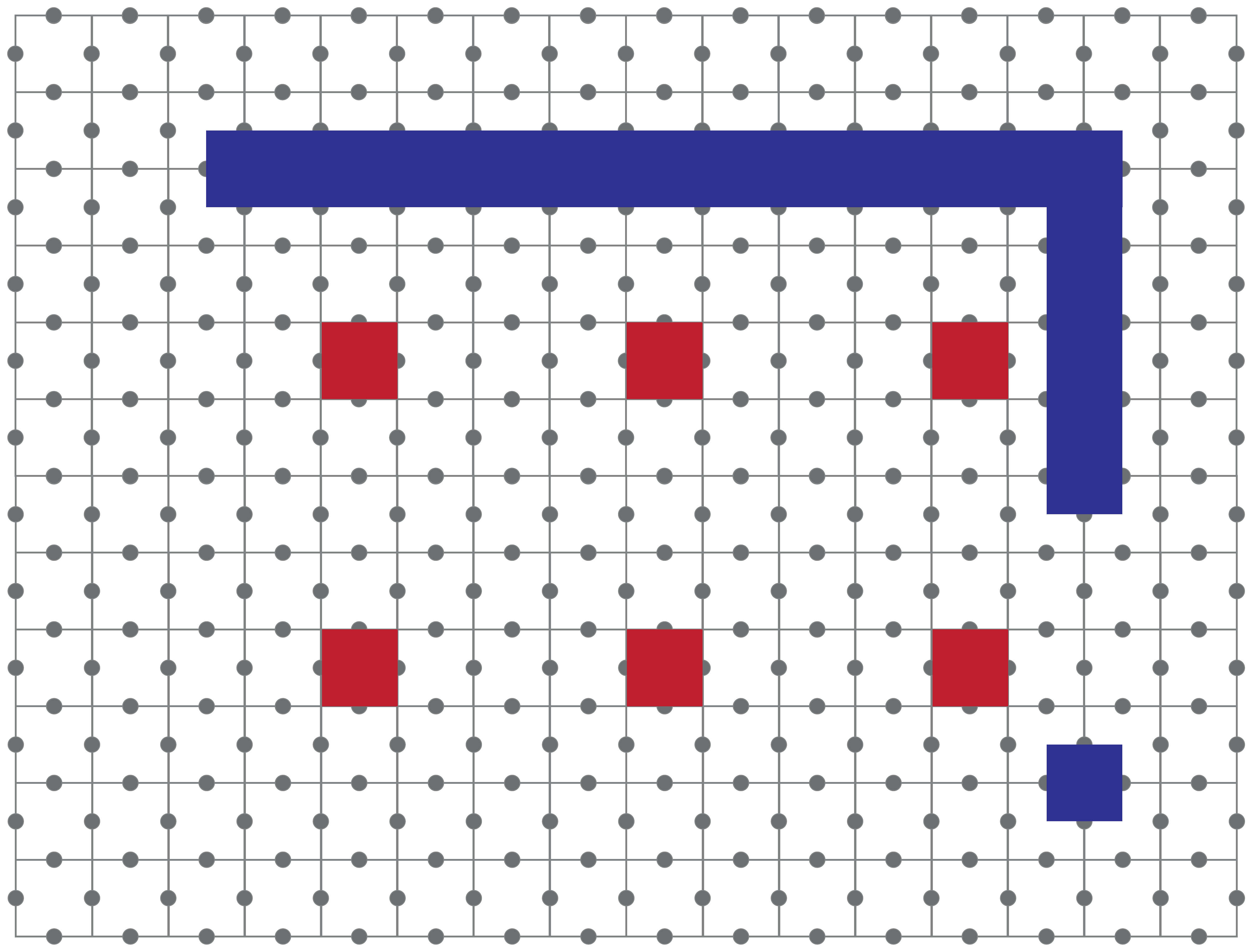}\\
  (e)\includegraphics[width=7cm]{fig/packed_cnot4.pdf}
  \caption[Example of optimazed braiding CNOT gates.]{
  CNOT gates between a rough qubit and three smooth qubits works as a CNOT gate between two smooth qubits.
  (a) The first situation.
  (b) First half expansion of the rough qubit for $8d$ steps.
  (c) First half shrink the rough qubit for $8d$ steps.
  (d) Second half expansion of the rough qubit for $8d$ steps.
  (e) Second half shrink the rough qubit for $8d$ steps. Logical CNOT gates between the rough qubit and all three smooth qubits are completed.
  $32d$ steps in total.
  }
  \label{packed_cnot}
 \end{figure}

Logical CNOT gate for defect-based surface code is achieved by ``braiding''.
A sequence of braiding CNOT gates can be merged and executed faster.
Figure \ref{packed_cnot} shows three rough-smooth CNOT gates
~\cite{1209.0510,srep04657,Polian:2015:DAC:2744769.2747921}.
Hence if a sequence of CNOT gates is a bottoleneck of the computation,
it can be executed faster by gathering logical qubits in defect-based area.

\subsection{Dynamic qubit loss during computation}
A dynamic loss is an escape of a quantum on which a qubit is encoded,
such as a photon loss and electron escape from quantum dot
~\cite{PhysRevA.71.042322,PhysRevLett.97.120501,PhysRevLett.96.020501,PhysRevA.75.010302,RevModPhys.79.135,QIC.8.0399,PhysRevA.82.022323,1409.4880}.
Generally a dynamic loss is treated as a measurement that we do not know the measured value,
therefore the lost qubit is traced out from the original density matrix.
To fix the lost qubit, initialize a new qubit at the lost location and measuring stabilizers involving the restored qubit,
then the code gets corrected.

\subsection{Static qubit loss due to fabrication difficulty}
There is a possibility that we have faulty devices each of which should but cannot hold a quantum for encoding a qubit.
There also be a chance that we may have qubits which have too worse fidelity to use than others.
Such qubits are treated as static losses during quantum computation~\cite{van-meter10:dist_arch_ijqi,1608.05764,1367-2630-19-2-023050}.
They are caused by technological difficulty of fabricating quantum devices.

\subsection{Physical realization of dynamic and static losses}
DiVincenzo offered an architecture of superconducting hardware for the surface code~\cite{1402-4896-2009-T137-014020},
in which a superconducting loop which does not show the appropriate quantum effect will be a static loss.
Jones et al. proposed an architecture for scalable quantum computation
with self-assembled quantum dots used
to trap electrons, which are used as qubits~\cite{Jones:2012Layered_Architecture_for_Quantum_Computing}.
There very likely will be defective quantum dots which cannot trap a single electron,
leaving static loss on the computation chip.

Lindner et al. shown an approach to creating a linear cluster state via a so-called ``photon machine gun''
~\cite{PhysRevLett.103.113602}.
Their idea is to have a quantum dot emit photons continuously followed by
certain operations that, if successful, form a cluster state for one way quantum computation~\cite{PhysRevLett.86.5188,Raussendorf20062242,raussendorf07:_topol_fault_toler_in_clust}.
If a quantum dot cannot emit a photon, there will be a static loss in the cluster state.
Devitt et al. proposed an efficient design for constructing photonic
topological cluster state with photon-photon coupling.
Static losses may occur in the cluster state because of the less-than-perfect probability/fidelity of the coupling~\cite{1367-2630-11-8-083032}.

Donor-based quantum computation chip is achieved by doping atoms such as phosphorus in sillicon~\cite{Hille1500707,PhysRevB.74.045311}.
Locations where doping failed will be static losses.

Static losses can be detected before the computation, by scanning the computation chip.

\subsection{Leakage errors}
The leakage error is that a quantum gets in a state which is not used as the computational basis~\cite{1367-2630-18-4-043021,1510.05653,Hille1500707}.
Methods to tolerate leakage errors are suggested which can work with quantum error correcting codes
~\cite{quant-ph/9705052,Lo-Spiller-Popescu:iqci,Aliferis:2007:FQC:2011706.2011715,PhysRevB.91.085419}.


 \section{Quantum Computer Architecture}
	Several quantum computer architectures have been proposed. Some of them are for a specific physical system and some are more general.
	Some are derived from classical computer architectures.
	
	Oskin et al. proposed a quantum computer architecture in which a quantum ALU and quantum memories are separately implemented~\cite{976922}. The quantum ALU has the ability to operate a universal set of quantum gates hence it operates both of the quantum computation and the quantum error correction. The quantum memories is designed to be dedicated for keeping quantum data, so it operates the quantum error correction more efficiently than the quantum ALU but it does not have the ability for universal quantum gates. Those blocks may employ different quantum error correcting codes. The transfer of quantum data between those blocks are implemented with the code teleportation that each half of a Bell pair is sent to each block and encoded into different error correcting codes then quantum teleportation is operated consuming the heterogeneously encoded Bell pair.
	
	Copsey et al. proposed a quantum computer architecture which has separated processor, cache and memory each of which employs different concatenated quantum error correcting code~\cite{copsey02:_quant-mem-hier}. Generally the concatenated codes have a tradeoff that a code which encodes more logical qubit on a certain number of physical qubits and holds a certain robustness against errors has larger computational overhead. The processor employs a code of less computational overhead and the memory employs a code with more efficiency of the number of physical qubits.
	
	Metodi et al. proposed the Quantum Logic Array (QLA) architecture, a 2-D microarchitecture of tiled logical qubits separated by data paths in which qubits (ions) go through, and Thaker et al. extended the architecture to utilize caches which employ the same error correcting code with the processor to avoid stalls in the processor~\cite{Metodi:2005:QLA:1099547.1100566,Thaker:2006:QMH:1150019.1136518}. Although they target ion trap qubits, the architecture can be partially applied to other physical technologies. They also showed that key quantum algorithms does not have large parallelism, hence such architectures which employ heterogeneous scheme on each component matched to its purpose would be efficient than the homogeneous design of ``sea of computational qubits''.
		
  	Whitney et al. evaluated such quantum processor microarchitectures for Shor's algorithm with the practical feasibility described by area of the computation circuit, duration of the circuit and error probability~\cite{Whitney:2009:FTA:1555815.1555802}. They found that carry-lookahead adders may be better than ripple-carry adders in feasibility. They also showed that a regular allocation of computation area and transfer components, and a regular allocation of computation area, memory area and transfer components are beaten by variable sized allocation of those areas and transfer components dependent on the circuit. This result infers that a dedicated design is superior to a general design for a purpose.
	
	Brun et al. extended the architecture composed from dedicated components; summarized the operations on multi-qubit large block codes for memories, arranged the ancilla resource factory and analyzed errors precisely~\cite{1504.03913}.
	
	Jones et al. arranged layers of a quantum computer architecture; layer 5: application, layer 4: logical, layer 3: quantum error correction, layer 2: virtual and layer 1: physical~\cite{Jones:2012Layered_Architecture_for_Quantum_Computing}.
	Though their analysis targets quantum dots, the physical implementation is concealed by the virtual layer therefore the layered architecture to manage the quantum computation to execute applications can be applied to other physical systems.
	
	Van Meter et al. proposed a quantum computer architecture assembled from semiconductor nanophotonics and quantum dots~\cite{van-meter10:dist_arch_ijqi}.
	Their architecture are designed considering static losses and separated qubits can have interaction via optics
	hence the lattice for surface code can be kept static loss-free regardless of the existence of faulty devices.

	Devitt et al. proposed photonic system for measurement-based quantum computation utilizing chip-based devices as fundamental building blocks~\cite{1367-2630-11-8-083032}.
	
	 Choi analyzed dual-code quantum computation model, which may partially determine the quantum computer architecture~\cite{choi2015}.
	 It is shown that the dual-code model for non-topological quantum correcting codes has advantages when the concatenation level
	 is low and the cost of non-transversal gates are high.
	 This is because the dual-code system gets free from non-transversal gates which require expensive state distillation 
 	 but deeper concatenation in the dual-code model requires more expensive cost for code conversion
	 which is achieved by quantum teleportation between heterogeneously encoded qubits.

  \section{Quantum Network and Distributed architecture}

Quantum network is a computer network for quantum computation
~\cite{kimble08:_quant_internet}.
The quantum repeater is a core infrastructure component of a quantum network,
tasked with constructing distributed quantum states or
relaying quantum information as it routes from the source to the destination
~\cite{briegel98:_quant_repeater,dur:PhysRevA.59.169,6246754,kimble08:_quant_internet}.
The quantum repeater creates new capabilities: end-to-end quantum communication,
avoiding limitations on distance and the requirement for trust in quantum key distribution networks~\cite{doi:10.1117/12.606489,1367-2630-11-7-075001,sasaki:11},
wide-area cryptographic functions~\cite{Ben-Or:2005:FQB:1060590.1060662},
distributed computation~\cite{RevModPhys.82.665,buhrman03:_dist_qc,buhrman1998quantum,PhysRevA.89.022317,Chien:2015:FOU:2810396.2700248,broadbent2010measurement,crepeau:_secur_multi_party_qc}
and possibly use as physical reference frames~\cite{jozsa2000qcs,komar14:_clock_qnet,chuang2000qclk,RevModPhys.79.555}.
To protect quantum data from losses and errors, sending a qubit consists of two steps;
first the quantum network and the end nodes cooperatively create a high fidelity Bell pair between the ends
and secondly the ends execute a quantum teleportation.

Several different classes of quantum repeaters have been proposed~\cite{munro2011designing,vanmeter2014book,takeoka2014fundamental} and these class distinctions often relate to 
how classical information is exchanged when either preparing a connection over multiple repeaters, or sending a piece of quantum information from source to destination.
The first class utilizes purification and swapping of physical Bell pairs~\cite{Nature:10.1038/35106500,PhysRevLett.96.240501,RevModPhys.83.33,Jiang30102007}.
First, neighboring repeaters establish raw (low fidelity) Bell pairs which are recursively used to purify a single pair to a desired fidelity.  
Adjacent stations then use entanglement swapping protocols to double the total range of the entanglement. In purify/swap protocols, classical information is exchanged 
continuously across the entire network path to herald both failures of purification protocols and entanglement swapping.  This exchange of information limits the speed of such a network 
significantly, especially over long distances.
The second class utilizes quantum error correction throughout the
end-to-end
communication~\cite{munro2010quantum,PhysRevA.79.032325,PhysRevLett.104.180503,1367-2630-15-2-023012,knill96:concat-arxiv} and limits the exchange of classical information to either two-way 
communications between adjacent repeaters or to ballistic communication, where the classical information flow is unidirectional from source to receiver.
These approaches depend on either high probability of success for
transmitting photons over a link with high fidelity, or build on top
of heralded creation of nearest neighbor Bell pairs and
purification, if necessary.  
If the probability of successful connection between adjacent repeaters is high enough we can use quantum
error correcting codes and relax constraints on the
technology, especially memory decoherence times and the need for large
numbers of qubits in individual repeaters, by sending logically
encoded states hop by hop in a quasi-asynchronous
fashion~\cite{munro2010quantum,munro2012quantum} or using
speculative or measurement-based
operations~\cite{1367-2630-15-2-023012,PhysRevA.85.062326,munro2012quantum}.

Several physical channel types have been suggested for quantum entanglement distribution
over long distance, notably, optical fiber, free-space,
satellite and sneakernet~\cite{Inagaki:13,Herbst15:free-space143km,PhysRevLett.115.040502,Devitt:arXiv:1410.3224}.
 The distributed quantum computation is the paradigm where many small or medium scale quantum computers are connected and composes a large cluster which can run large scale quantum operation~\cite{crepeau:_secur_multi_party_qc,buhrman03:_dist_qc,RevModPhys.82.665}.
 The distributed quantum computation can be a good solution for the difficulties of integrating quantum devices and for the large resource requirement for fault tolerant quantum computation. 
	Jiang et al. analyzed an extreme situation of the distributed quantum computation
	in which each node holds only a few ($\le 5$) qubits~\cite{jiang07:PhysRevA.76.062323}.
	Van Meter et al. analyzed distributed quantum computation by evaluating a distributed simple circuit~\cite{VanMeter:2006:DAQ:1150019.1136517,Meter:2008:ADQ:1324177.1324179}. They revealed that moving data by quantum teleportation is superior to teleportation-based gates, decomposing a quantum teleportation into several steps and managing the steps separately improves the system performance, a linear topology of computation nodes is efficient enough and I/O bandwidth dominates the performance rather than that a node holds only a few logical qubits.

	Monroe et al. proposed a quantum computer architecture in which there are many units of logical qubits that are connected to a optical crossbar hence the units can share a Bell pair via the optical crossbar~\cite{PhysRevA.89.022317}. They target atomic qubit, however, their computer architecture can be partially applied to other technologies. This architecture allows interactions among logical qubits separately memorized but actually requires longer execution time than the QLA architecture because of the probabilistic nature of the photonic network.
	
	Ahsan et al. proposed a quantum computer architecture in which there are many units each of which has components specialized for either memory, ancilla generation for non-Clifford gates, error correction or communication and interconnections are achieved by optical crossbars of two layers~\cite{ahsan2015designing}.
	
Nickerson et al. proposed a quantum computer architecture 
for the surface code in which 
several ancilla qubits are attached to each data qubit
and the ancilla qubits are networked~\cite{Nickerson2013apr}.
They first create GHZ state of high fidelity among ancilla qubits equipped on data qubits of a stabilizer utilizing purification. Next hey use the GHZ state for the stabilizer. They claim that the fault-tolerant threshold is $0.825\%$.


\clearpage
\chapter{Surface code on defective lattice overview}
\label{chap:defective}
\label{sec:solution}
This Chapter describes 
the adaptation of
the surface code 
to run
on defective lattices,
making the surface code capable of tolerating static losses caused by faulty devices.
In the proposed quantum computer architecture,
the surface code is employed in quantum CPUs, memories and would be used in the magic state generation area.
By this adaption, tolerances against quantum imperfections, state errors, dynamic losses and static losses, are completed.

The difficulty in working around faulty devices arises from the nearest neighbor architecture and the two separate roles for qubits.
Distant qubits have to interact around faulty devices but the nearest neighbor architecture does not provide the capability for such qubits to interact directly.
SWAP gates are brought in to solve this problem.
The solutions for faulty syndrome qubits and for faulty data qubits differ.
To tolerate faulty data qubits, we introduce the ``superunit'' that Stace, Barrett and Doherty called the ``superplaquette'' \cite{PhysRevLett.102.200501}.
The idea is to maintain error correction by modifying the shape of stabilizers around lost data qubits (faulty devices).
On the other hand, we do not have to modify the unit of stabilizers when syndrome qubits are faulty.
We can gather error syndromes onto another syndrome qubit instead of the faulty syndrome qubit, by using SWAP gates.
Stabilizers which do not involve faulty devices remain as normal stabilizers.

\section{Stabilizer reconfiguration}
\begin{figure}[t]
 \begin{center}
(a)
  \includegraphics[width=180pt]{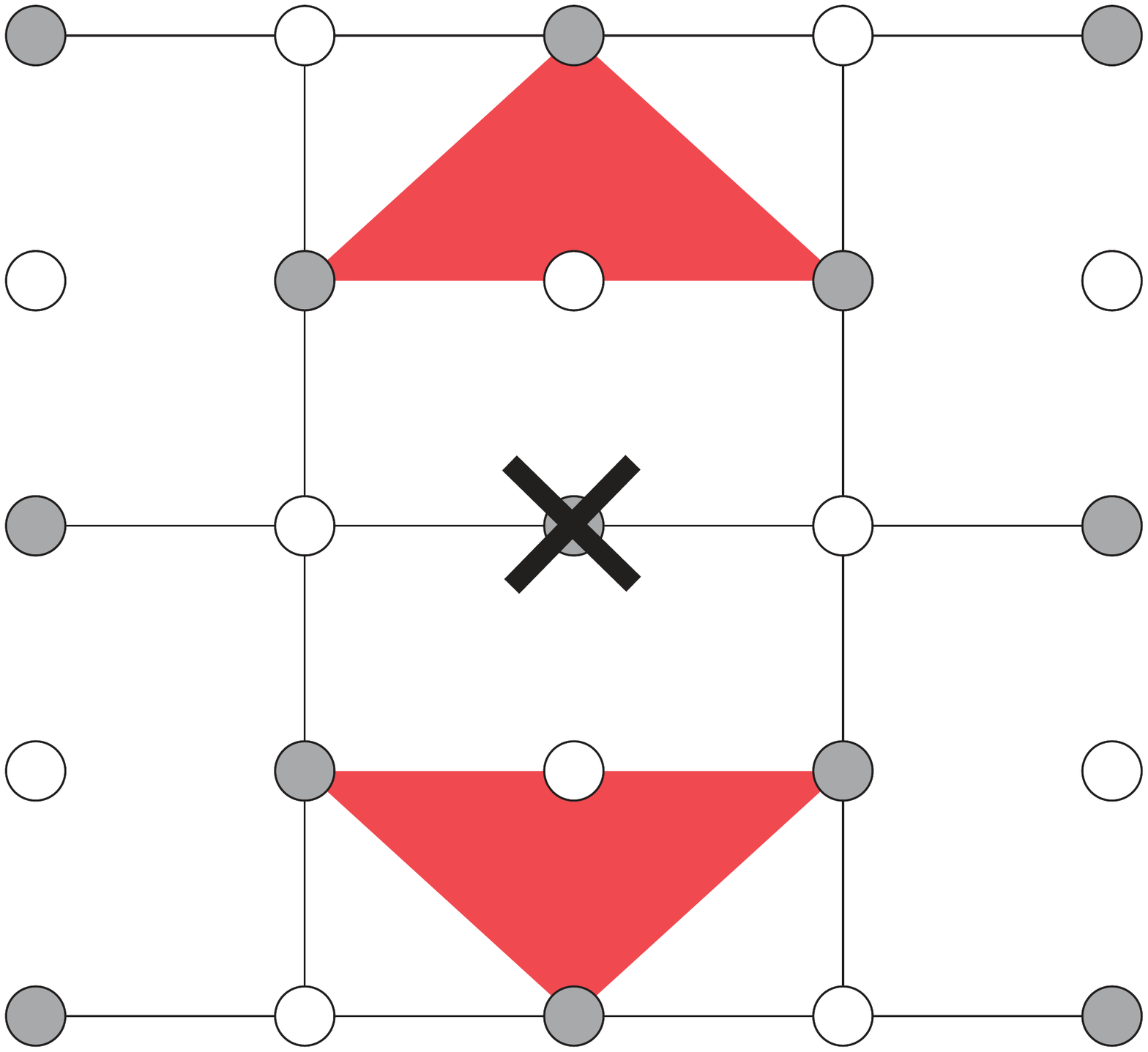}
(b)
  \includegraphics[width=180pt]{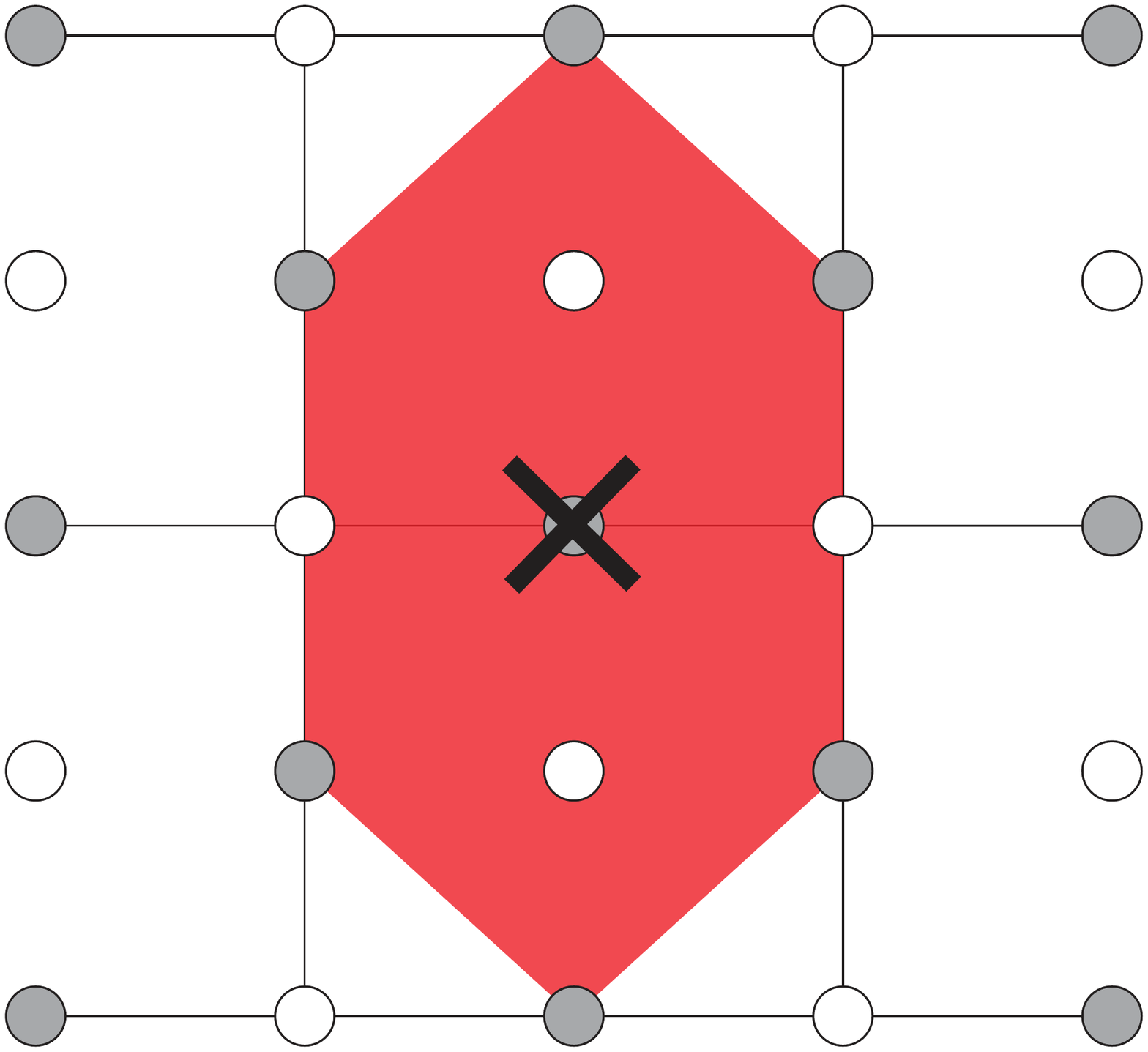}
  \caption[Modified stabilizers around a faulty device marked with the black cross.]
  {(a) A pair of Z triangular stabilizers. (b) A superunit Z stabilizer.}
  \label{fig:sol:modified_unit}
 \end{center}
\end{figure}

There are two ways to reconfigure around a faulty device.
The first is to introduce two triangular stabilizers by purging the broken qubit from stabilizers which involve the broken qubit, leaving two stabilizers composed of three data qubits and one syndrome qubit, as depicted in Figure \ref{fig:sol:modified_unit}(a). 
\begin{figure}[t]
 \begin{center}
  (a)
  \includegraphics[width=180pt]{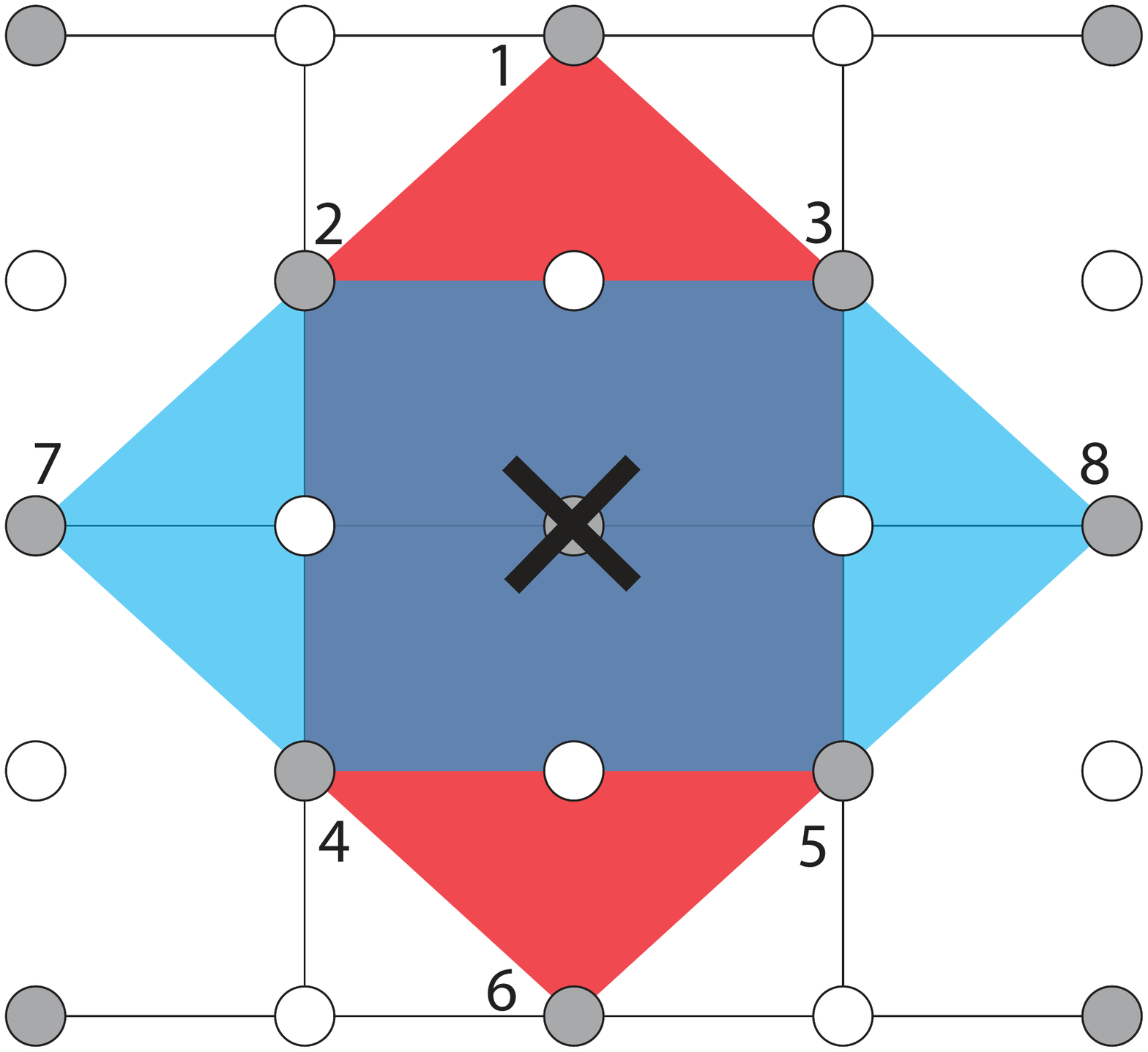}
  (b)
  \includegraphics[width=180pt]{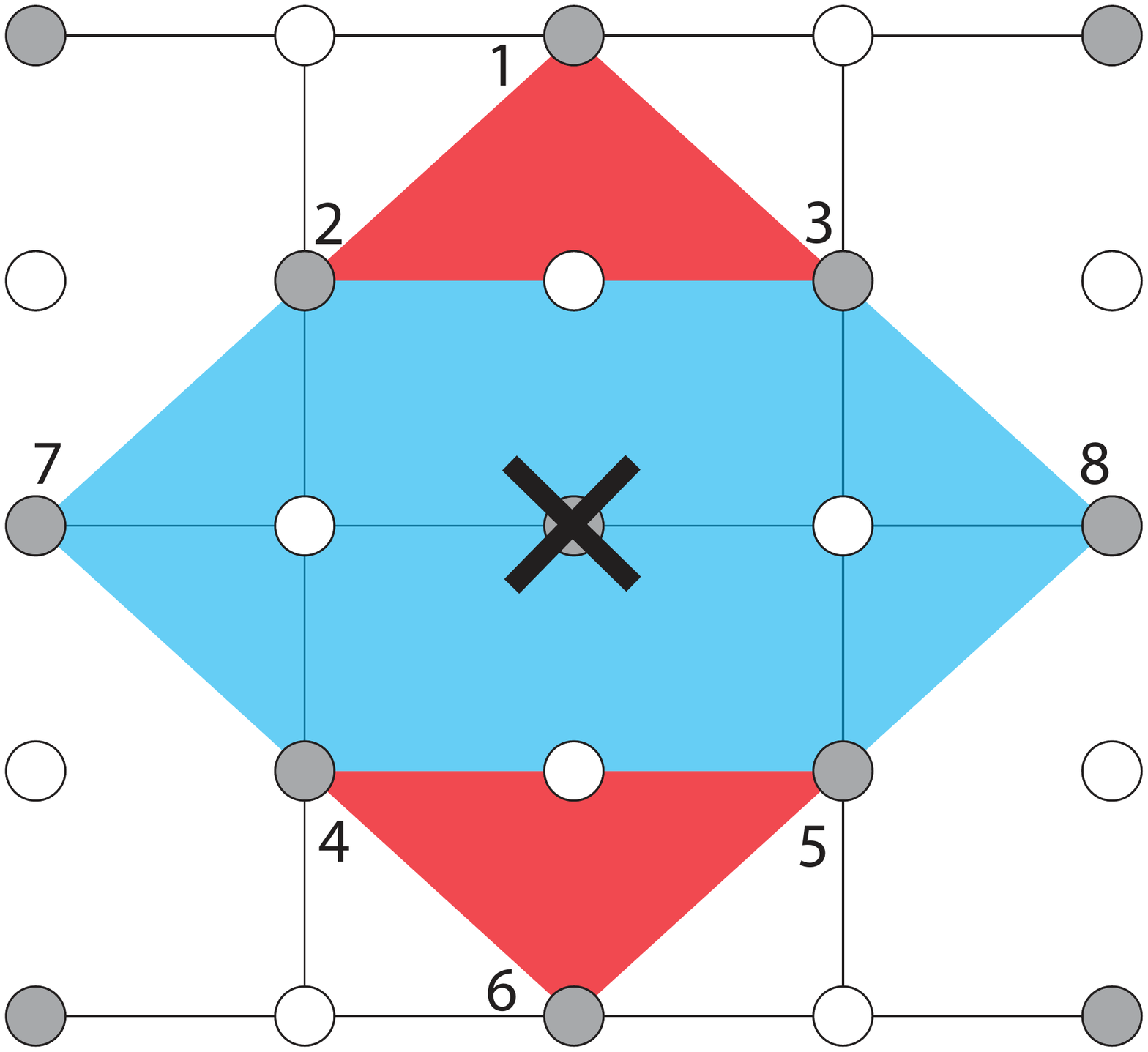}
  \caption[Two sets of modified stabilizers which commute.]
  {The corresponding stabilizers are shown in Table \ref{tab:sol:commute-stabilizers}. (a) Superunit stabilizer is adopted both to the Z stabilizer and to the X stabilizer. (b) Triangular stabilizer is adopted in Z stabilizer and superunit stabilizer is adopted to the X stabilizer. }
  \label{fig:sol:commute}
 \end{center}
\end{figure}
\begin{figure}[t]
 \begin{center}
  \includegraphics[width=180pt]{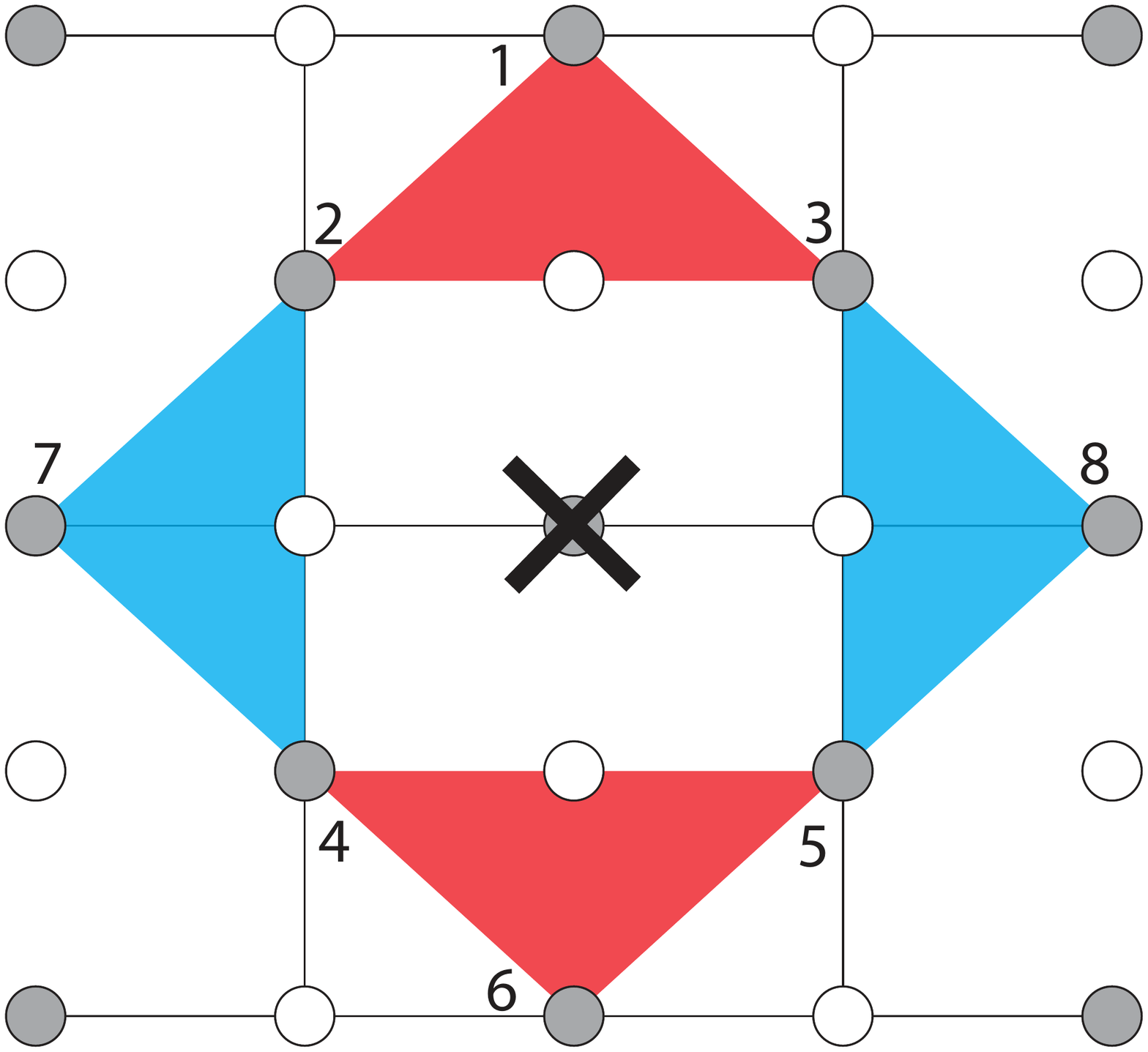}
  \caption[A set of modified stabilizers which anti-commute.]
  {The corresponding stabilizers are shown in Table \ref{tab:sol:anticommute-stabilizers}.}
  \label{fig:sol:anticommute}
 \end{center}
\end{figure}
\begin{table}[b]
  \begin{center}
   \caption[Stabilizers of two sets of modified unit stabilizers.]
   {(a) Superunit stabilizers are adopted both to the Z stabilizer and to the X stabilizer (Figure~\ref{fig:sol:commute}(a)).
   (b) A triangular stabilizers are adopted to the Z stabilizer and a superunit stabilizer is adopted to the X stabilizer (Figure~\ref{fig:sol:commute}(b)). }
   \label{tab:sol:commute-stabilizers}
   (a)
   \begin{tabular}[t]{cccccccc}
    1 &2 &3 &4 &5 &6 &7 &8 \\
    \hline
    Z &Z &Z &Z &Z &Z & & \\
    &X &X &X &X & &X &X \\
   \end{tabular}
   (b)
   \begin{tabular}[t]{cccccccc}
    1 &2 &3 &4 &5 &6 &7 &8 \\
    \hline
    Z &Z &Z & & & & & \\
    & & &Z &Z &Z & & \\
    &X &X &X &X & &X &X \\
   \end{tabular}
  \end{center}
\end{table}
\begin{table}[b]
  \begin{center}
   \caption{Stabilizers of four triangular unit stabilizers as in fig \ref{fig:sol:anticommute}. Some pairs anti-commute.}
   \label{tab:sol:anticommute-stabilizers}
   \begin{tabular}[t]{cccccccc}
    1 &2 &3 &4 &5 &6 &7 &8 \\
    \hline
    Z &Z &Z & & & & & \\
    & & &Z &Z &Z & & \\
    &X & &X & & &X & \\
    & &X & &X & & &X \\
   \end{tabular}
  \end{center}
\end{table}
It is impossible to adopt triangular stabilizers for both stabilizers around a faulty device since neighboring Z triangular stabilizers and X triangular stabilizers do not commute when they have only one qubit in common,
as shown in Figure~\ref{fig:sol:anticommute}.
Note that those four triangles cannot be stabilizers but can be gauge operators
for the subsystem code~\cite{bacon05:_operator-self-qec,PhysRevLett.95.230504,Bravyi:2013:SSC:2535639.2535643}.
We leave this solution for future work.

The second approach is to generate a superunit stabilizer by merging the two broken unit stabilizers, depicted in Figure \ref{fig:sol:modified_unit}(b).
At least one lattice unit must adopt a superunit stabilizer.
In this paper, we form superunit stabilizers for both stabilizers after Stace et al. and
Barrett et al.~\cite{PhysRevLett.102.200501,PhysRevLett.105.200502,PhysRevA.81.022317}.
To form superunit stabilizers for both stabilizer around a faulty device produces a degree of freedom which results in a logical qubit by code deformation~\cite{1751-8121-42-9-095302}, which is also called a ``junk qubit''~\cite{PhysRevLett.102.200501}.
However, operations specifically targeted at this qubit are required to execute two qubit gates
between this new logical qubit and the logical qubits of the planar code and defect-based code,
so that the effect of its presence in this dissertation is just reducing the minimum distance between the boundaries.

\section{Stabilizer circuits around faulty devices}
\begin{figure}[t]
 \begin{center}
  (a)
  \includegraphics[width=6cm]{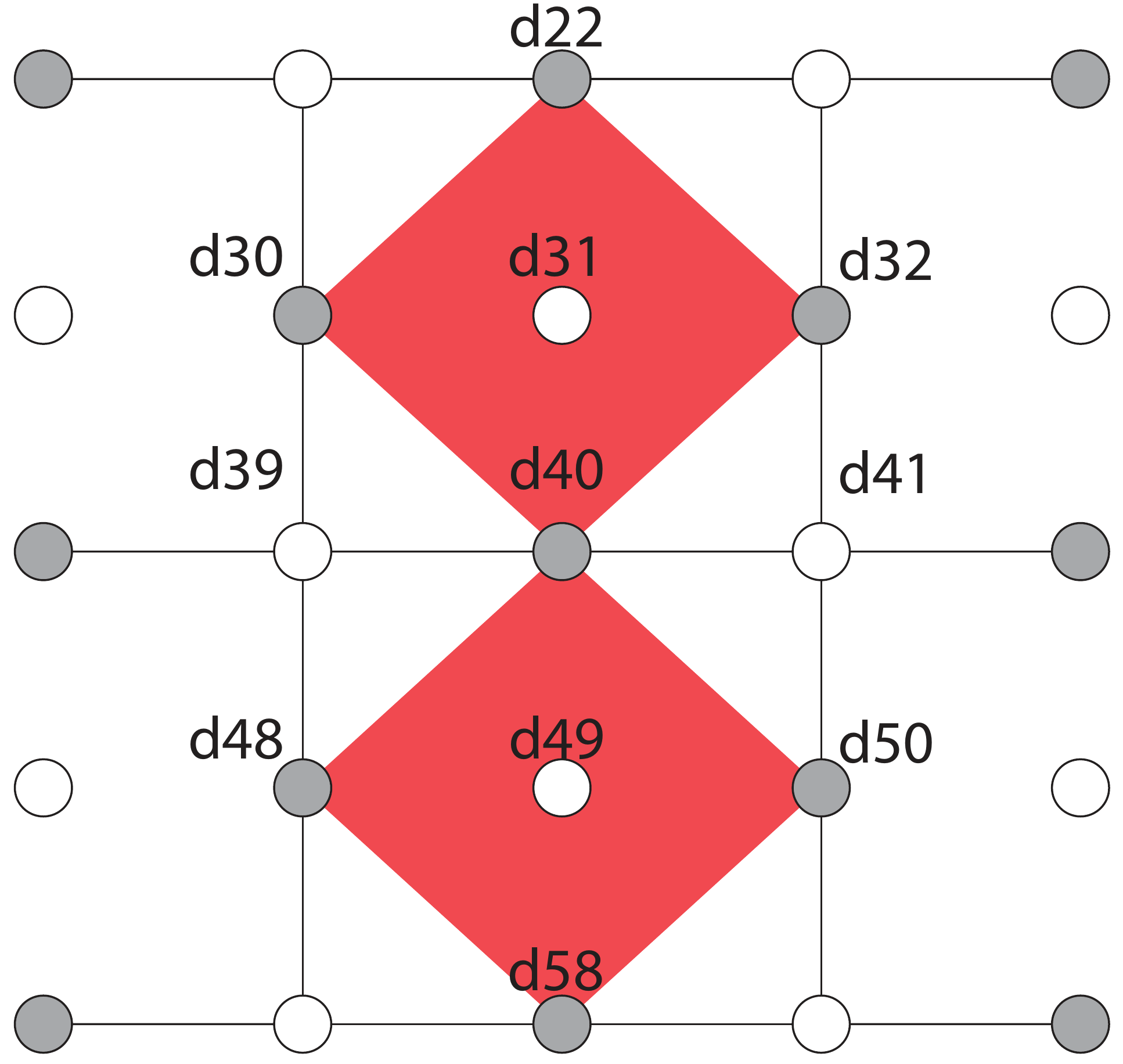}
  \includegraphics[height=5cm]{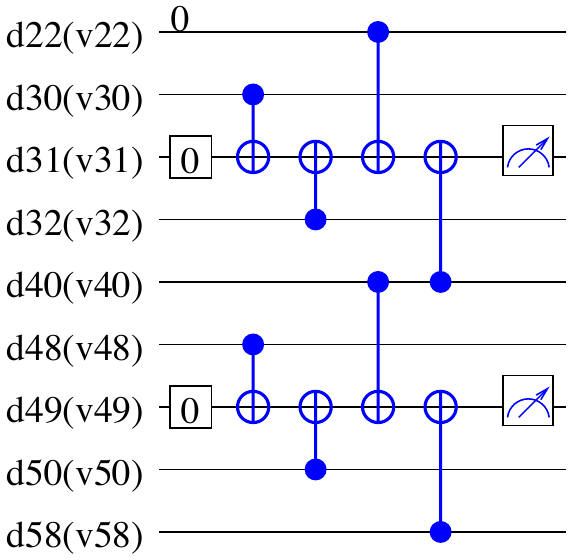}\\
  (b)
  \includegraphics[width=6cm]{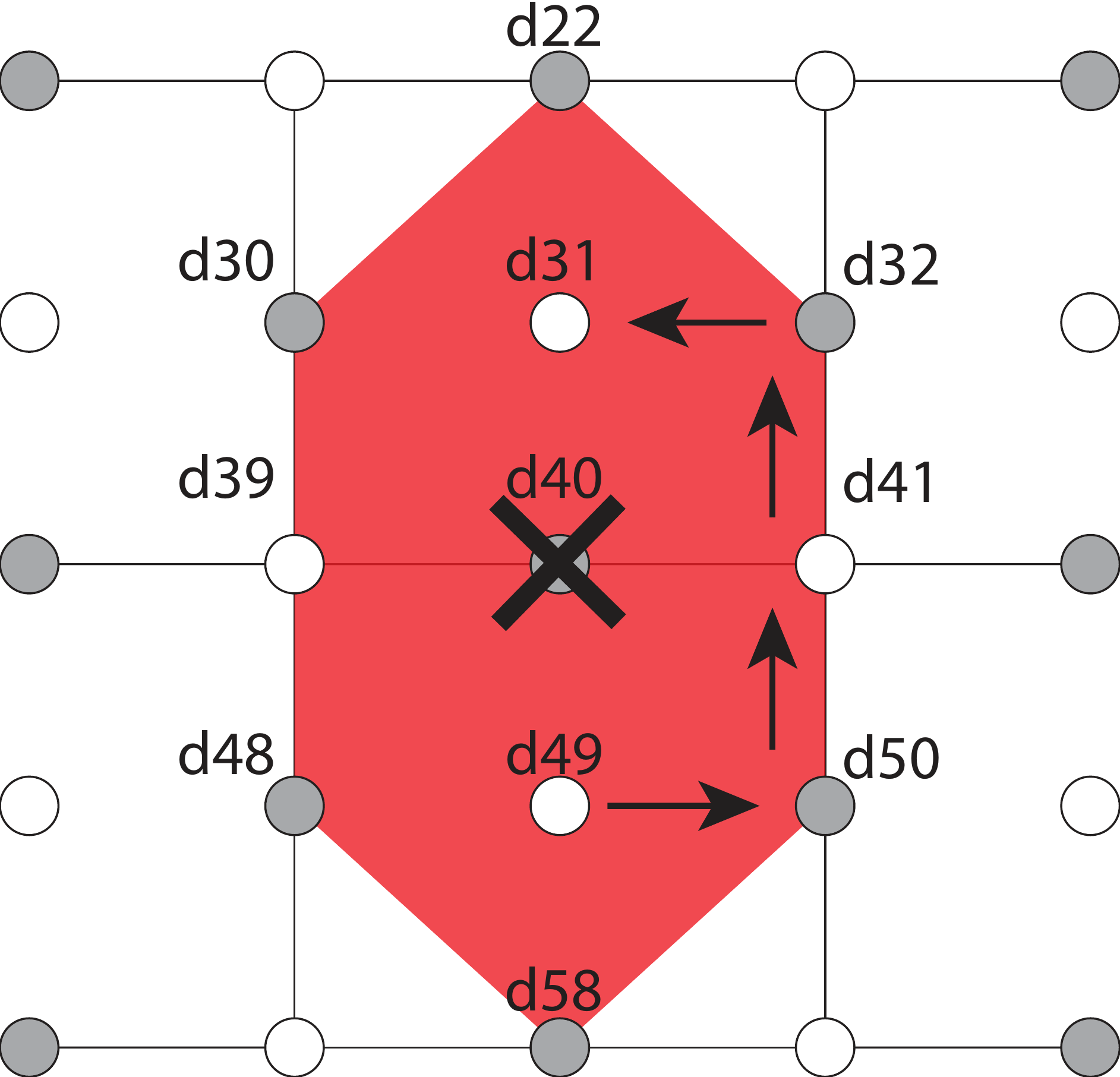}
  \includegraphics[height=5cm]{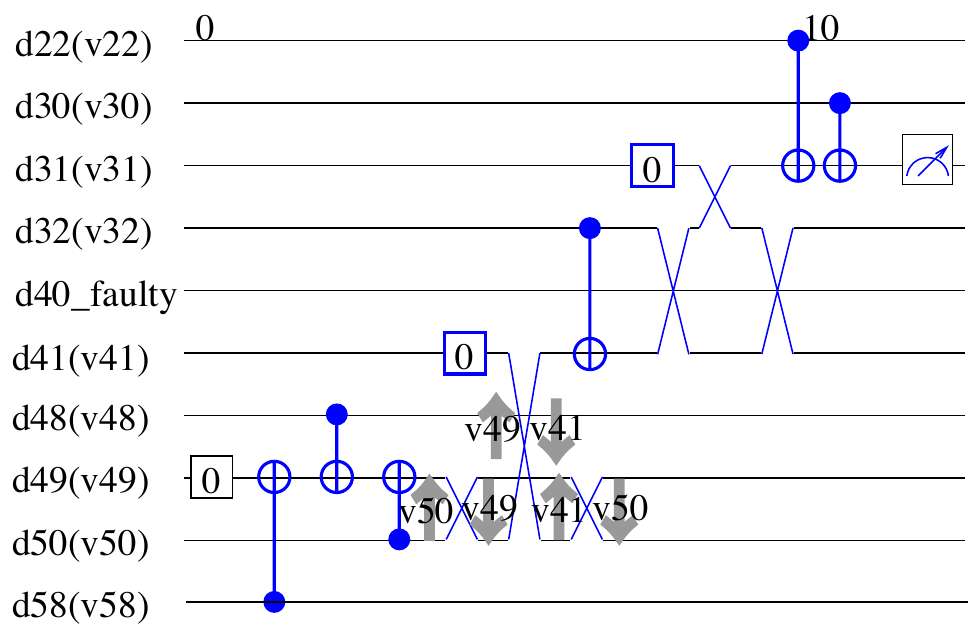}
  \caption[Stabilizers and their circuits.]
  {(a) A set of normal $Z$ stabilizer circuits around d40 for the case where d40 is properly functional. 
  The stabilizers are $Z_{v58}Z_{v48}Z_{v50}Z_{v40}$ and $Z_{v40}Z_{v32}Z_{v22}Z_{v30}$.
  (b) An example of a superunit stabilizer circuit.
  The qubit device d40 is faulty and two units are connected.
  The new stabilizer is $Z_{v58}Z_{v48}Z_{v50}Z_{v32}Z_{v22}Z_{v30}$.
  The circuit measuring only this stabilizer is shown on the right.
  This stabilizer circuit is isolated from the whole circuit shown in Figure \ref{fig:sol:whole_circuit_single_faulty_center}.
  }
  \label{fig:sol:a_2-units_superunit}
 \end{center}
\end{figure}

Stabilizer-measurement circuits working around faulty devices have different shape and depth from the circuits of normal stabilizers.
Figure \ref{fig:sol:a_2-units_superunit}(b) shows the shape of a superunit in which two units are connected by a faulty device and its circuit.
We call a circuit for an individual stabilizer a ``stabilizer circuit'' and the circuit for a complete lattice the ``whole circuit''.
We define two terms, ``qubit device'' and ``qubit variable''. A qubit device is the physical structure that holds the qubit variable, such as the semiconductor quantum dot or loop of superconducting wire. A qubit variable is the information encoded on a qubit device.
This distinction corresponds to the difference between a register or memory location in a classical computer, and the program variable held in that location.
In Figure \ref{fig:sol:a_2-units_superunit}(a), the horizontal lines correspond to qubit devices and qubit devices are distinguished with the labels (numbers).
The labels of qubit variables are tagged to the label of the qubit device, that is, we distinguish qubit variables with the label of the qubit device of the qubit variable's original position.

In Figure~\ref{fig:sol:a_2-units_superunit}(b), the qubit device labeled d40 is faulty,
hence the variable v40 does not exist.
The data qubit variables v58, v48, v50, v32, v22 and v30, initially held respectively in
the qubit devices d58, d48, d50, d32, d22 and d30, are stabilized by the red stabilizer.
The syndrome qubit variable v49 is initialized while residing in d49,
then moves around using SWAP gates to gather error syndromes of those data qubits.
After gathering three error syndromes from v58, v48 and v50, v49 moves into d41 via d50.
The data qubit variable v50 is moved onto d49 by the first SWAP gate between d49 and d50.
After moving v49 from d50 to d41, the data qubit v50 on d49 is moved back to d50 by the second SWAP gate.
v41, now in d49, is disentangled from other qubits, hence we can initialize d49 any time.
v49 is eventually moved to d31, finishes gathering all error syndromes and gets measured.
Figure \ref{fig:sol:a_2-units_superunit} (b) summarizes the move of v49 from d49 to d31.

To find the optimized stabilizer circuit,
first we search the smallest set of syndrome qubit devices
which are neighbored to all the data qubits of the stabilizer.
By moving a syndrome qubit variable around the syndrome
qubit devices in the set,
all data qubits are neighbored to the syndrome qubit variable
during the move
and error syndromes can be collected onto the syndrome qubit
variable with nearest-neighbor interaction.
For optimization, we solve a traveling salesman problem
to find the shortest move of the syndrome qubit variable of the stabilizer
among the syndrome qubit devices.
If there are several smallest sets of syndrome qubit devices,
we compare the moves of each set and adopt the set with the shortest move.
This procedure gives the optimized circuit for single stabilizer reconfiguration.

In Figure \ref{fig:sol:a_2-units_superunit}, we see that the superunit circuit is deeper than the normal unit stabilizer circuit.
In general, superunit stabilizers require more steps to gather error syndromes than normal stabilizers.
Obviously, the deeper stabilizer will have more physical errors during a stabilizing cycle, due to the increased opportunities to accumulate physical errors over a longer period of time.
Thus, an important engineering goal is to create stabilizer circuits that are as shallow as possible.

We present a basic algorithm to compose a stabilizer circuit shown in Algorithm~\ref{alg:sol:stabilizer_circuit}.
A syndrome qubit variable travels one way to gather error syndromes.
In this algorithm, we search for the shortest traversable path in which error syndromes can be gathered from all data qubits.

\IncMargin{1em}
 \begin{algorithm*}
\small
 \caption{Stabilizer Circuit Composition}
 \label{alg:sol:stabilizer_circuit}
 \SetKwInput{Input}{Input}
 \SetKwInOut{Output}{Output}
 \SetKwFunction{GatherSyndrome}{GatherSyndrome}
 \SetKwFunction{AddOp}{AddOp}
 \SetKwFunction{IsNotAlreadyGathered}{IsNotAlreadyGathered}
 \SetKwFunction{SolveTravelingSalesman}{SolveTravelingSalesman}
 \SetKwFunction{Combination}{Combination}
 \SetKwFunction{Neighbors}{Neighbors}
 \SetKwFunction{Num}{Num}
 \SetKwFunction{NextInMinPath}{NextInMinPath}
 \Input{$Dat$: Set of data qubits belonging to the stabilizer (typically 4 or 6)}
 \Input{$Anc$: Set of ancilla qubits around the stabilizer (typically 1 or 8)}
 \Input{$G$: Graph of qubits, describing qubits' neighbor relationships}
 \Output{Stabilizer circuit of shallowest depth}
 $MinCost = \_\_INT\_MAX\_\_$; $MinPath = None$\;
 /* Search for the smallest set of ancilla qubits which neighbor all data qubits. */\\
 \For{$n \in (1..\Num(Anc))$}{
   \For{$ancs \in \Combination(Anc,n)$}{
     \If{$d \in \Neighbors(ancs), \forall d \in Dat$}{
       /* Search for the shortest path involving data qubits. */\\
       $cost, path = \SolveTravelingSalesman(ancs)$\;
       \If{$cost < MinCost$}{
         $MinCost = cost$\;
         $MinPath = path$\;
       }
     }
   }
   \If{$MinPath \ne None$}{
     break\;
   }
 }
 /* Add operations along the path found. */\\
 \AddOp(Initializations(MinPath.ancillas))\;
 \For{$q \in MinPath$}{
   \If{$q \in Anc$}{
     \ForEach {$d \in [d | d \in Dat, d \in \Neighbors(q)]$}{
       \If{\IsNotAlreadyGathered($d$)}{
         \AddOp(\GatherSyndrome($d$, $q$))\;
       }
     }
   }
   /* SWAP gate which moves the ancilla qubit variable which holds error syndrome to the next hop in MinPath. */\\
   \If{$q.next \ne Null$}{
     \AddOp(SWAP($q$, $q.next$))\;
   }
   /* SWAP gate which returns the data qubit variable which was swapped to the previous hop to the original positions. */\\
   \If{$q \in Dat$}{
     \AddOp(SWAP($q.prev$, $q$))\;
   }
 }
 /* Measure the ancilla qubit which holds error syndrome. */\\
 \AddOp(Measurement($q$))\;

\end{algorithm*}

In our scheme, a single syndrome qubit is used to collect error syndromes from all data qubits in a stabilizer.
Hence, if a $Z$ ($X$) error occurs on the syndrome qubit in $Z$ ($X$) stabilizer,
the error propagates to data qubits whose error syndromes are collected after the error occurrence.
This error propagation is 
correctly incorporated in our model.

\section{Building a whole circuit from stabilizer circuits}
\label{subsec:whole}
On a perfect lattice, the stabilizer circuits are highly synchronous and easily scheduled efficiently.
The circuits for a defective lattice must be asynchronous on account of the different depth of stabilizers.
Such asynchronicity introduces a problem when several stabilizers try to access a qubit at the same time. We have to assign priorities to stabilizers.
Stabilizers with lower priority have to wait for other stabilizers, so that they have more opportunities to accumulate physical
errors on data qubits and may be blocked during stabilizing so that they have more opportunities to accumulate physical errors on ancilla qubits.
Therefore we give higher priority to stabilizers which have deeper stabilizer circuits to avoid error opportunities from concentrating there,
since a shorter error chain is obviously preferred for error correction.
The scheduling algorithm is shown in Algorithm~\ref{alg:sol:scheduling_algorithm}.

\IncMargin{1em}

\begin{algorithm*}
\caption{Scheduling algorithm}
\label{alg:sol:scheduling_algorithm}
\SetKwInput{Input}{Input}
\SetKwInOut{Output}{Output}
\SetKwFunction{Sort}{Sort}
\SetKwFunction{Head}{Head}
\SetKwFunction{AfterHead}{AfterHead}
 \SetKwFunction{Schedule}{Schedule}
 \SetKwFunction{Any}{Any}
\SetKwFunction{NotOverCurrentStep}{NotOver\_CurrentStep}
\Input{$SC$: The set of stabilizer circuits}
\Input{$MaxStep$: The number of time steps to output}
 \Output{$WholeCircuit$}
/* Sort stabilizers in order of depth, longest first. If they tie, stabilizers on top-left of the lattice have priority. */\\
$sortedSC = \Sort(SC)$\;
$deepest = \Head(sortedSC)$\;
$afterHead = \AfterHead(sortedSC)$\;

$WholeCircuit = NULL$\;
$wholeCeil = 0$\;
\While{$wholeCeil \leq MaxStep$}{ 
 /* wholeCeil is the step when the deepest stabilizer last scheduled finishes. */\\
 $deepest.ceil = wholeCeil = \Schedule(deepest,\textrm{ }WholeCircuit$)\; 
 \ForEach{$s \in afterHead$}{
  /* schedule every stabilizer once */\\
 $s.ceil = \Schedule(s)$\; 
 }
 /* loop until every s.Ceil $\rangle$ wholeCeil */\;
\While{$\Any(s.ceil \leq wholeCeil| s \in afterHead)$}{ 
  \ForEach{$s \in afterHead$}{
   \If{$s \leq wholeCeil$}{
    $s.ceil = \Schedule(s,\textrm{ }WholeCircuit)$\; 
   }
  }
 }
}

\end{algorithm*}

The scheduling algorithm is:
\begin{enumerate}
 \item Sort stabilizers in order of depth, longest first. If they tie, stabilizers in the upper left of the lattice have priority. (lines 1-2)
 \item The deepest stabilizer is scheduled. The step when the deepest stabilizer finishes is saved (currentCeil). (line 9)
 \item Each non-deepest stabilizer is scheduled once, in order of decreasing depth. (lines 10-13)
 \item Each non-deepest stabilizer which does not exceed the currentCeil is scheduled once again, in order of depth. 
	 Short ones may be scheduled twice or more before the loop terminates. (lines 14-21)
 \item If all of the non-deepest stabilizers exceed the currentCeil, return to step 2. Otherwise, return to step 4. (lines 21-22)
\end{enumerate}

Our algorithm must enforce important restrictions,
which the completely synchronous circuits of the perfect lattice fulfill without explicitly being stated.
Conflicts may occur in asynchronous scheduling since several stabilizers may attempt to use a qubit at the same time.
Additionally, different types of stabilizers which share an even number of data qubits must
access those qubits in the same order.
For example, if we have two stabilizers $X_1X_2$ and $Z_1Z_2$ on qubits $1$ and $2$,
we have to execute them as $X_1X_2$ then $Z_1Z_2$ (or reverse order).
$X_1Z_2$ then $Z_1X_2$ is not allowed because of the stabilizer commutivity.
We postpone stabilizers of low priority to resolve conflicts by simply adding identity gates.
\subsection{Example of solving conflicts in scheduling}
\label{sec:appendix:conflict}
\begin{figure}[t]
 \begin{center}
  (a)\includegraphics[width=6.5cm]{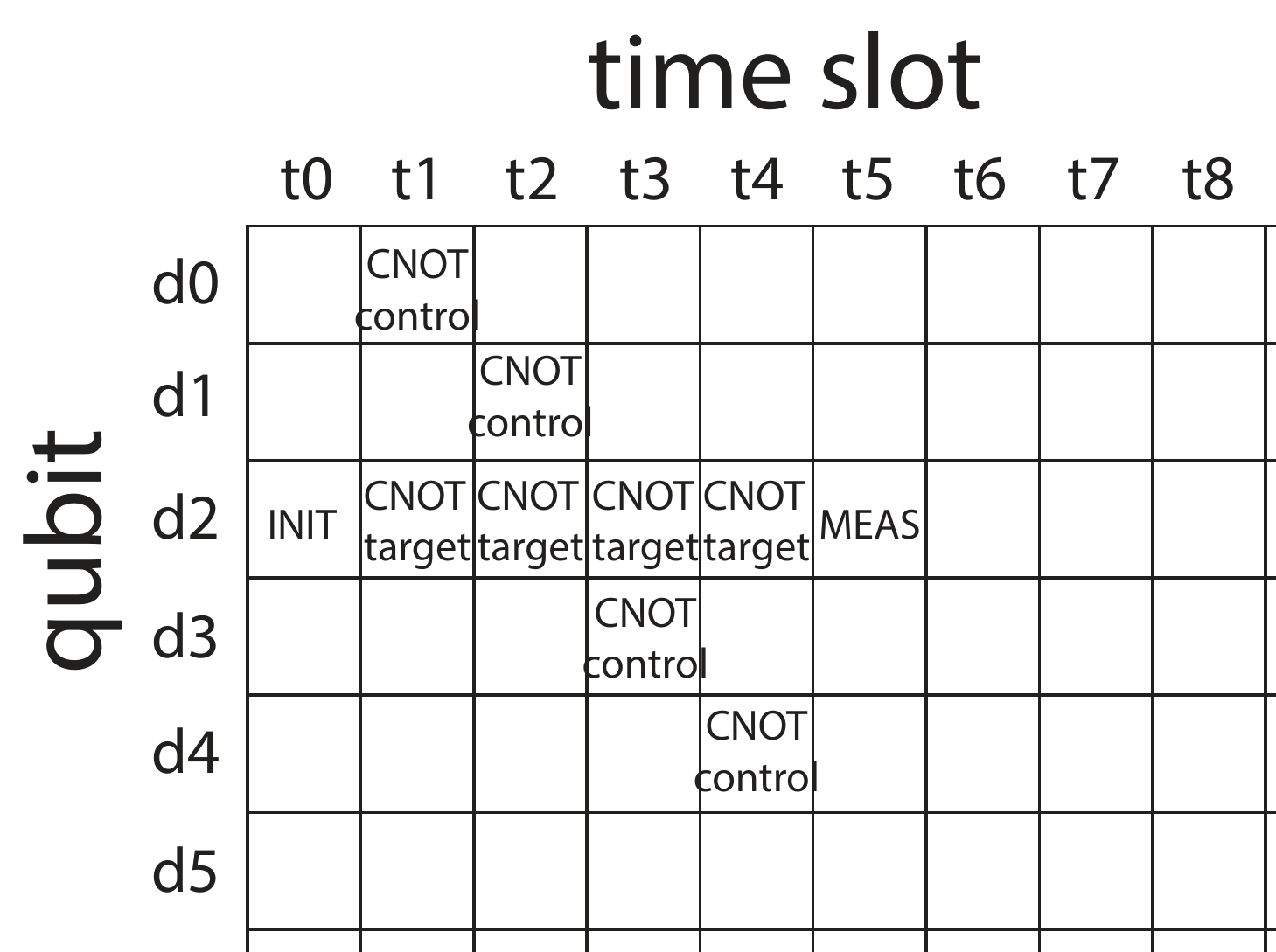}
  (b)\includegraphics[width=6.5cm]{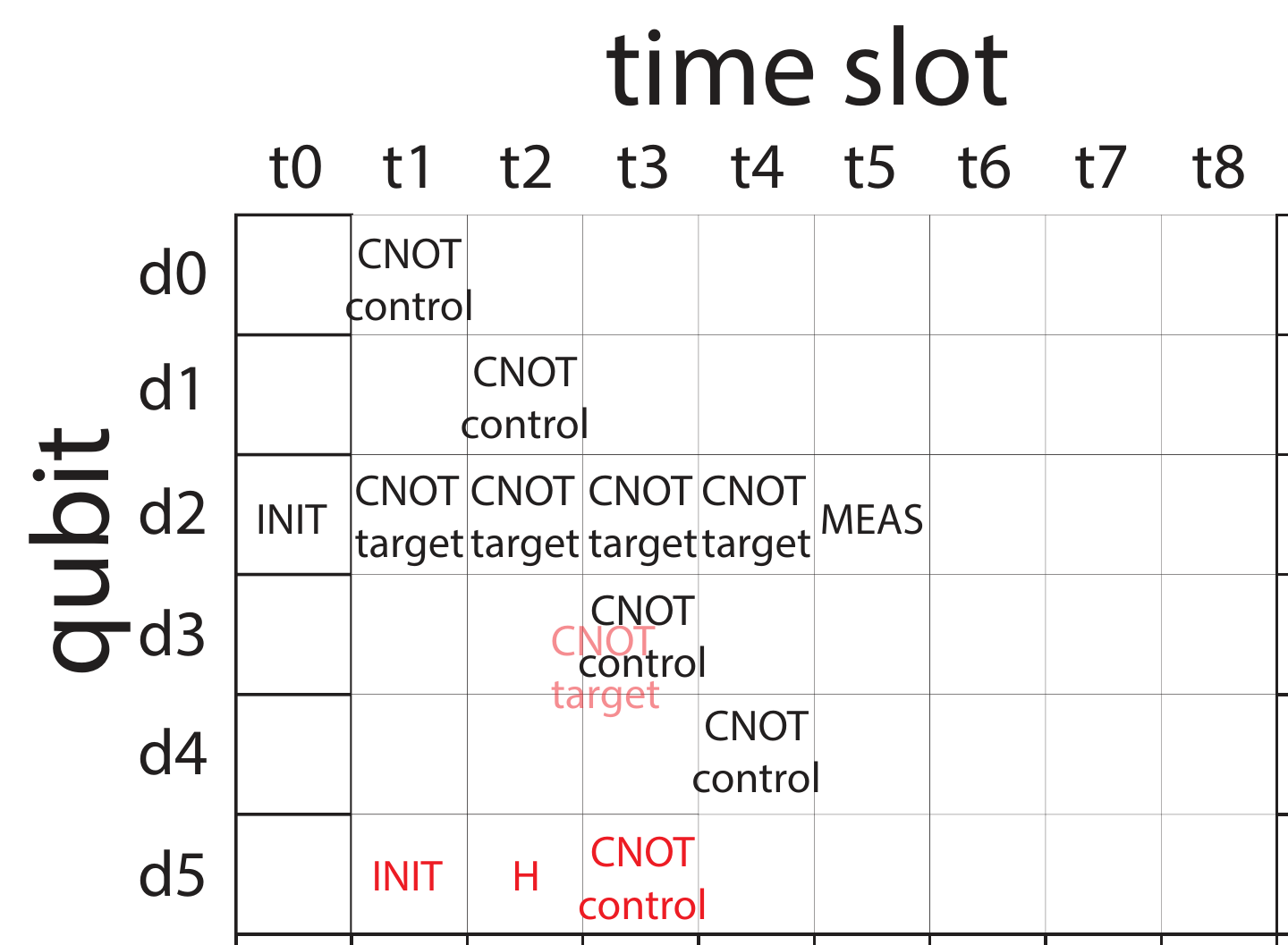}\\
  (c)\includegraphics[width=6.5cm]{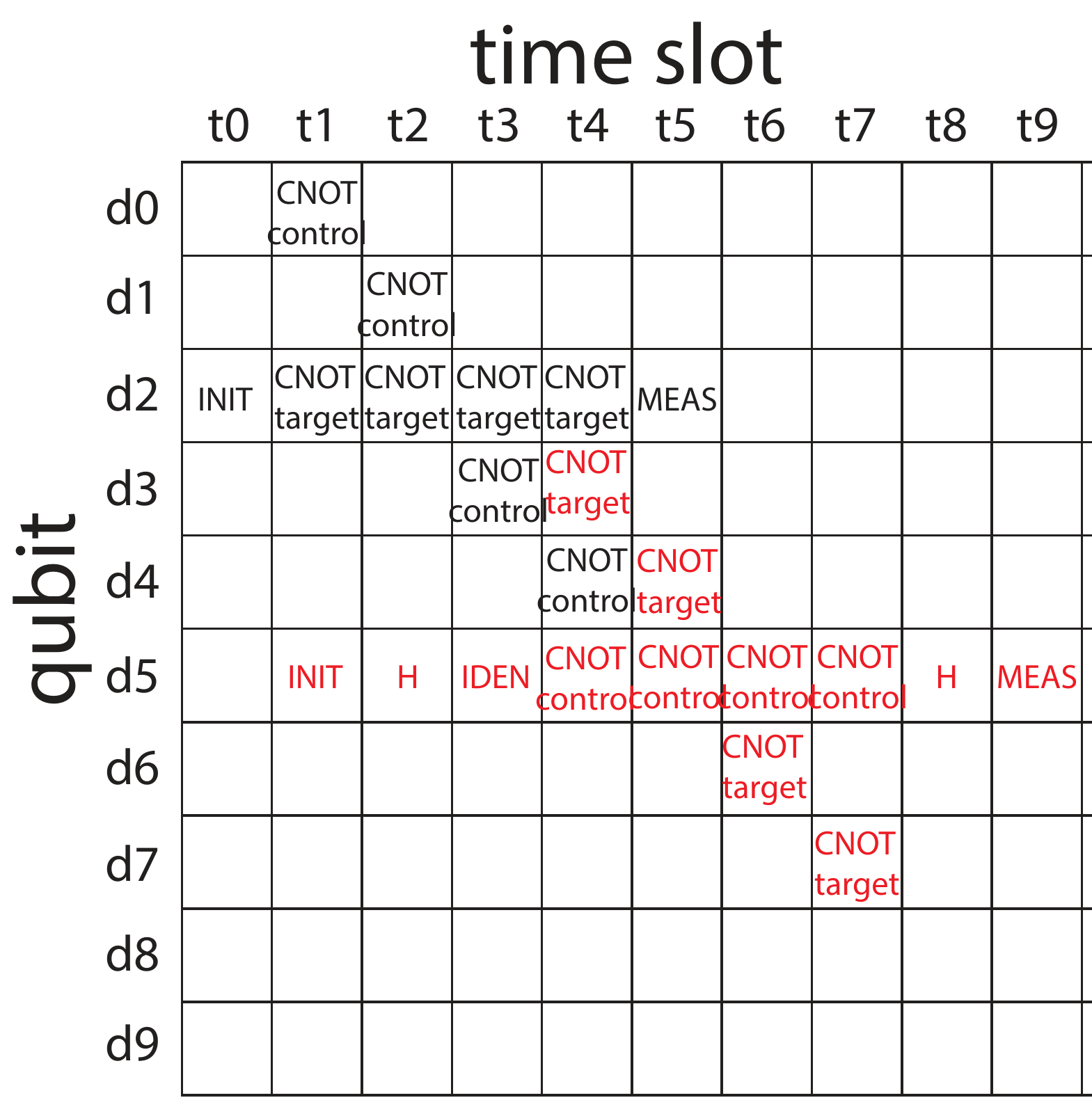}
  (d)\includegraphics[width=6.5cm]{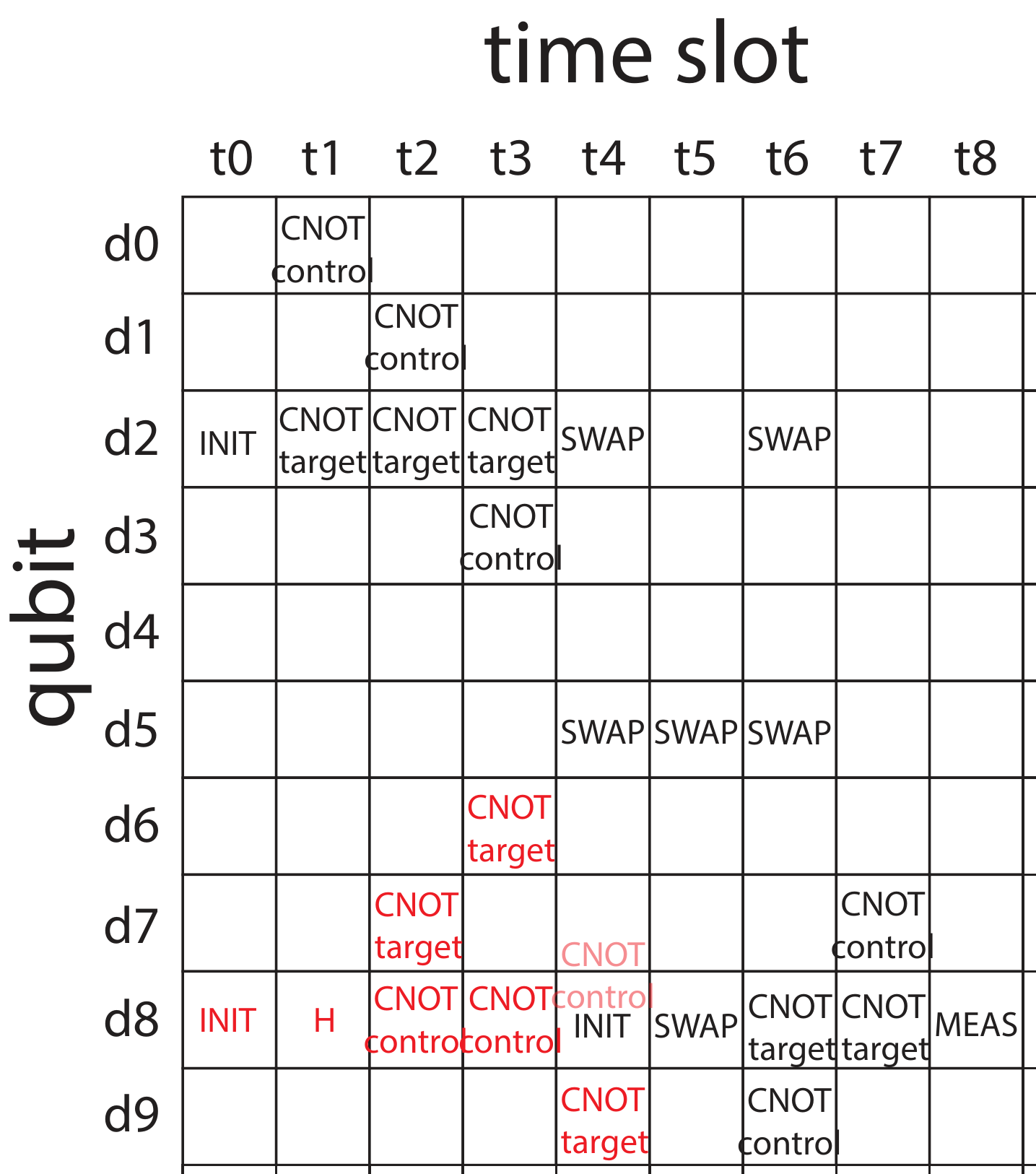}\\
  \caption[Example of the conflict of gates for data qubits in scheduling stabilizers.]{
  (a) Scheduling chart for the six 
  gates comprising a Z stabilizer circuit.
  Each box is called a slot and NULL slots are vacant.
  (b) Example of resource allocation conflict occurring between two stabilizers, with the higher priority one in black and the
  lower priority one in red.
  d3-t3 is allocated to the black stabilizer.
  The red stabilizer cannot lock the slot of d3-t3, as indicated by the light-colored CNOT target.
  (c) Solution to the contention for a data qubit. The red stabilizer waits for the slot of d3-t3 to become unlocked.
  (d) The slot of d8-t4 is allocated to the black stabilizer. The black SWAP gate already locks the slot of d8-t4 and the red CNOT cannot lock into it.
  }
  \label{fig:sol:slots1}
 \end{center}
\end{figure}

\begin{figure}[t]
 \begin{center}
  (e)\includegraphics[width=14cm]{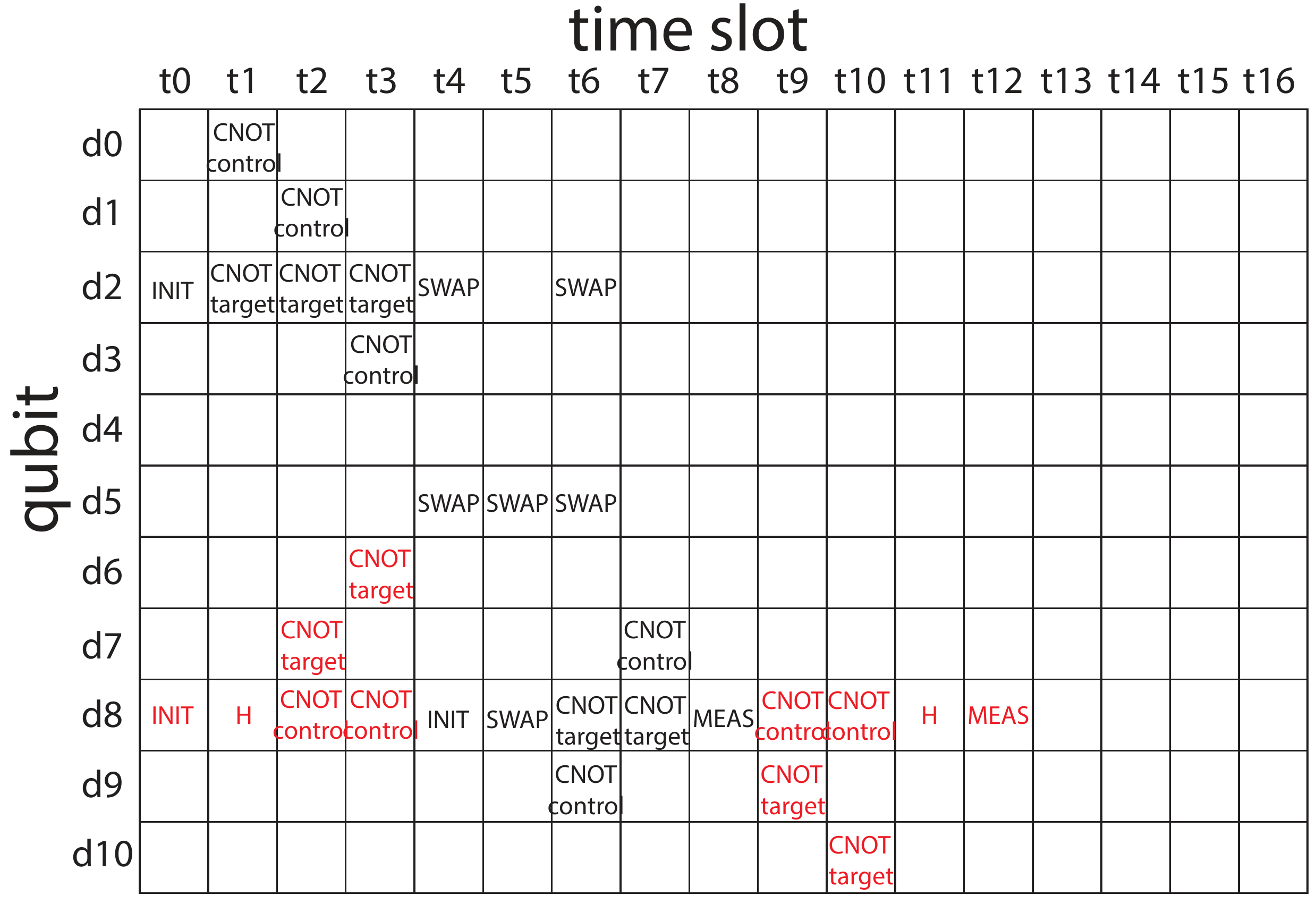}
  (f)\includegraphics[width=14cm]{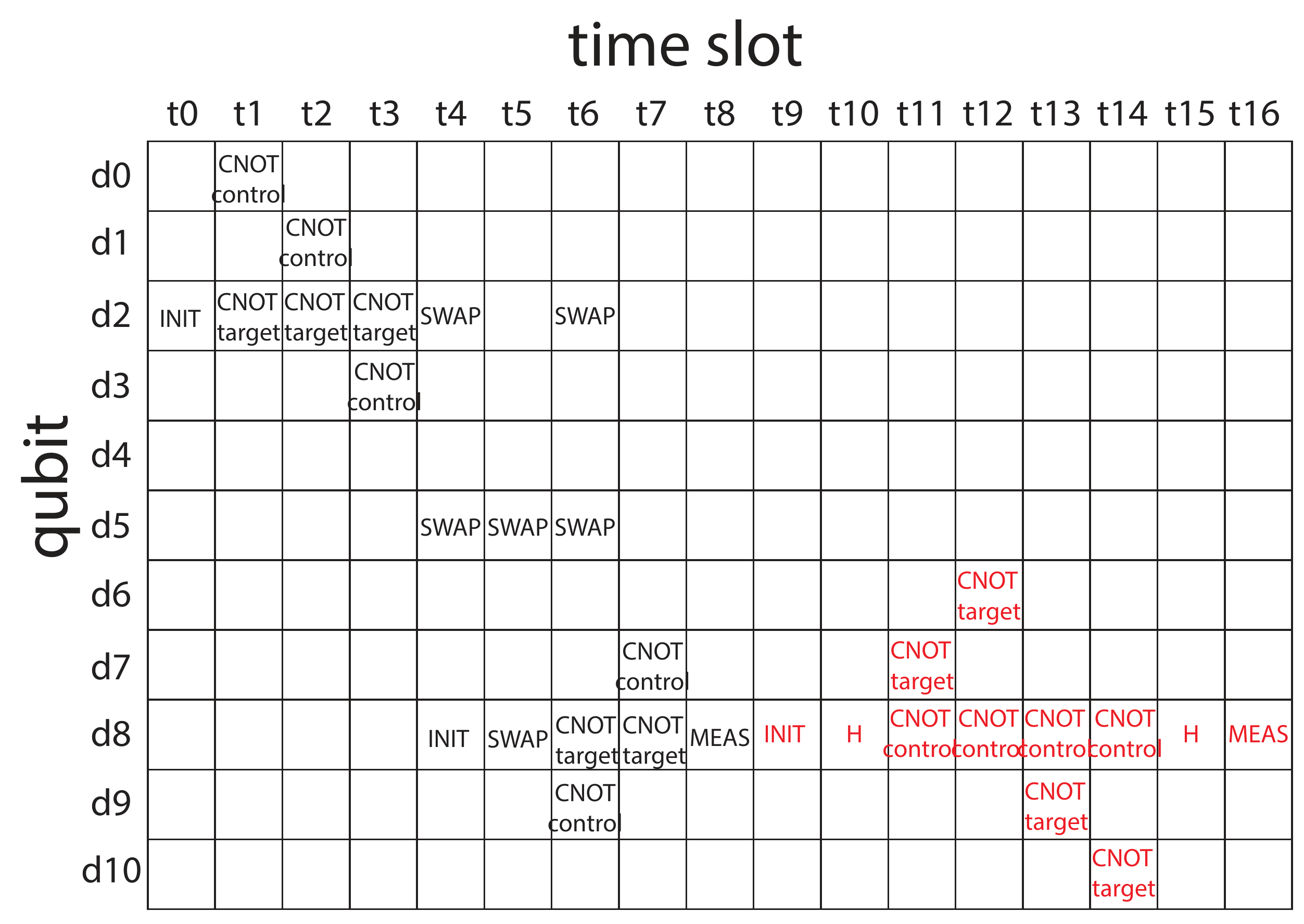}
  \caption[Example of the conflict of gates for syndrome qubits in scheduling stabilizers.]{
  (e) Attempting to solve the competition for a syndrome qubit by waiting. The red stabilizer is split and the former half of its error syndromes are deleted by the initialization gate in d8-t4. This is invalid.
  (f) Solution to the competition of a syndrome qubit. The red gates are all rescheduled after the black stabilizer.
  }
  \label{fig:sol:slots2}
 \end{center}
\end{figure}

Figures~\ref{fig:sol:slots1} and~\ref{fig:sol:slots2} illustrate various conflicts that occur during scheduling and our solutions.
Our scheduling is implemented to allocate ``slots'' to gates of stabilizer circuits, as shown in Figure \ref{fig:sol:slots1}(a).
Each qubit on each step has a slot and
only one gate can operate in a slot.
When a gate is set in a slot, the slot gets locked.
When a swap gate is set in the slot of a data qubit, the data qubit is locked
until the data qubit variable returns to the original data qubit device.
If conflicts occur, we add identity gates to resolve any conflicts
as shown in Figure \ref{fig:sol:slots1}(b) and Figure \ref{fig:sol:slots1}(c).
This method does not work for conflict on syndrome qubits.
This is because a data qubit may be unlocked after a single time step but a syndrome qubit
may not be unlocked for several steps, as shown in Figure \ref{fig:sol:slots1}(d).
Any sequence of gates on syndrome qubits in a stabilizer starts with an initialization gate,
which removes the error syndromes which have already gathered by the syndrome qubit,
as illustrated in d8-t4 in Figure \ref{fig:sol:slots1}(d).
If the stabilizer currently being scheduled (red gates) waits for the syndrome qubit to be unlocked,
the initialization gate deletes the syndrome qubit variable with some error syndromes of
the stabilizer as shown at d8-t4 in Figure \ref{fig:sol:slots2}(e).
To avoid this problem, the currently scheduled stabilizer is completely rescheduled
after the previously scheduled stabilizer finishes, as shown in Figure \ref{fig:sol:slots2}(f).

\subsection{Irregular whole circuit on account of a fault}
\begin{figure}[t]
 \begin{center}
  \includegraphics[width=300pt]{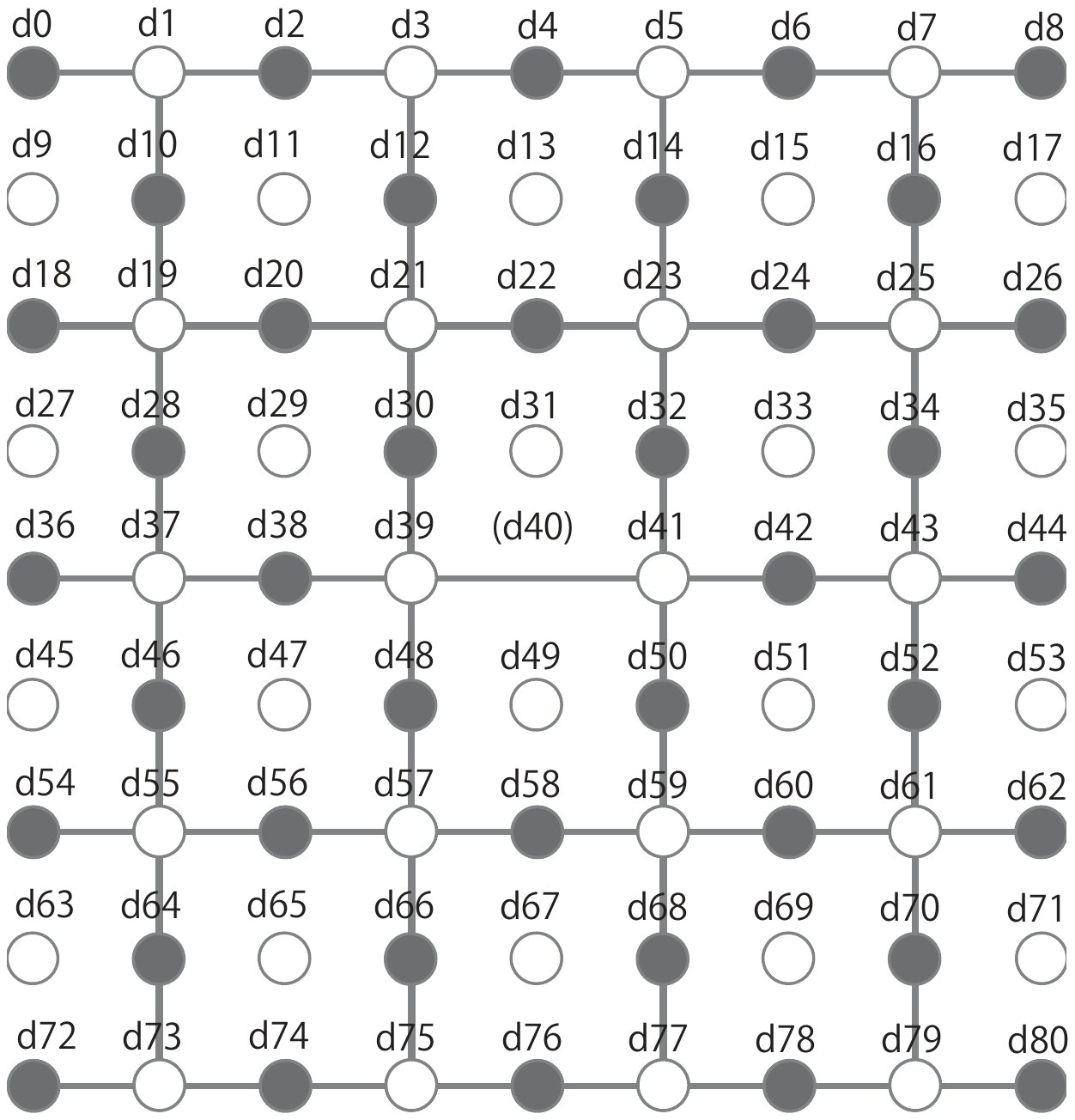}
  \caption[The picture of a lattice corresponding to Figure \ref{fig:sol:whole_circuit_single_faulty_center}.]
  {Gray dots are data qubits and white dots are syndrome qubits.
  The data qubit labeled (d40) is faulty.}
  \label{fig:sol:single_faulty_center_lattice}
 \end{center}
\end{figure}

\begin{figure}[t]
 \begin{center}
  \includegraphics[height=550pt]{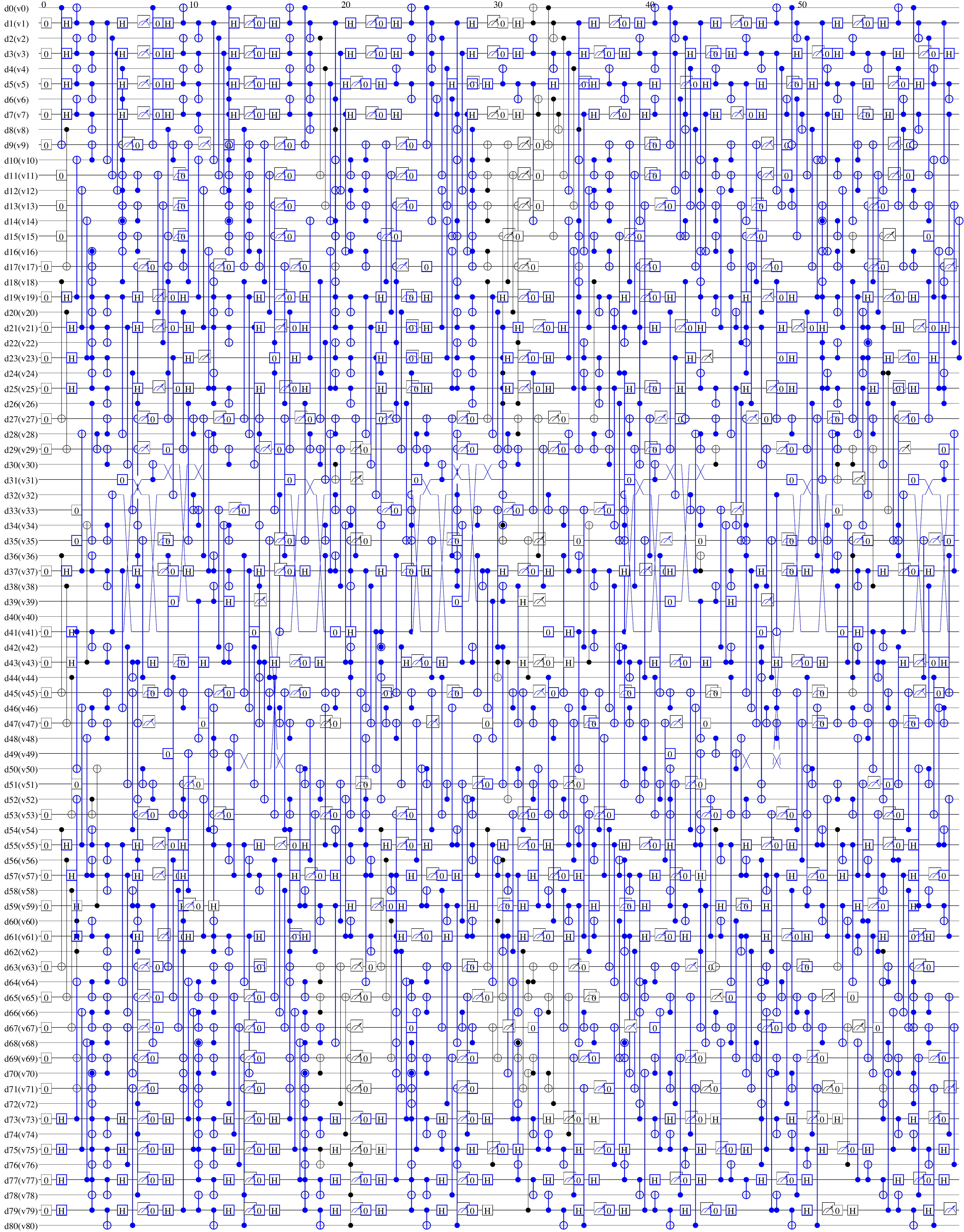}
  \caption
  [The first sixty steps of a whole circuit for a lattice of d=5 and in which the qubit device d40 in the center is faulty.]
  {The lattice condition is shown in Figure \ref{fig:sol:single_faulty_center_lattice}.
  We can see swap gates around d40.
  The detail of the irregular stabilizer circuit is shown in Figure \ref{fig:sol:a_2-units_superunit}.}
  \label{fig:sol:whole_circuit_single_faulty_center}
 \end{center}
\end{figure}

Figure~\ref{fig:sol:single_faulty_center_lattice} show an example of
a defective lattice in which the central qubit d40 is faulty.
Figure \ref{fig:sol:whole_circuit_single_faulty_center} shows the first tens of steps of the whole circuit of the lattice.
We can see that the circuit becomes irregular around the faulty device.

\section{Adapting matching to asynchronous operation}
\begin{figure}[t]
 \begin{center}
  \includegraphics[width=15cm]{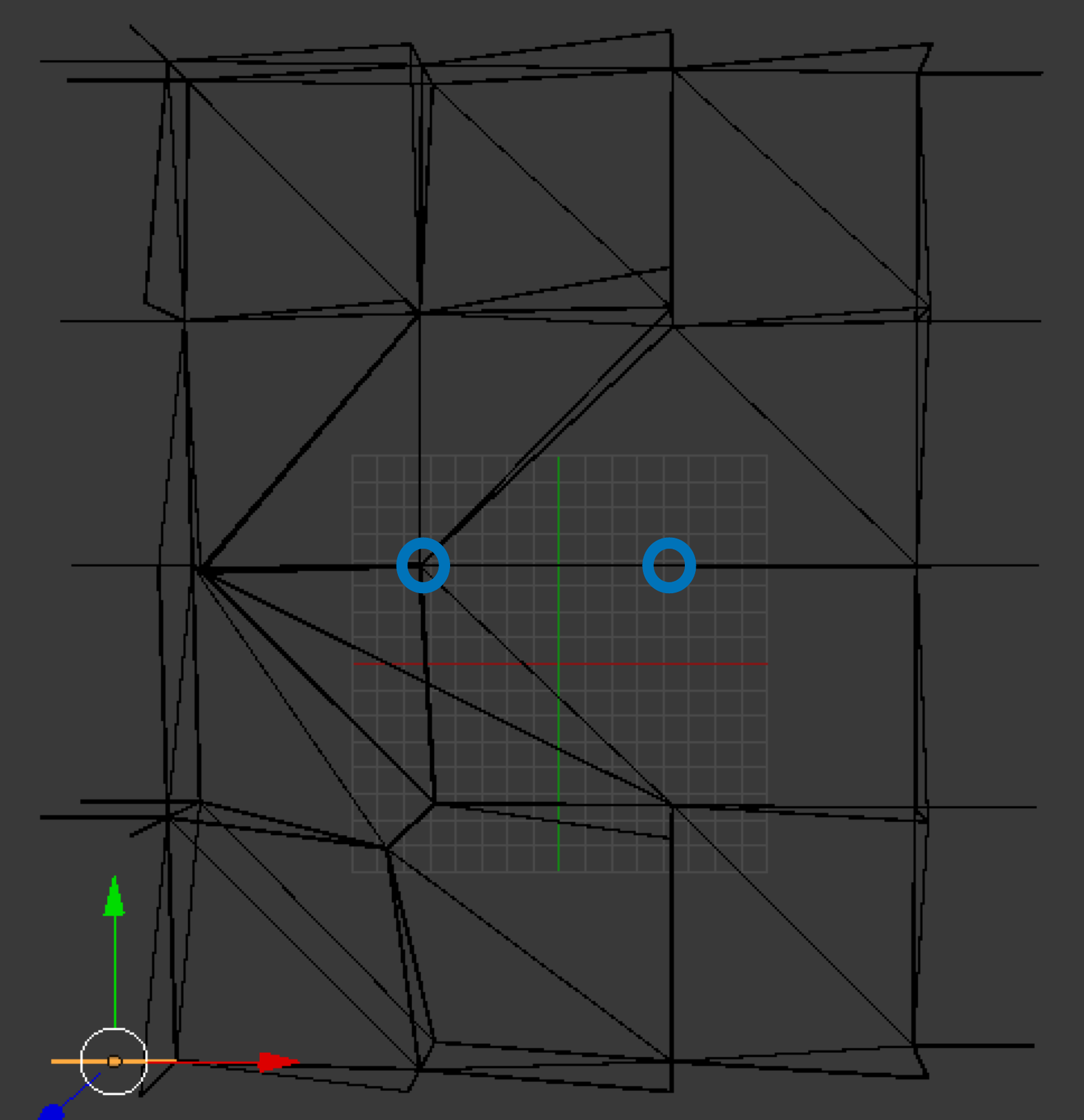}
  \caption[A top-view of a visualization of an asynchronous nest.]
  {Each vertex describes a syndrome measurement and each edge connects two syndrome measurements which might be changed by same errors. The two blue circles indicate the positions where two unit stabilizers originally existed and now they are merged into a superunit stabilizer placed in the left blue circle.}
  \label{fig:sim:nest-topview}
 \end{center} 
\end{figure}

\begin{figure}[t]
 \begin{center}
  \includegraphics[width=15cm]{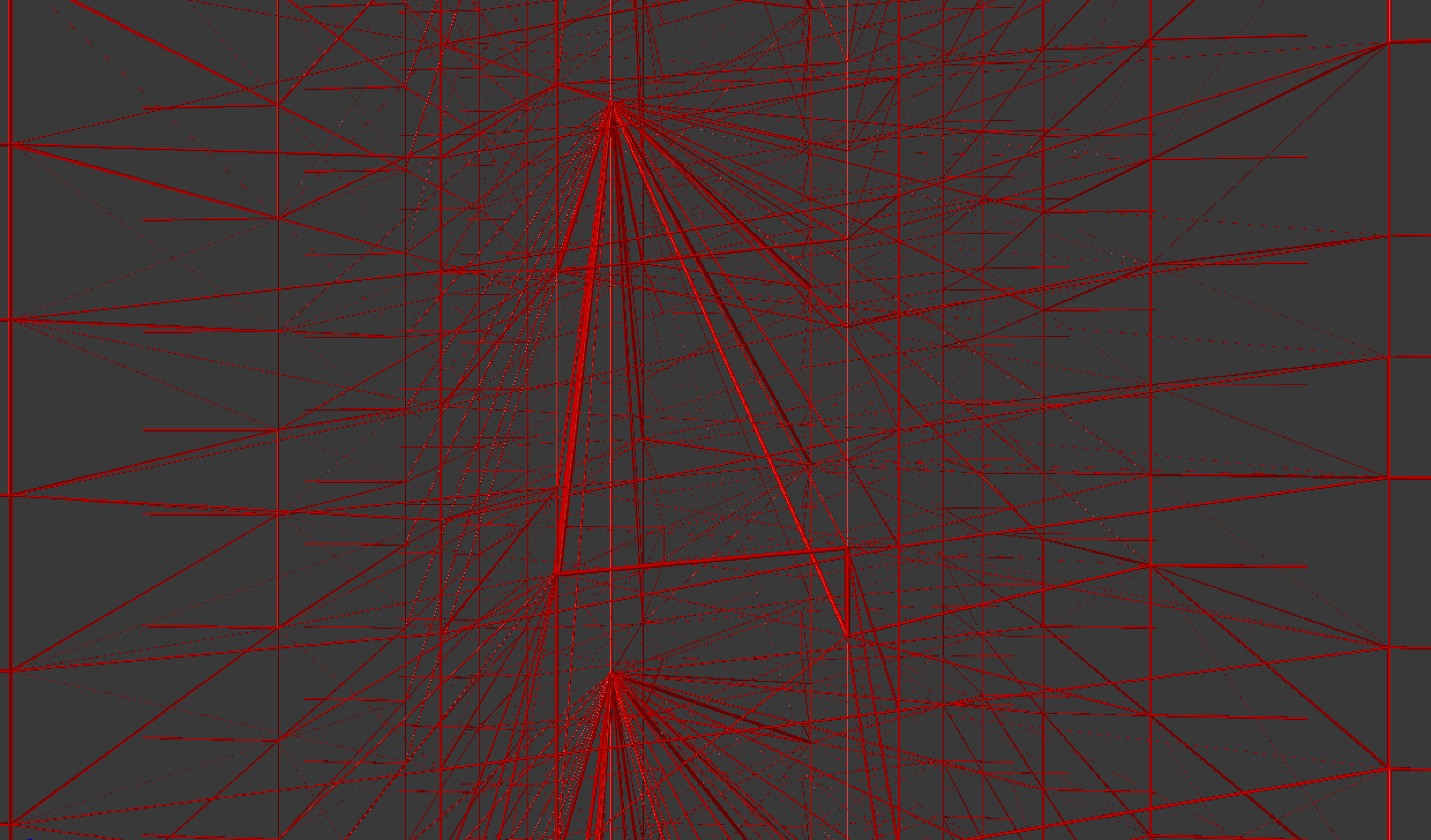}
  \caption[A frontview of a visualization of an asynchronous nest.]
  {The diameter of the edge indicates the weight of the matching.}
  \label{fig:sim:nest-frontview}
 \end{center}
\end{figure}

Irregular stabilizer circuits degrade the parallelism of stabilizer measurements of
the whole circuit, so that the surface code on a defective lattice has irregular error matching nests, as shown in Figures \ref{fig:sim:nest-topview} and \ref{fig:sim:nest-frontview}.
A superunit stabilizer is measured in a longer cycle than normal stabilizers and
a vertex corresponding to a superunit stabilizer has many edges.
These figures are output by Autotune during simulation of our surface code on a defective lattice~\cite{Fowler:2012autotune}.
We can see the irregularity of the nest.
A superunit stabilizer is measured in a longer cycle than normal stabilizers and
the vertex of a superunit stabilizer has many edges, some of which are thick.
This thickness is proportional to the error probability.
Additionally, the weights of edges are generated by tracing propagation, along the stabilizer circuits, of virtually created errors on every qubit at every physical step. 
Therefore the weights of edges reflect the possibilities of errors accurately, allowing Blossom V to achieve a result likely to correctly match the error pairs and return the state to the correct one.

\section{System architecture}
\begin{figure}[t]
 \begin{center}
 \includegraphics[width=400pt]{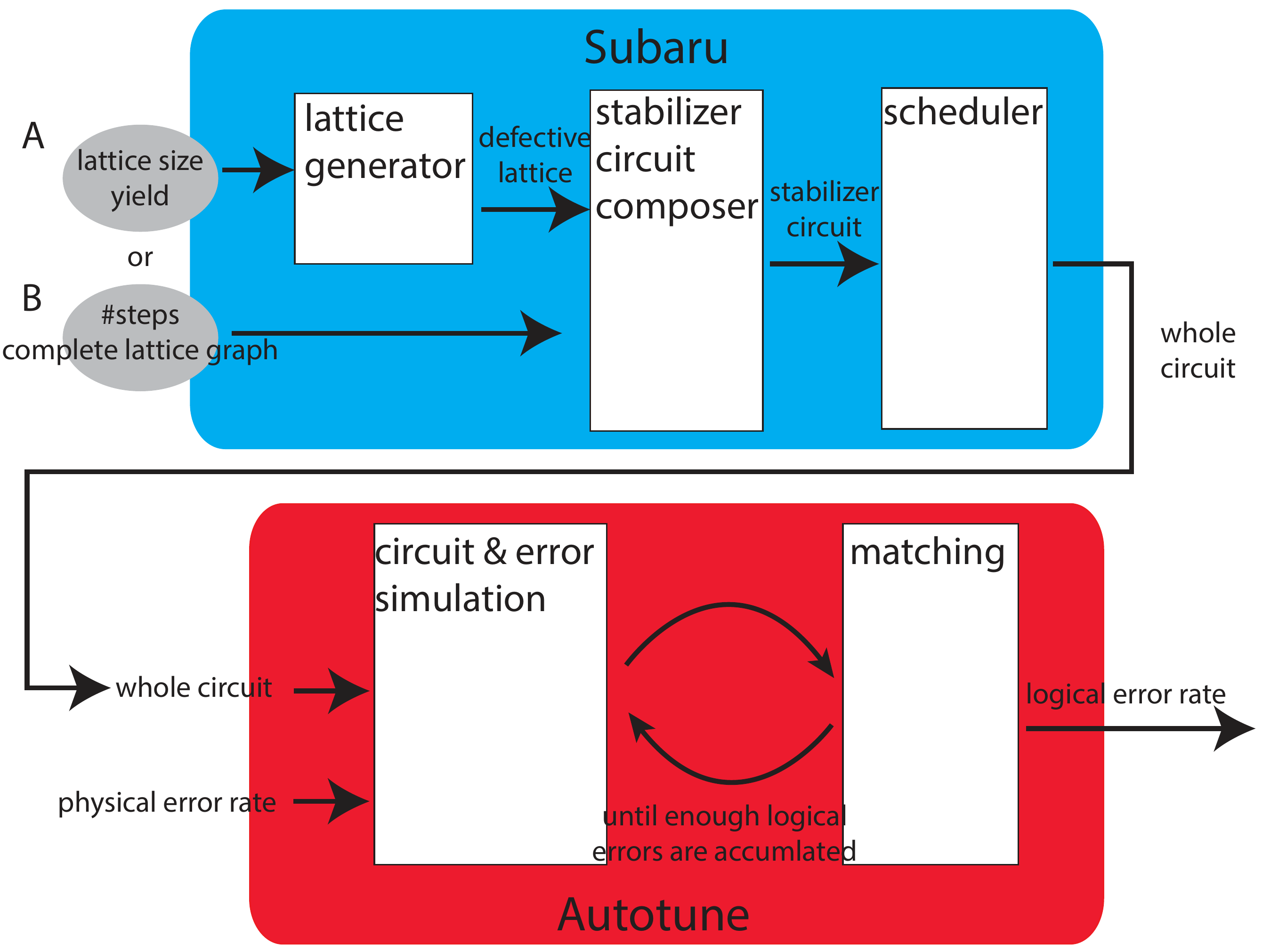}
  \caption[Simulation system design.]{
  Subaru can have one of two mutually exclusive inputs -- a pair of yield and a lattice size (labeled ``A'') or a full description of a lattice (labeled ``B'').
  With the former input, it randomly generates a defective lattice, then builds circuits to suit.
  With the latter input, it builds circuits for the provided defective lattice. It outputs a whole circuit of the requested number of steps.
  The circuit and physical error rate are input into Autotune, and Autotune outputs the logical error rate corresponding to the inputs.}
  \label{fig:sim:system_design}
 \end{center}
\end{figure}

Figure \ref{fig:sim:system_design} shows the major software components for compiling a circuit for the surface code and simulating its behavior on a defective lattice.
My circuit generator Subaru produces a whole circuit for a defective lattice and Autotune by Fowler et al. simulates surface code along the circuit \cite{Fowler:2012autotune}.
Subaru performs the tasks described in section \ref{sec:solution}.
Subaru can have alternative inputs -- yield or a lattice.
The yield is the probability of fabricating qubits which work properly.
Instead of a yield, a complete lattice can be input into Subaru.
This enables us to investigate particular conditions using hand-constructed lattices.

\section{Summary}
The surface code has large redundancy of encoded information.
This redundancy is generated by many equivalent logical operators
produced by many stabilizers spread vertically and horizontally.
This redundancy is originally designed to tolerate state errors.
Later this redundancy gets proposed to be used to tolerate losses, especially dynamic losses.
This Chapter gives a practical and concrete method to work around static losses,
utilizing the redundancy.

A lost data qubit is worked around by merging stabilizers the lost qubit originally belongs to into a superstabilizer.
This merge causes two overheads.
The first is that the merge may reduce the shortest hops of stabilizers between the boundaries, hence the error tolerance gets reduced.
The second is that the superstabilizer has more data qubits than normal stabilizers and the data qubits do not neighbor to single ancilla qubits so we need SWAP gates to move an ancilla qubit to collect error syndromes.
This procedure lengthen the depth of the stabilizer circuit.
This lengthened depth increases the chance to accumulate physical errors during error correction cycle.
Therefore those overheads affect the logical error rate of the lattice worse.

A lost ancilla qubit is worked around by utilizing another ancilla qubit near the stabilizer, originally prepared to measure another stabilizer.
This working around also requires SWAP gates to move the ancilla qubit to collect error syndromes.
Additionally, the cycle time of the stabilizer the utilized ancilla qubit is originally used for is lengthen.
Those overheads also affect the logical error rate of the lattice worse.

Those overheads are quantitatively evaluated in Section \ref{sec:defective:evaluation} by numerical simulation.

\clearpage
\chapter{Deformation-based surface code}
\label{chap:deformation}
This Chapter presents the construction of the densely packed surface code.
In fault-tolerant quantum computation, there are two potential ways to achieve significant resource reduction:
develop a new quantum error correcting code with universal logical gates for fault-tolerant quantum computation,
and or utilize multiple quantum error correcting codes as complementary systems
in which each can execute some operations more efficiently than others~\cite{976922,copsey02:_quant-mem-hier,choi2015}.
The new code 
presented here packs logical qubits more densely than other forms of the surface code.
It can be realized as an extension of Bombin's deformation-based surface code~\cite{1751-8121-42-9-095302}.

\section{Conventional deformation-based surface code and the new deformation-based surface code}
Bombin and Delgado introduced another way to produce a qubit on the surface code, the deformation-based surface code~\cite{1751-8121-42-9-095302}.
The deformation-based code produces a logical qubit by cutting a hole in a big/infinite regular lattice
in which the boundary of the unused region is composed from both types of boundaries,
hence it is like turning a planar code qubit inside out.
They showed Clifford gates and initialization to $|0\rangle$ and $|+\rangle$.
They demonstrated a CNOT gate by braiding, which can be executed between
two logical qubits in the deformation-based code and
even between the deformation-based code and the defect-based code.
Since a SWAP gate can be implemented with three CNOT gates,
arbitrary state injection to the deformation-based code can be achieved
utilizing this heterogeneous CNOT gate.
First, use the standard state injection method in the
defect-based code, then swap into the deformation-based code.
However, this method is an indirect way to achieve state injection to the deformation-based code.

We find a conversion from the defect-based code to the deformation-based code
that enables the deformation-based code to hold an arbitrary state,
and demonstrate that a crossed pair of an $X$ superstabilizer and a $Z$ superstabilizer produces a deformation-based qubit,
without sacrificing the surface code's advantages.
Far from killing the advantages, the lattice of my deformation-based surface code is fulfilled by stabilizers
so that denser packing of logical qubits can be achieved.
We employ the fault-tolerant stabilization utilizing a cat state generated by parallel $ZZ$ stabilization.
Additionally, we demonstrate a lattice surgery-like CNOT gate for the deformation-based code~\cite{Horsman:2012lattice_surgery}.
Lattice surgery is a non-transversal hence scalable means of executing a CNOT gate on the planar code that
requires fewer resources than the braiding of the defect-based code.
Our lattice surgery-like CNOT gate for the deformation-based code requires fewer qubits than the conventional braiding.
Nevertheless, the error suppression ability is similar to conventional surface code
since the logical state is protected by normal stabilizers.
My proposals may reduce the resource requirements of the surface code
in spatial accounting hence reducing the size of a practical quantum computer.

\section{Overview of the new deformation-based surface code}
\label{sec:logical_qubit}
\begin{figure}[t]
 \begin{center}
  \includegraphics[width=8cm]{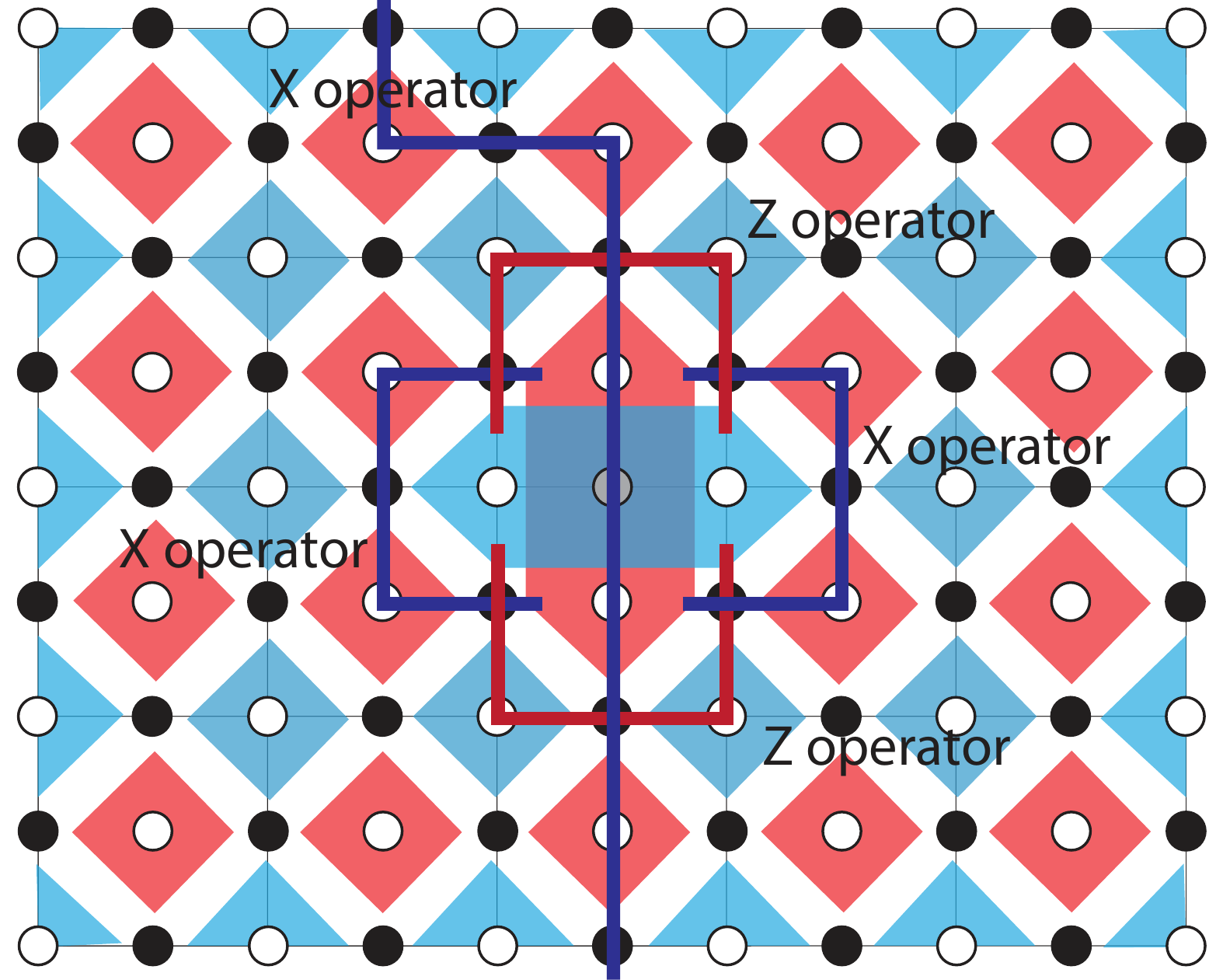}
  \caption[The deformation-based qubit of distance 3.]
  {Black dots depict data qubits and white dots are ancilla qubits.
  Each red diamond describes a $Z$ stabilizer and each blue diamond describes an $X$ stabilizer.
  The gray dot in the center depicts the unused data qubit, and
  the two 4-qubit $Z$ stabilizers the unused data qubit originally belonged to are merged to form the 6-qubit $Z$ stabilizer shown.
  The two 4-qubit $X$ stabilizers the unused data qubit originally belonged to are also merged to form the 6-qubit $X$ stabilizer shown.
  The thick lines are logical operators of the superstabilizer qubit.
   Any of the blue or the red paths serves as a logical $X$ operator
  or a logical $Z$ operator, respectively.
  } 
  \label{fig:integrated_qubit}
 \end{center}
\end{figure}

Figure \ref{fig:integrated_qubit} shows a distance 3 deformation-based qubit,
existing on the surface code lattice.
The surface code uses physical qubits placed on a 2D lattice.
The black dots are data qubits, and the white dots are ancilla qubits.
The lattice is separated into plaquettes as shown by black lines in the Figure.
A stabilizer $U$ is an operator which does not change a state,
\begin{equation}
 U\vert \psi \rangle = \vert \psi \rangle.
\end{equation}
An ancilla qubit in the center of a plaquette is used to measure the eigenvalue of a $Z$ stabilizer such as
$Z_aZ_bZ_cZ_d$ where $a\sim d$ denotes the surrounding four data qubits.
Recall that an ancilla qubit on the vertex is used for an $X$ stabilizer.

The number of logical qubits $k$ on a state of $n$ physical qubits is $k=n-s$ where $s$ is the number of independent stabilizers.
In Figure \ref{fig:integrated_qubit}, there are 48 data qubits, 19 $Z$ stabilizers
and 28 independent $X$ stabilizers, 
since any of the $X$ stabilizers is the product of all the others,
leaving a single degree of freedom for one logical qubit.

Two $Z_L$ operators of a deformation-based qubit
are shown in Figure \ref{fig:integrated_qubit}, either of which acts
on the logical qubit.
Three $X_L$ operators are shown in the figure, also working on the same logical qubit.
Two of the $X_L$ operators are the same shape as the described $Z_L$ operators,
while the third crosses the $Z$ superstabilizer ends at the boundaries of the lattice.
Those two $Z_L$ and two $X_L$ logical operators surrounding the superstabilizers correspond to the logical operators shown in
Figure 5 (a) in~\cite{1751-8121-42-9-095302}, except that our deformation-based qubit employs superstabilizers.
As with other surface code qubits, the products of a logical operator and stabilizers produce the redundancy for measurements
of logical operators.

\begin{figure}[t]
 \begin{center}
  \includegraphics[width=8cm]{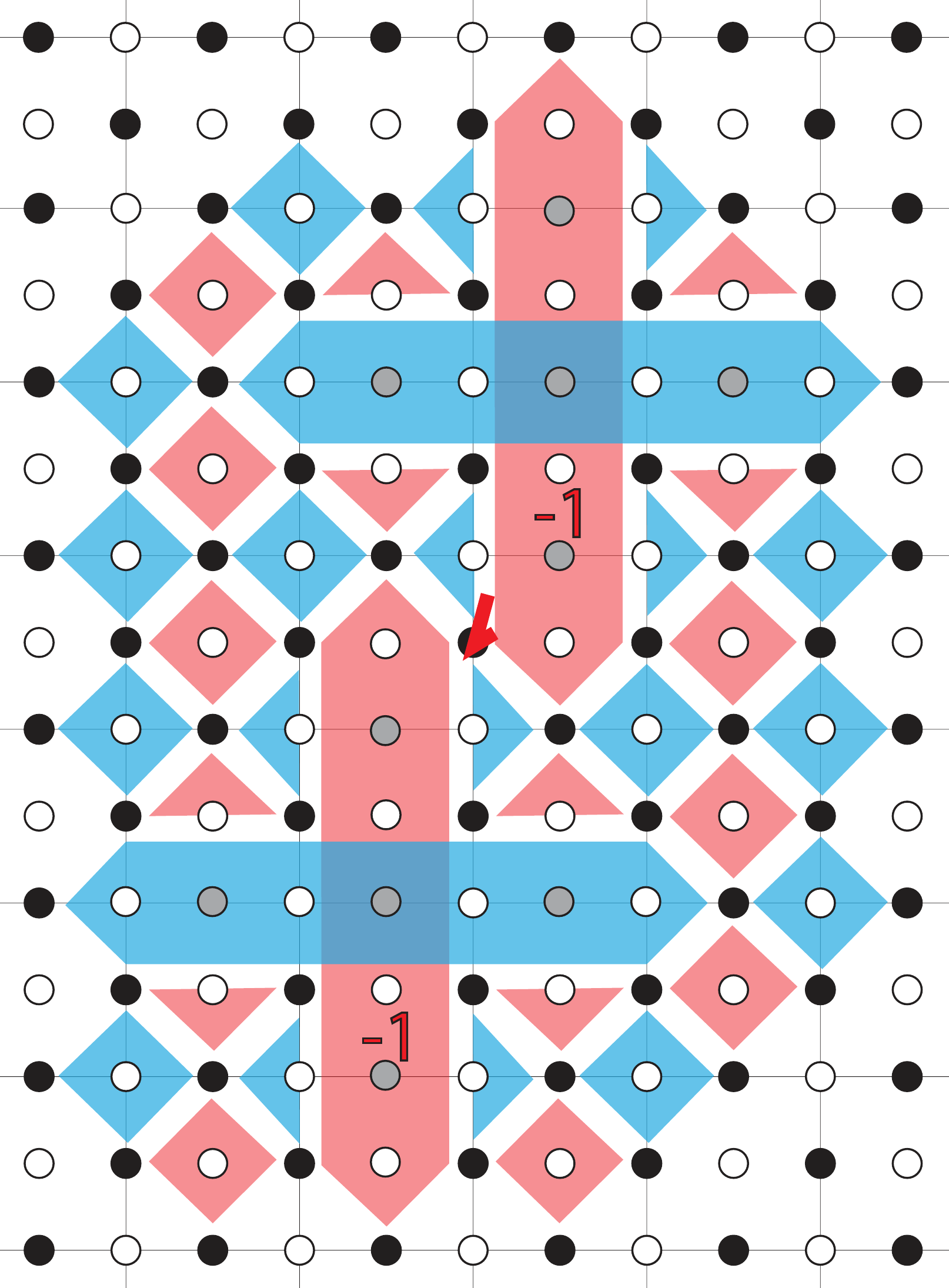}
  \caption[Neighboring distance 5 deformation-based qubits.]
  {Placement code distance apart from the boundary of the lattice is assumed.
  An $X$ error on the marked qubit results in $-1$ eigenvalues of the two red $Z$ superstabilizers.
  The two-defect surface code cannot correct an $X$ error on a data qubit which belongs to two defects,
  but the superstabilizers of the deformation-based code can.
  }
  \label{fig:correction}
 \end{center}
\end{figure}

Figure~\ref{fig:integrated_qubit} shows another important characteristic of the deformation-based qubit,
how to count its code distance.
Each logical operator consists of operations on three physical qubits, therefore the code distance of this deformation-based qubit is three.
An example of a longer code distance is shown in Figure~\ref{fig:correction},
which depicts two deformation-based qubits of distance five.

Figure~\ref{fig:correction} shows an advantage of deformation-based qubits compared to defect-based surface code qubits.
The deformation-based qubit exists at the junction of two superstabilizers,
so that every data  qubit alive in the lattice belongs to two $X$ stabilizers and two $Z$ stabilizers.
The two $Z$ superstabilizers find the $X$ error on the marked qubit in Figure~\ref{fig:correction},
hence the deformation-based qubits can be placed close to each other without being susceptible to logical errors,
though other surface code qubits must be placed
far enough away to maintain the code distance.

\section{Transformation}
\begin{figure}[t]
 \begin{center}
  \includegraphics[width=8cm]{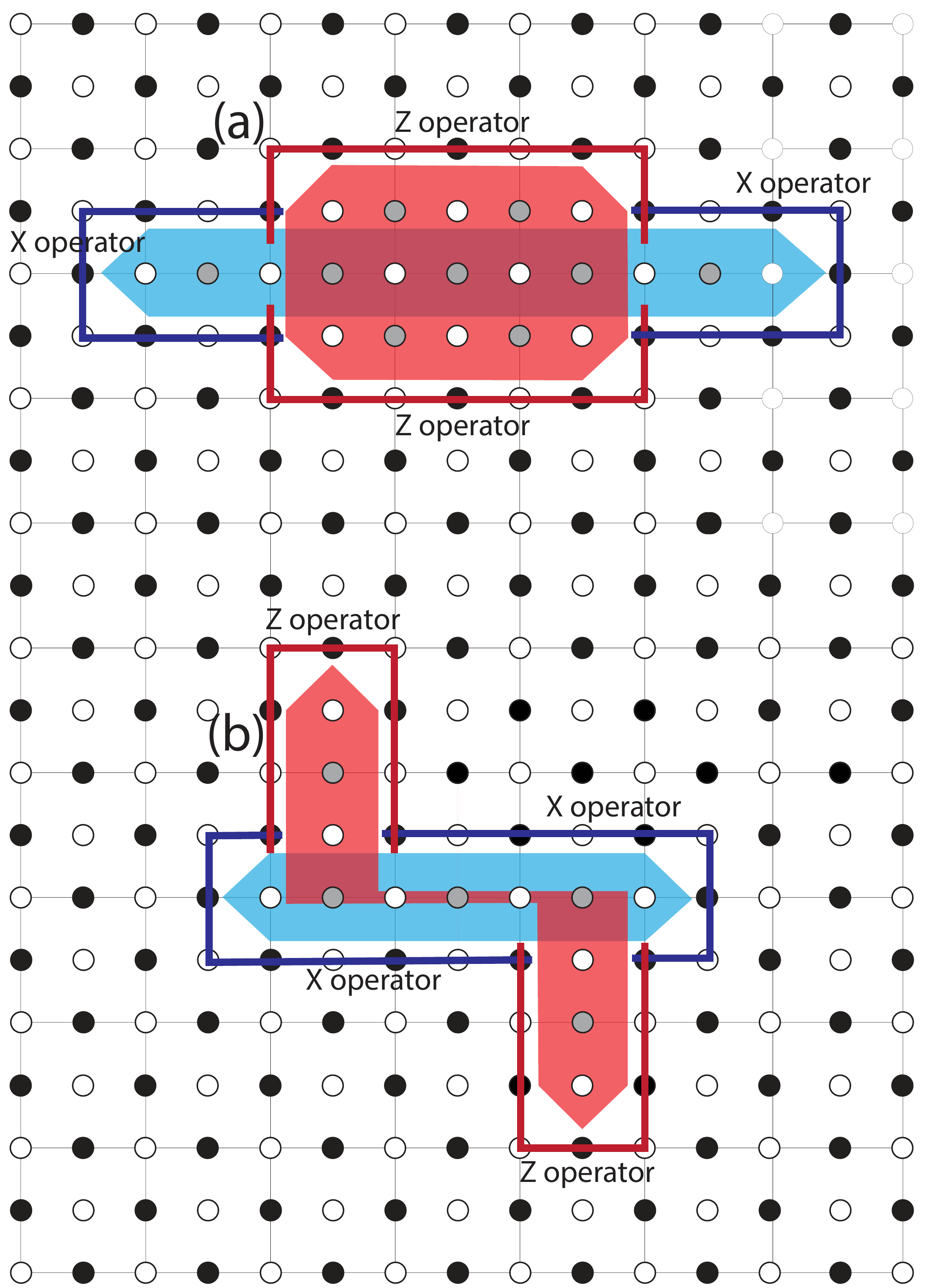}
  \caption[Transformation examples of deformation-based qubits.]
  {(a) ''Bar form'' deformation-based qubit, which has code distance 5.
  The $Z$ ($X$) logical operators exists between halves of the $X$ ($Z$) superstabilizer separated by the $Z$ ($X$) superstabilizer.
  (b) A deformation-based qubit of code distance 5, that has ``skew fin''.
  The $Z$ ($X$) logical operators exists between halves of the $X$ ($Z$) superstabilizer separated by the $Z$ ($X$) superstabilizer.
  }
  \label{fig:transform1} 
 \end{center}
\end{figure}

We have shown the ``four fin'' style deformation-based qubits.
Figure \ref{fig:transform1} shows two transformed deformation-based qubits of distance 5.
The deformation-based qubit in Figure \ref{fig:transform1} (a) is extended in the horizontal direction and
compressed in the vertical direction.
The perimeter of the $Z$ ($X$) superstabilizer can be considered to be separated by the $X$ ($Z$) superstabilizer.
The logical $Z$ ($X$) operator exists at any path connecting the separated halves.
The deformation-based qubit in Figure \ref{fig:transform1} (b) has a single, skewed $Z$ superstabilizer.
This transformation is achieved with more or less the defect-moving operations
of the defect-based surface code~\cite{Fowler:2009High-threshold_universal_quantum_computation_on_the_surface_code}.
The only difference is that the defect 
that does not have a stabilizer measurement
is replaced with the superstabilizer here.

\section{Conversion from a two-defect-based qubit}
\label{sec:state_injection}
\begin{figure}[t]
 \begin{center}
  \includegraphics[width=6cm]{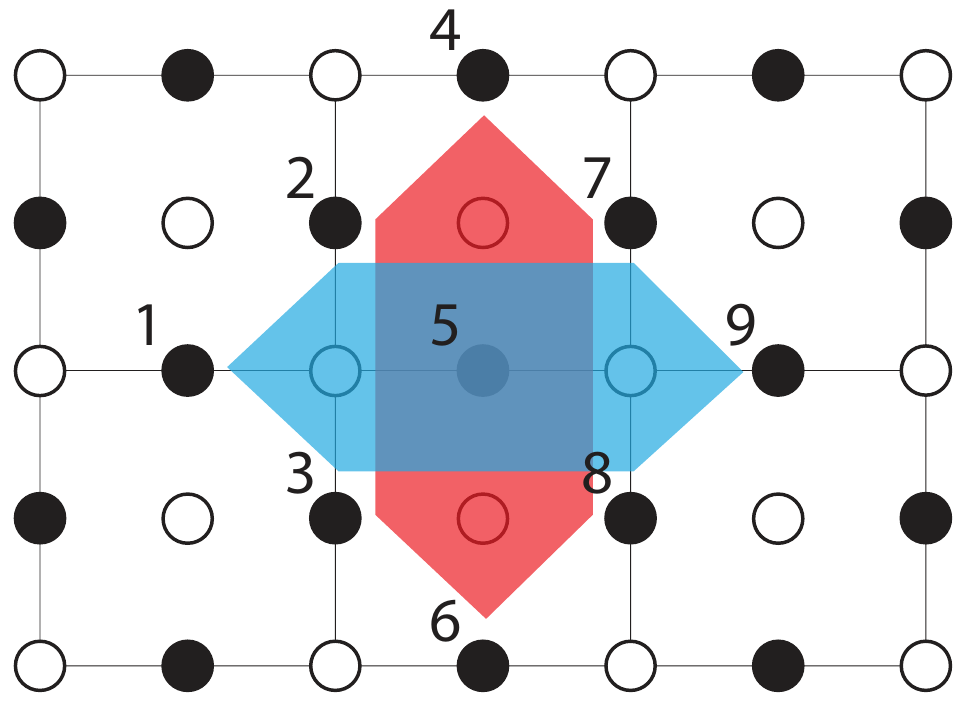}
  \caption[Surface code fragment to inject an arbitrary deformation-based qubit.]
  {
  The lattice has only normal stabilizers at first.
  The shown superstabilizers are introduced in several steps, as described in Section~\ref{sec:state_injection}.
  }
  \label{fig:state_injection}
 \end{center}
\end{figure}

Direct conversion from a two-defect surface code qubit to a deformation-based qubit can be achieved.
This conversion works as the state injection for the deformation-based qubit and
e.g. to support
networking among multiple quantum computers that employ heterogeneous error correcting codes
~\cite{PhysRevA.93.042338}.
To complete universality of the deformation-based surface code, we demonstrate the arbitrary state injection in this section.
We first inject an arbitrary qubit to a two-defect surface code following Fowler et al.
~\cite{Fowler:2009High-threshold_universal_quantum_computation_on_the_surface_code},
as depicted on a fragment of surface code in Figure \ref{fig:state_injection}.
The surface begins in normal operation, using qubit 5 and
measuring all 4-qubit stabilizers,
\begin{equation}
\begin{array}{c|ccccccccc}
 &1&2 &3 &4 &5 &6 &7 &8 &9 \\
 \hline
 &X&X &X &  &X & & & & \\
 & &  &  &  &X &  &X &X &X \\
 & &Z &  &Z &Z &  &Z & & \\
 & &  &Z &  &Z &Z &  &Z & \\
\end{array}
\end{equation}
where each number corresponds to the number in Figure~\ref{fig:state_injection}.
First, we measure qubit 5 in the X basis,
disentangling it from the larger state where $M_a^b$ denotes a measured value where $a$ is the measurement basis and $b$ is the qubit index.
\begin{equation}
\begin{array}{c|ccccccccc}
 & 1&2 &3 &4 &5 &6 &7 &8 &9 \\
 \hline
 & X&X &X &  &X & & & & \\
 &  &  &  &  &X &  &X &X &X \\
(-1)^{M_X^5} &  &  &  &  &X &  &  & & \\
 &  &Z &Z &Z &  &Z &Z &Z & \\
\end{array}
\end{equation}
If the -1 eigenvalue is measured, apply either $Z_2Z_4Z_5Z_7$ or $Z_3Z_5Z_6Z_8$
to restore $X_1X_2X_3$ and $X_7X_8X_9$ to +1 eigenvalues,
\begin{equation}
\begin{array}{c|ccccccccc}
 & 1&2 &3 &4 &5 &6 &7 &8 &9 \\
 \hline
 & X&X &X &  & & & & & \\
 &  &  &  &  & &  &X &X &X \\
 &  &  &  &  &X &  &  & & \\
 &  &Z &Z &Z &  &Z &Z &Z &
\end{array}.
\end{equation}
Next, qubit 5 is rotated to the arbitrary desired state~\footnote{\emph{Note:} Eqs. 5-9 and 14-21 describe \emph{states} that are stabilized by the corresponding terms, but do not correspond directly to stabilizer measurements conducted for error correction purposes.  In particular, the last line of Eq. 8 and 9 represents the newly introduced superstabilizer itself, while the two stabilizers just above illustrate the degree of freedom representing our logical qubit.  The stabilizers demarking the degree of freedom are labeled in the leftmost column inside the parentheses with $+$ or $-$ as appropriate.}, $\alpha (Z) + \beta (-Z)$,
 \begin{eqnarray}
  \alpha \left(\begin{array}{c|ccccccccc}
 & 1&2 &3 &4 &5 &6 &7 &8 &9 \\
 \hline
 & X&X &X &  & & & & & \\
 &  &  &  &  & &  &X &X &X \\
 +&  &  &  &  &Z &  &  & & \\
 &  &Z &Z &Z &  &Z &Z &Z & \\
	  \end{array}  \right)& \nonumber\\
 + \beta \left(\begin{array}{c|ccccccccc}
 & 1&2 &3 &4 &5 &6 &7 &8 &9 \\
 \hline
 & X&X &X &  & & & & & \\
 &  &  &  &  & &  &X &X &X \\
 -&  &  &  &  &Z &  &  & & \\
 &  &Z &Z &Z &  &Z &Z &Z & \\
	  \end{array}  \right).
 \end{eqnarray}
 Then we measure $Z_2Z_4Z_5Z_7$ and $Z_3Z_5Z_6Z_8$,
  \begin{eqnarray}
  \alpha \left(\begin{array}{c|ccccccccc}
 & 1&2 &3 &4 &5 &6 &7 &8 &9 \\
 \hline
 & X&X &X &  & & &X &X &X \\
 +&  &  &  &  &Z &  &  & & \\
 (-1)^{M_Z^{2457}} &  &Z & &Z &Z & &Z & & \\
 (-1)^{M_Z^{3568}} &  & &Z & &Z &Z & &Z & \\
	  \end{array}  \right)& \nonumber\\
 + \beta \left(\begin{array}{c|ccccccccc}
 & 1&2 &3 &4 &5 &6 &7 &8 &9 \\
 \hline
 & X&X &X &  & & &X &X &X \\
 -&  &  &  &  &Z &  &  & & \\
 (-1)^{M_Z^{2457}} &  &Z & &Z &Z & &Z & & \\
 (-1)^{M_Z^{3568}} &  & &Z & &Z &Z & &Z & \\
	  \end{array}  \right).
 \end{eqnarray}
 If the -1 eigenvalue is measured, apply either $X_1X_2X_3$ or $X_7X_8X_9$
 to give the desired state.
 The two defects exist at $X_1X_2X_3X_5$ and $X_5X_7X_8X_9$,
 a minimal logical qubit of distance 1,
\begin{eqnarray}
  \alpha \left(\begin{array}{c|ccccccccc}
 & 1&2 &3 &4 &5 &6 &7 &8 &9 \\
 \hline
 & X&X &X &  & & &X &X &X \\
 +&  &  &  &  &Z &  &  & & \\
 &  &Z & &Z &Z & &Z & & \\
 &  &  &Z & &Z &Z & &Z & \\
	  \end{array}  \right)& \nonumber\\
 + \beta \left(\begin{array}{c|ccccccccc}
 & 1&2 &3 &4 &5 &6 &7 &8 &9 \\
 \hline
 & X&X &X &  & & &X &X &X \\
 -&  &  &  &  &Z &  &  & & \\
 &  &Z &  &Z &Z & &Z & & \\
 &  &  &Z &  &Z &Z & &Z & \\
	  \end{array}  \right).
\end{eqnarray}

So far we have the logical qubit of the two-defect surface code.
Next we start to convert this logical qubit 
to the deformation-based surface code.

For pedagogical clarity,
we omit writing the stabilizers that do not change over the course of this operation,
depicted in white in the figures, and
we write $Z_2Z_4Z_5Z_7 \otimes Z_3Z_5Z_6Z_8 = Z_2Z_3Z_4Z_6Z_7Z_8$, which is a product of two stabilizers
and which can be measured as a stabilizer without breaking the logical state.
We again measure qubit 5 in the $X$ basis,
merging the two minimal defects into one superstabilizer,
    \begin{eqnarray}
 \alpha \left(\begin{array}{c|ccccccccc}
 & 1&2 &3 &4 &5 &6 &7 &8 &9 \\
 \hline
 & X&X &X &  & & &X &X &X \\
(-1)^{M_X^5} &  &  &  &  &X &  &  & & \\
 +&  &Z &  &Z & & &Z & & \\
 +&  &  &Z &  & &Z & &Z & \\
 &  &Z &Z &Z & &Z &Z &Z & \\
	  \end{array}  \right)& \nonumber\\
 + \beta \left(\begin{array}{c|ccccccccc}
 & 1&2 &3 &4 &5 &6 &7 &8 &9 \\
 \hline
 & X&X &X &  & & &X &X &X \\
(-1)^{M_X^5} &  &  &  &  &X &  &  & & \\
- &  &Z &  &Z & & &Z & & \\
- &  &  &Z &  & &Z & &Z & \\
 &  &Z &Z &Z & &Z &Z &Z & \\
	  \end{array}  \right).
    \end{eqnarray}
    If the -1 eigenvalue is obtained,
    apply either $Z_2Z_4Z_5Z_7$ or $Z_3Z_5Z_6Z_8$
    to preserve the parity of the logical $X$ operator such as $X_1X_2X_3X_5$ into $X_1X_2X_3$, giving
    
        \begin{eqnarray}
  \alpha \left(\begin{array}{c|ccccccccc}
 & 1&2 &3 &4 &5 &6 &7 &8 &9 \\
 \hline
 & X&X &X &  & & &X &X &X \\
 &  &  &  &  &X &  &  & & \\
 +&  &Z &  &Z & & &Z & & \\
 +&  &  &Z &  & &Z & &Z & \\
 &  &Z &Z &Z & &Z &Z &Z & \\
	  \end{array}  \right)& \nonumber\\
 + \beta \left(\begin{array}{c|ccccccccc}
 & 1&2 &3 &4 &5 &6 &7 &8 &9 \\
 \hline
 & X&X &X &  & & &X &X &X \\
 &  &  &  &  &X &  &  & & \\
- &  &Z &  &Z & & &Z & & \\
- &  &  &Z &  & &Z & &Z & \\
 &  &Z &Z &Z & &Z &Z &Z & \\
	  \end{array}  \right).
    \end{eqnarray}

    Now $Z_2Z_4Z_7$ and $Z_3Z_6Z_8$ share the desired state.
    We can now begin measuring $Z_2Z_3Z_4Z_6Z_7Z_8$ as our superstabilizer.
    As is common with state injections,
    because the process begins with a raw qubit,
    state distillation on the logical qubit is required after this process.

\section{CNOT gate}
\label{sec:cnot}
A CNOT gate can be performed utilizing lattice surgery~\cite{Horsman:2012lattice_surgery}.
The basic concept of the CNOT gate by lattice surgery is
\begin{enumerate}
 \item prepare a control (C) qubit in $\alpha \vert 0_{C} \rangle + \beta \vert 1_{C} \rangle$ and
	 a target (T) qubit in $\alpha' \vert 0_{T} \rangle + \beta' \vert 1_{T} \rangle$.
 \item prepare an intermediate (INT) qubit in $\vert +_{I} \rangle$.
	 The initial state is
\begin{equation}
	 \vert \psi ^{init} \rangle = (\alpha \vert 0_{C} \rangle + \beta \vert 1_{C} \rangle) \otimes \vert +_{I} \rangle \otimes (\alpha' \vert 0_{T} \rangle + \beta' \vert 1_{T} \rangle).
\end{equation}
 \item measure $Z_CZ_{I}$ and get
\begin{equation}
	 \vert \psi' \rangle = (\alpha \vert 0_{C}0_{I} \rangle + \beta \vert 1_{C}1_{I} \rangle) \otimes (\alpha' \vert 0_{T} \rangle + \beta' \vert 1_{T} \rangle)
\end{equation}
	 by applying $X_{I}$ if the -1 eigenvalue is observed.
 \item measure $X_{I}X_T$ and get
 \begin{eqnarray}
  \vert \psi '' \rangle = 
	 \alpha \vert 0_{C} \rangle (
	 \alpha' \vert 0_{I}0_{T} \rangle + \beta' \vert 0_{I}1_{T} \rangle + \beta' \vert 1_{I}0_{T} \rangle + \alpha' \vert 1_{I}1_{T} \rangle) \nonumber\\
	 + \beta \vert 1_{C} \rangle (
	  \beta' \vert 0_{I}0_{T} \rangle + \alpha' \vert 0_{I}1_{T} \rangle + \alpha' \vert 1_{I}0_{T} \rangle + \beta' \vert 1_{I}1_{T} \rangle	 ) \nonumber\\
	  \label{equ:ls}
 \end{eqnarray}
	 if the +1 eigenvalue is observed, and get
	 \begin{eqnarray}
 \vert \psi ''' \rangle = 
	 \alpha \vert 0_{C} \rangle (
	 \alpha' \vert 0_{I}0_{T} \rangle + \beta' \vert 0_{I}1_{T} \rangle - \beta' \vert 1_{I}0_{T} \rangle - \alpha' \vert 1_{I}1_{T} \rangle	 ) \nonumber\\
	 + \beta \vert 1_{C} \rangle (
	 - \beta' \vert 0_{I}0_{T} \rangle - \alpha' \vert 0_{I}1_{T} \rangle + \alpha' \vert 1_{I}0_{T} \rangle + \beta' \vert 1_{I}1_{T} \rangle) \nonumber\\
	 \end{eqnarray}
	 if the -1 eigenvalue is observed. Apply $Z_CZ_I$ and get Equation \ref{equ:ls} when -1 is observed.
	 Merging $I$ and $T$ by the lattice surgery, the Z operators are XORed and finally we get
	 \begin{eqnarray}
 \vert \psi ^{final} \rangle = 
	 	 \alpha \vert 0_{C} \rangle (
		  \alpha' \vert 0_{m} \rangle + \beta' \vert 1_{m} \rangle) 
		  + \beta \vert 1_{C} \rangle (\beta' \vert 0_{m} \rangle + \alpha' \vert 1_{m} \rangle) \nonumber\\
	 \end{eqnarray}
	 where \textit{m} stands for \textit{merged}, indicating the merged qubit of $I$ and $T$.
\end{enumerate}

\begin{figure}[t]
 \begin{center}
  \includegraphics[width=8cm]{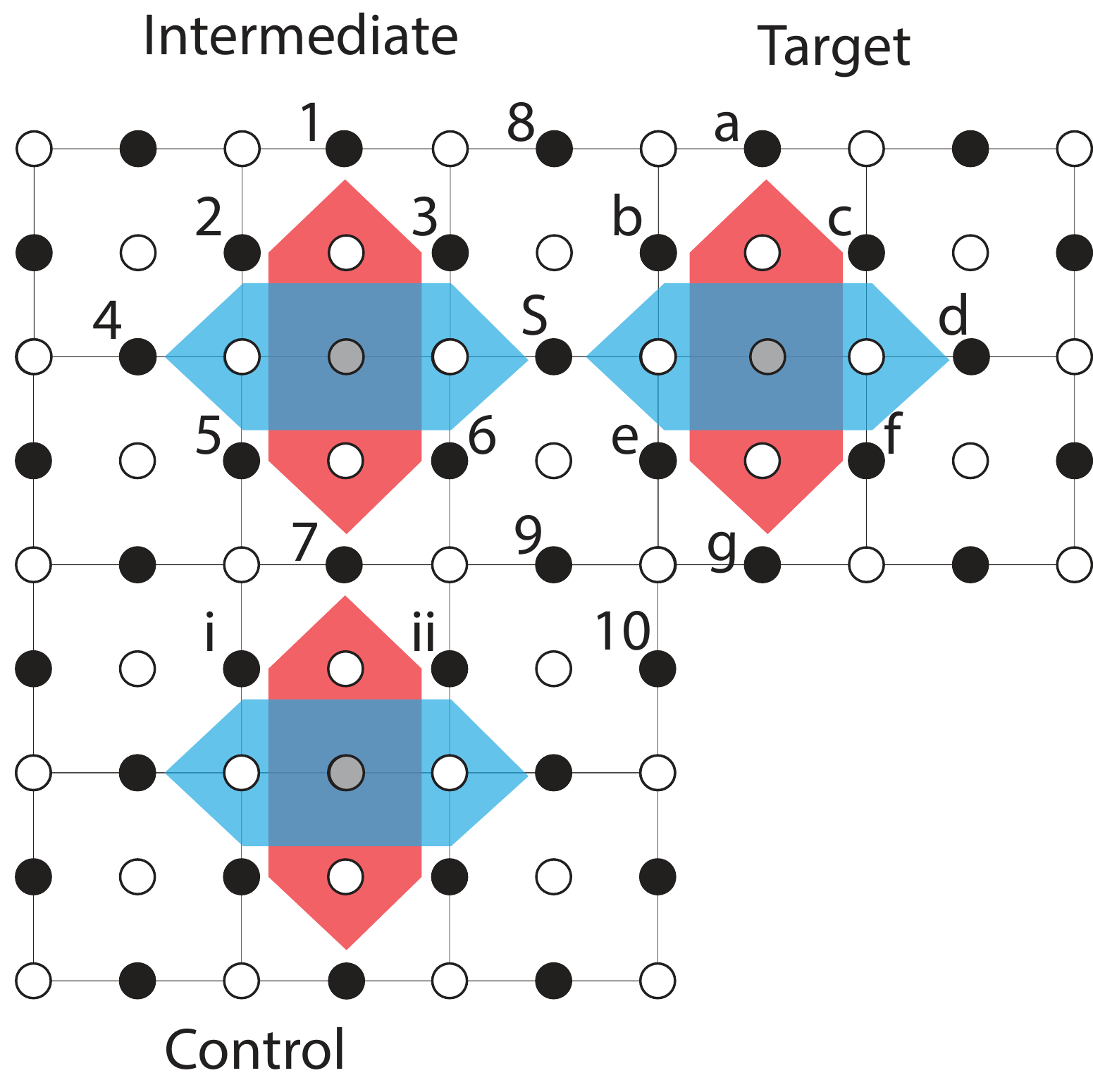}
  \caption[An example of lattice-surgery of deformation-based qubits.]
  {Three deformation-based qubits to demonstrate CNOT gate between the control qubit and the target qubit by lattice-surgery like operations
  in Section \ref{sec:cnot}.
  The intermediate qubit is initialized in $\vert + \rangle$.
  The code distance for those logical qubits is still 3 during lattice surgery.
  }
  \label{fig:lattice_surgery}
 \end{center}
\end{figure}

Figure~\ref{fig:lattice_surgery} depicts the logical CNOT gate of the deformation-based qubit by lattice surgery.

To measure $Z_CZ_{I}$, we measure $Z_5Z_6Z_{i}Z_{ii}$. This is achieved by swapping qubit $7$ with a neighboring ancilla qubit and
using the fault-tolerant stabilizer measurement described in Section~\ref{sec:fast_stabilizer}.
This measurement is repeated $d$ times for majority voting to correct errors, where $d$ is the code distance.
If the -1 eigenvalue is observed from the $Z_CZ_{I}$ measurement, $X_{I}$ is applied.
During the measurement of $Z_CZ_{I}$, we cannot measure the $Z$ superstabilizers of the intermediate qubit and the control qubit, meanwhile normal $Z$ stabilizers can be measured.
Hence, error chains connecting the two $Z$ superstabilizers, such as $X_7$ and $X_6X_9X_{ii}$,
may be caused. (Figure~\ref{sec:cnot} shows distance 3 code, therefore we should not allow an error chain of length 3 to go undetected.)
However, those error chains do not matter since they are stabilizers for $Z_5Z_6Z_{i}Z_{ii}$.

Next, we measure $X_{I}X_T$ and merge the intermediate qubit and the target qubit.
Here we describe the merge operation of deformation-based qubits.
The original state is
\begin{eqnarray}
&&(\alpha \vert 0_{C}0_{I} \rangle + \beta \vert 1_{C}1_{I} \rangle) \otimes (\alpha' \vert 0_{T} \rangle + \beta' \vert 1_{T} \rangle) \nonumber \\
 &=&\alpha\alpha'\vert 0_{C}0_{I}0_{T} \rangle +\alpha\beta'\vert 0_{C}0_{I}1_{T} \rangle \nonumber \\
 &&+\beta\alpha'\vert 1_{C}1_{I}0_{T} \rangle +\beta\beta'\vert 1_{C}1_{I}1_{T} \rangle
\label{equ:merge_orig_state}
 .
\end{eqnarray}
The first term of Equation~\ref{equ:merge_orig_state} is

         \begin{equation}
  \alpha\alpha' \vert0_C\rangle \left(\begin{array}{p{1.5mm}|p{1.5mm}p{1.5mm}p{1.5mm}p{1.5mm}p{1.5mm}p{1.5mm}p{1.5mm}p{1.5mm}p{1.5mm}p{1.5mm}p{1.5mm}p{1.5mm}p{1.5mm}p{1.5mm}p{1.5mm}p{1.5mm}p{1.5mm}}
	    &1&2&3&4&5&6&7&8&9&S&a&b&c&d&e&f&g\\
		    \hline
&$Z$&$Z$&$Z$&&$Z$&$Z$&$Z$&&&&&&&&&&\\
&&$X$&$X$&$X$&$X$&$X$&&&&$X$&&&&&&&\\
&&&&&&&&&&&$Z$&$Z$&$Z$&&$Z$&$Z$&$Z$\\
&&&&&&&&&&$X$&&$X$&$X$&$X$&$X$&$X$&\\
&&&$Z$&&&&&$Z$&&$Z$&&$Z$&&&&&\\
&&&&&&$Z$&&&$Z$&$Z$&&&&&$Z$&&\\
$+$&&&&&$Z$&$Z$&$Z$&&&&&&&&&&\\
$+$&&&&&&&&&&&&&&&$Z$&$Z$&$Z$\\
	  \end{array}  \right)
	   \end{equation}

where the logical state of two qubits exists in $Z_1Z_2Z_3$ and $Z_aZ_bZ_c$.
The two bottom lines are the logical operator states.
Measure qubit $S$ in the $Z$ basis, giving
         \begin{equation}
  \alpha\alpha' \vert0_C\rangle \left(\begin{array}{c|p{1.5mm}p{1.5mm}p{1.5mm}p{1.5mm}p{1.5mm}p{1.5mm}p{1.5mm}p{1.5mm}p{1.5mm}p{1.5mm}p{1.5mm}p{1.5mm}p{1.5mm}p{1.5mm}p{1.5mm}p{1.5mm}p{1.5mm}}
	    &1&2&3&4&5&6&7&8&9&S&a&b&c&d&e&f&g\\
		    \hline
&$Z$&$Z$&$Z$&&$Z$&$Z$&$Z$&&&&&&&&&&\\
&&$X$&$X$&$X$&$X$&$X$&&&&&&$X$&$X$&$X$&$X$&$X$&\\
&&&&&&&&&&&$Z$&$Z$&$Z$&&$Z$&$Z$&$Z$\\
(-1)^{M_Z^S}&&&$Z$&&&&&$Z$&&&&$Z$&&&&&\\
(-1)^{M_Z^S}&&&&&&$Z$&&&$Z$&&&&&&$Z$&&\\
(-1)^{M_Z^S}&&&&&&&&&&$Z$&&&&&&&\\
+&&&&&$Z$&$Z$&$Z$&&&&&&&&&&\\
+&&&&&&&&&&&&&&&$Z$&$Z$&$Z$\\
	  \end{array}  \right).
	   \end{equation}

If -1 eigenvalue is obtained, apply either $X_2X_3X_4X_5X_6X_S$ or $X_bX_cX_dX_eX_fX_S$ and get

         \begin{equation}
  \alpha\alpha' \vert0_C\rangle \left(\begin{array}{p{1.5mm}|p{1.5mm}p{1.5mm}p{1.5mm}p{1.5mm}p{1.5mm}p{1.5mm}p{1.5mm}p{1.5mm}p{1.5mm}p{1.5mm}p{1.5mm}p{1.5mm}p{1.5mm}p{1.5mm}p{1.5mm}p{1.5mm}p{1.5mm}}
	    &1&2&3&4&5&6&7&8&9&S&a&b&c&d&e&f&g\\
		    \hline
&$Z$&$Z$&$Z$&&$Z$&$Z$&$Z$&&&&&&&&&&\\
&&$X$&$X$&$X$&$X$&$X$&&&&&&$X$&$X$&$X$&$X$&$X$&\\
&&&&&&&&&&&$Z$&$Z$&$Z$&&$Z$&$Z$&$Z$\\
&&&$Z$&&&&&$Z$&&&&$Z$&&&&&\\
&&&&&&$Z$&&&$Z$&&&&&&$Z$&&\\
&&&&&&&&&&$Z$&&&&&&&\\
$+$&&&&&$Z$&$Z$&$Z$&&&&&&&&&&\\
$+$&&&&&&&&&&&&&&&$Z$&$Z$&$Z$\\
	  \end{array}  \right).
	   \end{equation}

	   Next, we measure $X_3X_bX_6X_e$ for the third step of lattice surgery.
	   We can measure $X_3$, $X_b$, $X_6$ and $X_e$ both to execute our merge and to measure $X_3X_bX_6X_e$.
	   Measure qubit $3$ in the $X$ basis. If -1 is obtained, apply either $Z_3Z_8Z_b$ or $Z_1Z_2Z_3Z_5Z_6Z_7$.

         \begin{equation}
  \alpha\alpha' \vert0_C\rangle \left(\begin{array}{p{1.5mm}|p{1.5mm}p{1.5mm}p{1.5mm}p{1.5mm}p{1.5mm}p{1.5mm}p{1.5mm}p{1.5mm}p{1.5mm}p{1.5mm}p{1.5mm}p{1.5mm}p{1.5mm}p{1.5mm}p{1.5mm}p{1.5mm}p{1.5mm}}
	    &1&2&3&4&5&6&7&8&9&S&a&b&c&d&e&f&g\\
		    \hline
&$Z$&$Z$&&&$Z$&$Z$&$Z$&$Z$&&&&$Z$&&&&&\\
&&$X$&&$X$&$X$&$X$&&&&&&$X$&$X$&$X$&$X$&$X$&\\
&&&&&&&&&&&$Z$&$Z$&$Z$&&$Z$&$Z$&$Z$\\
&&&&&&$Z$&&&$Z$&&&&&&$Z$&&\\
$+$&&&&&$Z$&$Z$&$Z$&&&&&&&&&&\\
$+$&&&&&&&&&&&&&&&$Z$&$Z$&$Z$\\
	  \end{array}  \right)
	   \end{equation}

Measure qubit $b$ in the $X$ basis. If the -1 is obtained, apply either $Z_1Z_2Z_5Z_6Z_7Z_8Z_b$ or $Z_aZ_bZ_cZ_eZ_fZ_g$.

         \begin{equation}
  \alpha\alpha' \vert0_C\rangle \left(\begin{array}{p{1.5mm}|p{1.5mm}p{1.5mm}p{1.5mm}p{1.5mm}p{1.5mm}p{1.5mm}p{1.5mm}p{1.5mm}p{1.5mm}p{1.5mm}p{1.5mm}p{1.5mm}p{1.5mm}p{1.5mm}p{1.5mm}p{1.5mm}p{1.5mm}}
	    &1&2&3&4&5&6&7&8&9&S&a&b&c&d&e&f&g\\
		    \hline
&$Z$&$Z$&&&$Z$&$Z$&$Z$&$Z$&&&$Z$&&$Z$&&$Z$&$Z$&$Z$\\
&&$X$&&$X$&$X$&$X$&&&&&&&$X$&$X$&$X$&$X$&\\
&&&&&&$Z$&&&$Z$&&&&&&$Z$&&\\
$+$&&&&&$Z$&$Z$&$Z$&&&&&&&&&&\\
$+$&&&&&&&&&&&&&&&$Z$&$Z$&$Z$\\
	  \end{array}  \right)
	   \end{equation}

Measure qubit $6$ in the $X$ basis, and apply either $Z_1Z_2Z_5Z_6Z_7Z_8Z_aZ_cZ_eZ_fZ_g$ if the -1 eigenvalue is observed.

         \begin{equation}
  \alpha\alpha' \vert0_C\rangle \left(\begin{array}{p{1.5mm}|p{1.5mm}p{1.5mm}p{1.5mm}p{1.5mm}p{1.5mm}p{1.5mm}p{1.5mm}p{1.5mm}p{1.5mm}p{1.5mm}p{1.5mm}p{1.5mm}p{1.5mm}p{1.5mm}p{1.5mm}p{1.5mm}p{1.5mm}}
	    &1&2&3&4&5&6&7&8&9&S&a&b&c&d&e&f&g\\
		    \hline
&$Z$&$Z$&&&$Z$&&$Z$&$Z$&$Z$&&$Z$&&$Z$&&&$Z$&$Z$\\
&&$X$&&$X$&$X$&&&&&&&&$X$&$X$&$X$&$X$&\\
$+$&&&&&$Z$&&$Z$&&$Z$&&&&&&$Z$&&\\
$+$&&&&&&&&&&&&&&&$Z$&$Z$&$Z$\\
	  \end{array}  \right)
	   \end{equation}

Measure qubit $e$ in the $X$ basis and apply both $Z_5Z_7Z_9$ as $Z_I$ and $Z_iZ_7Z_{ii}$ as $Z_C$ if the -1 eigenvalue is obtained.

Alternately, we can measure $X_3$, $X_b$, $X_6$ and $X_e$ in parallel.
After the parallel measurements, if an even number of $-1$ eigenvalues is observed, as in normal error correction,
a physical $Z$ operator chain connecting the remaining $X$ stabilizers with $-1$ eigenvalues is executed.
If an odd number of $-1$ eigenvalues is observed,
we execute the physical $Z$ operator chain and there still remains an $X$ stabilizer with $-1$ eigenvalue.
The $X$ superstabilizer of the merged qubit actually has the $-1$ eigenvalue in this case, hence we connect
the remaining $X$ stabilizer and the intermediate qubit side of the $X$ superstabilizer.
This operation keeps the eigenvalues of the lattice $+1$ and works as $Z_I$, like $Z_5Z_7Z_9$ was used in the sequential form above.
We execute $Z_iZ_7Z_{ii}$ as $Z_C$ when an odd number of $-1$ eigenvalue is observed.

Those measurements work for connecting the superstabilizers.
Therefore, those measurements are allowed to be non-fault-tolerant since the remaining stabilizers confirm the correctness of the measurements;
when qubit $e$ is measured in the $X$ basis, regardless of whether a measurement error occurs, if the remaining stabilizer $X_9X_gX_{10}$ outputs -1 repeatedly, we can conclude the correct measurement of qubit $e$ to be -1.

Now we have code space for only one qubit and the two qubits are merged into a qubit
whose logical operator state is the product of the first two, shown in the bottom line,
         \begin{equation}
  \alpha\alpha' \vert0_C\rangle \left(\begin{array}{p{1.5mm}|p{1.5mm}p{1.5mm}p{1.5mm}p{1.5mm}p{1.5mm}p{1.5mm}p{1.5mm}p{1.5mm}p{1.5mm}p{1.5mm}p{1.5mm}p{1.5mm}p{1.5mm}p{1.5mm}p{1.5mm}p{1.5mm}p{1.5mm}}
	    &1&2&3&4&5&6&7&8&9&S&a&b&c&d&e&f&g\\
		    \hline
&$Z$&$Z$&&&$Z$&&$Z$&$Z$&$Z$&&$Z$&&$Z$&&&$Z$&$Z$\\
&&$X$&&$X$&$X$&&&&&&&&$X$&$X$&&$X$&\\
$+$&&&&&$Z$&&$Z$&&$Z$&&&&&&&$Z$&$Z$\\
	  \end{array}  \right).
	   \end{equation}

By similar operations, Equation~\ref{equ:merge_orig_state} is rewritten to
 \begin{flushleft}
  \begin{eqnarray}
  \alpha\alpha' \vert0_C\rangle \left(\begin{array}{p{1.5mm}|p{1.5mm}p{1.5mm}p{1.5mm}p{1.5mm}p{1.5mm}p{1.5mm}p{1.5mm}p{1.5mm}p{1.5mm}p{1.5mm}p{1.5mm}p{1.5mm}p{1.5mm}p{1.5mm}p{1.5mm}p{1.5mm}p{1.5mm}}
	    &1&2&3&4&5&6&7&8&9&S&a&b&c&d&e&f&g\\
		    \hline
&$Z$&$Z$&&&$Z$&&$Z$&$Z$&$Z$&&$Z$&&$Z$&&&$Z$&$Z$\\
&&$X$&&$X$&$X$&&&&&&&&$X$&$X$&&$X$&\\
$+$&&&&&$Z$&&$Z$&&$Z$&&&&&&&$Z$&$Z$\\
	  \end{array}  \right)\nonumber\\
+  \alpha\beta' \vert0_C\rangle \left(\begin{array}{p{1.5mm}|p{1.5mm}p{1.5mm}p{1.5mm}p{1.5mm}p{1.5mm}p{1.5mm}p{1.5mm}p{1.5mm}p{1.5mm}p{1.5mm}p{1.5mm}p{1.5mm}p{1.5mm}p{1.5mm}p{1.5mm}p{1.5mm}p{1.5mm}}
	    &1&2&3&4&5&6&7&8&9&S&a&b&c&d&e&f&g\\
		    \hline
&$Z$&$Z$&&&$Z$&&$Z$&$Z$&$Z$&&$Z$&&$Z$&&&$Z$&$Z$\\
&&$X$&&$X$&$X$&&&&&&&&$X$&$X$&&$X$&\\
$-$&&&&&$Z$&&$Z$&&$Z$&&&&&&&$Z$&$Z$\\
	  \end{array}  \right)\nonumber\\
+  \beta\alpha' \vert1_C\rangle \left(\begin{array}{p{1.5mm}|p{1.5mm}p{1.5mm}p{1.5mm}p{1.5mm}p{1.5mm}p{1.5mm}p{1.5mm}p{1.5mm}p{1.5mm}p{1.5mm}p{1.5mm}p{1.5mm}p{1.5mm}p{1.5mm}p{1.5mm}p{1.5mm}p{1.5mm}}
	    &1&2&3&4&5&6&7&8&9&S&a&b&c&d&e&f&g\\
		    \hline
&$Z$&$Z$&&&$Z$&&$Z$&$Z$&$Z$&&$Z$&&$Z$&&&$Z$&$Z$\\
&&$X$&&$X$&$X$&&&&&&&&$X$&$X$&&$X$&\\
$-$&&&&&$Z$&&$Z$&&$Z$&&&&&&&$Z$&$Z$\\
	  \end{array}  \right)\nonumber\\
+  \beta\beta' \vert1_C\rangle \left(\begin{array}{p{1.5mm}|p{1.5mm}p{1.5mm}p{1.5mm}p{1.5mm}p{1.5mm}p{1.5mm}p{1.5mm}p{1.5mm}p{1.5mm}p{1.5mm}p{1.5mm}p{1.5mm}p{1.5mm}p{1.5mm}p{1.5mm}p{1.5mm}p{1.5mm}}
	    &1&2&3&4&5&6&7&8&9&S&a&b&c&d&e&f&g\\
		    \hline
&$Z$&$Z$&&&$Z$&&$Z$&$Z$&$Z$&&$Z$&&$Z$&&&$Z$&$Z$\\
&&$X$&&$X$&$X$&&&&&&&&$X$&$X$&&$X$&\\
$+$&&&&&$Z$&&$Z$&&$Z$&&&&&&&$Z$&$Z$\\
	  \end{array}  \right).\nonumber\\
\label{eqn:merge_2}
 \end{eqnarray}
 \end{flushleft}
 Using a new definition, we now have
 \begin{eqnarray}
  \vert 0_{m} \rangle = Z_5Z_7Z_9Z_fZ_g\\
  \vert 1_{m} \rangle = -Z_5Z_7Z_9Z_fZ_g
 \end{eqnarray}
 where $m$ stands for $merged$. Equation~\ref{eqn:merge_2} can be written as
 \begin{equation}
  \alpha\vert0\rangle(\alpha'\vert0_{m}\rangle + \beta'\vert1_{m}\rangle) + \beta\vert1\rangle(\beta'\vert0_{m}\rangle + \alpha'\vert1_{m}\rangle)
 \end{equation}
therefore now we have a complete CNOT gate.
From this point in the operation,
we start to measure the new superstabilizers.

\section{Arbitrary size stabilizer measurement}
\label{sec:fast_stabilizer}
We suggest using a cat state of an arbitrary length to measure superstabilizers.
%
%
In this section, we first discuss fault-tolerant preparation, then generic use of cat states for constant-time stabilizer measurement, before addressing superstabilizers in our system.
Finally, we return to the issue of errors.

\subsection{Arbitrary length cat state preparation}
\begin{figure}[t]
  \begin{center}
   \includegraphics[width=6cm]{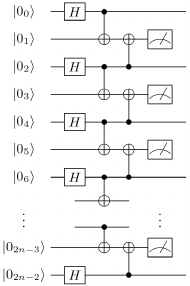}
  \end{center}
  \caption{Non-fault-tolerant circuit to make a $n$-size cat state in 5 steps.}
  \label{fig:cat_state_creation}
 \end{figure}

The non-fault-tolerant circuit to prepare an arbitrary length cat state in constant time is depicted in Figure~\ref{fig:cat_state_creation}.
In the circuit, many qubits in $\vert + \rangle$ are created and entangled by measuring $ZZ$ of every pair of neighboring qubits.
Here, we prepare two qubits in $\vert +_0+_2\rangle$ and a third qubit in $\vert 0_1 \rangle$,
\begin{equation}
 \vert \psi_{012} \rangle = \vert +_0 0_1 +_2 \rangle,
\end{equation}
with this order corresponding to the physical placement.
Dispensing with normalization, as every term has the same amplitude,
apply $CNOT$ for $Z_0Z_2$ measurement:
\begin{align}
 \vert \psi_{012}' \rangle &= CNOT[2,1]CNOT[0,1] \vert +_0 0_1 +_2 \rangle \nonumber \\
 &= \vert 0_00_10_2 \rangle + \vert 0_01_11_2 \rangle + \vert 1_01_10_2 \rangle + \vert 1_00_11_2 \rangle 
\end{align}
where $CNOT[a,b]$ denotes that qubit $a$ is the control qubit and $b$ is the target.
Measure the ancilla qubit $1$ in the $Z$ basis and if the $-1$ eigenvalue is obtained, apply $X_1$ to get
\begin{equation}
\vert\psi_{02}''\rangle = \vert 0_00_2 \rangle + \vert 1_01_2 \rangle.
\end{equation}
We can entangle another qubit in $\vert + \rangle$ to this state in the same way and 
we can make a cat state of arbitrary length.
However, this procedure is not fault-tolerant and there is a chance of getting a problematic state
such as $\vert 00001111 \rangle + \vert 11110000 \rangle$. Using this state for a stabilizer measurement may produce a logical error
because the logical operator of the deformation-based qubit is a half of a superstabilizer.
Therefore we need to confirm that we have a proper cat state.
It is well-known that measuring $ZZ$ of every pair of qubits comprising the cat state
is good enough for this proof~\cite{nielsen-chuang:qci}.
Since measuring $ZZ$ of every pair of qubits requires many SWAP gates and a lot of steps,
we suggest repeating the $ZZ$ measurement of every pair of neighboring qubits $d$ times,
which guarantees the probability that the state is in a problematic state is $O(p^{\lceil\frac{d}{2}\rceil})$,
where $d$ is the code distance and $p$ is the physical error rate,
which is same as the error rate the fault tolerant quantum computation achieves.
(The state yet could be an imperfect cat state such as $\vert 00100000 \rangle + \vert 11011111 \rangle$
due to individual physical errors, which is tolerable.)
 
 \subsection{Stabilizer measurement in constant time using cat state}
A three qubit cat state can be rewritten as
 \begin{align}
  \vert \psi_{cat}\rangle = &\vert 000 \rangle + \vert 111 \rangle \nonumber\\
  =& (\vert + \rangle + \vert - \rangle)(\vert + \rangle + \vert - \rangle)(\vert + \rangle + \vert - \rangle)\nonumber\\
  & + (\vert + \rangle - \vert - \rangle)(\vert + \rangle - \vert - \rangle)(\vert + \rangle - \vert - \rangle)\label{equ:cat_0}\\
  =& \vert +++ \rangle + \vert +-- \rangle + \vert -+- \rangle + \vert --+ \rangle.
  \label{equ:cat_1}
 \end{align}

 The $\vert 000 \rangle$ and $\vert 111 \rangle$ are rewritten in symmetric fashion
 except that the signs of factors involving an odd number of $\vert-\rangle$ differs, as shown in Equation~\ref{equ:cat_0}.
 From this fact and the binomial expansion, a cat state of any length involves an even number of $\vert-\rangle$.
 Applying a $Z$ to any qubit in the cat state, the state in Equation~\ref{equ:cat_1} is changed to 
\begin{equation}
\vert \psi_{cat}' \rangle = \vert -++ \rangle + \vert --- \rangle + \vert ++- \rangle + \vert +-+ \rangle.
\end{equation}
Applying a $Z$ to any qubit again, this state returns to the state in Equation~\ref{equ:cat_1}.
To observe whether we have the ``even'' cat state or the ``odd'' cat state,
we need to measure all ancilla qubits in the $X$ basis and calculate the product of the measured values.

Let us assume that
we have as many ancillae for the cat state as we have 
data qubits to stabilize, and we can assign a qubit in the cat state to each data qubit,
then apply $CNOT$ from each cat state qubit to the corresponding data qubit.
This set of $CNOT$s is equivalent to the syndrome propagation for the $X_1X_2...X_n$ stabilizer.
The cat state starts from the ``even'' state and if an odd number of flips is performed the cat state results in the ``odd'' state.
The $CNOT$s can be applied simultaneously and the measurement can be performed simultaneously,
therefore this procedure requires three steps ($CNOT$, Hadamard and measurement in $Z$ basis).
  
\subsection{Superstabilizer implementation}
  To suppress the probability of having an improper cat state to $O(p^{\lceil\frac{d}{2}\rceil})$,
  a linear placement requires $d$ cycles of $ZZ$ stabilizers,
  but a circular placement requires only $\lceil\frac{d}{2}\rceil$ cycles.
  Let us assume that an example of the problematic states, $\vert 0_00_10_20_31_41_51_61_7 \rangle + \vert 1_01_11_21_30_40_50_60_7 \rangle$
  has developed.
  In a linear arrangement, we have $ZZ$ stabilizers only between neighboring qubits.
  After $d$ cycles of $ZZ$ stabilizers,
  the problematic state generation is caused
  by $p^{\lceil\frac{d}{2}\rceil}$ errors at the $Z_3Z_4$ stabilizer.
  In a circular arrangement, we have another $Z_7Z_0$ stabilizer.
  Hence, after $d$ cycles of $ZZ$ stabilizers,
  even though we have $p^{\lceil\frac{d}{2}\rceil}$ errors at the $Z_3Z_4$ stabilizer,
  $d$ cycles of the $Z_7Z_0$ stabilizer tell us that we have an improper cat state.
  Therefore for instance
  $p^{\lceil\frac{d}{2}\rceil}$ errors at the $Z_3Z_4$ stabilizer and
  $p^{\lceil\frac{d}{2}\rceil}$ errors at the $Z_7Z_0$ stabilizer
  are required to generate a problematic cat state after $d$ cycles of $ZZ$ stabilizers,
  suppressing the improper cat state generation probability to $O(p^{d})$.
  Hence, to suppress the error probability to $O(p^{\lceil\frac{d}{2}\rceil})$,
  circular fashion cat state generation requires only $\lceil\frac{d}{2}\rceil$ cycles of $ZZ$ stabilizers.

\begin{figure}[t]
   \begin{center}
    \includegraphics[width=8cm]{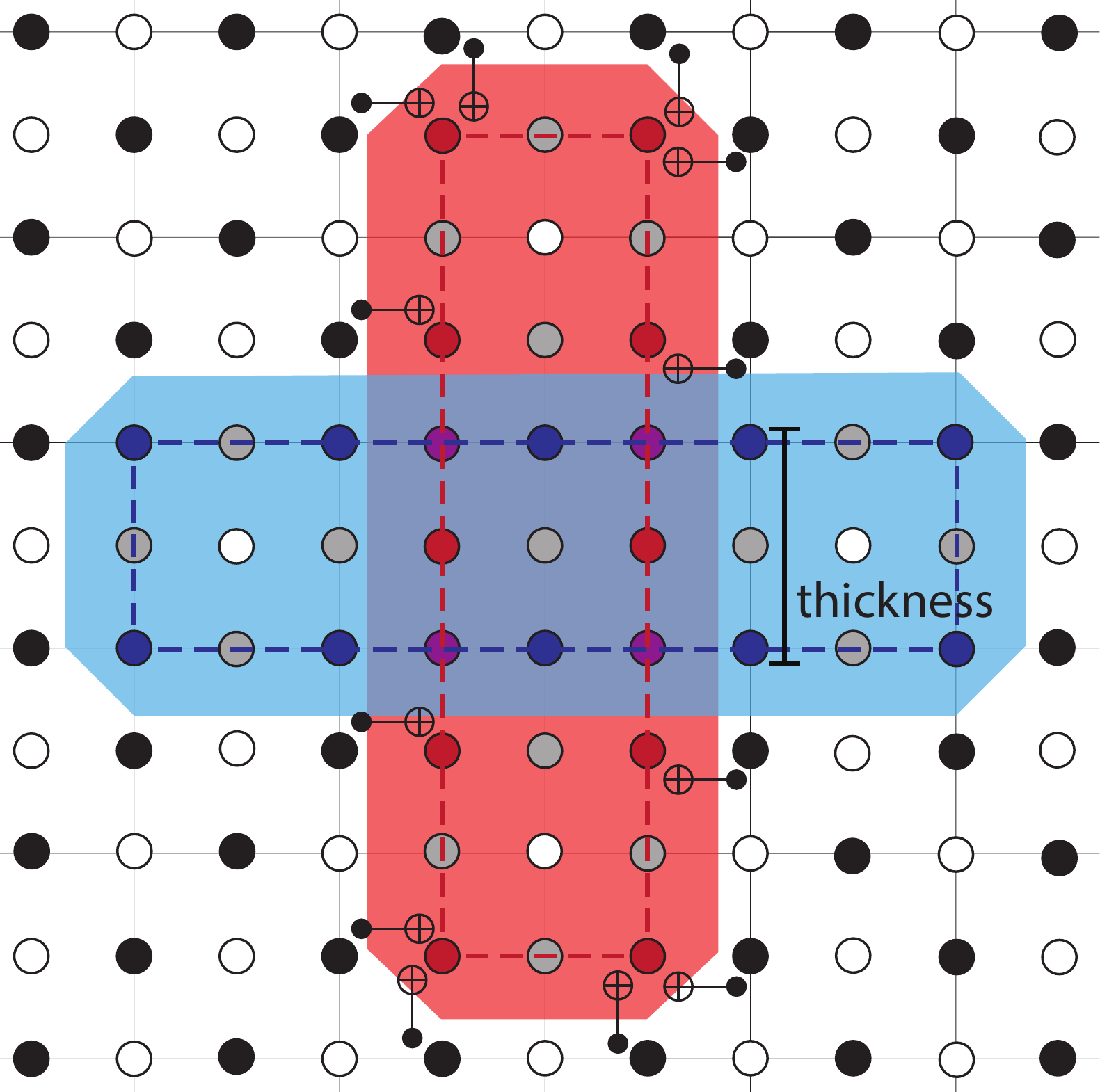}
    \caption[Implementation of two cat states for the two superstabilizers of a deformation-based qubit.]
    {The red dots are ancilla qubits prepared in a cat state for the $Z$ superstabilizer.
    The red dashed loop describes the pairs for $ZZ$ stabilizers to create and confirm the cat state.
    The $ZZ$ stabilizer on each pair of neighboring red dots in this red dashed loop is executed.
    The gray qubits under the red dashed circle are qubits with odd indices in Figure~\ref{fig:cat_state_creation},
    used to measure $ZZ$ stabilizers.
    So as the blue dots and the blue dashed circle for the $X$ superstabilizer.
    The dots under the crosses of the dashed circles are used for both cat state creation alternately.
    The ``thickness'' of this deformation-based qubit is 2.
    The $CNOT$ gates of the $Z$ superstabilizer are shown. Each ancilla qubit on the corner of the loop
    handles two data qubits and those along the sides handle one.
    }
    \label{fig:superstab}
   \end{center}
  \end{figure}

  Figure ~\ref{fig:superstab} depicts the placement of two sets of ancilla qubits,
  each of which is prepared in a cat state for the $X$ superstabilizer and for the $Z$ superstabilizer.
  The dashed lines describe the cat state qubits; red (blue) dots are qubits
  composing the cat state for the $Z$ ($X$) superstabilizer and
  gray dots are ancilla qubits to create and confirm the cat state (the ancillas' ancilla).
  The qubits under both dashed lines are used for the $Z$ and $X$ ancilla qubits alternately.
  Therefore we need $\frac{d}{2} \times 2 = d$ cycles to measure both the $Z$ superstabilizer and the $X$ superstabilizer.
  The ``thickness'' of the deformation-based qubit in Figure~\ref{fig:superstab} is 2 to allow us to have the loop cat state.
  Greater thickness requires fewer cycles of repeating $ZZ$ stabilizer to confirm the cat state.
  We assume that the thickness is 2 through the rest of this dissertation to show the basic idea of our architecture.

  The depth of the circuit to initialize a cat state is five.
  A cycle of the following $ZZ$ measurements for the proof requires four steps.
  The maximum number of \textsc{CNOT}s to propagate error syndromes from data qubits to an ancilla qubit is 2,
  as shown in Figure ~\ref{fig:superstab}, at the corners of the superstabilizers.
  The total number of steps to measure a superstabilizer is the sum of 
  $5 + 4 (d-1)=4d+1$ steps for cat state creation and the proof,
  $1$ step for a Hadamard gate for $Z$ superstabilizer,
  $2$ steps for syndrome propagation,
  $1$ step for a Hadamard gate for $X$ superstabilizer,
  $1$ step for measurements, where $d$ is the $code{\ }distance$.
  Therefore the number of steps to measure a superstabilizer is $4 d + 5$.

  \begin{figure}[t]
   \begin{center}
    \includegraphics[width=8cm]{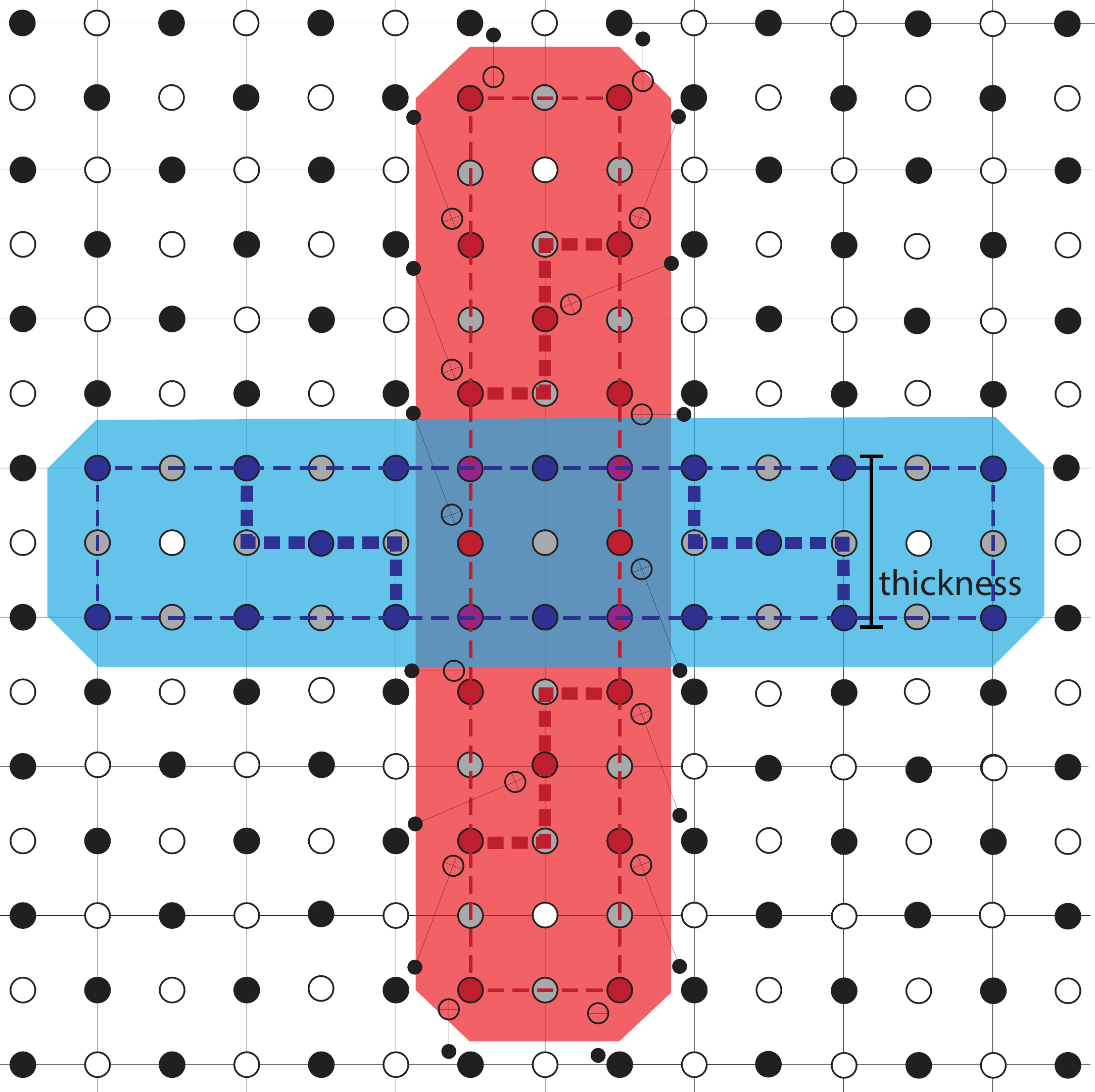}
    \caption[$Z$ superstabilizer in which a cat state qubit stabilizes a data qubit.]
    {Qubits on the thick dashed lines are newly added to the cat state qubits.
    It does not matter that cat state qubit on the cross of a thick dashed line and a thin dashed line
    is stabilized by three stabilizers for the proof of the correctness of the cat state
    since one cycle of stabilization for the proof takes four steps.
    Non-neighbor $CNOT$ gates are executed after SWAP gates to neighbor the control and the target qubits.
    }
    \label{fig:superstab2}
   \end{center}
  \end{figure}
However, in Figure \ref{fig:superstab}, two data qubits neighboring a corner of a loop cat state execute CNOT gates with the corner qubit
so that an error on the corner qubit may propagate to the two data qubits,
which may reduce the error suppression ability of the code.
By judicious use of the unused qubits, we can recover the code distance lost, as shown in Figure \ref{fig:superstab2}.
In Figure \ref{fig:superstab}, for simplicity we show thickness $t=2$ employing a cat state forming a complete loop,
in which each corner cat state qubit stabilizes two data qubits, resulting in reducing the effective code distance by $2$.
Figure \ref{fig:superstab2} shows that, by utilizing unused physical qubits inside a superstabilizer,
we can add more qubits to the cat state and can allow every cat state qubit to stabilize a data qubit.
This improvement can be applied with code distance $8$ or higher.
This process is the same as the previous one, except that only one step is required for propagation.
The first SWAP gates overlap with the measurements, then we add
$1$ step for the second SWAP gates,
  $1$ steps for syndrome propagation of ranged pairs,
  $1$ step for a Hadamard gate for $X$ superstabilizer,
  $1$ step for measurements.
to replace a corner cat state qubit with one made inside the superstabilizer, followed by error syndrome propagation and measurement.
In total, $4d+9$ steps are required.

\section{Errors}
\label{sec:errors}
\begin{figure}[t]
   \begin{center}
     \includegraphics[width=8cm]{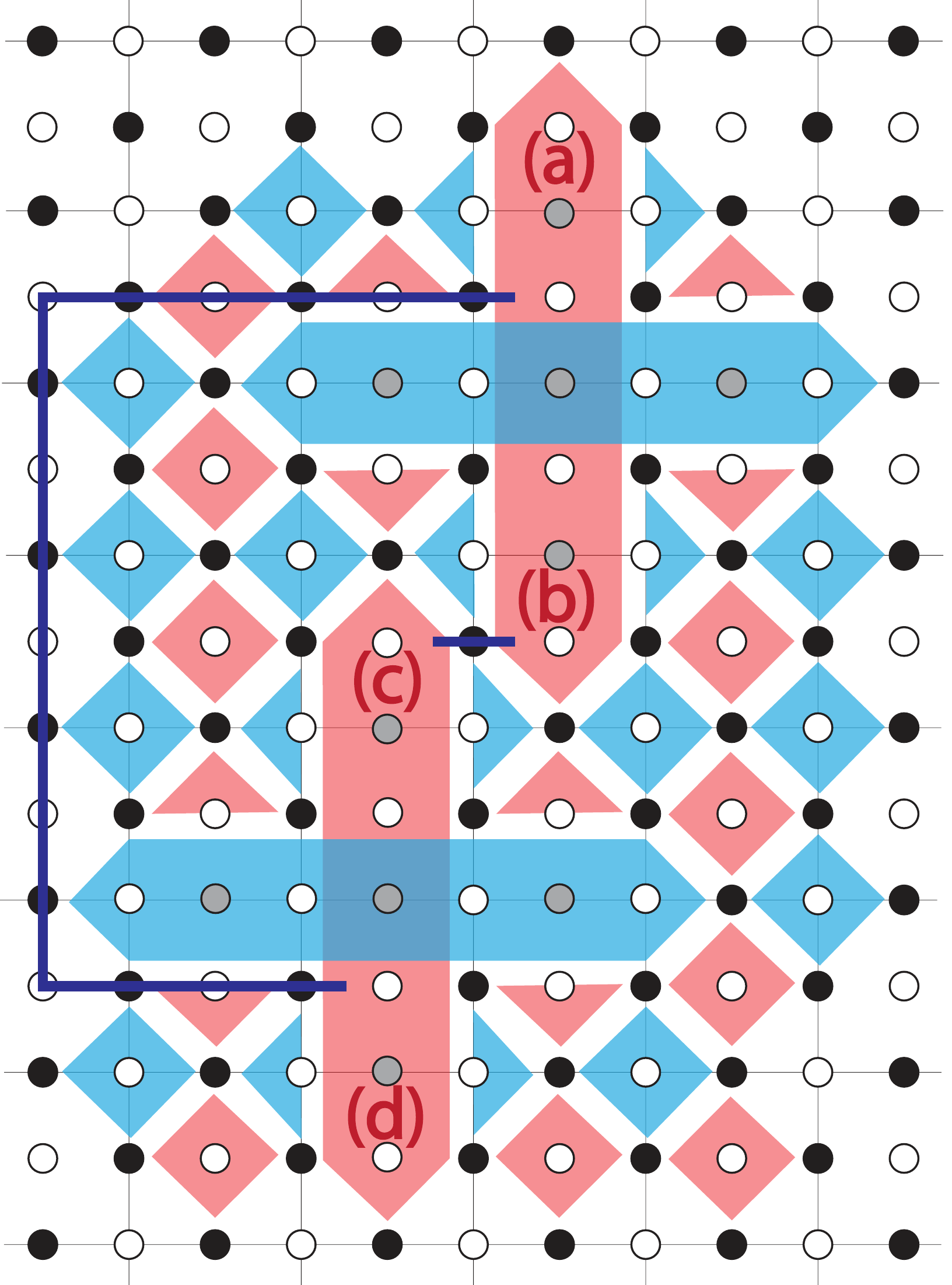}
    \caption[Errors on deformation-based qubits due to the long execution time to measure superstabilizers.]{
    Either (a) or (b) is a half of a $Z$ superstabilizer.
    A physical $X$ error chain connecting those halves results in a logical $X$ error for this deformation-based qubit.
    So are (c) and (d).
    The set of blue lines describes the product of logical $X$ operators of the two deformation-based qubits.
    }
    \label{fig:operator6}
   \end{center}
  \end{figure}

  Though it might be thought that
  the deep circuit of the superstabilizer measurement results in a
  higher logical error rate than another surface code in which
  any stabilizer requires 8 steps,
  we argue that the deformation-based surface code will exhibit a similar logical error rate with the conventional surface code.
  Figure~\ref{fig:operator6} shows an example of two deformation-based qubits.
  Obviously, any single logical operator is protected by code distance 5,
  as shown in Figure~\ref{fig:correction}.
  Any single operator is protected by normal stabilizers at every $8$ physical steps.
  Therefore conventional error analysis for surface code can be applied.

  The pair of blue lines in Figure~\ref{fig:operator6} indicates the product of
  the two logical qubits' logical $X$ operators.
  In order for a logical error to arise undetected,
  both error chains must occur.
  The short fragment of the operator product between (b) and (c) may occur easily
  and will be detected only by superstabilizer measurements,
  which are completed at every $4d + 5$ physical steps.
  The long fragment of the operator product between (a) and (d) should occur only rarely,
  because the long fragment is protected by normal stabilizers
  and has a longer length than the code distance.
  Therefore the probability that this product operator happens to be executed by errors is
  strongly suppressed, though (b) and (c) are close and $4 d + 5$ physical steps are required to measure superstabilizers.

\begin{figure}[t]
   \begin{center}
    \includegraphics[width=8cm]{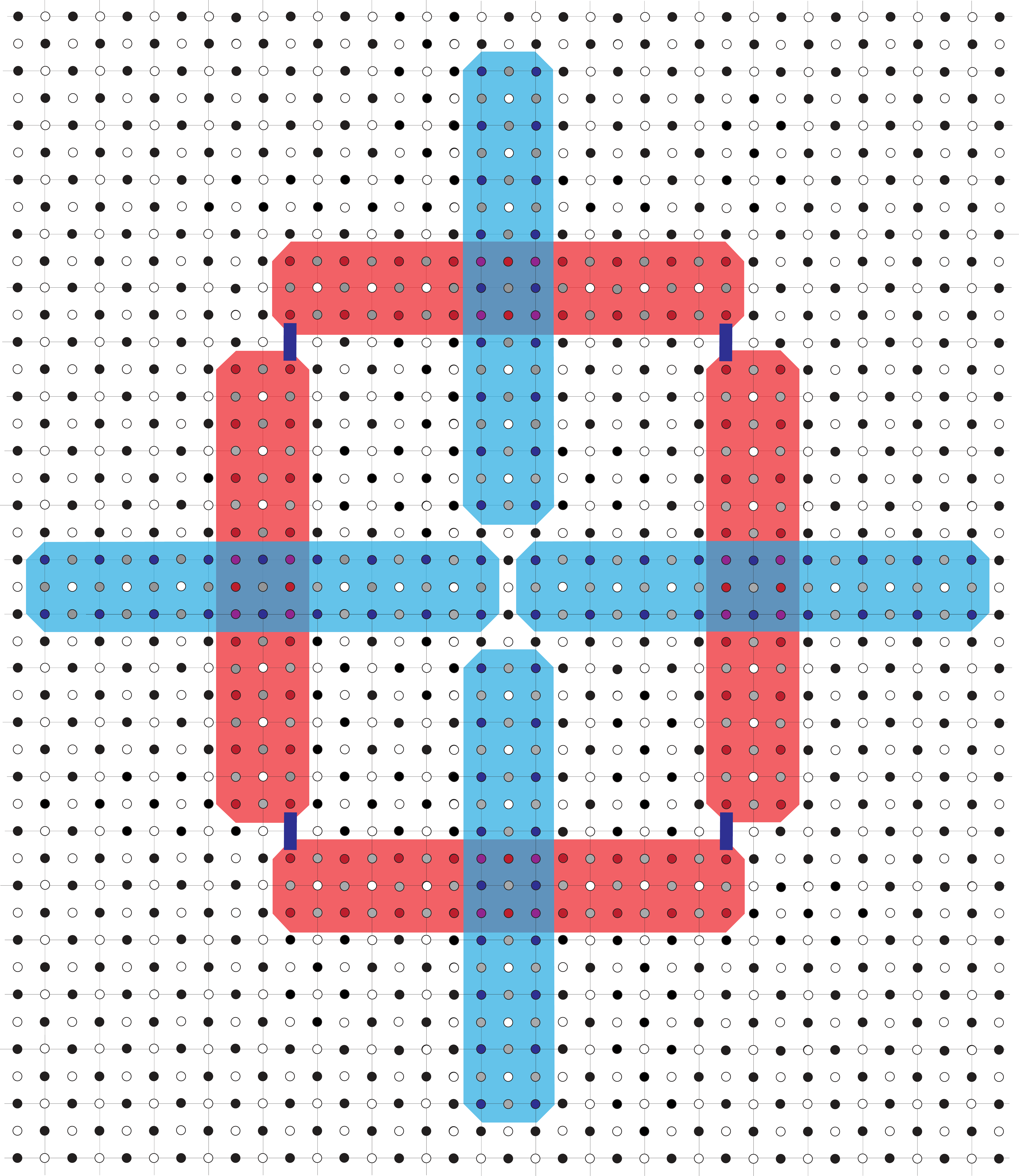}
    \caption[Problematic placement of deformation-based qubits.]{
    Each deformation-based qubit has code distance 10.
    The shortest combined logical $X$ operator for those four logical qubits
    is only 4, the combination of the shown blue lines.
    }
    \label{problematic_placement}
   \end{center}
  \end{figure}

  Figure~\ref{problematic_placement} shows a problematic placement of deformation-based qubits.
  The code distance of each deformation-based qubit is 10.
  However, the product of the four logical $X$ operators of those deformation-based qubits results in
  the combination of the four blue lines,
  each of which exists between two neighboring $Z$ superstabilizers,
  consisting of only four physical qubits,
  reducing our minimum error chain to $4$.
  Deformation-based qubits must be placed so that their superstabilizers do not form a loop.

  In the next section, we present dense packing that meets these constraints, then continue the discussion of errors.

\section{Summary}
This Chapter gives the mechanism of the new deformation-based code.
By utilizing the combination of triangular stabilizers and superstabilizers,
we can design the shapes of shortest logical operators on the 2-D lattice.

In this Chapter I give the concept of the new deformation-based code and the denser packing.
I also give how to transform and move the new code in the surface code lattice.
State injection by conversion from defect-based code
and $CNOT$ gate by lattice surgery
are also described hence a universal gate set is fulfilled.
Concrete and practical method to measure large superstabilizers is demonstrated too.

In Section \ref{sec:deformation:eval},
I analyze
the resource requirements of the new code
and show the advantages than other surface codes.

\clearpage
\chapter{Scalable Bell Pair delivery with code interoperability}
\label{chap:hetero}
This Chapter shows
the scheme designed to create heterogeneously encoded Bell pairs.
This scheme is utilized at interconnections between components in the proposed quantum computer architecture.
As discussed in Chapter \ref{chap:sdqcarch},
quantum CPUs, memories and other components have different requirements depending on their roles.
Hence the most suitable quantum error correcting code for each component must differ
since quantum error correcting codes have different characteristics, advantages and disadvantages.
Heterogeneously encoded Bell pairs are used to establish interoperability.

The scalability of interconnection will hit the performance of a quantum computer.
If communication between a set of components blocks communication between another set of components,
operations in the latter components have to stall.
Hence the scalable method to create heterogeneously encoded Bell pairs between arbitrary two components is required.

\section{Overview of the heterogeneously encoded Bell pairs}
The optical crossbar switch demonstrated by Kim et al. can be a core module for switching internal connections of a quantum computer~\cite{PhysRevA.89.022317,ahsan2015designing,kim03:_1100_port_mems}.
Such an architecture for physical systems which have direct interaction with optics
or which have conversion via several hops, such as optical photon - ferromagnet magnon - microwave photon - superconducting qubit
~\cite{Tabuchi405,PhysRevLett.113.083603,PhysRevLett.116.223601,PhysRevB.93.174427}.
In Kim's architecture, many logical qubit devices are connected to a crossbar switch
and the crossbar switch switches interconnections among those logical qubits for multiple-qubit gates.
This architecture is scalable
since connections do not block each other and
since we can enlarge the crossbar and can install new chips and new network interface cards.

A concern is that there are components of different error correcting codes.
The main quantum computation chips, the complementary computation chips and the network links hence the network interface cards
may employ different codes.
To bridge components of different codes, 
either converting a logical qubit from one code to another,
or building entanglement between two logical qubits in separate codes is required.
Direct code conversion transforms an encoded state
$\vert \psi \rangle_L$ into an encoded state $\vert \psi \rangle_{L'}$ where $L$ and $L'$ indicate two distinct codes.
Since this change operates on valuable data, the key point is
to find an appropriate fault-tolerant sequence that will convert the
stabilizers from one code to the other
~\cite{PhysRevLett.113.080501,Hill:2013:FQE:2481614.2481619,PhysRevA.77.062335}.
Entanglement spanning two separate codes allows us to perform code
teleportation. We employ a heterogeneously encoded Bell pair, in which
each half of the pair is encoded in a separate QEC code. Therefore,
the key point is the method for preparing such a state.
The concern of this architecture is now the fault-tolerance of the heterogeneously encoded Bell pairs.

We give the detailed analysis of the generalized approach to create heterogeneously encoded Bell pairs
for interoperability of quantum components of different error correcting codes.
We evaluate this approach between the Steane [[7,1,3]] code, a distance three surface code,
and unencoded (raw) physical qubits.
We chose those two codes because they have simple structure, are well-investigated and will clearly demonstrate the principle of interconnection.

We have studied three possible schemes to increase the fidelity of the heterogeneously encoded Bell pairs:
{\it purification before encoding}, {\it purification after encoding} and {\it purification after encoding with strict post-selection}.
{\it Purification before encoding} does entanglement purification at the level of physical Bell pairs.
{\it Purification after encoding} does entanglement purification at the level of encoded Bell pairs.
{\it Purification after encoding with strict post-selection} also does entanglement purification at the level of encoded Bell pairs.
The difference from the previous scheme is that encoded Bell pairs in which any eigenvalue (error syndrome) of -1 is measured
in the purification stage are discarded and the protocols restarted.
(Because stabilizers are only measured during logical purification, the fourth combination of physical
purification with strict post-selection does not exist.)
We determine the error probability and the resource efficiency of these schemes by Monte Carlo simulation
with the Pauli error model of circuit level noise~\cite{landahl:arXiv:1108.5738}.

\section{Heterogeneously encoded Bell pairs}
There are two methods for building heterogeneously encoded Bell pairs for code teleportation.
The first is to inject each qubit of a physical Bell pair to a different code
~\cite{copsey02:_quant-mem-hier}.
The second is to prepare a common cat state for two codes to check the ZZ parity of two logical qubits
~\cite{976922,Thaker:2006:QMH:1150019.1136518,doi:10.2200/S00331ED1V01Y201101CAC013}.
It has been shown that code teleportation utilizing a cat state is better than direct 
code conversion because the necessary stabilizer checking for the latter approach is too expensive~\cite{6675561}.
Direct code conversion and code teleportation utilizing a cat state are specific for a chosen code pair as the specific 
sequence of fault-tolerant operations has to match the two codes chosen.
In contrast, code teleportation by injecting a physical Bell pair can be used for any two codes, and provided encoding circuits 
are available for the two codes in question, the protocol can be generalized to arbitrary codes.

Putting things together, heterogeneous Bell pairs of long distance can be created
by entanglement swapping (physical or logical) or a method appropriate to each network, allowing
an arbitrary quantum state encoded in some code 
to be moved onto another code by teleportation
~\cite{PhysRevA.79.032325,PhysRevLett.104.180503}.
In a single computer, code conversion has been proposed for memory hierarchies and for cost-effective fault tolerant quantum computation
~\cite{copsey02:_quant-mem-hier,Thaker:2006:QMH:1150019.1136518,PhysRevLett.113.080501,PhysRevLett.112.010505,PhysRevLett.111.090505,choi2015}.

\begin{figure}[t]
 \begin{center}
  \includegraphics[width=9cm]{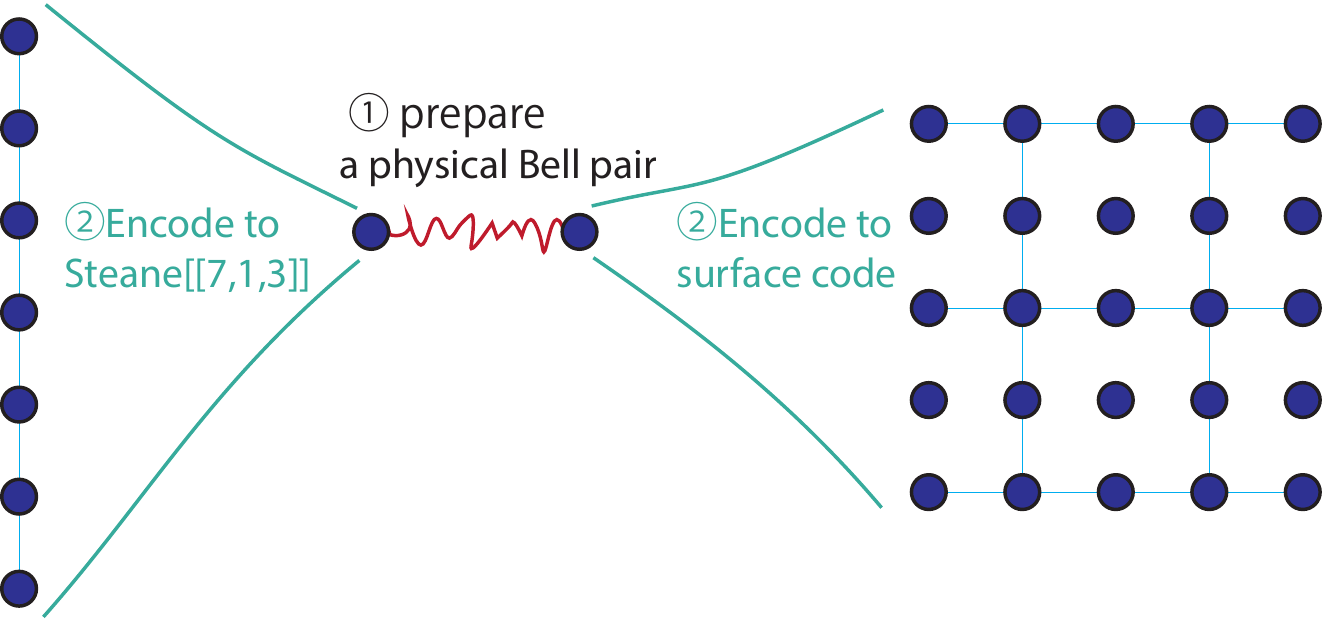}
  \caption[Procedure for encoding a Bell pair heterogeneously.]
  {A qubit of a Bell pair is encoded onto Steane[[7,1,3]] on the left side of the figure, adding 6 ancilla qubits..
  The other qubit of the Bell pair is encoded onto the surface code of distance 3, adding 24 ancilla qubits.
  on the right side of the figure. Eventually, a heterogeneously encoded logical Bell pair
  is achieved.}
  \label{fig:heterogeneousEncode}
 \end{center}
\end{figure}

Figure \ref{fig:heterogeneousEncode} shows the basic procedure for creating
a heterogeneously encoded logical Bell pair. Each dot denotes a
physical qubit and thin blue lines connecting those dots demark the set of physical qubits
comprising a logical qubit.
Each qubit of a Bell pair is processed separately and encoded onto its respective code
through non-fault-tolerant methods to create arbitrary encoded states.

\begin{figure}[t]
   \begin{center}
    \includegraphics[width=9cm]{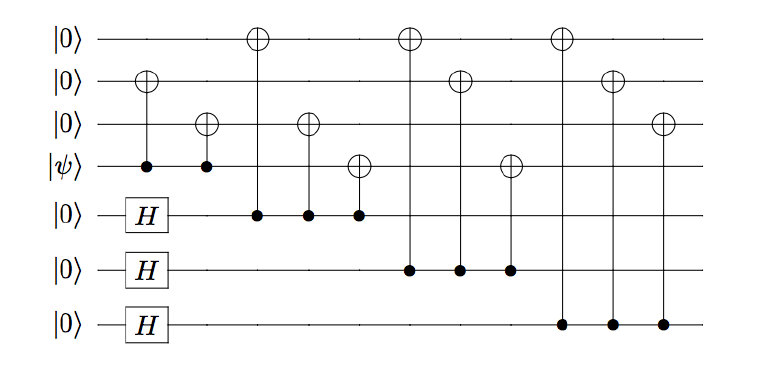}
    \caption[{Circuit to encode an arbitrary state to the Steane [[7,1,3]] code}~\cite{Sidney:10.1007:s11128-012-0414-7}.]
    {$\vert \psi \rangle$ is the state to be encoded.
  This circuit is not fault-tolerant. The KQ of this circuit is 42 because some gates can be performed simultaneously.}
  \label{fig:CSS7qubitEncode}
   \end{center}
\end{figure}

\begin{figure}[t]
   \begin{center}
    \includegraphics[width=9cm]{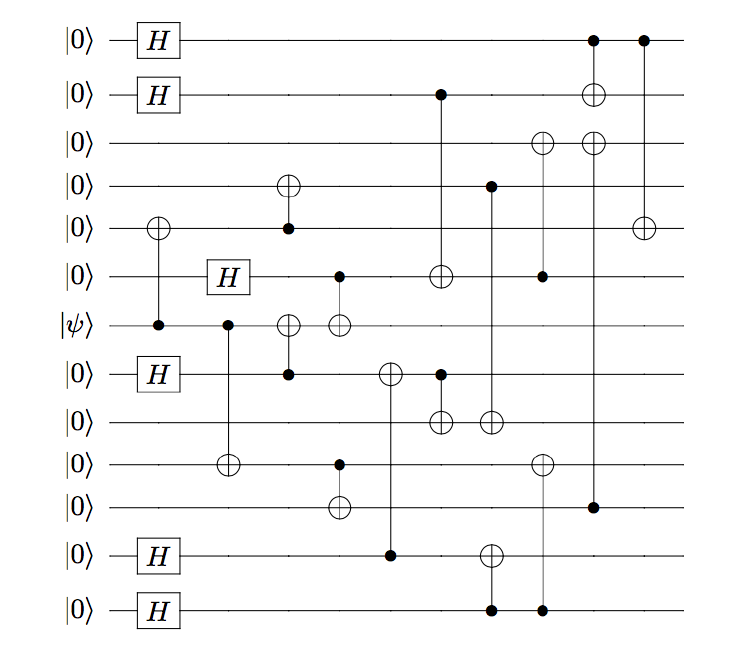}
    \caption[Circuit to encode an arbitrary state $\vert \psi \rangle$ to a distance three surface code
  ~\cite{:/content/aip/journal/jmp/43/9/10.1063/1.1499754}.]
    {This circuit is not fault-tolerant.
  The KQ of this circuit is 250 if some gates are performed simultaneously.
  }
  \label{fig:SurfaceCodeDist3Encode}
   \end{center}
\end{figure}

Figure \ref{fig:CSS7qubitEncode} shows the circuit to encode an arbitrary quantum state
in the Steane[[7,1,3]] code~\cite{steane:10.1098/rspa.1996.0136,steane:10.1038/20127}.
Figure \ref{fig:SurfaceCodeDist3Encode} shows the circuit to encode an arbitrary quantum state
in the surface code~\cite{:/content/aip/journal/jmp/43/9/10.1063/1.1499754}.

The KQ of a circuit is the number of qubits times the circuit depth, giving an estimate of the number of opportunities for errors to occur~\cite{steane:10.1098/rspa.1996.0136}.
Note that those circuits are not required to be fault-tolerant because the state being purified is generic,
rather than inreplaceable data.
If the fidelity of the encoded Bell pair is not good enough
(e.g. as determined operationally using quantum state tomography),
entanglement purification is performed~\cite{bravyi2005uqc,PhysRevA.73.062309}.

 \section{Three Methods to Prepare a Heterogeneously Encoded High Fidelity Bell Pair}
Entanglement purification is performed to establish high fidelity entanglement
~\cite{duer03:_pur-qc,dur2007epa}.
Entanglement purification can be viewed as a distributed procedure for testing a proposition about a distributed state~\cite{vanmeter2014book}.

\begin{figure}[t]
   \begin{center}
    \includegraphics[width=9cm]{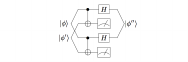}
 \caption[Circuit for entanglement purification~\cite{dur:PhysRevA.59.169}.]{The two measured values are compared.
  If they disagree, the output qubits are discarded.
  If they agree, the output qubits are treated as a new Bell pair.
  At this point, the X error rate of the output Bell pair is suppressed from the input Bell pairs.
  The Hadamard gates exchange the X and Z axes, so that the following round of purification suppresses the Z error rate.
  As the result, entanglement purification consumes two Bell pairs and
  generates a Bell pair of higher fidelity stochastically.
 }
 \label{sc:circ_purification}
   \end{center}
\end{figure}
Figure \ref{sc:circ_purification} shows the circuit for the basic form of entanglement purification where $\vert\phi\rangle$ is a noisy Bell pair.
The input is two low fidelity Bell pairs and on success the output is a Bell pair of higher fidelity.
 One round of purification suppresses one type of error, X or Z.
If the initial Bell pairs are Werner states, or approximately Werner states, then to suppress both types, two rounds of purification are required.
The first round makes the resulting state into a binary state with only one significant error term but not a significantly improved fidelity.
The second round then strongly suppresses errors if the gate error rate is small.
Thus, the overall fidelity tends to improve in a stair step fashion.
After two rounds of purification, the distilled fidelity will be, in the absence of local gate error,
\begin{equation}
 F'' \sim \frac{F^2}{F^2 + (1-F)^2}
\end{equation}
where the original state is the Werner state
\begin{equation}
 \rho = F|\Phi^+\rangle\langle\Phi^+| + \frac{1-F}{3}(|\Phi^-\rangle\langle\Phi^-| + |\Psi^+\rangle\langle\Psi^+|+ |\Psi^-\rangle\langle\Psi^-|)
\end{equation}
and $F$ is the fidelity $F=\langle\phi|\rho|\phi\rangle$ if $|\phi\rangle$ is the desired state.
The probability of success of a round of purification is
\begin{equation}
 p = F^2 + 2F\frac{1-F}{3} + 5\left(\frac{1-F}{3}\right)^2.
\end{equation}
Table \ref{tab:purificationresult} in the appendix provides the numerical data for this to compare with our protocols.
Our simulation assumptions are detailed in section \ref{sec:hetero:sim}.

\subsection{{\it Purification before encoding}}
\begin{figure}[t]
  \begin{center}
   \includegraphics[width=9cm]{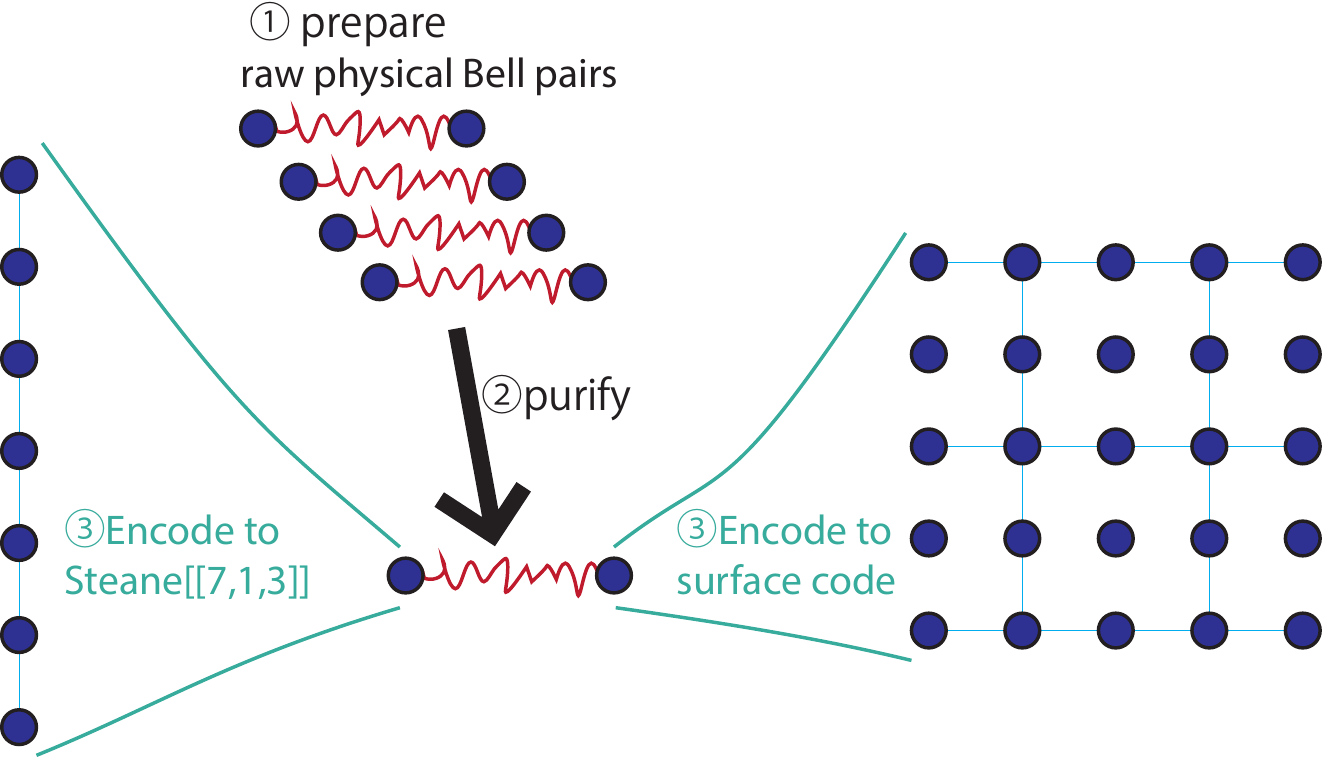}
   \caption[Overview of the scheme which purifies physical Bell pairs to generate an encoded Bell pair of high fidelity.]
   {First, entanglement purification is conducted between physical Bell pairs an arbitrary number of times.
   Second, each qubit of the purified physical Bell pair is encoded to heterogeneous error correcting code.}
   \label{fig:overview_purify_before_encode}
  \end{center}
\end{figure}
Figure \ref{fig:overview_purify_before_encode} shows the overview of the scheme to make heterogeneously encoded Bell pairs
that are purified before encoding.
To create an encoded Bell pair of high fidelity,
entanglement purification is repeated the desired number of times.
Next, each qubit of the purified Bell pair is encoded to its respective error correcting code.
To estimate the rate of {\em logical} error after encoding, we perform a perfect 
syndrome extraction of the system to remove any residual correctable errors.
After the whole procedure finishes, we check whether logical errors exist.
 Table \ref{tab:purify_before_encode} in the appendix presents the details of the simulated error probability and resource efficiency of {\it purification before encoding}.

\subsection{{\it Purification after encoding}}
\begin{figure}[t]
  \begin{center}
   \includegraphics[width=9cm]{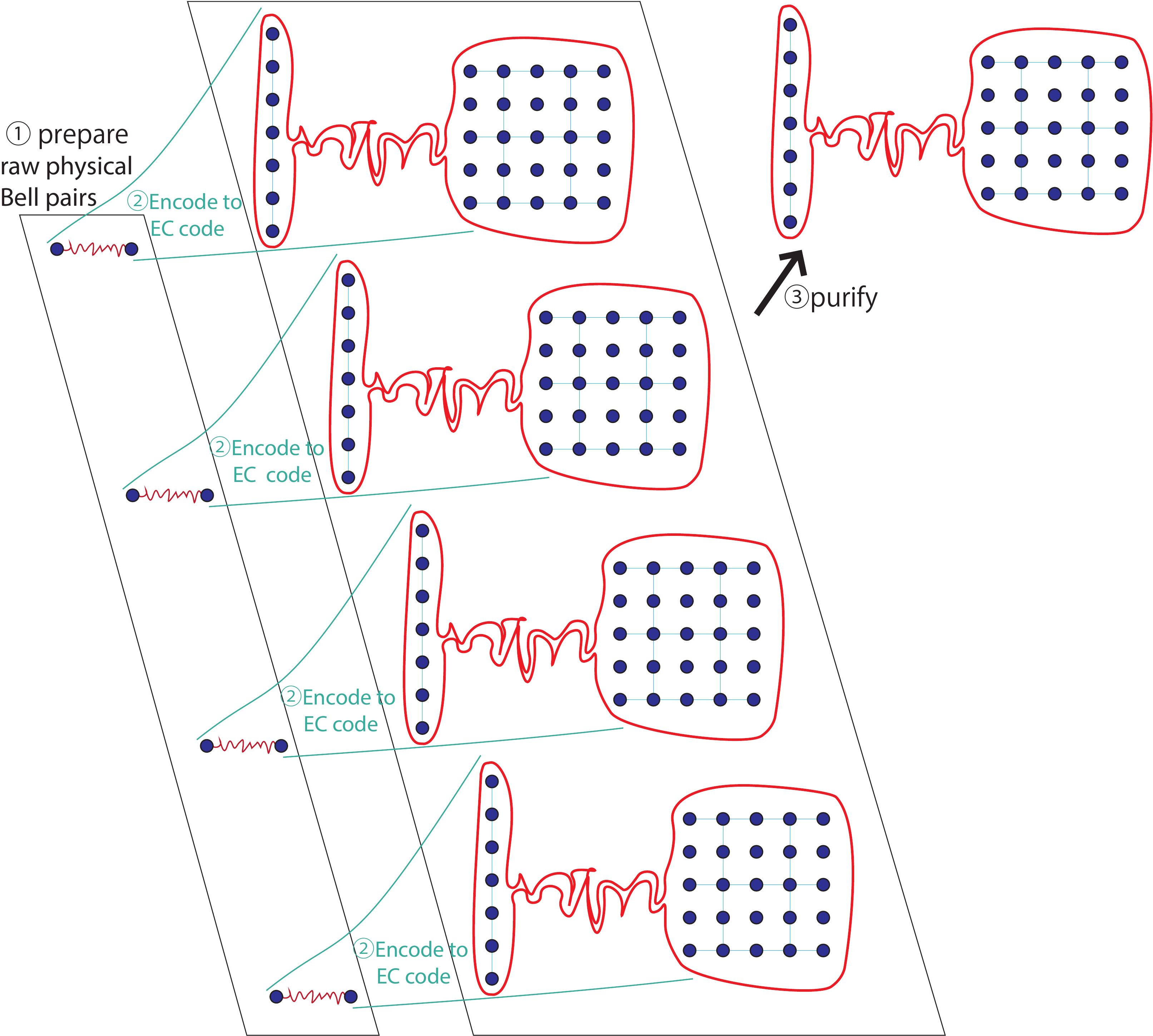}
   \caption[Overview of the scheme which purifies encoded Bell pairs to achieve an encoded Bell pair of high fidelity.]
   {In this method, first, raw physical Bell pairs are encoded into our heterogeneous error correcting code,
   Secondly, those heterogeneously encoded Bell pairs are purified
   directly at the logical level.}
   \label{fig:overview_purify_after_encode}
  \end{center}
\end{figure}
Figure \ref{fig:overview_purify_after_encode} shows the overview of the scheme to make heterogeneously encoded Bell pairs
that are purified after encoding.
In this scheme, to create an encoded Bell pair of high fidelity,
heterogeneously encoded Bell pairs are generated first by encoding each qubit of a raw physical Bell pair
to our chosen heterogeneous error correcting codes.
Next, those encoded Bell pairs are purified at the logical level the desired number of times, via transversal CNOTs and logical measurements.
Logical purification is also achieved by the circuit shown in Figure \ref{sc:circ_purification}, operating on logical rather than physical qubits.
 Table \ref{tab:purify_after_encode} presents the details of the simulated error probability and resource efficiency of {\it purification after encoding}.

\subsection{{\it Purification after encoding with strict post-selection}}
\begin{figure}[t]
  \begin{center}
   \includegraphics[width=9cm]{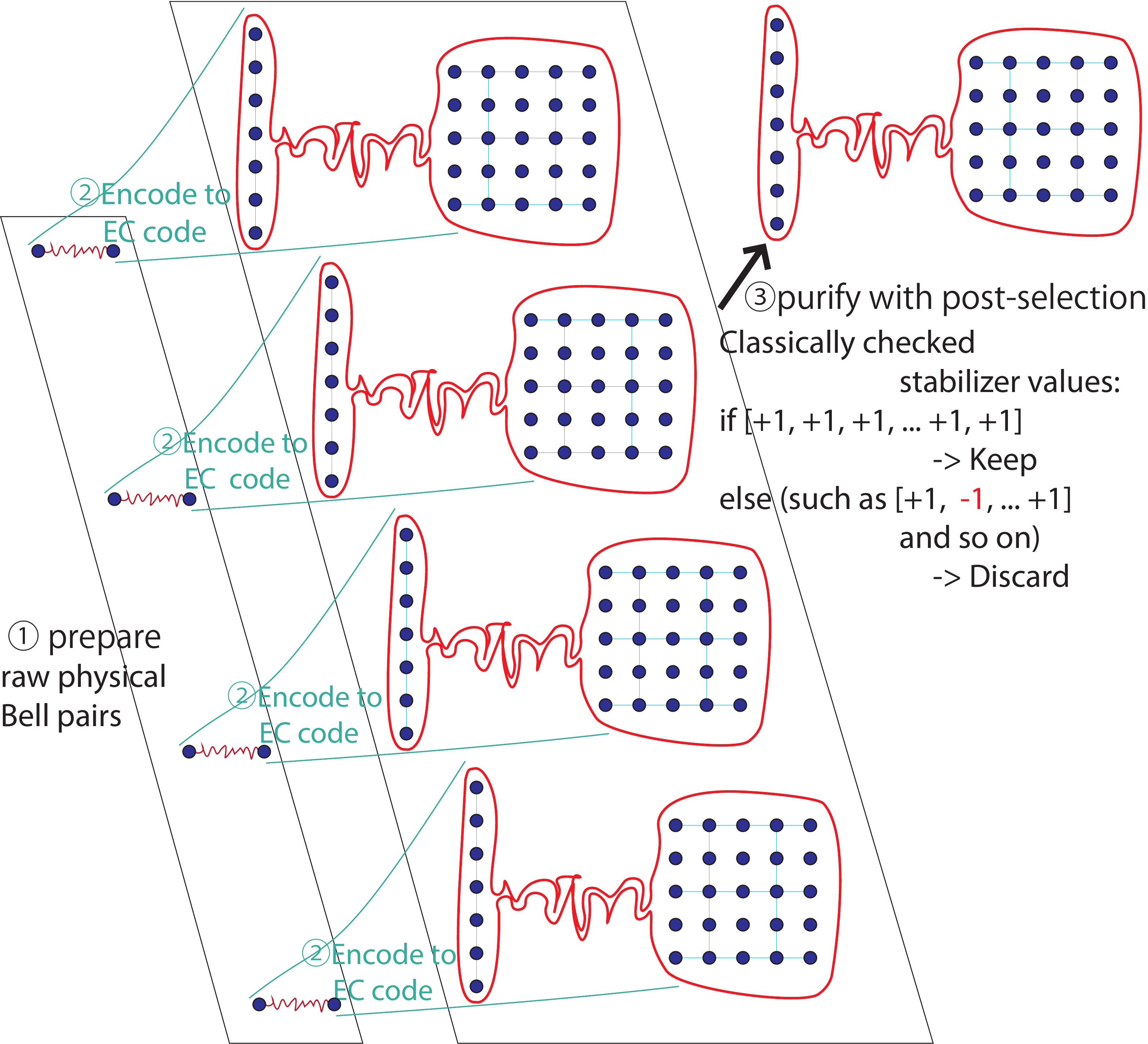}
   \caption[Overview of the scheme which purifies encoded Bell pairs to achieve an encoded Bell pair of high fidelity with strict post-selection.]
   {First, raw physical Bell pairs are encoded to heterogeneous error correcting code, same as {\it purification after encoding}.
   Secondly, at measurement in purification, eigenvalues of each stabilizer are checked classically.
   If any eigenvalue of -1 is measured, the output Bell pair is discarded (in a similar manner to if the overlying purification 
   protocol failed).
   }
   \label{fig:overview_purify_after_encode_strict}
  \end{center}
\end{figure}
Figure \ref{fig:overview_purify_after_encode_strict} shows the overview of the scheme to make encoded Bell pairs,
purified after encoding with strict post-selection protocols to detect errors.
This scheme uses a procedure similar to {\it purification after encoding}.
In this scheme, to create an encoded Bell pair of high fidelity,
heterogeneously encoded Bell pairs are generated first by encoding each qubit of a raw physical Bell pair
to our chosen heterogeneous error correcting codes. We then run purification protocols at the logical level, similarly to the previous 
protocol. However, when we perform a logical measurement as part of this protocol, we also calculate (classically) the 
eigenvalues of all code stabilizers. If any of these eigenvalues are found to be negative, we treat the operation as a failure
(in a similar manner to odd parity logical measurements for the purification)
and the output Bell pair of the purification is discarded. This simultaneously performs purification and 
error correction using the properties of the codes.
 Table \ref{tab:purify_after_encode_strict} presents the details of the numerically calculated error probability and
 resource efficiency of {\it purification after encoding with strict post-selection}.

\section{Summary}
In this Chapter I proposed three schemes to create heterogeneously encoded Bell pairs
between arbitrary two components with scalability.
The architecture employs a method
that
physical Bell pairs are distributed by optical connections
and
each half of a Bell pair is encoded into heterogeneous quantum error correcting code at distributed components.
This method suffers from state errors caused at creation of physical Bell pairs and at the non-fault-tolerant encoding.

To manage such state errors, I propose three schemes to make the method fault-tolerant,
{\it Purification before encoding},
{\it Purification after encoding} and
{\it Purification after encoding with strict post-selection}.

In Section \ref{sec:hetero:sim},
I give the numerical analysis of those error management schemes.

\clearpage
\chapter{Scalable distributed QC architecture utilizing benefits of various codes}
\label{chap:sdqcarch}
This chapter proposes the quantum computer architecture for ``distributed quantum computing utilizing multiple codes on imperfect hardware''.
The most basic requirements to achieve quantum computation in theory are
\begin{itemize}
 \item a quantum universal gate set, and
 \item system size large enough to run quantum algorithms of practical size.
\end{itemize}
However, physical qubits have error rates higher than we can tolerate, hence
\begin{itemize}
 \item fault-tolerance
\end{itemize}
is also required. 
Quantum error correcting codes which achieve the fault-tolerance require excessive resource consumption and have limited ability to achieve a quantum universal gate set. To realize a quantum universal gate set within the limited abilities, more resources are required. See Chap. ~\ref{chap:intro} and ~\ref{chap:qcbasic} for details of those three elements.

This chapter organizes those problems and requirements for the quantum computer architecture and proposes a quantum computer architecture,
starting from the fault-tolerance, then the universal gate set and lastly addressing adequate system size.

\section{Requirements and means to achieve quantum computation}
This section discusses why a distributed quantum computer based on the surface code is
preferred, and the requirements and means to achieve it.

\subsection{For fault tolerance}
\label{subsec:proarch:surface}
Qubits everywhere always suffer from the quantum imperfections discussed in Section~\ref{sec:qc_imperfect}.
The best solution to the quantum imperfections would be to improve the quantum engineering technology.
The required error rate to successfully complete a large scale quantum computation properly is $10^{-15}\sim 10^{-16}$,
however, the physical state error rate today is around $10^{-2} \sim 10^{-3}$~\cite{0034-4885-74-10-104401}.
 Therefore, this solution is impractically far away.

The second solution to the quantum imperfections is fault-tolerant quantum computation executed 
everywhere we use quantum information in the computer.
From the point of view of physical feasibility,
the surface code is the most promising quantum error correcting code.
As stated in Subsec. \ref{surface_code},
the surface code has the highest threshold of physical state error rate in quantum error correcting codes found so far.
It requires interactions only between nearest neighbor qubits on a 2-D lattice
and is robust against dynamic loss. The surface code is robust against static loss, as shown in Chap. ~\ref{chap:defective}.

Therefore the proposed architecture employs the surface code as the fundamental quantum error correcting code.

\subsection{For universal quantum computation}
\label{subsec:fuqc}
A universal quantum computer requires a universal set of quantum gates.
As noted in Subsec. \ref{subsec:qgates},
by Solovay-Kitaev decomposition,
the set of the $CNOT$ gate, $H$ gate and $T$ gate is a universal set.
No quantum error correcting code which is practically feasible in 2-D fashion
and which produces a universal gate set has been discovered~\cite{1367-2630-17-8-083002}.

As shown in Subsec. ~\ref{surface_code},
since the surface code does not support the $T$ gate in a transversal, fault-tolerant fashion,
the surface code requires an ancilla-supported $T$ gate utilizing an $\vert A \rangle$ state.
The ancilla-supported $T$ gate problematically requires the application of an $S$ gate as a correction
and the surface code also does not support $S$ gate in a transversal, fault-tolerant manner,
hence the surface code needs the ancilla-supported $S$ gate utilizing a $\vert Y \rangle$ state, too.

There are several ways to generate the ancilla states $\vert A \rangle$ and $\vert Y \rangle$ in the surface code.
The first way is state injection from physical qubits followed by the magic state distillation~\cite{bravyi2005uqc,PhysRevA.86.052329}.
Though the magic state distillation guarantees adequate fidelity of injected $\vert A \rangle$ and $\vert Y \rangle$,
its process is stochastic and its cost is high both in space and in time.
To achieve a stable supply of the ancilla states generated probabilistically and
to distribute the required number of ancilla states 
which are generated in groups in a magic state distillation~\cite{PhysRevA.86.052329}
to the required locations,
division of components used to generate ancilla states from components used to execute quantum gates would be preferred.

The second way is to get support from other error correcting codes by code teleportation (conversion)~\cite{1504.03913,choi2015}.
The Reed-Muller code can create $\vert A \rangle$ transversally.
The Steane [[7,1,3]] code can create $\vert Y \rangle$ transversally.
As stated in \cite{choi2015}, this approach is advantageous only
when the concatenation level of supporting codes is low
and when the cost of magic state injection and distillation is relatively expensive.
Such complementary code may prefer a different physical architecture, such as
physical qubit layout, from the surface code, hence separated components for the complementary code
from those for the surface code is preferred.

Either way may prefer separated components to generate ancilla states from those for the surface code.
Therefore the proposed architecture employs separate, dedicated components to generate ancilla states,
leaving the selection flexible.

\subsection{For system size large enough}
As mentioned in Chap. ~\ref{chap:intro}, 
Shor's algorithm to factor a number described with $N$ bits requires at least $2N+2$ high-quality qubits.
Additionally, as noted in Subsec. ~\ref{surface_code} and ~\ref{subsec:proarch:surface}, the surface code requires excessive resource overhead.
Installing all quantum computational resource on a single quantum chip would not be feasible, as noted in Chap. ~\ref{chap:intro}.
Hence, to execute quantum algorithms to solve practically meaningful large problems, 
we need to connect components of small performance to build up a large scale quantum computer.

Connecting many components leads to the idea of division of roles among components~\cite{976922,copsey02:_quant-mem-hier}.
The roles in computation are broadly divided into logic operations and storage.
By dividing roles, we can combine the two benefits,
executing quantum operations in fast fashion in quantum CPUs and
storing quantum information in space-saving fashion in quantum memories.
Therefore, different choices may be made in different components,
such as choice of physical technologies~\cite{0034-4885-74-10-104401} and error correcting codes~\cite{copsey02:_quant-mem-hier,1504.03913}.
Unlike the CPUs of the conventional computer of today, quantum logic gates are not relatively high-level instructions~\cite{Metodi:2005:QLA:1099547.1100566,Thaker:2006:QMH:1150019.1136518,Whitney:2009:FTA:1555815.1555802}.
Quantum CPUs with the surface code are more like executing logic gates directly on memory than integration of registers, pipelined devices, ALUs and so on~\cite{Jones:2012Layered_Architecture_for_Quantum_Computing,van-meter10:dist_arch_ijqi,1367-2630-11-8-083032}.
Hence the first difference of quantum CPUs and quantum memories is the choice of quantum error correcting code.
The physical implementation of the components will vary depending on the quantum error correcting code;
this dissertation leaves physical technologies employed by each component flexible.
Different quantum error correcting codes have different duration for quantum logic gates and different resource consumption.
Therefore the proposed architecture employs the defect-based surface code in quantum CPUs and
employs the deformation-based surface code in quantum memories.

Connecting components requires internal quantum connections in a quantum computer.
Such connections must support heterogeneous encoding and must be fault-tolerant.
Oskin et al. and Copsey et al. proposed architectures utilizing code teleportation based on heterogeneously encoded Bell pairs~\cite{976922,copsey02:_quant-mem-hier}. Their means are not robust against errors during encoding.
Monroe et al. and Ahsan et al. proposed an internal connection backplane utilizing optical crossbar switches
for scalability of parallelism of connections~\cite{PhysRevA.89.022317,ahsan2015designing}.
Extending their means to involve purification, discussed in Chap. ~\ref{chap:hetero},
this system realizes a
scalably fault-tolerant interconnection supporting heterogeneous encoding.
Components need to support quantum teleportation using the Bell pairs shared by the backplane.

These connections will resemble quantum networks.
This architecture can naturally be extended to installing quantum network interface cards
hence making possible distributed quantum computation for more scalability ~\cite{crepeau:_secur_multi_party_qc,buhrman03:_dist_qc,RevModPhys.82.665}.

\section{Requirements of building blocks and internal connections of the proposed architecture}
In my design, the roles for quantum computation are divided to three types of building blocks
and a backplane, the internal connections of building blocks.
The three types of building blocks are quantum CPUs which execute quantum algorithms,
quantum memories which maintain quantum data that is not operated at that time,
and magic state generation areas which produce magic states for non-Clifford gates.

Table \ref{tab:sdqa_req} summarizes the must-have requirements of building blocks of the quantum computer architecture for ``name will be added''.
Other desirable but less critical features are discussed in the following section.

The requirements for quantum CPUs can be summarized as processing
quantum data without any problems.
Therefore, quantum CPUs must tolerate quantum imperfections;
they must support a universal gate set; and
they must support transferring quantum data 
to and from the internal connection point.

The requirements for quantum memories can be summarized as maintaining quantum data without any problems.
Therefore, quantum memories must tolerate quantum imperfections;
and 
they must support transferring quantum data 
to and from the internal connection point.

The requirements for magic state generation area can be summarized as producing
high fidelity magic states for non-Clifford gates.
Hence, magic state generation areas must tolerate quantum imperfections;
they must create magic states of high fidelity, such as $\vert A \rangle$ and $\vert Y \rangle$; and 
they must support transferring quantum data
to and from the internal connection point.

The requirements for internal connections can be summarized as
transfer quantum data from one building block to another with adequate fidelity.
Hence, internal connections must tolerate quantum imperfections;
they must support routing of data transfer; and
they must bridge heterogeneous error correcting codes.

 \begin{table}[b]
  \begin{center}
   \caption{Requirements of building blocks and internal connections of the quantum computer architecture for ``Distributed quantum computing utilizing multiple codes on imperfect hardware''.}
   \label{tab:sdqa_req}
	\begin{tabular}{|c|l|}
	 \hline
	element & requirements \\
	\hline
     \hline
     Quantum CPUs & tolerate quantum imperfections \\
                 & support an universal gate set \\
                & support transferring quantum data \\
     \hline
     Quantum Memories & tolerate quantum imperfections \\
                    & support transferring quantum data \\
	   \hline
     Magic State  & tolerate quantum imperfections \\
     Generation   & create magic states of high fidelity, such as $\vert A \rangle$ and $\vert Y \rangle$ \\
	 area       & support transferring quantum data \\
	   \hline
Internal connections & tolerate quantum imperfections \\
 & support routing of data transfer \\
 & bridge heterogeneous error correcting codes \\
	 \hline
	\end{tabular}
  \end{center}
 \end{table}

\subsection{Desired characteristics of building blocks and internal connections}
For the division of roles, 
building blocks and internal connections have different desirable characteristics.
Table \ref{tab:sdqa_des} summarizes the desired characteristics of building blocks and internal connections.

Quantum CPUs desire fast quantum gates to minimize computation time.
When employing fast gates, high consumption of spatial resources would be 
an acceptable compromise
as long as the quantum memories are spatially efficient so that the overall system size is feasible.

Quantum memories desire spatially small resource requirement to minimize the overall system size.
As long as quantum CPUs have fast gates to process the quantum computation rapidly,
slow operations on quantum memories would be an acceptable compromise;
by analogy with classical computers, access time to memories 
that is hundreds of times slower than registers in CPU
would be acceptable~\cite{Hennessy:2011:CAF:1999263}.

For the magic state generation area,
as noted in Subsec. \ref{subsec:fuqc}, 
there are two candidates for the design
and the better design differs depending on the environment and on the performance of base technologies.
This area should employ the design which prepares ancilla states more efficiently in terms of the product of space and time, when building the quantum computer.
I leave the design of this area flexible.

Internal connections hopefully have scalable parallelism of connections.
With scalable parallelism of connections, stall of computation caused by queued data transfer would be reduced.

 \begin{table}[b]
\begin{center}
   \caption[Desired characteristics of building blocks and internal connections]{ of the quantum computer architecture for ``Distributed quantum computing utilizing multiple codes on imperfect hardware''.}
   \label{tab:sdqa_des}
	\begin{tabular}{|c|l|}
	 \hline
	element & desired characteristics \\
	\hline
     \hline
     Quantum CPUs & fast gates \\
     \hline
     Quantum Memories & space-saving design\\
	   \hline
     Magic State  & space-saving design \\
     Generation   & fast preparation of magic states \\
	 area       & \\
	   \hline
Internal connections & scalable data transfer\\
	   \hline
	\end{tabular}
\end{center}
 \end{table}

\section{The proposed quantum computer architecture}
  \begin{figure}[t]
   \begin{center}
  \includegraphics[height=18cm]{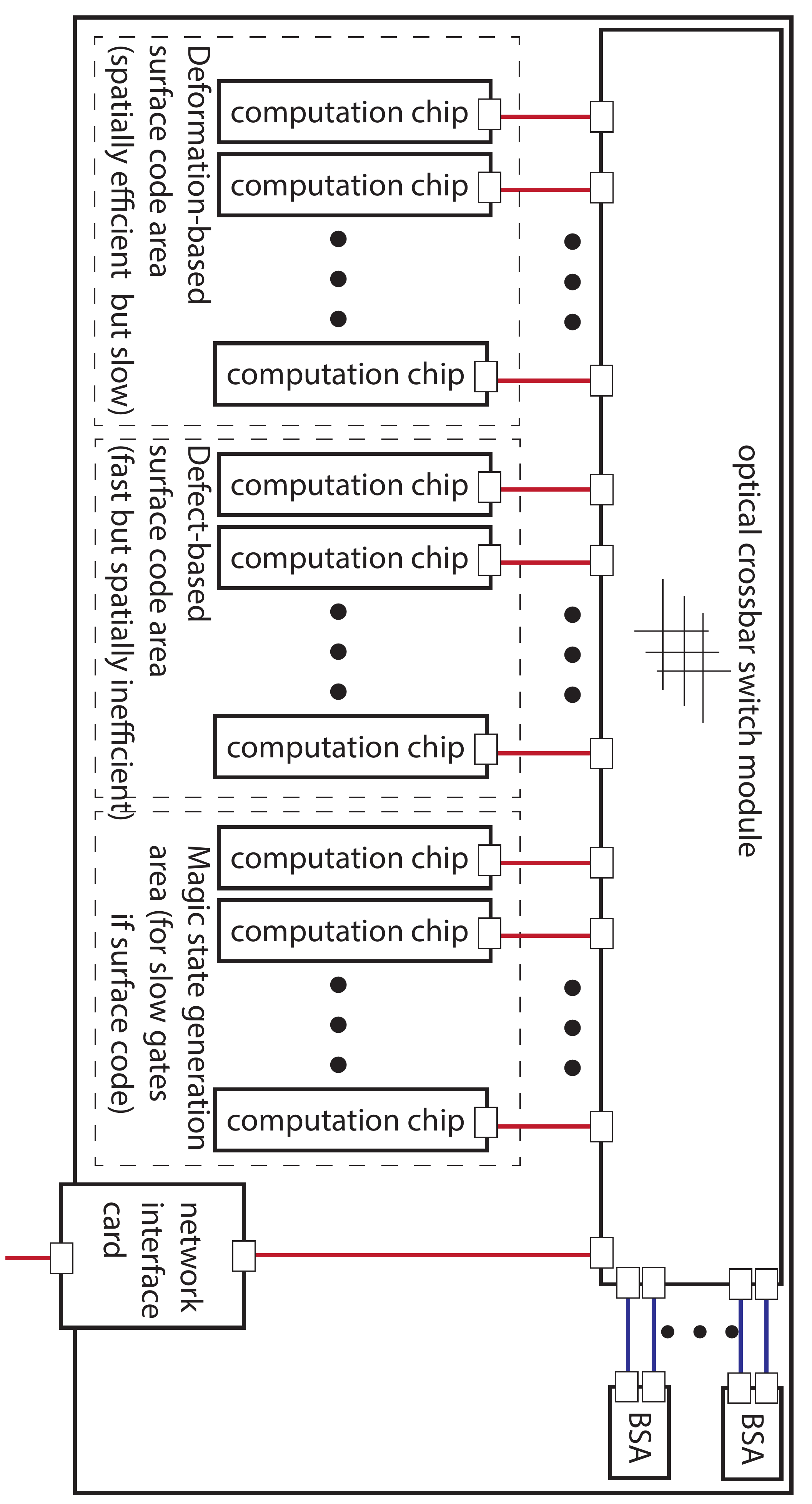}
  \caption[The proposed quantum bus architecture supporting the multiple code model and networking for distributed computation.]
    {Colored lines are optical connections. The red lines are input to the crossbar switch. The blue lines are output from the crossbar switch.
  The BSA stands for Bell state analyzer~\cite{PhysRevLett.59.2044}.
  }
  \label{fig:prop_arch}
   \end{center}
  \end{figure}
 Figure \ref{fig:prop_arch} shows the proposed quantum bus architecture, extended from~\cite{ahsan2015designing,kim03:_1100_port_mems}.
 Colored lines are optical connections or internal photon path of the optical crossbar.
 The red lines are inputs to the crossbar switch. The blue lines are outputs from the crossbar switch.
 Each input source component has links with the crossbar and holds a solid qubit to emit a photon.
 Two output lines lead to a
 Bell state analyzer (BSA), which executes Bell measurement for
 entanglement swapping~\cite{PhysRevLett.59.2044}.

 To achieve quantum communication between two components,
 each of the two components emit a photon to the crossbar switch so that we have each stationary qubit entangled with a photon,
 leaving a solid qubit behind in the component.
 The crossbar switch switches the connections to route the two photons
 to two links to an assigned BSA.
 BSA executes Bell measurement on the two photons for entanglement swapping,
 leaving the two stationary qubits in the two components entangled.

 Each of the two solid qubits is then encoded to the error correcting code used in the component.
 This homo- or hetero-encoded Bell pair is created as the resource for quantum teleportation
 at the logical level.

 The deformation-based surface code area and the defect-based surface code area
 work as quantum memories and as CPUs, respectively.
 Their physical implementation may be same.
 Both of them have physical qubits placed on a 2-D lattice,
 since the surface code runs in both area.
 However, the encoding of logical qubits on the surface code lattice differs,
 therefore the way of using the devices differs depending only on software.

\subsection{Defect-based surface code area}
This area employs the defect-based surface code~\cite{PhysRevA.86.032324}.
the $CNOT$ gate by braiding requires $32d$ steps, for
$8d$ steps for each of first half expansion, first half shrink, second half expansion and second half shrink respectively.
 
\subsection{Densely packed deformation-based surface code area}
This area employs the deformation-based surface code proposed in Chapter~\ref{chap:deformation}.
The deformation-based surface code packs logical qubits denser than other surface codes.
Logical $X$ and logical $Z$ can be executed efficiently.
However, logical $CNOT$ gate by lattice surgery requires $4d^2+9d$ steps
where $d$ is the code distance while $24d$ steps are required for the standard planar code.
This fact would make the deformation-based surface code area hundreds of times slower than the defect-based code area.
On top of this, the logical Hadamard gate of the deformation-based qubit is three times slower.
Hence this area should be a storage, or a memory, to keep logical qubits for a long time in a spatially efficient fashion.
A concern is whether the hundreds slower memory than the CPU is usable or not.
In my design, the memory access time is proportional to the duration of $CNOT$ gate
because data transfers are implemented by quantum teleportation.
I would conclude that the much slower memory is acceptable from this fact in classical computers~\cite{Hennessy:2011:CAF:1999263}.

$CNOT$ gate for the deformation-based surface code is expensive,
hence teleportation based on $ZZ$ stabilizer is more preferred to send a logical qubit from this memory.
Such as, the first state is
\begin{eqnarray}
 (\alpha \vert 0_q \rangle + \beta \vert 1_q \rangle) (\frac{1}{\sqrt{2}}\vert 0_{b0}0_{b1} \rangle + \vert 1_{b0}1_{b1} \rangle) \vert 0_a \rangle,
\end{eqnarray}
where qubit $q$ is the state we want to send and qubits $b0$ and $b1$ are a Bell pair and qubit $a$ is an ancilla qubit for the $ZZ$ stabilizer.
By applying $CNOT(q,a)$ and $CNOT(b0,a)$, we get
\begin{eqnarray}
 \frac{1}{\sqrt{2}}(\alpha \vert 0_q0_{b0}0_{b1}0_a \rangle + \beta \vert 1_q1_{b0}1_{b1}0_a \rangle 
  + \alpha \vert 0_q1_{b0}1_{b1}1_a \rangle + \beta \vert 1_q0_{b0}0_{b1}1_a \rangle),
\end{eqnarray}
by measuring $Z_a$ and applying $X_{b0}X_{b1}$ if $-1$ eigenvalue is observed, then we get
\begin{eqnarray}
 \alpha \vert 0_q0_{b0}0_{b1} \rangle + \beta \vert 1_q1_{b0}1_{b1} \rangle,
\end{eqnarray}
by measuring $X_q$ and $X_{b0}$ and applying $Z_{b1}$ for each measurement if $-1$ eigenvalue is observed, then eventually we get
\begin{eqnarray}
 \alpha \vert 0_{b1} \rangle + \beta \vert 1_{b1} \rangle.
\end{eqnarray}

  \subsection{Magic state generation area}
  This area may consist of quantum chips of several codes, such as the Steane [[7,1,3]] code and the Reed-Muller 15 qubit code.
  This is because the Steane [[7,1,3]] code can create an ancilla state for an S gate efficiently and
  the Reed-Muller 15 qubit code can create an ancilla state for a T gate efficiently~\cite{PhysRevLett.112.010505}.
  
  Another possible implementation is of the surface code, to create logical ancilla states by magic state distillation
  ~\cite{bravyi2005uqc,Fowler:2009High-threshold_universal_quantum_computation_on_the_surface_code}.
  It is not yet sure which is better for this purpose the defect-based code with braiding of the planar code with lattice surgery.

  Created ancilla states are sent to the computation areas by quantum teleportation.
  With any codes, the chip must have dedicated design for distillation.
 
  \subsection{Network interface card}
  The network interface card has one or more attachments for optical fibers to connect to other quantum nodes,
  after the classical network interface card.
  By utilizing this component, the computation chips are enabled to cooperate with computation chips installed in other machines.

  External links have much larger loss/error rate than internal links because of the environment and the longer fiber~\cite{Inagaki:13}.
  External links can achieve Bell pair creation at a lower rate than internal links do,
  hence internal Bell pairs may go unused as we wait for an external Bell pair.
  Figure \ref{fig:prop_arch} has only one network interface card and only one external link from the card.
  If more throughput is required, we can employ a network interface card which has two or more fiber attachments instead
  and use two or more fibers for the link.
  The new fiber should be connected to the same repeater/router as the first one,
  then the quantum links are aggregated after the classical link aggregation~\cite{7297898}.
  
  Another means to increase throughput is inserting another network interface card.
  This means allows the quantum computer to be connected to two quantum repeaters/routers/computers.
  This may be useful for complex use of quantum network.
  This work targets the bus architecture and more consideration about network interface cards is an open problem.

 \section{Hardware and Software of building blocks}
 Both quantum CPUs and quantum memories employ the surface code,
 so their physical implementation are the same;
 they have 2-D lattice of physical qubits in which nearest neighbor interactions are implemented;
 and on the edges of the components, optical-solid conversion
 is implemented for internal connections.
 Therefore the difference of the two areas will be
 how the physical qubits are used by software; how logical qubits are encoded;
 how logical gates are executed; and how logical qubits are transferred.
 
 If magic state generation area employs the surface code with
 state injection and magic state distillation,
 its hardware implementation will be same as the quantum CPUs and quantum memories.
 Then, the software may be the same as the quantum CPUs.
 If it employs the other, 
 the hardware implementation will depend on the employed error correcting code.

 The hardware of internal connections are composed from optical fibers, the optical cross bar switch and BSAs.
 The number of optical fibers connected to a component directly affects the
 performance of the bandwidth of the internal communication.
 One BSA can support one internal connection at a time.
 Hence the parallelism of internal connections is determined by the number of BSAs~\cite{ahsan2015designing,kim03:_1100_port_mems}.

\clearpage
\chapter{Evaluation}
\label{chap:eval}
\section{Performance of the surface code on defective lattice}
\label{sec:defective:evaluation}
We assume a circuit-based error model, summarized by Landahl et al. \cite{landahl:arXiv:1108.5738}.
This circuit-based error model assumes that each gate acts ideally, followed by a noisy channel, and that each measurement acts ideally, after a noisy channel.
Errors may occur at every gate in the circuit.
Our error channel for a single-qubit gate has error probability $p$, meaning that each error (X, Z or Y) occurs with probability $p$/3.
In a similar fashion, for two-qubit gates, our error model has probability $p$/15 for each two-qubit error (IX, IZ, IY, XI, XX, XZ, XY, ZI, ZX, ZZ, ZY, YI, YX, YZ, YY).
We assume that the set of physical gates available includes CNOT, SWAP and Hadamard gates.
We assume that INIT and measurement in Z basis have X error probability $p$.
All operations require one time step.

Our circuit is asynchronous in the sense that stabilizers are measured at different frequencies.
Stabilizers whose circuits have shallower depth may be measured more times than those whose circuits have deeper depth.
To achieve proper syndrome matching,
the surface code requires that the lattice be covered by stabilizers.
Otherwise, an unstabilized area works as a defect-based qubit which may serve as an end of error chains, leading to undetectable logical errors.
Hence, after all stabilizers covering the lattice
have been measured at least once since the last execution of the matching algorithm, the matching algorithm is re-executed.
Typically, this timing is dependent on the deepest stabilizer circuit.
From the result of matching, we make a map of Pauli frames which describes where Pauli frame corrections should be applied for error correction.
Because our circuit is asynchronous and there might be SWAP gates, we must keep track of the location of data qubits to combine the error information about data qubits and the map to check the result of error correction.

We have conducted extensive simulations, beginning with a perfect lattice, then extending to imperfect ones.
First we show the numerical result of several basic test cases:
only a single faulty device exists, in the center of the lattice;
only a single faulty device exists, in the west of the lattice; and
only a single faulty device exists, in the northwest of the lattice
for the distances 5, 7, 9 and 13.
Our simulation holds $d$ temporal rounds of measured stabilizer values for error correction.
Hence $d$ measurements are saved for the stabilizer with the deepest circuit and
more measurements are saved for normal stabilizers, because of the scheduling algorithm shown in Subsection~\ref{subsec:whole}.
After finishing an error correction cycle, the oldest round is discarded, a new round is created by new measurements and error correction is re-executed.
Next, we show the numerical result for random generated lattices for different yields, 80\%, 90\% and 95\%.
We generated 30 lattices for each pair of yield and code distance of 5, 7, 9, 13, 17 and 21.
Some defective lattices cannot encode a logical qubit for the code distance becomes 0 as a result of merging stabilizers, so that eventually
we simulate 478 lattices (details described in subsection~\ref{subsec:num_res_random})
for each physical error rates of $0.1\%$, $0.2\%$, $0.3\%$, $0.4\%$, $0.5\%$, $0.6\%$, $0.7\%$, $0.8\%$, $0.9\%$, $1\%$ and $2\%$.
It is hard to collect enough logical errors in Monte Carlo simulation
as the logical error rate is exponentially suppressed,
therefore we choose $0.1\%$ as the lowest physical error rate
for our simulation.
Therefore we simulated 5258 parameter combinations.

The computational resource devoted to circuit simulation, excluding chip generations and circuit constructions, was more than 100,000 CPU days, executed on the 
StarBED project testbed~\cite{Miyachi:2006:SSL:1190095.1190133}.
Each preparation of stabilizer circuits which solves the traveling salesman problem
required up to 1 CPU day.
After construction of the nest, for example, the simulation of $d=5$ of single-faulty-northwest for $p=10^{-3}$
consisted of 370945 rounds of error correction to find 500 logical $X$ errors in 1424.98 seconds.
The simulation of $d=13$ of single-faulty-northwest for $p=10^{-3}$
consisted of 315550 rounds of error correction in 5.8 days but found 0 logical $X$ errors.

Peak memory sizes are estimated to be 30GB for 320 lattices,
63GB for 133 lattices and
more than 100GB for 25 lattices.
The greatest memory consumption is during nest building, shown in Figure B5, B6 and B7.
To give accurate weights to the edges of the ``nests'', Autotune virtually creates errors on every qubit at every physical step, and traces their propagation.
Roughly speaking, the size of the error structure is 136 bytes.
A lattice includes 1089 qubits for distance 17.
Let us assume: 200 physical steps per error correction cycle due to the asynchronous stabilizers;
each error propagates to 10 physical qubits on average;
each error remains for 100 physical steps on average.
Then memory consumption is
$136 * 1089 * 200 * 10 * 100 = 29620800000$ bytes, roughly 30 GB.
Several factors affect this rough estimate.
Faulty devices reduce the number of qubits
and
other structures generated to create the nests,
but the memory consumption remains on the order of tens giga bytes.

Those peak memory sizes are big, however,
they do not affect the quantum computation in practice.
This is because the heavy operations that
Autotune virtually creates errors on every qubit at every physical step,
and traces their propagation by the circuits
to give accurate weights to the edges of the nests
can be executed preliminarily.
To avoid redundant execution of heavy creation of nests,
all 11 physical gate error rates for a single lattice are simulated in parallel on a single simulation node, allowing us to share a single in-memory copy of the nest.
%
We attempted to simulate distance 21, but failed because
we cannot accumulate enough logical errors
to have valid data points, for one of several reasons:
good lattices have strong tolerance against errors;
even bad lattices have strong error tolerance at lower physical error rate;
at higher physical error rates, simulating an error correction cycle takes too much computation time because
many physical errors occur in our extended asynchronous error correction cycle,
taxing the scalability of the matching algorithm;
or because the simulation requires more than 128GB memory, the maximum available in our system.

\subsection{perfect lattice}
\begin{figure}[t]
 \begin{center}
  \includegraphics[width=13cm]{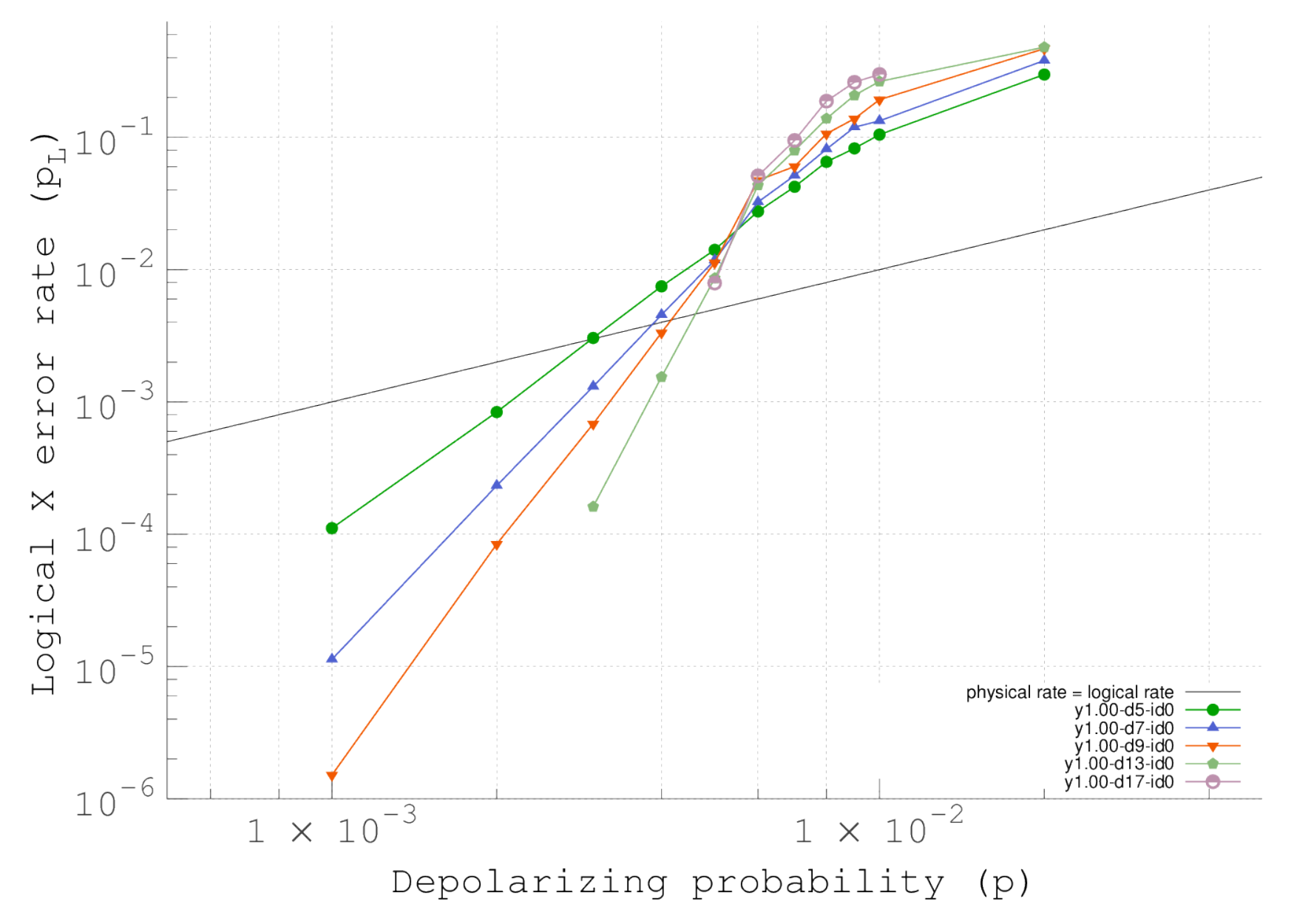}
  \caption[Results of baseline simulations of perfect lattices of code distance 5, 7, 9, 13 and 17.]
  {The average number of steps per error correction cycle for every code distance is
  $8.1$, $8.0$, $8.0$, $8.0$ and $8.0$ respectively.
  The black line is the break-even line.
  The threshold seems to be around $0.58\%$.
  Each data point has $50\sim 1500$ logical errors.
  The irregularity of the point at $p=0.6\%$ of distance 9 may come from statistical variance.
  }
  \label{fig:graph_perfect_lattice}
 \end{center}
\end{figure}

Figure \ref{fig:graph_perfect_lattice} depicts the results of simulation of perfect lattices, used as our baseline for comparison.
Each curve represents a set of simulations for a lattice of a particular code distance, for varying physical gate error rates.
Points below the break-even line are conditions in which the logical error rate
in the logical state is below that of a bare, unencoded physical qubit.
The break-even line indicates whether the error correcting code bears fruit at the physical error rate in the sense of the error probability,
with lengthening the execution time of the quantum computation since a logical gate consists of many physical gates.
Distance 9 achieves break-even at $p=0.3\%$.
The crossing point of the curves, each of which describes a code distance, is called the \emph{threshold}, the physical error rate below which
the larger code distance has the lower logical error rate.
Above the threshold, the error correction process introduces more errors than it corrects,
and the higher code distance has the higher logical error rate.

The threshold indicated by this simulation is around $0.58\%$, similar to the $0.60\%$ reported in
~\cite{Fowler:2009High-threshold_universal_quantum_computation_on_the_surface_code}.
This related work employs the assumptions most similar to our perfect lattice simulation,
other than the asynchronous scheduling of stabilizers.
Our error correction circuits are designed to omit identity gates to shrink the asynchronous circuit
depth, whereas circuits of related work achieve perfect synchronization and parallelism through careful insertion of identity gates.
For example, identity gates on the qubit d17 in Fig \ref{fig:sc:stabilizers} (b) between the initialization and the CNOT gates or between the CNOT gates and the measurement are omitted in our simulation.
We infer that our baseline simulation follows the related work, our baseline simulation is valid and the effect of asynchronicity to the perfect lattice is small.

\subsection{lattice with a single faulty device}
\begin{figure}[t]
 \begin{center}
  (a)
  \includegraphics[width=9cm]{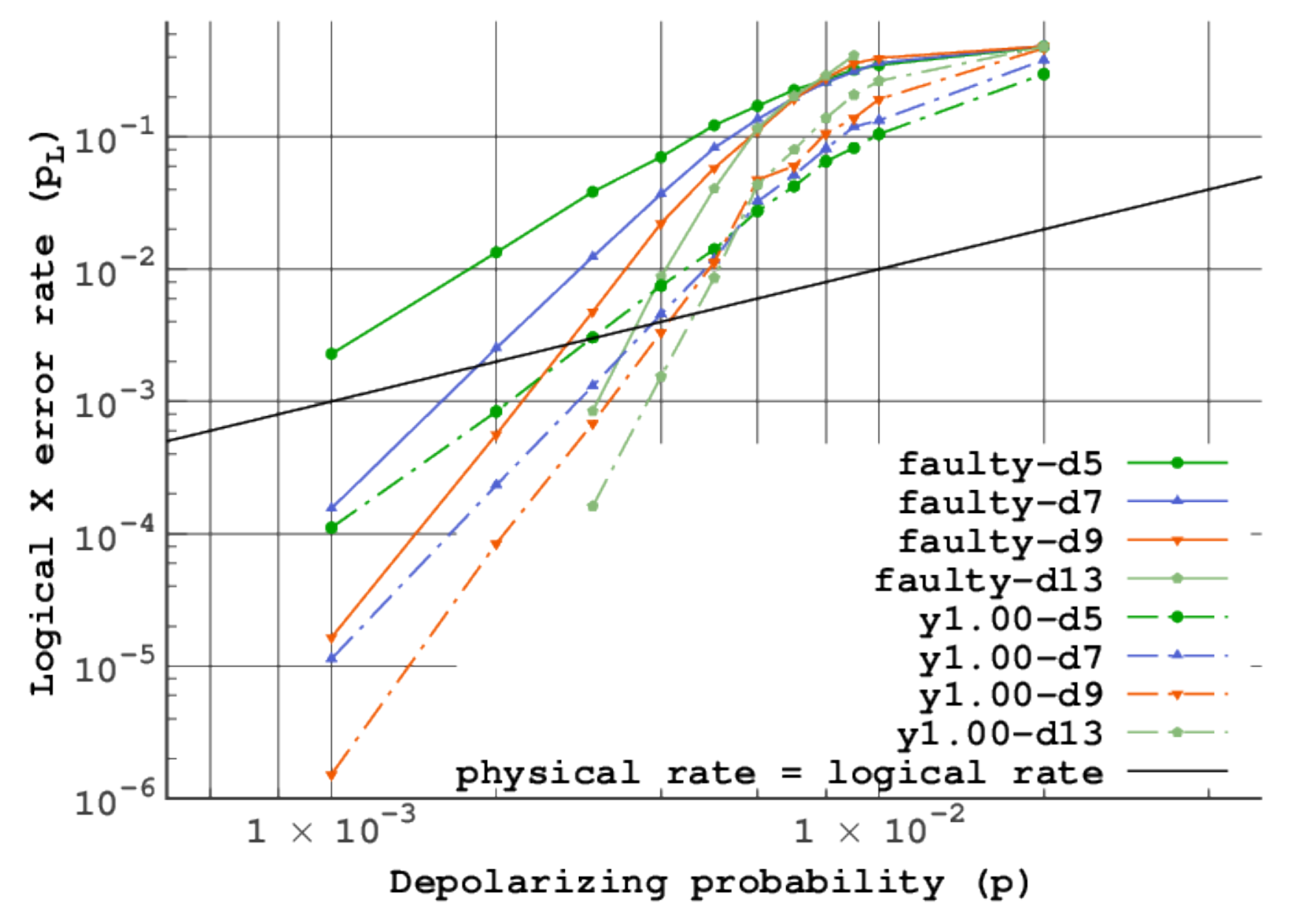}\\
  (b)
  \includegraphics[width=9cm]{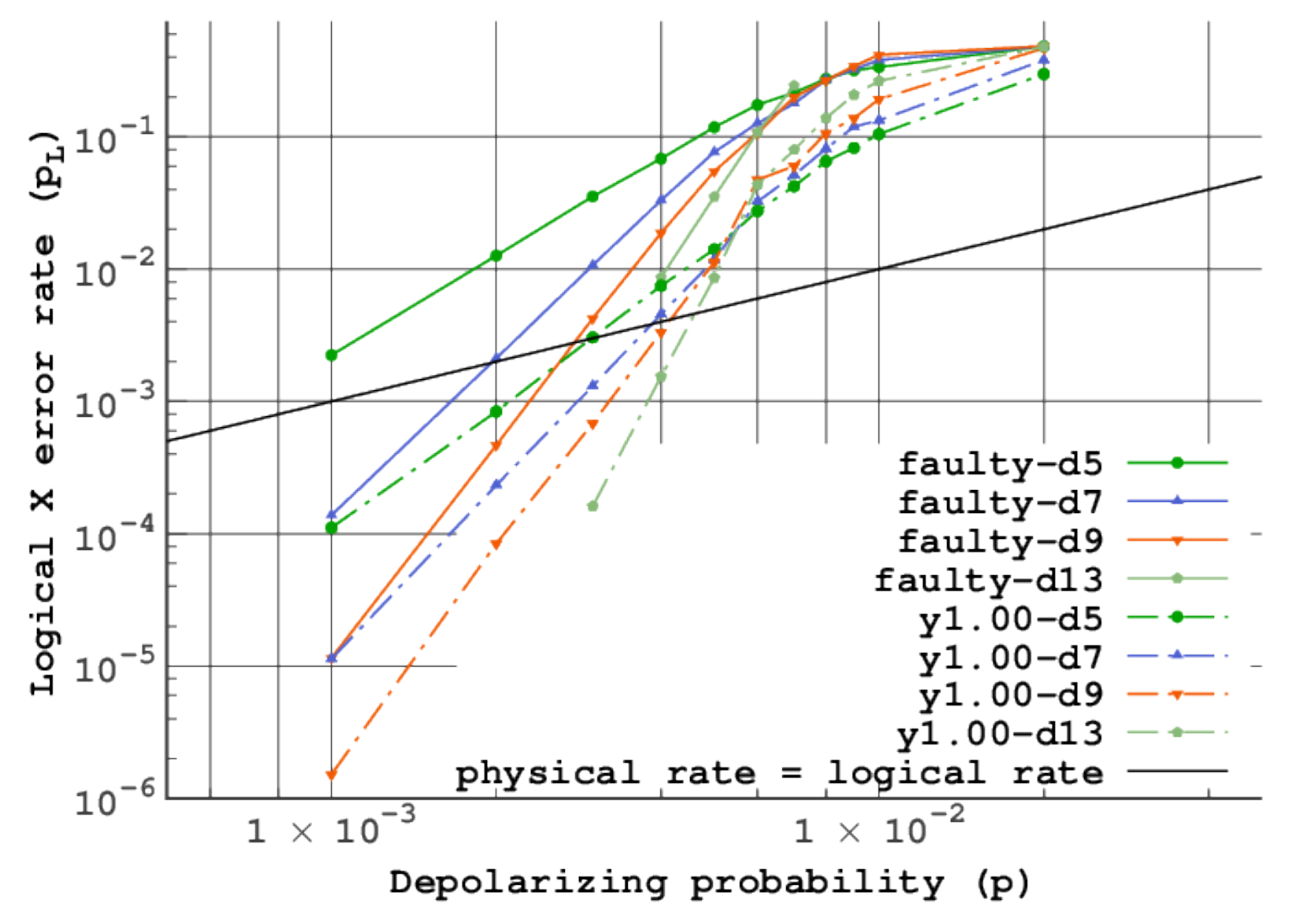}\\
  (c)
  \includegraphics[width=9cm]{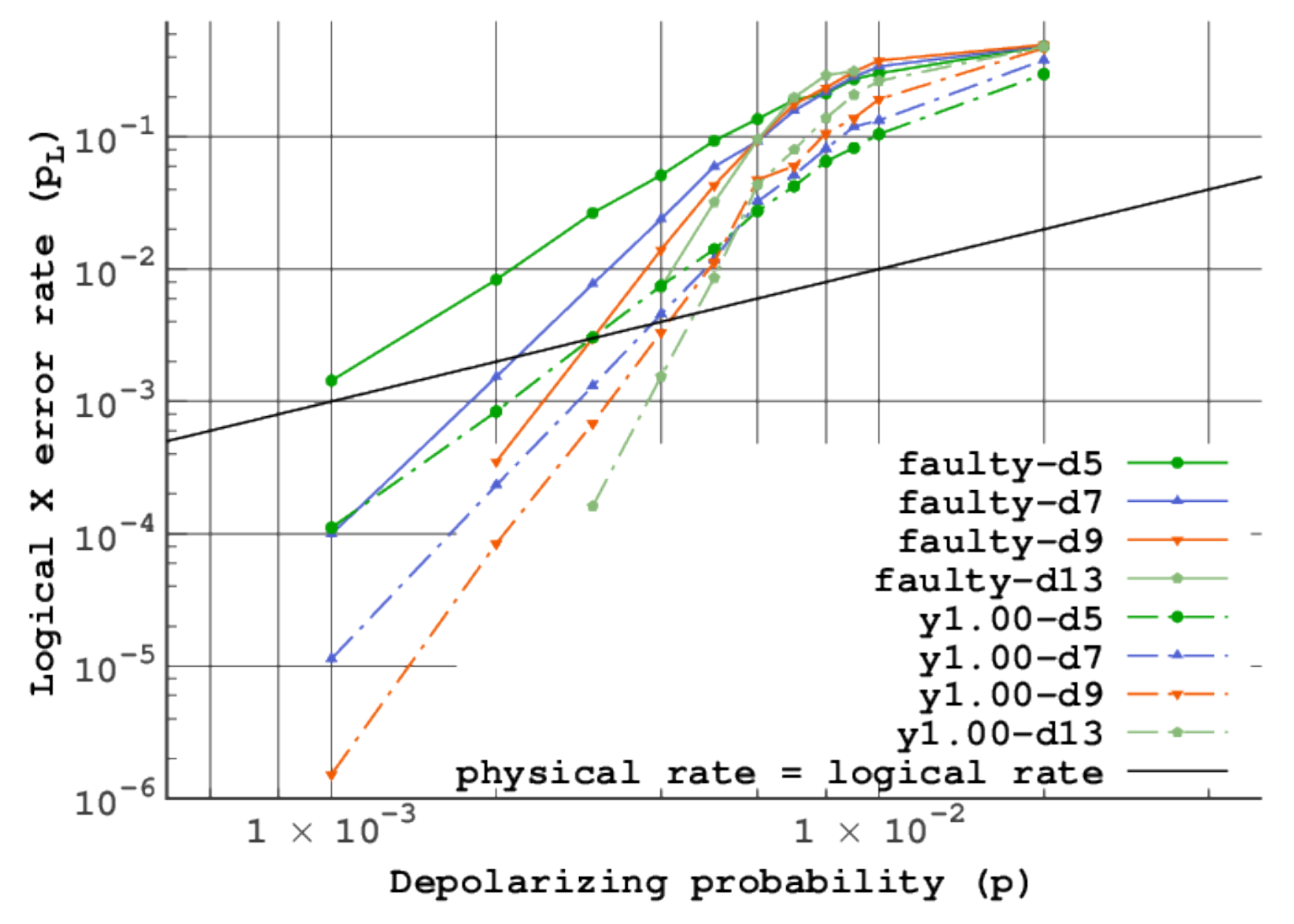}\\
  \caption[Results of simulations of defective lattices which have a single faulty device]{
  (a) in the center of the lattice,
  (b) in the west of the lattice
  and
  (c) in the northwest of the lattice respectively.
  Dashed lines are of the perfect lattices for reference.
  The code distances are 5, 7, 9 and 13.
  The average numbers of steps per error correction cycle are all $32.5$ for every code distance of every fault location.
  }
  \label{fig:graph_single_faulty}
 \end{center}
\end{figure}

Figure \ref{fig:graph_single_faulty}(a), (b) and (c)
depict the results of simulations
to investigate the effect of a single faulty device in the center, on the west edge
and on the northwest corner of the lattice, respectively.
The plots show that our approach works properly
because the larger code distance has the lower logical error rate at lower physical error rates.

Each single-fault residual error rate is worse than that of the corresponding perfect lattice.
The slope of each code distance of single-fault chips is lower than that of the corresponding perfect lattice.
The gap grows slightly as the physical error rate is reduced, visible as the less-steep curve for the defective lattice.

There are differences depending on the single-fault location.
Comparing the points $d=9$ of the perfect lattice with those of single-faulty-center, single-faulty-west and single-faulty-northwest at $p=0.1\%$,
faulty lattices are $10.9\times$, $7.60\times$ and $7.20\times$ worse than the perfect lattice, respectively.
Single-faulty-northwest has a lower residual error rate than the others.
This may be because the big stabilizer which causes asynchronous scheduling of stabilizers is
on the periphery, so that the number of stabilizers which are close to the big stabilizer and hence
which have stronger scheduling restrictions than more remote stabilizers is smaller than other single-faulty chips.
Across the range of our simulations, the negative impact is $6\times\sim 11\times$ depending on location, distance and error rate.

From the point of view of absolute logical error rate,
the penalty for having a defect is greater at lower physical error rates.
An ``effective'' code distance is the code distance at the same physical error rate of the perfect lattice which has the closest logical error rate to the defective lattices. 
For single-faulty-center,
at $p=0.3\%$, faulty $d=9$ is $1.5\times$ worse than perfect $d=5$, hence the effective code distance of faulty $d=9$ at $p=0.3\%$ is $\approx 5$.
The effective code distance is useful when considering the resource overhead of modifications.
In the example above, to achieve a logical error rate 
equivalent to that of $d=5$ on the perfect lattice at $p=0.3\%$,
we at least need $d=9$ for the defective lattice. This indicates that $3.5 \times$ the number of physical qubits are required.

From the point of view of the effective code distance,
the penalty for having a defect is smaller at lower physical error rates.
At $p=0.3\%$, faulty $d=9$ is $1.5\times$ worse than perfect $d=5$, and
at $p=0.1\%$, faulty $d=9$ is $14.9\times$ better than perfect $d=5$ while
faulty $d=7$ is $1.4\times$ worse than perfect $d=5$.
Hence, to exceed the effective code distance 5, $p=0.3\%$ requires us to use $d=11$
while $p=0.1\%$ only requires us to use $d=9$.
The trend of the penalty of the effective code distance and that of the absolute logical error rate differ.
This difference is caused by the difference of slopes of each code distance of each lattice.
We have to be mindful of those trends when designing a quantum computer
to achieve an adequate logical error rate.
%

Because the proportional impact of a single fault should lessen as the code distance increases,
the crossing point of the curves is not a good measure of performance here.
Figure \ref{fig:graph_single_faulty} shows that the crossing points of two distances would differ.
The crossing point of distance 9 and 13 appears to be around $0.6\%$ which is
the threshold for the perfect lattice as shown in Figure \ref{fig:graph_perfect_lattice},
whereas the crossing point of distance 5 and 7 is around $0.8\%$.

Table \ref{tab:num_static_result} shows the data of the single-faulty lattice simulations.
The reduced code distance is the minimum distance between corresponding boundaries shortened by merging stabilizers.
The naive hypothesis would be that reduced code distance is a good metric to predict the logical error rate of the lattice,
since the number of physical errors required to cause a logical error is a minimum on the shortest logical operator,
which is the minimum distance between corresponding boundaries.
 However, the effect is more complex. We will explore this further in Section~\ref{subsec:num_res_random} and Chapter~\ref{chap:conclusion}.

 \begin{landscape}
 \begin{table}[b]
  \baselineskip=2mm
  \caption[The $X$ error rate of single-faulty lattices and corresponding $Z$ stabilizer data.]
  {Averages here are arithmetic means.
  ``Faulty location'' is the location of the faulty device (static loss).
  ``\#X and Z stabs'' stands for the total number of X stabilizers and Z stabilizers.
  ``Reduced distance'' is the minimum distance between corresponding boundaries shortened by merging stabilizers.
  ``\#Z stabs'' is the number of Z stabilizers.
  ``Biggest \#dataq of Z stabs'' is the largest number of data qubits in a Z stabilizer.
  ``Ave. $CDQ$ of Z stabs'' is the average of $CDQ$s (metric is the space-time product of an error correction circuit: the number of data qubits $DQ$ involved, multiplied by the ``cycle'', the sum of the circuit depth and the waiting time for next stabilization $C$, after~\cite{steane02:ft-qec-overhead}) of Z stabilizers.
  ``Biggest Z $CDQ$'' is the largest $CDQ$ for any stabilizer circuit of the chip.
  ``Ave. \#dataq of Z stabs'' is the average of the number of data qubits in Z stabilizers.
  }
\renewcommand{\baselinestretch}{1.5}
\baselineskip=1.2\normalbaselineskip
  \label{tab:num_static_result}
 \begin{tabular}[t]{c|c|c|c|c|c|c|c|c|c|c|c}
  faulty &code    &\#X and  &residual&reduced &\#Z stabs&biggest&steps per &ave. $CDQ$&biggest&ave. steps &ave. \#dataq\\
location &distance&Z        &X error &distance&         &\#dataq&error     &of Z      &Z $CDQ$&per error  &of Z stabs\\
         &        &stabs    &rate    &        &         &of Z   &correction&stabs     &       &correction & \\
         &        &         &        &        &         &stabs  &cycle     &          &       &of Z stabs & \\
  \hline
center&5&38&7.038E-02&4&19&6&32&35.871&195.194&8.844&3.684\\
west&5&38&6.843E-02&4&19&6&32&35.803&195.194&8.822&3.684\\
  northwest&5&38&3.335E-02&4&19&6&32&35.648&195.194&8.791&3.684\\
  \hline
center&7&82&3.704E-02&6&41&6&32&32.249&195.194&8.149&3.756\\
west&7&82&3.319E-02&6&41&6&32&32.270&195.194&8.153&3.756\\
northwest&7&82&1.969E-02&6&41&6&32&32.505&195.194&8.228&3.756\\
  \hline
center&9&142&2.224E-02&8&71&6&32&31.215&195.194&7.943&3.803\\
west&9&142&1.873E-02&8&71&6&32&31.456&195.194&8.016&3.803\\
northwest&9&142&1.046E-02&8&71&6&32&31.539&195.194&8.033&3.803\\
  \hline
center&13&310&8.776E-03&12&155&6&32&30.711&195.194&7.824&3.858\\
west&13&310&8.662E-03&12&155&6&32&31.028&195.194&7.910&3.858\\
northwest&13&310&4.869E-03&12&155&6&32&31.084&195.194&7.925&3.858\\
 \end{tabular}
 \end{table}
 \end{landscape}

 \subsection{Random multiple faulty devices}
 \label{subsec:num_res_random}
We generated 30 randomly defective lattices for each combination of three yields, $80\%$, $90\%$ and $95\%$
and of 5 code distances, 5, 7, 9, 13, 17, so that
we generated 450 lattices.
Table \ref{tab:num_simulated} shows the number of defective lattices generated and simulated.
On some defective lattices, by chance the faulty qubit placement
results in a lattice for which we are unable to build an effective circuit for 
encoding a logical qubit, so they are not simulated.
Our software successfully built circuits for almost all lattices at $y=0.90$ and above,
but only about two-thirds at $y=0.80$.
  \begin{table}[b]
   \caption[The number of defective lattices generated and simulated.]{}
   \label{tab:num_simulated}
 \begin{tabular}[t]{c||c|c|c|c|c|c|c|c|c|c|c|c|c|c|c}
  yield         & \multicolumn{5}{|c|}{0.80}& \multicolumn{5}{|c|}{0.90}& \multicolumn{5}{|c}{0.95}\\
  code distance& 5& 7& 9&13&17& 5& 7& 9&13&17& 5& 7&9& 13& 17\\ \hline  \hline
  \#encodable        &     20&     24&     22&     19&     19&     29&     29&     30&     30&     30&     28&     30&     30&     30&     30\\
  \#unencodable      &     10&      6&      8&     11&     11&      1&      1&      0&      0&      0&      2&      0&      0&      0&      0\\
 \end{tabular}
  \end{table}

  \begin{table}[b]
   \caption[The average number of faulty qubits in all generated lattices, in 50\%-culled lattices and in 90\%-culled lattices, respectively.]
   {The numbers of qubits of perfect lattice of distance 5, 7, 9, 13 and 17 are 81, 169, 289, 625 and 1089, respectively.}
   \label{tab:num_faulty}
   \scalebox{0.8}[1.0]{
 \begin{tabular}[t]{c||c|c|c|c|c|c|c|c|c|c|c|c|c|c|c}
  y         & \multicolumn{5}{|c|}{0.80}& \multicolumn{5}{|c|}{0.90}& \multicolumn{5}{|c}{0.95}\\ \hline \hline
  d& 5& 7& 9&13&17& 5& 7& 9&13&17& 5& 7&9& 13& 17\\ \hline  \hline  
all &13.9&29.7&54.0&122.6&215.6&8.4&16.2&29.6&61.0&107.3&3.8&8.0&14.7&31.3&53.7\\
50\%&13.2&28.9&55.1&122.6&214.5&7.3&14.2&26.7&56.3&101.7&2.9&7.1&11.9&28.5&51.3\\
90\%&12.3&29.0&55.0&121.3&221.7&5.7&9.3&23.7&48.0&96.0&1.7&4.7&9.0&25.3&51.3
 \end{tabular}
}
  \end{table}
  
  This unencodable condition occurs when a defective data qubit chain stretches from a boundary of the lattice
  to the other boundary of the same type (south and north for Z stabilizer boundary, or west and east for X stabilizer boundary).
  For instance, if a faulty qubit is on a boundary, say, the $qubit 1$ which is stabilized by $Z_1$ of the stabilizer $Z_1Z_2Z_3Z_4$ and
  the qubit is not stabilized by another Z stabilizer, then $Z_1Z_2Z_3Z_4$ cannot be merged with another stabilizer to work around $Z_1$.
  Hence we remove $Z_1Z_2Z_3Z_4$ with $qubit 1$ and eventually $qubit 2$, $qubit 3$ and $qubit 4$ become a part of the boundary instead.
  In general, this adaptation reduces the code distance (shown in Tables \ref{tab:linear_correlation} and \ref{tab:exponential_correlation}).
  Therefore, a lattice of lower yield and of lower code distance may be unencodable with higher probability.
  Though only 30 instances for each condition are too few to get explicit statistics, table \ref{tab:num_simulated} shows
  this trend at yields of 90\% and 95\%.
  $Y=80\%$ might be saturated in encodable rates
  because the code distances do not show meaningful differences.

\begin{landscape}
\begin{figure}[t]
 \begin{center}
  \includegraphics[width=6cm]{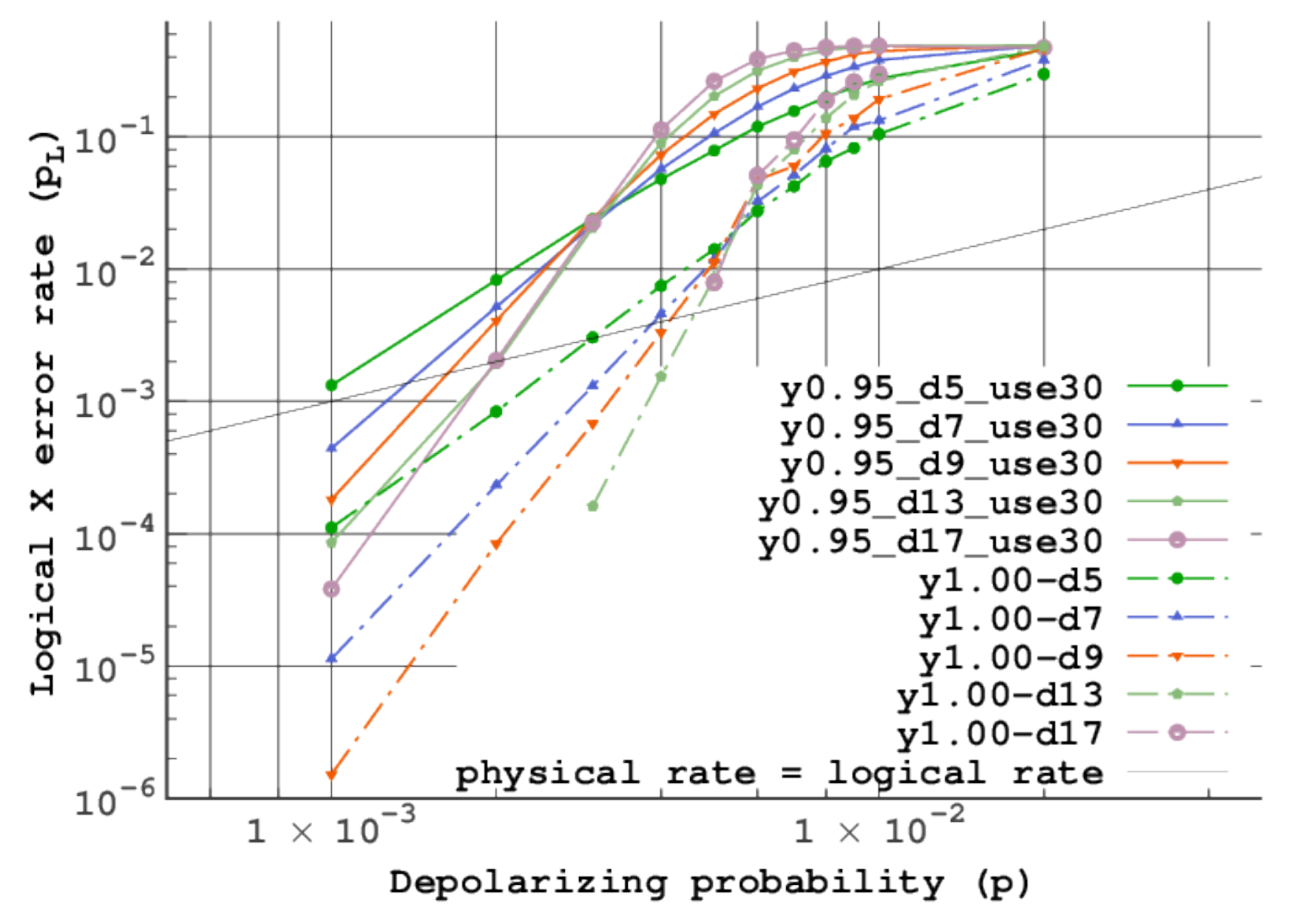}
  \includegraphics[width=6cm]{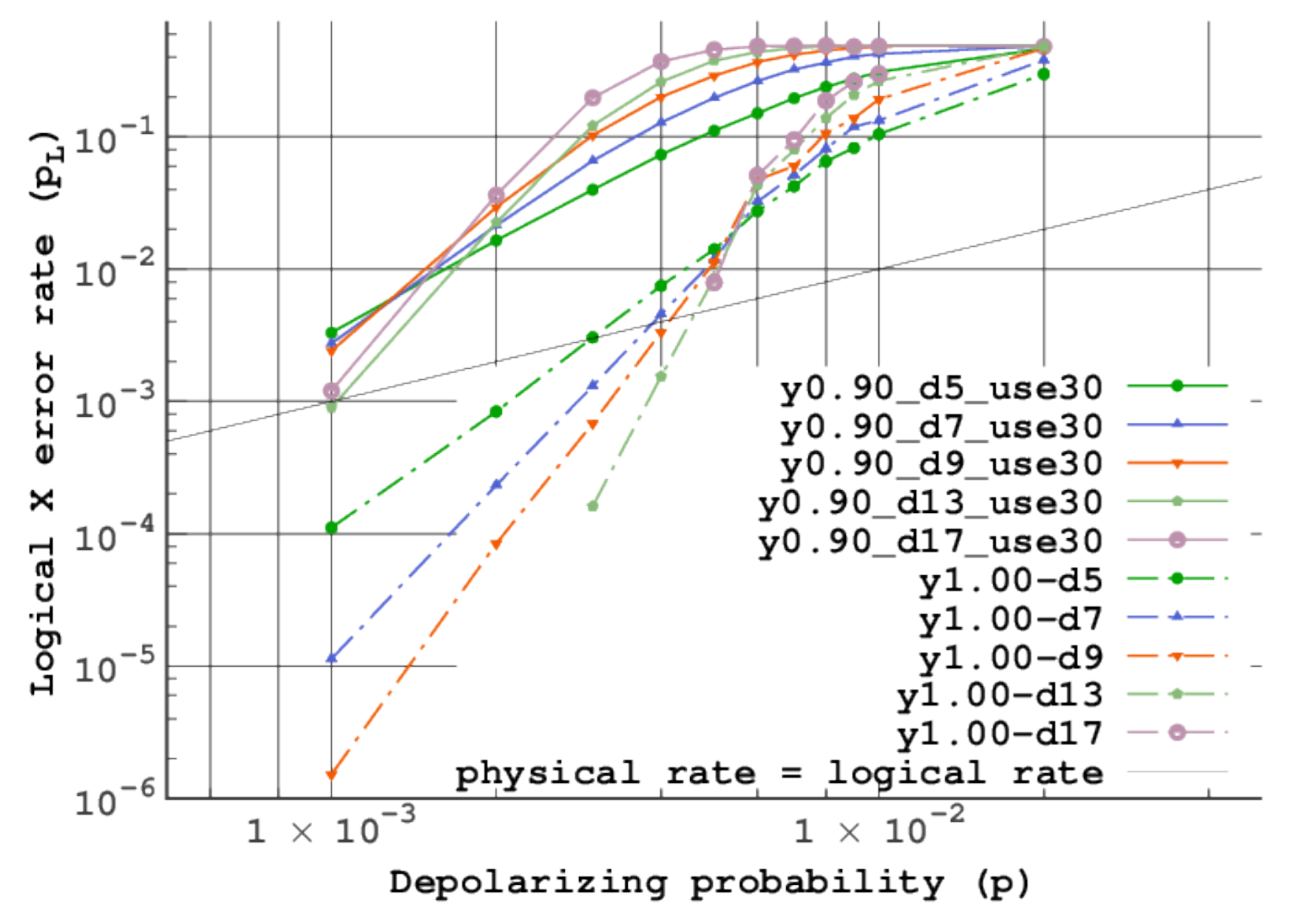}
  \includegraphics[width=6cm]{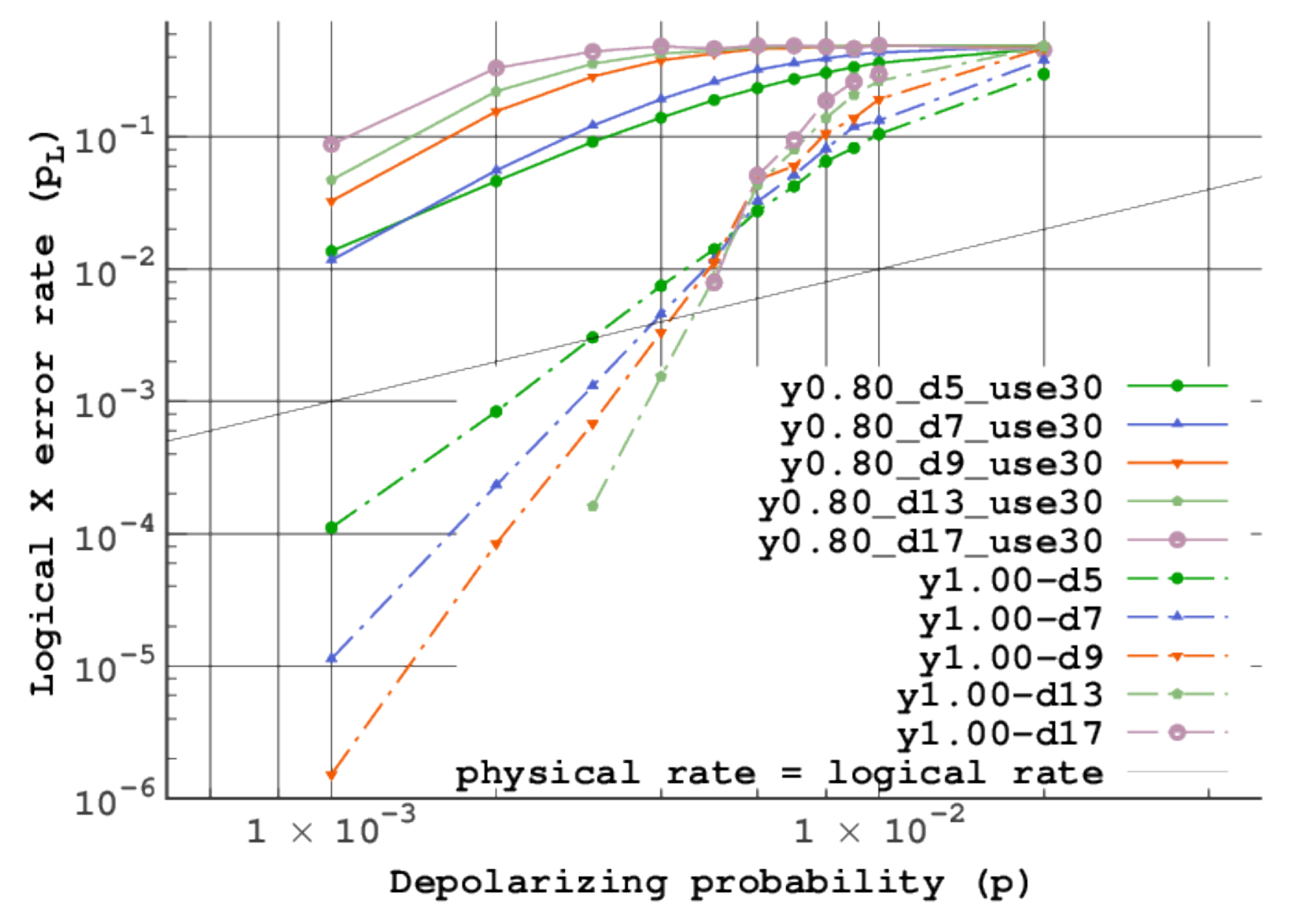}\\
  \includegraphics[width=6cm]{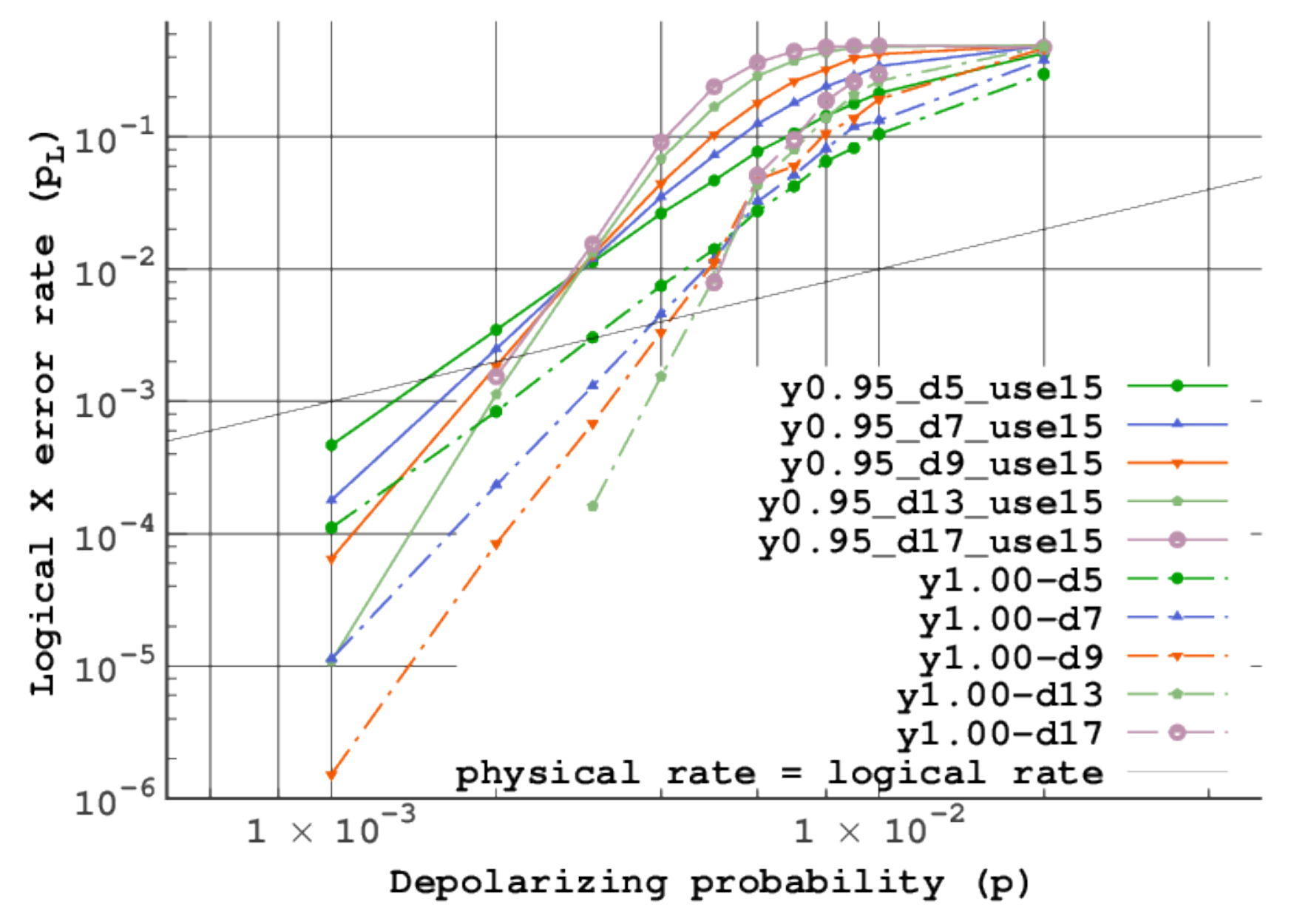}
  \includegraphics[width=6cm]{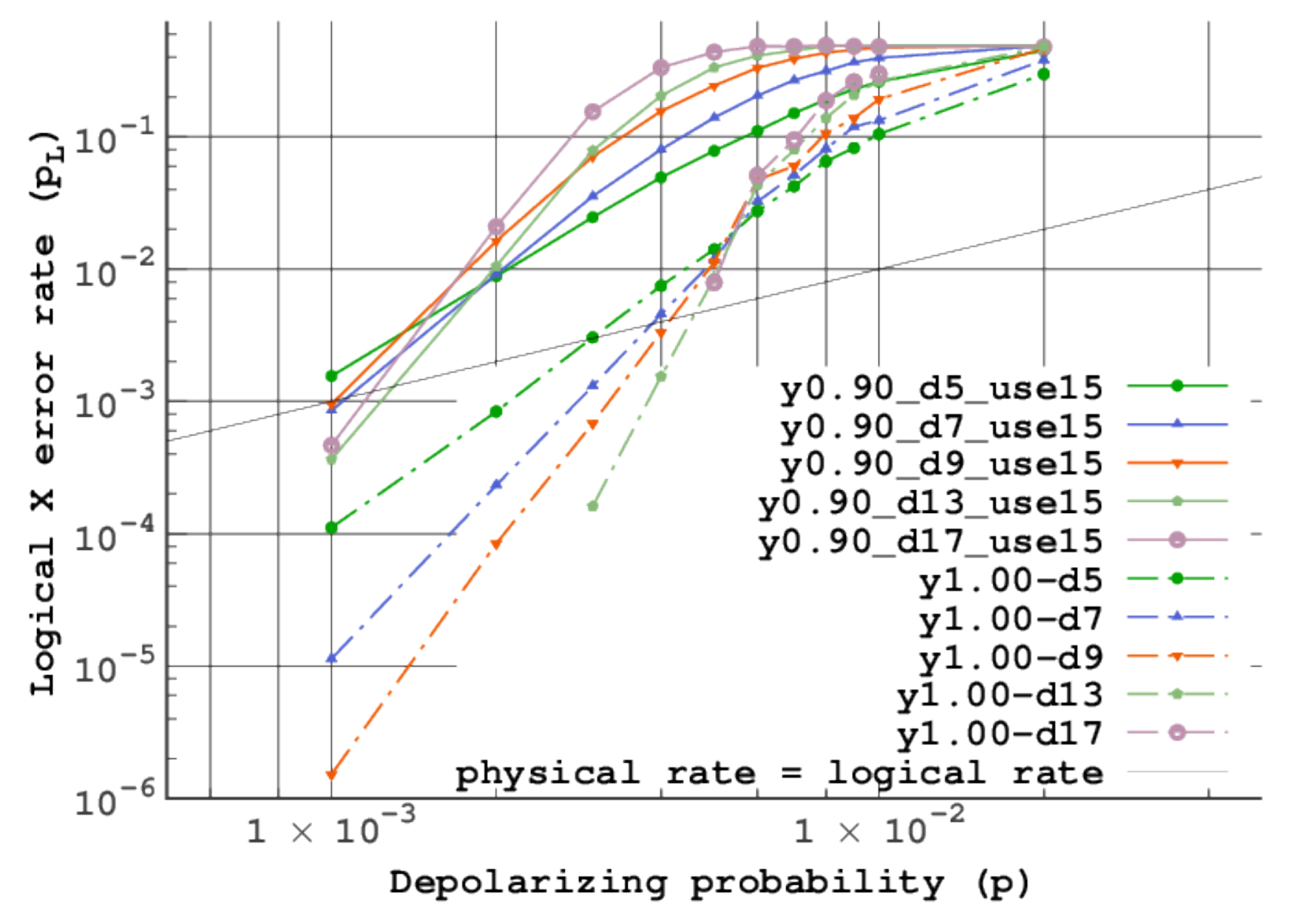}
  \includegraphics[width=6cm]{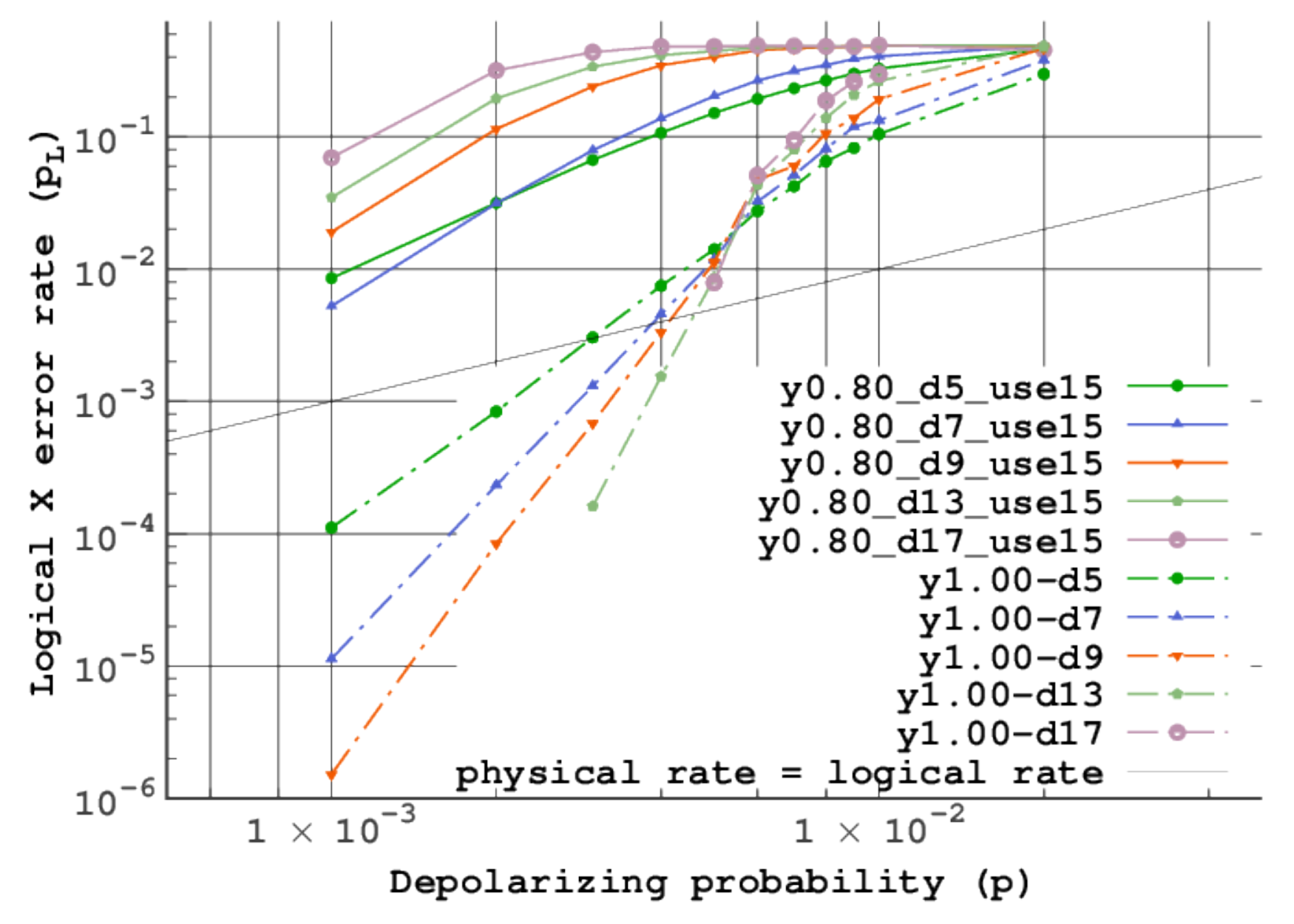}\\
  \includegraphics[width=6cm]{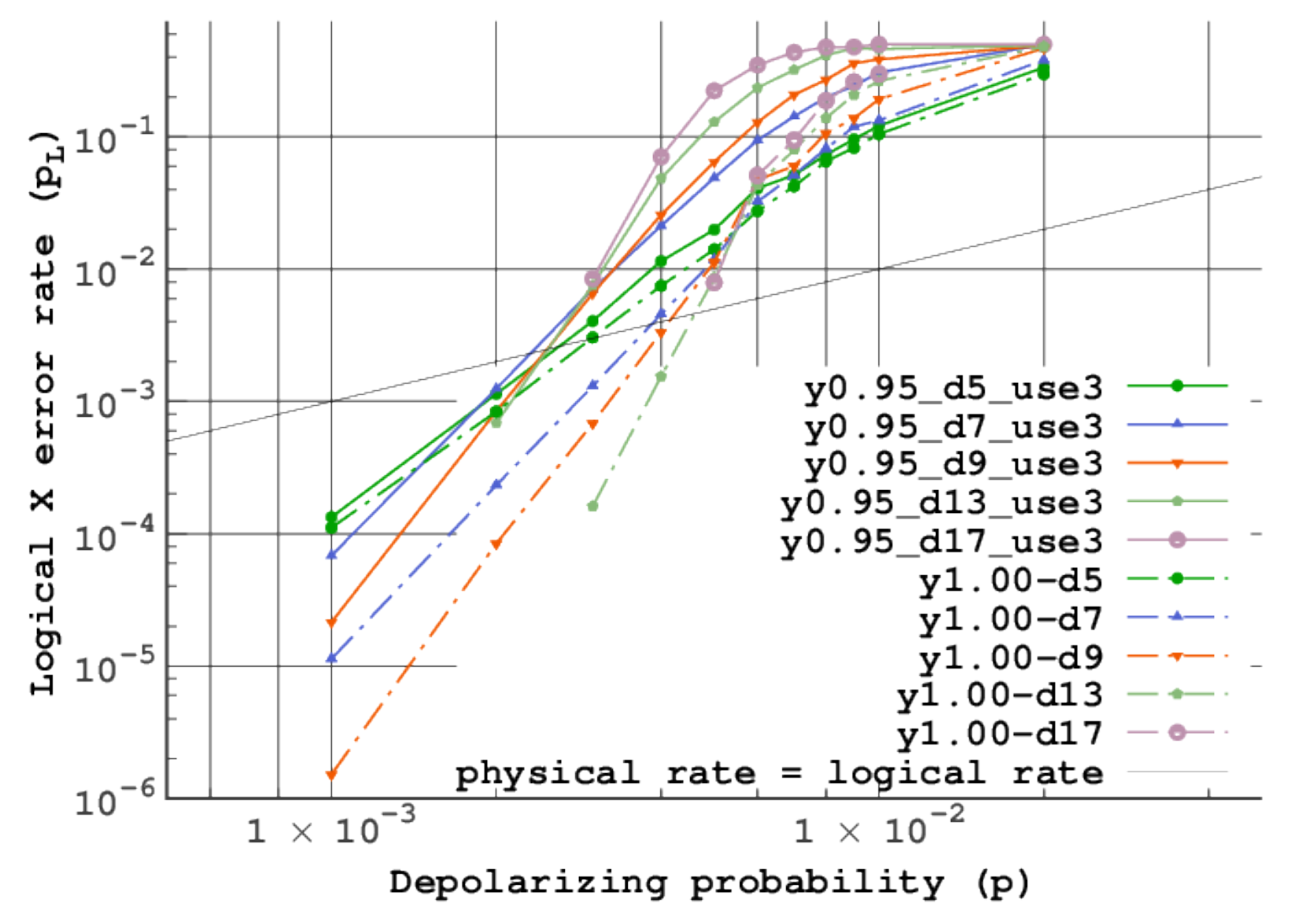}
  \includegraphics[width=6cm]{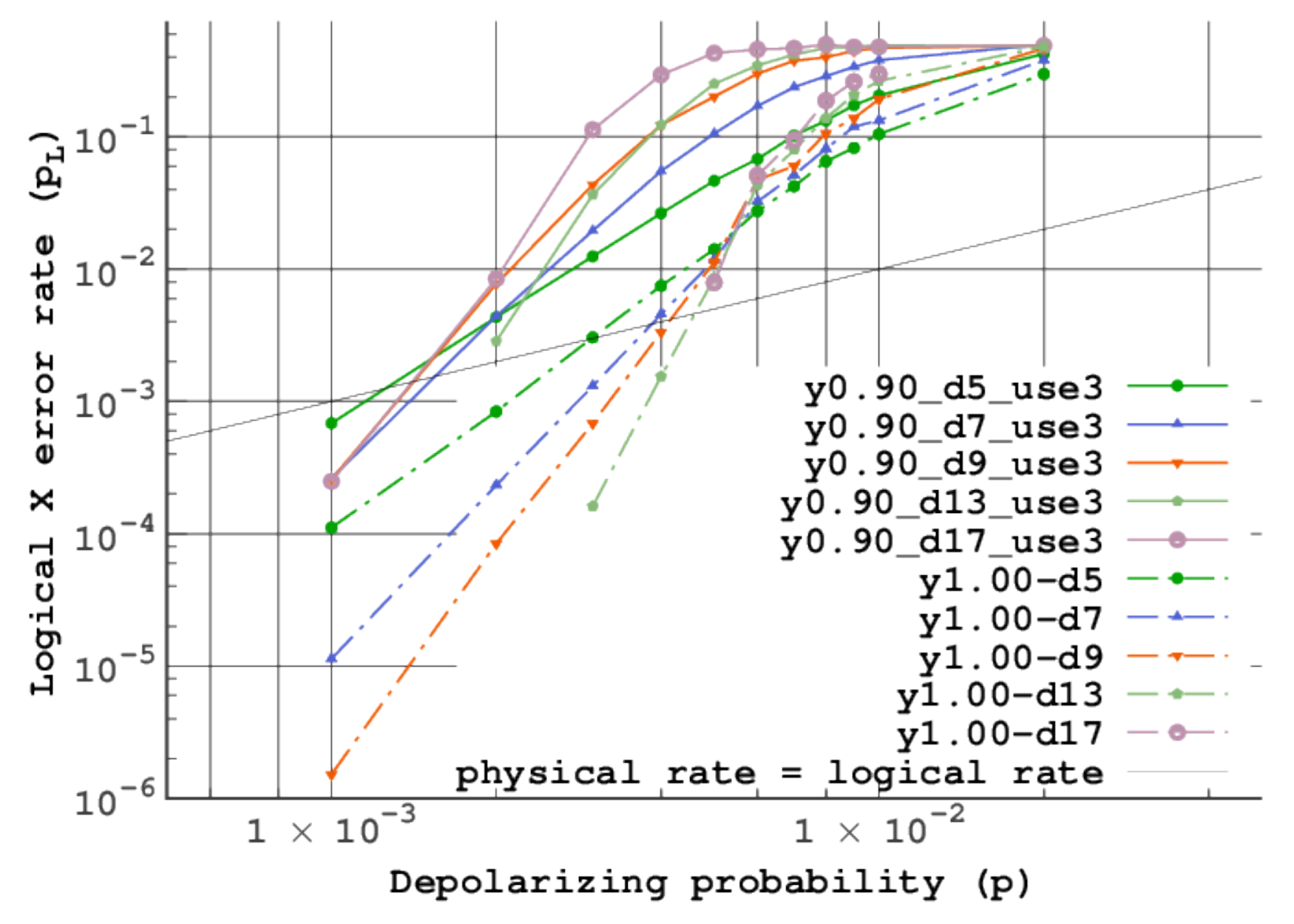}
  \includegraphics[width=6cm]{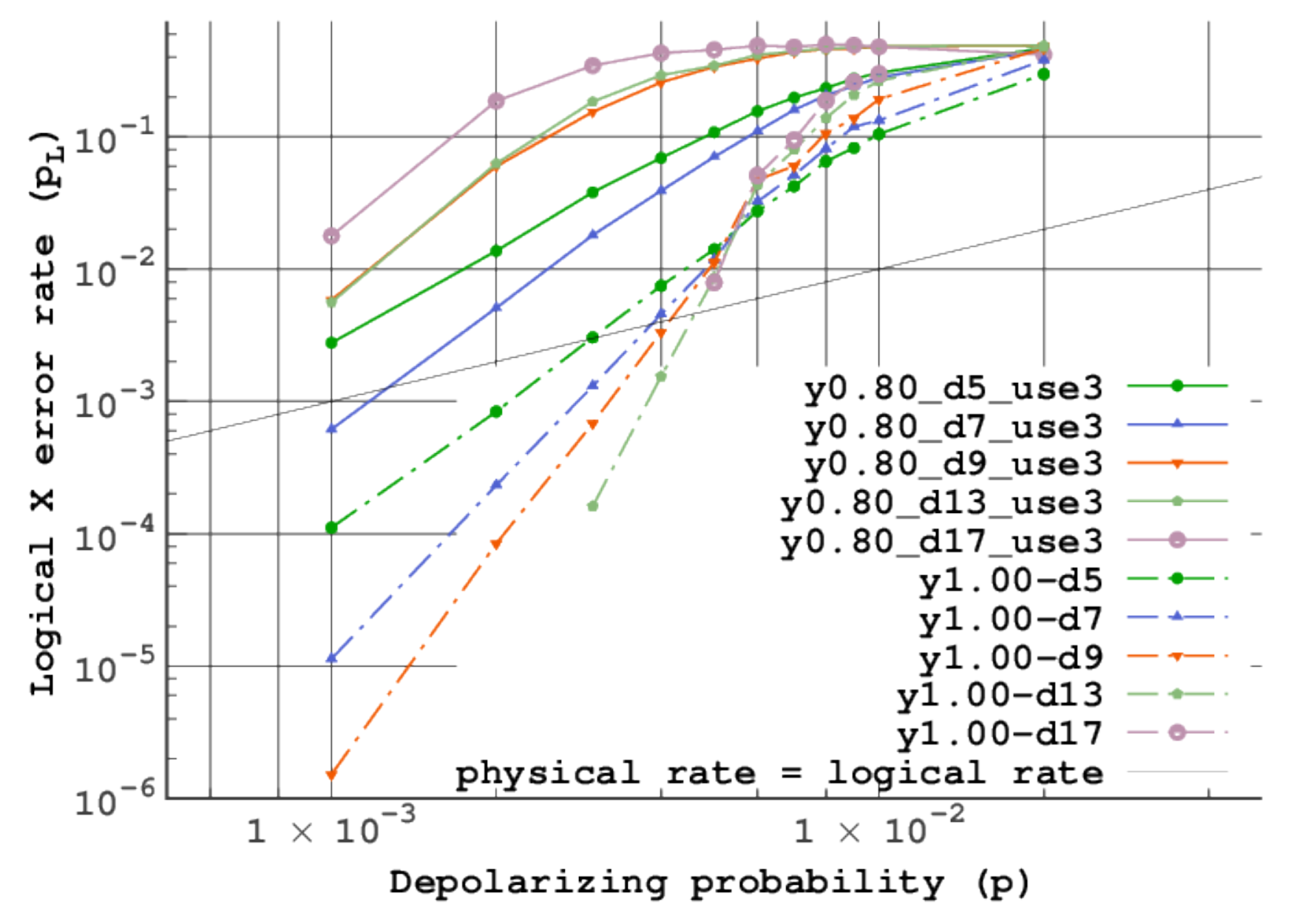}
 \end{center}
\caption[Graphs of randomly defective lattice.]
{Dashed lines are of the perfect lattices for reference.
The left column is of $y=95\%$,
the middle column is of $y=90\%$
and the right column is of $y=80\%$.
The top row is of all generated lattices,
the middle row is of culling worse 50\%
and the bottom row is of culling worse 90\%.
}
\label{fig:graphs_random}
\end{figure}
\end{landscape}

  Figure \ref{fig:graphs_random} shows the geometric mean of sets of encodable lattices, plotting physical error rates versus logical error rates.
  Appendix \ref{subsec:scatterplots} shows the scatter plots of raw data.

  The left column in Figure~\ref{fig:graphs_random} show the graphs of $y=95\%$, describing the geometric mean of all encodable lattices,
  of the better 50\%, and of the best 10\% of generated lattices, from the top, respectively.
  Note that those cull percentages are based on the original set of 30 generated lattices,
  not the smaller number of the encodable lattices.
  Some points of longer distance at lower physical error rate are not plotted
  since not enough logical errors are accumulated because of the very low logical error rates.
  
  At $95\%$ functional qubit yield, we see many chips beating break-even at $p=10^{-3}$.
  The threshold is about $0.3\%$, about half of the threshold error rate for a perfect lattice.
  The significant
  penalty in both threshold and residual error rate can be dramatically reduced by culling poorer chips and discarding them.
  At $50\%$ cull at $p=10^{-3}$, the residual error rate for $d=7$
  is about that of $d=5$ with a perfect lattice, and $d=13$ is about that of a perfect $d=7$.  

Naturally, the logical error rates get better as we discard more of the
poorest lattices.  At $p=0.2\%$, 
unculled $y=95\%$ shows that even
distance 17 is just on the break-even line and 90\%-culled (c)
shows that all 5 distances exceed break-even.  The steepness of the
slope of the curves of culled defective lattices exceeds that of the
curves of lower code distances on the perfect lattice, though it does
not match the perfect lattice of the same distance.  Thus, an
appropriate culling strategy reduces the penalty for a 5\% fault rate
to a manageable level, allowing us to achieve a desired level of error
suppression by using a slightly larger code distance.
  At $p=0.1\%$, by culling $90\%$, the penalty against the perfect lattices changes
  from $12.0\times$ to $1.2\times$ at $d=5$,
  from $39.0\times$ to $6.1\times$ at $d=7$, and
  from $119.9\times$ to $14.2\times$ at $d=9$.
  We do not have data points for $90\%$-culled of $d=13$ and of $d=17$ since
  not enough logical errors are accumulated on their best $10\%$ of the lattices.
  The smaller code distance gets closer to the perfect lattice
  because it has fewer qubits, therefore
  good outliers may be generated with higher probability,
  as shown in Table~\ref{tab:num_faulty}.
  Table~\ref{tab:num_faulty} summarizes the average number of static losses on all the generated lattices,
  on the 50\%-culled lattices and on the 90\%-culled lattices.
  The remaining 3 lattices of 90\%-culled distance 5 have 1, 2 and 2 static losses respectively.
  Table~\ref{tab:num_faulty} also allows us to see that the importance of static loss placement
  because the numbers of static losses of longer distances do not decrease much
  but all the logical error rates gets better.

  The middle column in Figure~\ref{fig:graphs_random} is the graphs of $y=90\%$,
  under the same conditions with those of $y=95\%$.
  The threshold is about $0.15\%$, about a quarter of the threshold for a perfect lattice.
  At $90\%$ cull (in the middle-bottom of Figure~\ref{fig:graphs_random}), 
  at $p=10^{-3}$, the residual error rates for $d=7$, $d=9$ and $d=17$
  are about twice those of $d=5$ with a perfect lattice. $d=13$ would be better than that of perfect $d=5$,
  but it is missing since the logical error rate may be too low to accumulate enough number of logical errors.
  At $p=0.1\%$ of $y=90\%$,
  unculled (middle-top) shows that only distance 13 exceeds the break-even,
  but 90\%-cull (middle-bottom) shows all five distances exceed the break-even.

  The right column in Figure~\ref{fig:graphs_random} is the graphs of $y=80\%$.
  At $y=80\%$, we have already seen that only two-thirds of the chips can even be encoded.
  Our simulations indicate that even those chips for which we could create a circuit are unusable.
  Even at $p=10^{-3}$, there is no evidence of a correctable threshold, and although residual error rates
  do decline as the physical error rate is reduced, only a single data point reaches break-even.
  We conclude that $y=80\%$ is not good enough to build a computer.

Unculled $y=95\%$ shows that distances 13 and 17 are approximately identical at $p=0.2\%$, while other distances show that longer is better.
Unculled distance 13 and 17 for $y=90\%$ do not show that longer is better, though other distances do.
Unculled $y=80\%$ shows that distance 7 exceeds distance 5 at $p=0.1\%$, while other distances do not show an improvement for the longer distance.
Those indicate that the longer code distances cross at lower physical error rate.
We need to consider this fact when deciding the code distance to use.  

\subsection{Metrics for selecting good chips}
\label{subsec:metrics}
  Both to improve our understanding of the root causes of the error rate
  penalty and to provide a simple means of selecting good chips, we evaluated the
  correlation between a set of easy-to-calculate metrics and the simulated residual error rate.
  Tables \ref{tab:linear_correlation} and \ref{tab:exponential_correlation} describe the correlations 
  between eighteen metrics and logical error rates or log(logical error rates), respectively,
  for $p=0.002$ for each combination of yield and code distance.

  The simplest possible metrics, just counting numbers of qubits in various categories, show only modest correlation.

  Steane's $KQ$ metric is the space-time product of a circuit: the number of qubits $Q$ involved, multiplied by the
  circuit depth $K$~\cite{steane02:ft-qec-overhead}.

  The $CDQ$ and $CQ$ is the product of the ``cycle'', which is the average 
  number of steps in a stabilizer measurement
  and the number of data qubits or the total number of qubits 
  including ancillae involved in the stabilizer, respectively.
  The $CDQ$ and $CQ$ reflect the total probabilities of possible physical errors which occur 
  in a measurement of the stabilizer.
  Both Tables \ref{tab:linear_correlation} and \ref{tab:exponential_correlation} 
  indicate that 
  the average of the $CDQ$ and the average of the $CQ$ of $Z$ stabilizers 
  have the strongest and the second strongest correlations with the logical $X$ error rate.
  The average number of qubits in a $Z$ stabilizer and
  the average ``cycle'' of $Z$ stabilizers show the next strongest correlations.
  Those mean that the accumulation of possible errors in a stabilizer may be the factor most 
  strongly correlated to the logical error rate.

  Somewhat to our surprise, both the $KQ$ of the largest stabilizer and the average across the entire lattice 
  do not have good correlations.
  This may be because this form of $KQ$ does not correctly capture the total probabilities of possible physical errors which occur 
  in a measurement of the stabilizer.

  Table \ref{tab:num_faulty} implies that the number of faulty devices is correlated
  with the logical error rate. 
  By culling bad lattices, 
  Table \ref{tab:num_faulty} shows that the average number of faulty devices on a lattice is reduced
  and Figure~\ref{fig:graphs_random}
  shows that the logical error rate gets better.
  However, the average $CDQ$ of $Z$ stabilizers has significantly higher correlation with
  logical $X$ error rate, 0.76, than that of the number of faulty devices, 0.43.
  We calculated the cross-correlation of elements for $y=0.95$ and $d=9$.
  The correlation between the number of faulty devices and the average $CDQ$ of $Z$ stabilizers is 0.79.

  The number of faulty ancilla qubits is the most weakly correlated to the logical error rate.
  This fact indicates that even if the number of faulty ancilla qubit increases,
  the logical error rate does not decline rapidly.
  For a given yield, the placement of faults matters more than the exact number.

\begin{landscape}
   \begin{center}
 \begin{table}[t]
   \begin{center}
     \scalebox{0.6}[0.8]{
  \begin{tabular}[t]{c|c||c|c|c|c||c|c||c|c|c|c||c|c||c|c|c|c||c|c||c|c|c|c||c}
yield& code  &\#stab&\#flty&\#flty&\#flty & $Z$   &\#$Z$& bgst  & ave.  & bgst  & ave.  & dpst  & ave.  & bgst & ave. & bgst & ave. & bgst & ave. & bgst & ave. & bgst & ave. & ave.\\
     & dist. &      & qubit& data & syn.  & redu- &stab& \#qubit&\#qubit&\#data &\#data & depth & depth & $KQ$ & $KQ$ & $KDQ$& $KDQ$& $Z$  & $Z$  & $CQ$ & $CQ$ & $CDQ$& $CDQ$& \#$Z$ \\
     &       &      &      & qubit& qubit & ced   &     & of $Z$& of $Z$& qubit & qubit & of $Z$& of $Z$& of   & of   & of   & of   & cycle& cycle& of   & of   & of   & of   & stab\\
     &       &      &      &      &       & dist. &     & stabs & stabs & of $Z$& of $Z$& stabs & stabs & $Z$  & $Z$  & $Z$  & $Z$  &      &      & $Z$  & $Z$  & $Z$  & $Z$  & msmts\\
     &       &      &      &      &       &       &     &       &       & stabs & stabs &       &       & stabs& stabs& stabs& stabs&      &      & stabs& stabs& stabs& stabs& /step\\
\hline
\hline
0.95&5&-0.64&0.59&0.70&0.08&-0.67&-0.71&0.81&0.83&0.79&0.71&0.26&-0.01&0.24&0.37&0.23&0.16&0.69&0.71&0.74&\textbf{0.88}&0.76&0.86&-0.69\\
0.95&7&-0.40&0.45&0.65&-0.13&-0.62&-0.46&0.65&0.44&0.70&0.44&0.05&0.00&0.07&0.11&0.08&0.04&0.78&0.55&0.76&0.78&\textbf{0.79}&\bf{0.79}&-0.44\\
0.95&9&-0.50&0.68&0.72&0.33&-0.60&-0.48&0.63&0.81&0.64&0.79&0.42&-0.07&0.46&0.48&0.40&0.24&0.62&0.81&0.67&0.88&0.68&{\bf 0.89}&-0.69\\
0.95&13&-0.28&0.48&0.55&0.11&-0.19&-0.27&0.36&0.56&0.37&0.52&-0.02&-0.47&0.02&-0.14&0.03&-0.25&0.30&0.59&0.33&\textbf{0.71}&0.34&0.69&-0.40\\
0.95&17&-0.18&0.34&0.27&0.25&-0.10&-0.18&0.20&0.39&0.15&0.40&0.05&-0.32&0.17&-0.06&0.05&-0.16&0.13&0.50&0.11&\textbf{0.56}&0.10&0.55&-0.40\\
\hline
0.90&5&-0.19&0.35&0.26&0.29&-0.48&-0.24&\textbf{0.62}&0.20&0.55&0.16&-0.23&-0.21&0.06&-0.13&-0.11&-0.16&0.49&0.28&0.69&0.48&\textbf{0.62}&0.46&-0.21\\
0.90&7&-0.31&0.37&0.43&0.07&-0.59&-0.33&0.50&0.53&0.49&0.45&0.20&-0.33&0.59&0.23&0.43&-0.01&0.51&0.58&0.71&0.71&0.70&\textbf{0.74}&-0.46\\
0.90&9&-0.39&0.40&0.54&0.09&-0.65&-0.42&0.67&0.65&0.63&0.58&-0.29&-0.54&0.01&-0.12&-0.18&-0.32&0.28&0.39&0.43&\textbf{0.68}&0.43&0.66&-0.45\\
0.90&13&-0.36&0.47&0.63&-0.01&-0.30&-0.37&0.47&0.77&0.52&\textbf{0.81}&-0.26&-0.25&-0.10&0.24&-0.28&0.04&0.46&0.57&0.60&\textbf{0.81}&0.62&\textbf{0.81}&-0.51\\
0.90&17&-0.55&0.58&0.62&0.10&-0.62&-0.56&0.56&0.76&0.56&0.74&-0.05&-0.25&0.34&0.33&0.10&0.08&0.17&0.65&0.35&0.86&0.37&\textbf{0.87}&-0.74\\
\hline
0.80&5&-0.20&0.30&0.22&0.17&-0.43&-0.22&\textbf{0.78}&0.59&0.46&0.40&-0.11&-0.13&0.43&0.30&-0.19&0.05&0.64&0.72&\textbf{0.78}&0.75&0.63&0.67&-0.49\\
0.80&7&0.22&0.35&0.10&0.40&-0.51&0.13&0.56&0.81&0.44&0.75&0.07&-0.34&0.29&0.58&0.03&0.24&0.56&0.74&0.64&0.79&0.69&\textbf{0.83}&-0.48\\
0.80&9&-0.39&0.37&0.31&0.15&-0.47&-0.38&0.63&0.88&0.68&0.68&-0.42&-0.45&0.41&0.58&0.12&-0.01&0.50&0.77&0.58&0.86&0.66&\textbf{0.92}&-0.53\\
0.80&13&0.02&0.34&0.18&0.20&-0.43&-0.05&0.65&0.79&0.66&0.80&0.21&0.09&0.45&0.68&0.22&0.52&0.70&0.63&0.65&0.79&0.70&\textbf{0.84}&-0.35\\
0.80&17&0.39&0.25&0.19&0.19&-0.24&0.34&0.37&0.60&0.29&0.57&-0.15&-0.41&0.28&0.37&0.04&0.07&0.51&0.55&0.34&0.61&0.33&\textbf{0.62}&0.16\\
\hline
\multicolumn{2}{c||}{average}&-0.25&0.42&0.42&0.15&-0.46&-0.28&0.56&0.64&0.53&0.59&-0.02&-0.25&0.25&0.25&0.06&0.04&0.49&0.60&0.56&0.74&0.56&\textbf{0.75}&-0.44
  \end{tabular}
  }
\renewcommand{\baselinestretch}{0.5}
\baselineskip=1.0\normalbaselineskip
  \caption[Linear Correlation between candidate metrics for lattice quality factors and X logical error rates for each combination of yield and code distance.]
    {\small Boldface is the strongest correlation in each row.
    ``Dist.'', ``flty'', ``bgst'', ``syn.'', ``stab'' and ``msmt'' stand for distance, faulty, biggest, syndrome, stabilizer and measurement, respectively.
    Averages here are arithmetic means.
    ``Reduced dist.'' is the minimum distance between corresponding boundaries shortened by merging stabilizers.
    ``\#Z stabs'' is the number of Z stabilizers.
    ``Bgst \#qubit of Z stabs'' and
    ``ave. \#qubit of Z stabs'' is the biggest and the average number of qubits involving data qubits and syndrome qubits in Z stabilizer circuits.
    ``Bgst \#data qubit of Z stabs'' and
    ``ave. \#data qubit of Z stabs'' is the biggest and the average number of data qubits in a Z stabilizer.
    ``Dpst depth of Z stabs'' is the depth of the deepest Z stabilizer circuit.
    ``Ave. depth of Z stabs'' is the average depth of Z stabilizer circuits.
    ``Bgst $KQ$ of Z stabs'' and
    ``Ave. $KQ$ of Z stabs'' is the biggest and average $KQ$ (metric is the space-time product of a circuit: the number of qubits $Q$ involved, multiplied by the circuit depth $K$~\cite{steane02:ft-qec-overhead}) of Z stabilizer circuits.
    ``Bgst $KDQ$ of Z stabs'' and
    ``Ave. $KDQ$ of Z stabs'' is the biggest and average $KDQ$ which is the product of the number of data qubits $DQ$ involved and of the circuit depth $K$ of Z stabilizer circuits.
    ``Cycle'' here indicates every how many steps a stabilizer is measured, including waiting time by scheduling of stabilizers.
    ``Bgst Z cycle'' and
    ``Ave. Z cycle'' is the biggest and the average cycle of Z stabilizers.
    ``CQ'' is the product of the cycle $C$ and the number of qubits $Q$, similar to $KQ$.
    ``Bgst $CQ$'' and
    ``Ave. $CQ$'' is the biggest and the average $CQ$ of stabilizers.
    ``Bgst $CDQ$'' and
    ``Ave. $CDQ$'' is the biggest and average $CDQ$ which is the product of the number of data qubits $DQ$ involved and of the cycle $C$ of Z stabilizer circuits.
    ``Ave. \#Z stabs per step'' is how many stabilizers are measured in a step on average.
  }
\renewcommand{\baselinestretch}{1.5}
\baselineskip=1.2\normalbaselineskip
  \label{tab:linear_correlation}
  \end{center}
 \end{table}
  \end{center}
 \end{landscape}

\begin{landscape}
   \begin{center}
 \begin{table}[t]
   \begin{center}
     \scalebox{0.6}[0.9]{
  \begin{tabular}[t]{c|c||c|c|c|c||c|c||c|c|c|c||c|c||c|c|c|c||c|c||c|c|c|c||c}
yield& code  &\#stab&\#flty&\#flty&\#flty & $Z$   &\#$Z$& bgst  & ave.  & bgst  & ave.  & dpst  & ave.  & bgst & ave. & bgst & ave. & bgst & ave. & bgst & ave. & bgst & ave. & ave.\\
     & dist. &      & qubit& data & syn.  & redu- & stab&\#qubit&\#qubit&\#data &\#data & depth & depth & $KQ$ & $KQ$ & $KDQ$& $KDQ$& $Z$  & $Z$  & $CQ$ & $CQ$ & $CDQ$& $CDQ$& \#$Z$ \\
     &       &      &      & qubit& qubit & ced   &     & of $Z$& of $Z$& qubit & qubit & of $Z$& of $Z$& of   & of   & of   & of   & cycle& cycle& of   & of   & of   & of   & stab\\
     &       &      &      &      &       & dist. &     & stabs & stabs & of $Z$& of $Z$& stabs & stabs & $Z$  & $Z$  & $Z$  & $Z$  &      &      & $Z$  & $Z$  & $Z$  & $Z$  & msmts\\
     &       &      &      &      &       &       &     &       &       & stabs & stabs &       &       & stabs& stabs& stabs& stabs&      &      & stabs& stabs& stabs& stabs& /step\\
\hline
\hline
0.95&5&-0.63&0.66&0.71&0.18&-0.74&-0.68&0.87&0.76&\textbf{0.88}&0.64&0.44&-0.05&0.40&0.33&0.39&0.12&0.72&0.75&0.72&0.82&0.76&0.83&-0.75\\
0.95&7&-0.49&0.62&0.79&-0.05&-0.67&-0.53&0.57&0.59&0.65&0.55&0.27&0.06&0.29&0.26&0.30&0.15&0.79&0.69&0.69&0.85&0.73&\textbf{0.87}&-0.54\\
0.95&9&-0.63&0.81&0.83&0.42&-0.63&-0.62&0.70&0.77&0.69&0.72&0.39&-0.15&0.43&0.38&0.36&0.14&0.66&0.83&0.69&0.87&0.70&\textbf{0.88}&-0.79\\
0.95&13&-0.31&0.58&0.66&0.15&-0.29&-0.31&0.43&0.71&0.43&0.68&-0.05&-0.30&0.01&0.09&0.03&-0.03&0.31&0.70&0.41&\textbf{0.82}&0.40&0.81&-0.52\\
0.95&17&-0.16&0.18&0.17&0.12&-0.08&-0.16&0.25&0.24&0.18&0.29&-0.02&-0.24&0.08&-0.08&-0.02&-0.13&0.15&0.34&0.17&\textbf{0.41}&0.16&0.40&-0.28\\
\hline
0.90&5&-0.10&0.41&0.28&0.36&-0.42&-0.15&0.62&0.32&0.53&0.19&-0.14&-0.32&0.13&-0.17&-0.01&-0.24&0.52&0.45&\textbf{0.66}&0.60&0.59&0.58&-0.24\\
0.90&7&-0.34&0.45&0.39&0.30&-0.65&-0.35&0.51&0.54&0.46&0.40&0.29&-0.39&0.51&0.16&0.39&-0.11&0.51&0.65&0.64&0.72&0.61&\textbf{0.73}&-0.50\\
0.90&9&-0.49&0.51&0.58&0.19&-0.67&-0.52&0.66&0.63&0.63&0.53&-0.22&-0.58&0.11&-0.15&-0.09&-0.36&0.42&0.52&0.46&\textbf{0.74}&0.48&0.73&-0.58\\
0.90&13&-0.43&0.69&0.74&0.21&-0.48&-0.46&0.46&0.87&0.49&0.86&-0.14&-0.28&0.03&0.29&-0.15&0.04&0.47&0.77&0.54&0.92&0.55&\textbf{0.93}&-0.69\\
0.90&17&-0.50&0.75&0.67&0.30&-0.63&-0.51&0.49&0.76&0.49&0.67&0.05&-0.29&0.40&0.31&0.18&0.03&0.25&0.73&0.35&\textbf{0.87}&0.36&0.86&-0.76\\
\hline
0.80&5&-0.20&0.29&0.24&0.14&-0.51&-0.26&0.69&0.65&0.49&0.55&0.01&0.07&0.39&0.48&-0.09&0.28&0.77&\textbf{0.81}&0.75&0.79&0.68&0.77&-0.63\\
0.80&7&0.19&0.36&-0.06&0.51&-0.62&0.03&0.65&0.78&0.56&\textbf{0.83}&-0.04&-0.31&0.31&0.59&0.01&0.33&0.44&0.67&0.58&0.69&0.60&0.75&-0.62\\
0.80&9&-0.36&0.27&0.24&0.09&-0.54&-0.35&0.59&0.80&0.58&0.60&-0.35&-0.40&0.19&0.46&-0.07&-0.06&0.55&0.77&0.61&0.82&0.64&\textbf{0.87}&-0.49\\
0.80&13&0.13&0.20&0.01&0.19&-0.36&0.05&0.59&0.78&0.64&\textbf{0.80}&0.27&0.17&0.50&0.73&0.32&0.58&0.56&0.61&0.54&0.71&0.60&0.79&-0.28\\
0.80&17&0.43&0.19&0.14&0.15&-0.13&0.40&0.41&0.59&0.33&0.54&-0.18&-0.40&0.29&0.35&0.01&0.05&0.49&0.50&0.36&\textbf{0.61}&0.35&\textbf{0.61}&0.25\\
\hline
\multicolumn{2}{c||}{average}&-0.26&0.47&0.43&0.22&-0.50&-0.30&0.57&0.65&0.54&0.59&0.04&-0.23&0.27&0.27&0.10&0.05&0.51&0.65&0.54&0.75&0.55&\textbf{0.76}&-0.49
  \end{tabular}
  }
  \caption[Logarithmic Correlation between candidate metrics for lattice quality factors and X logical error rates for each combination of yield and code distance.]
    {See Table \ref{tab:linear_correlation} for column heading definitions.
  }
  \label{tab:exponential_correlation}
  \end{center}
 \end{table}
  \end{center}
 \end{landscape}

\section{Dense Placement}
\label{sec:deformation:eval}
\begin{figure}[t]
 \begin{center}
  \includegraphics[width=8cm]{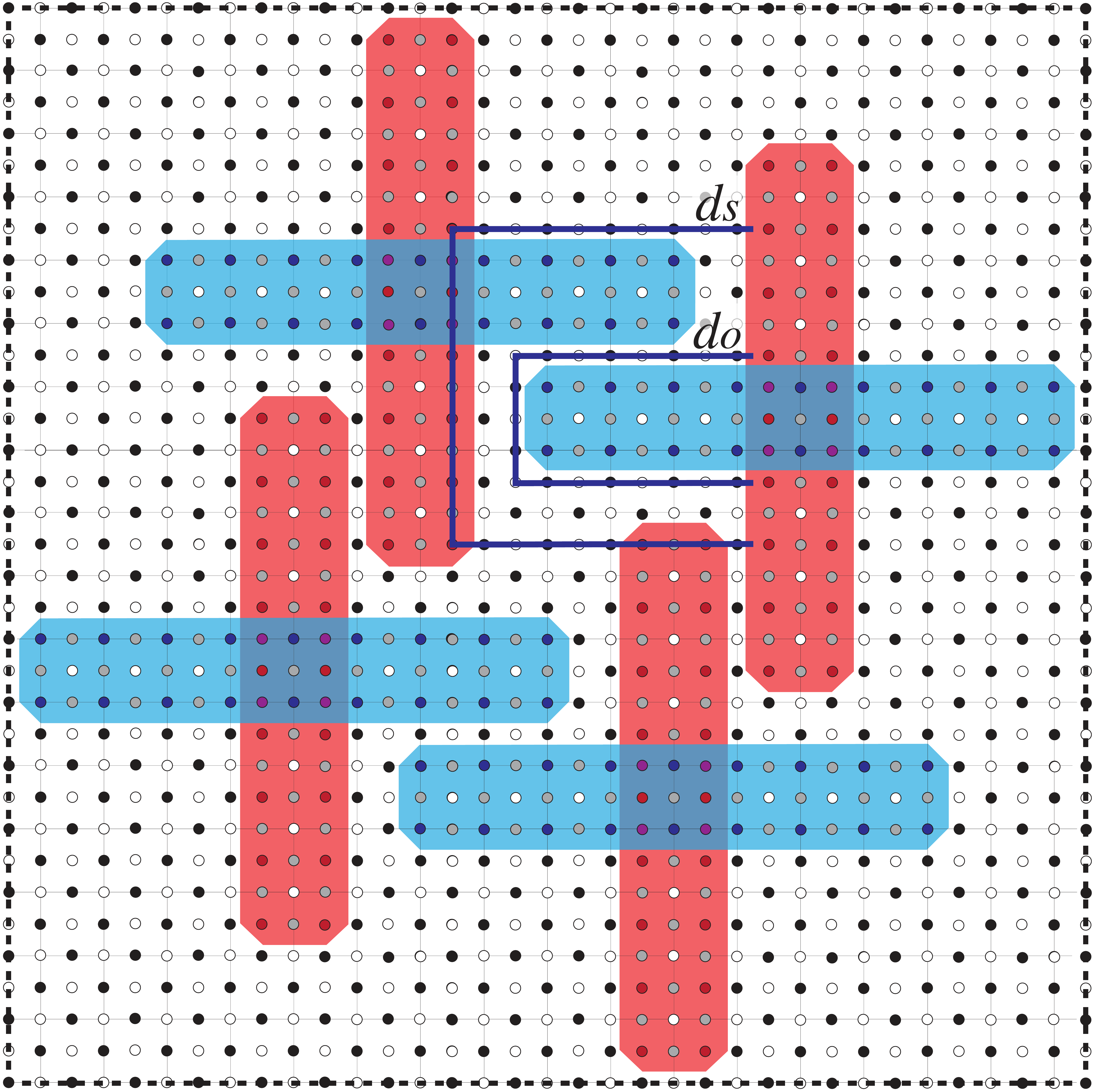}
  \caption[Local placement of the deformation-based surface code.]
  {There are four logical qubits of distance 10 ($d_o$), however, since the thickness of superstabilizers shorten others' code distance by $1$,
  the shortened code distance $d_s$ is 9.
  This placement enables the four logical qubits to have lattice surgery-like CNOT with other logical qubits.
  For thickness $t=2$, each row and column has $3d_s + 8$ physical qubits
  and $(3d_s+8)^2 = 9d_s^2 + 48d_s + 64$
  physical qubits are required for four logical qubits.
  The dashed box corresponds to the dashed box in Figure \ref{fig:placement_global}.
  The blue lines describe minimal $X$ error chains from the point of view of $d_o$ and $d_s$ respectively, for the top-right logical qubit.
  To downgrade the error tolerance in this analysis
  because of the two superstabilizers which have longer cycle time,
  we introduce the effective code distance $d_e$ where $d_e = d_s - 2$ and
  $(3d_s+8)^2 = (3d_e+14)^2 = 9d_e^2 + 84d_e + 196$
  physical qubits are required for four logical qubits.
  }
  \label{fig:placement_local} 
 \end{center}
\end{figure}

\begin{figure*}[t]
 \begin{center}
  \includegraphics[width=15cm]{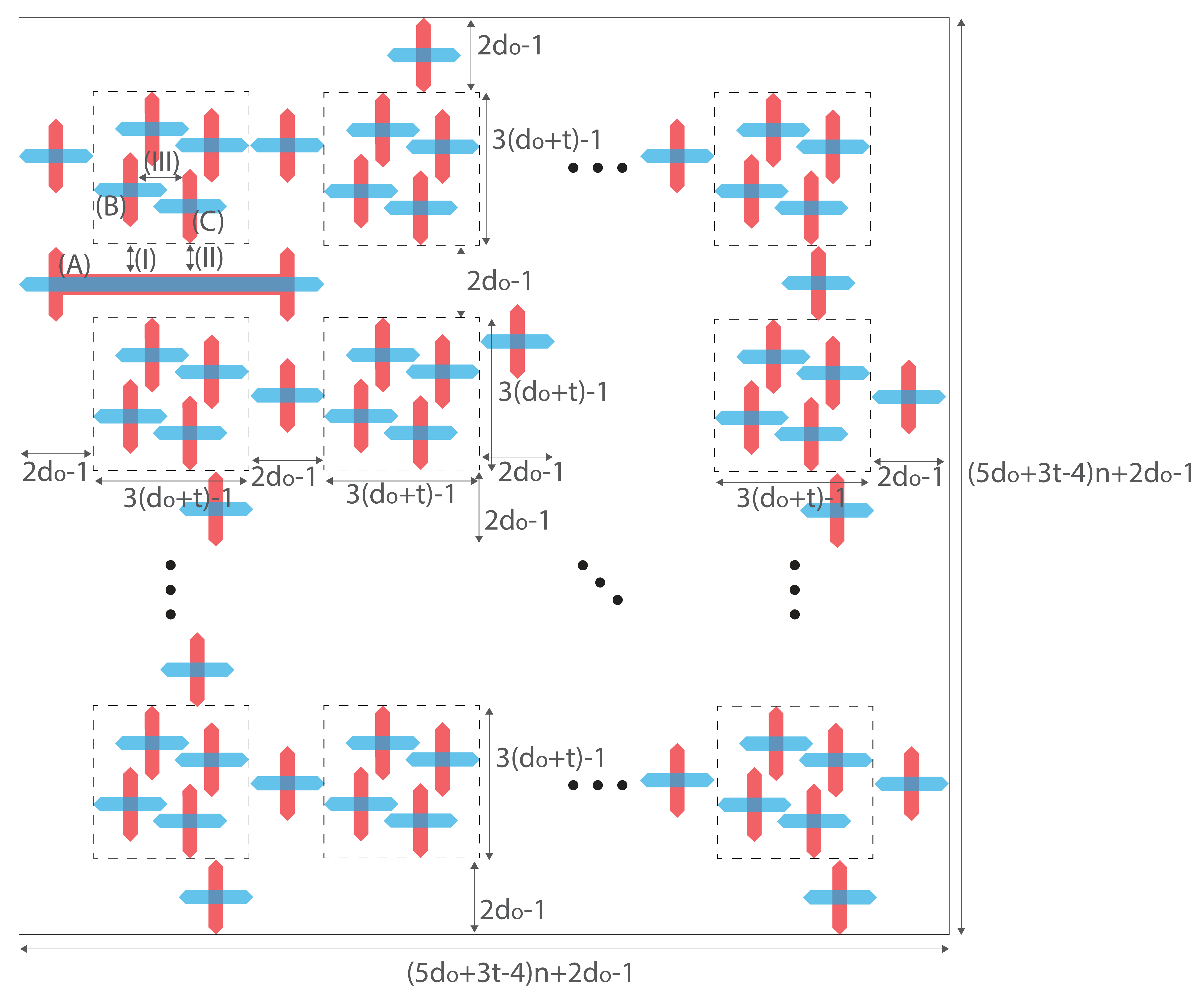}
  \caption[Global placement of the deformation-based surface code.]
  {Each dashed box is the dashed box shown in Figure \ref{fig:placement_local}.
  The spaces between the boxes are paths to move logical qubits and intermediate qubits.
  The deformation-based qubits outside of the dashed boxes are examples of intermediate qubits.
  There are $n$ by $n$ sets of the local placement.
  The lengths include both data qubits and ancilla qubits, hence
  $2d$ in this figure corresponds to the code distance $d$.
  The stretched qubit indicated with (A) is being routed from location to location.
  To retain the fault-tolerance of (B) and of (C), $(I) + (II)$ needs to be $\frac{d}{2}$ or more,
  therefore (A) is transformed.
  The qubits on the boundary between a local placement set and a path are included both in $3(d_o+t)-1$ and $2d_o-1$,
  hence there are $(5d_o+3t-4)n+2d_o-1$ rows and $(5d_o+3t-4)n+2d_o-1$ columns.
  The total number of physical qubits is $((5d_o+3t-4)n+2d_o-1)^2$,
  for $4n^2$ logical qubits excluding intermediate qubits.
  This placement requires
  $(\frac{5d_o+3t-4}{2})^2 = \frac{25d_o^2 +30d_ot + 9t^2 -40d_o -24t + 16 }{4}$
  physical qubits per logical qubit for large enough $n$.
  This corresponds to $(\frac{5d_e+17}{2})^2 = \frac{25d_e^2+170d_e+289}{4}$ physical qubits are required for a logical qubit
  for $t = 2$.
  }
  \label{fig:placement_global} 
 \end{center}
\end{figure*}

Because of the restrictions described in Section \ref{sec:errors},
we locally set four deformation-based qubits as a box, as shown in Figure~\ref{fig:placement_local},
and globally place the boxes apart to maintain fault-tolerance and to have free space available for routing intermediate qubits
as shown in Figure~\ref{fig:placement_global}.
This local placement actually achieves dense packing,
however, placing the logical qubits close together in this fashion results in
error paths that shorten the effective distance.

To qualitatively
analyze this effect, we define a hierarchy of distances: First, $d_o$
is the original code distance of a single logical qubit, corresponding
to the length around one of the arms, as in Fig. 1.  Next, $d_s$ is
the shortened code distance, where the presence of neighboring
superstabilizers may result in an error path of fewer hops.  Finally,
$d_e$ is the effective code distance: the superstabilizers' longer
cycle time results in higher vulnerability to errors, so we downgrade
their ability to protect our data in this analysis by creating this
artificially shortened distance.  We want this final $d_e$ to give us
protection equivalent to or better than the protection of a
planar code qubit of distance $d$, leading to the relation
\begin{equation}
d_o \ge d_s \ge d_e \ge d.
\end{equation}
In the rest of this section, we explore this relationship in detail by
comparing the number of error paths of the minimum length in several
scenarios.

In Figure \ref{fig:placement_local}, the path labeled $d_s$ crosses two superstabilizers,
one laterally and the other longitudinally.
An error chain can cross a superstabilizer in a single hop.
Thus, although the path $d_s$ covers more ground than $d_o$, the number of errors in an undetected error chain is $d_s = d_o - t + 1$,
where $t$ is the thickness of the superstabilizer crossed laterally.

Worse, the deeper circuit of the superstabilizer increases the likelihood of error,
so we choose to treat the superstabilizers as having no positive effect on error suppression.
Removing them, our effective distance is $d_e = d_s - 2 = d_o - t - 1$.

Finally,
setting $t=2$,
this leads us to this relationship for the dense packing of
Figure ~\ref{fig:placement_local},
\begin{align}
d_e = d \\
d_s = d_e + 2\\
d_o = d_s + 1 = d_e + 3 = d + 3.
\end{align}
With this layout, our four-fin logical qubits begin with a distance
three longer than the defect-based qubits to achieve comparable
logical error rates.
As a result, $\frac{(5d_e+17)^2}{4} = \frac{25d_e^2+170d_e+289}{4}$ physical qubits are required for a logical qubit.

The global placement is shown in Figure~\ref{fig:placement_global}.
  The transformed qubit indicated with (A) is being routed.
  (A) is transformed during moving from one crossroads to another.
  %
  %
  Since the surface code places data qubits and ancilla qubits alternately,
  $2d$ columns/rows are required to have code distance $d$.
  To avoid the situation shown in Figure \ref{problematic_placement},
  $(I)+(II)+(III) \ge 2d$ must be satisfied to guarantee code distance $d$ of (B) and (C).
  Since (III) is $d$, $(I) + (II)$ needs to be $d$ or more hence each of $(I)$ and $(II)$ must be $\frac{d}{2}$ or more.
  Therefore (A) is transformed.
  
  This placement design requires
  $(\frac{5d_o+3t-4}{2})^2 = \frac{25d_o^2 +30d_ot + 9t^2 -40d_o -24t + 16 }{4}$
  physical qubits per logical qubit for enough large $n$.
  Choosing $t=2$,
  $(\frac{5d_s+7}{2})^2 = (\frac{5d_e+17}{2})^2 = \frac{25d_e^2+170d_e+289}{4}$
  physical qubits are required for a logical qubit,
  including ancilla qubits.

\begin{figure}[t]
 \begin{center}
  \includegraphics[width=8cm]{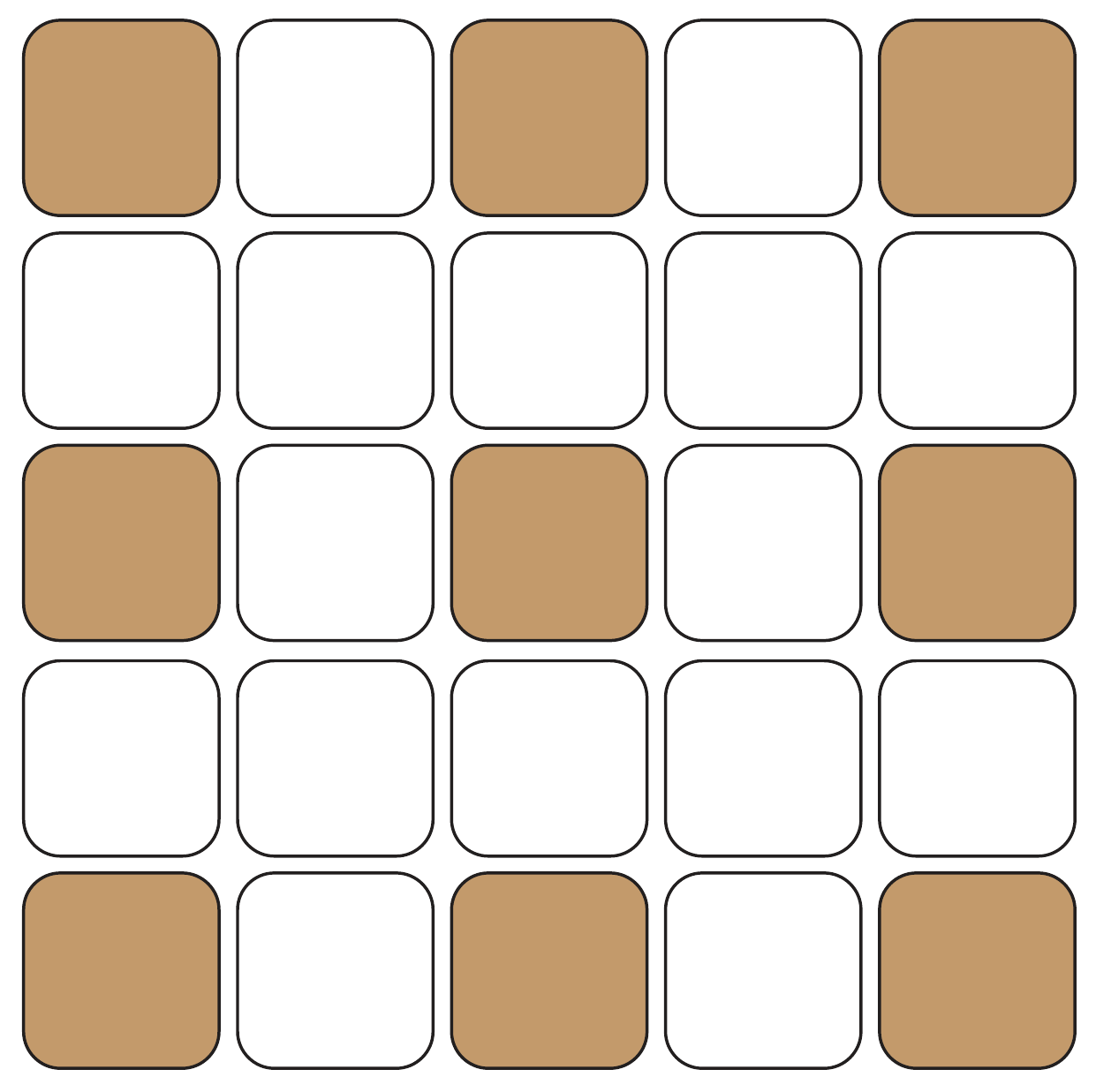}
  \caption[Planar code placement for comparison, after Figure 12 in ~\cite{Horsman:2012lattice_surgery}.]
  {Each shaded area holds a logical data qubit and
  blank areas are available for intermediate qubits for CNOT gate by lattice surgery.
  Each area has $2d-1$ by $2d-1$ physical qubits, including ancillae.
  }
  \label{fig:planar}
 \end{center}
\end{figure}

  In contrast, the planar code's placement for lattice surgery-based operation,
  shown in Figure \ref{fig:planar}, requires $(4d-2)^2 = 16d^2 -16d + 4$ physical qubits per logical qubit.
  %
  %
  As a result the deformation-based surface code requires $50\%$ fewer physical qubits than the planar code.
  Horsman et al. showed that 
  the number of required qubits for the defect-based surface code is similar to that of the planar code
  in large scale quantum computation,
  so deformation-based surface code also requires fewer physical qubits than the defect-based surface code~\cite{Horsman:2012lattice_surgery}.

  The complex interactions during syndrome extraction and the difficulties of the error matching processing make direct calculation of residual logical error rates infeasible, but we can make a qualitative comparison by examining
  the number of redundant logical operators,
  each of which may be potentially a logical error.
  For a code of distance $d$, logical operators of length $d$ dominate.
  A planar code qubit has $d$ redundant logical operators of length $d$, for each type of logical operator.
  A defect-based code qubit has $\frac{d}{4}$ redundant logical operators of length $d$ between the two defects.
  In contrast, a deformation-based qubit which is distant from the lattice boundary and distant from other logical qubits
  has exactly two redundant logical operators of length $d_o$ for each type.
  Hence the isolated deformation-based code has fewer potential error operators.
  With the placement shown in Figure~\ref{fig:placement_local}, because of the presence of a superstabilizer of a neighboring logical qubit,
  the number of potential logical error operators increases to
  $\frac{d_e^3-11d_e^2+35d_e-25}{8}$ redundant logical operators of length $d_e$.
  This is because, for example,
  a horizontal error chain on the top-left of the lattice and a horizontal error chain on the bottom-right do not result in a logical error
  since they would be distant if the surface consisted only of normal stabilizers.
  But, if there is a vertical long superstabilizer in the middle of the lattice which makes the error chains close,
  those two error chain may cause a logical error.

  For distance 9, 19 and 29, the deformation-based code may have 2, 23 and 70 times as many potential logical error operators
  of length $d_e$ as the planar code.
  (Obviously a deformation-based code of effective code distance $d_e$ does not have error operators of geometric length $d_e$,
  since the shortened code distance $d_s$ is the geometric distance and $d_e = d_s - 2$;
  hence we counted error operators by the shortened code distance corresponding to the effective code distance we want.)
  The effective code distance $d_e$ is defined stringently
  therefore we may not need to be concerned about this overhead to compare the deformation-based qubit and other surface code qubits.
  Otherwise, this overhead can be tolerated by employing one greater code distance.

A surface code qubit utilizing the rotated lattice requires
only $2d^2 - 1$ physical qubits to encode a logical qubit~\cite{Horsman:2012lattice_surgery}.
However, since the rotated planar code does not directly support lattice surgery because of its irregular boundaries,
either transversal CNOT gate or conversion to a standard planar code is required to achieve practical quantum computation.
Implementing transversal CNOT gate may kill the surface code's advantage on feasibility.
Conversion to the standard planar code requires a memory area large enough for the standard planar code
and requires the paths for logical qubit transfer wide enough to transfer standard planar code qubits,
eventually killing the rotated lattice qubit's advantage on the resource requirement.
Therefore we focus on the standard planar code for comparison rather than the rotated lattice planar code.

We employed the thickness $t=2$ in this example for simplicity.
Using thickness $t=3$ instead 
will shorten the columns and the rows
of a deformation-based qubit.
Because an even code distance has the same error suppression capability as the odd distance just below it,  a $t=2$ logical qubit and a $t=3$ logical qubit should have $2d + 1$ or $2d - 1$ columns/rows, respectively. This allows us to slightly narrow the inter-block channels in Figure~\ref{fig:placement_global}.

  As in the defect-based code, the Hadamard gate is executed by isolating the logical qubit from the rest of the surface, exchanging X and Z stabilizers, then reconnecting it to the surface. With the dense packing, there is not room around the qubit to disconnect it from the surface, so the qubit first should be moved out into the channel before performing the Hadamard.

\section{Performance of creating heterogeneously encoded Bell pairs in fault-tolerant way}
\label{sec:hetero:sim}
We calculate the error probability and estimate resource requirements by Monte Carlo simulation.
The physical Bell pairs' fidelity is assumed to be 0.85; the state is assumed to be, following N\"olleke \emph{et al.}~\cite{PhysRevLett.110.140403},
\begin{equation}
\rho = 0.85 |\Phi^+\rangle\langle\Phi^+| + 0.04 |\Phi^-\rangle\langle\Phi^-| + 0.055|\Psi^+\rangle\langle\Psi^+| + 0.055|\Psi^-\rangle\langle\Psi^-|.
  \label{equ:input_bell_pair}
\end{equation}
We have chosen to model our interface-to-interface coupling as an
optical coupling, based on the experimental values of N\"olleke et
al.~\cite{PhysRevLett.110.140403}.
This organization corresponds to a classical Internet router
architecture in which separate network interface cards connect to each other
through a crossbar switch on a backplane as shown in figure \ref{fig:router_arch},
which in our case is assumed to be an
intermediate-fidelity optical connection~\cite{kim09:_integ_optic_ion_trap}.
Although the exact
numerical results will of course vary, the principles described in
this paper are independent of the exact numbers.
For comparison, Tables \ref{tab:lg_phys_phys}-\ref{tab:lg_purify_after_encode_strict} in Appendix~\ref{sec:appendixB} present results of simulations in which raw Bell pairs are created using local gates, an approach that could be used with a simpler but less scalable repeater architecture.

Our error model is the Pauli model of circuit level noise~\cite{landahl:arXiv:1108.5738}.
This model consists of memory error, 1-qubit gate error, 2-qubit gate error, and measurement error
each of which occurs with the error probability $p$.
Memory, 1-qubit gates and measurement are all vulnerable to X, Y and Z errors and we assume a balanced model, where probabilities are $\frac{p}{3}$ respectively.
Similarly, 2-qubit gates are vulnerable to all fifteen possibilities, 
each with a probability of $\frac{p}{15}$.
Errors propagate during all circuits after the initial distribution of Bell pairs.

 Figure \ref{fig:graph_CSS7_CSS7} shows a baseline homogeneous simulation
 creating logical Bell pairs
 of Steane [[7,1,3]] code using our physical-to-logical mechanism.
The figure plots the number of consumed raw physical Bell pairs versus
 logical error rate in the output state.
\begin{figure*}[t]
  \begin{center}
   \includegraphics[width=15cm]{fig/./PlotCSS7_CSS7.pdf}
   \caption[Results of a baseline simulation of creation of a Steane {[[7,1,3]]}-Steane {[[7,1,3]]} homogeneous Bell pair, showing
   residual logical error rate versus physical Bell pairs consumed.]
   {The three schemes plus the baseline case of purification of
   physical Bell pairs are each represented by a line. Each point
   along a line corresponds to the number of rounds of purification.
   The leftmost point represents no purification, the second point is
   one round of purification, and the rightmost point represents four
   rounds of purification a.-c. Improving values of
   local gate error rate.  d.  The three cases with residual error
   rate of $10^{-3}$ or less.}
   \label{fig:graph_CSS7_CSS7}
  \end{center}
\end{figure*}
 Figure \ref{fig:graph_SCD3_SCD3} shows a similar baseline homogeneous simulation
 for surface code distance 3.
\begin{figure*}[t]
  \begin{center}
   \includegraphics[width=15cm]{fig/./PlotSCD3_SCD3.pdf}
\caption[Results of a baseline simulation of creation of a surface code distance 3-surface code distance 3 homogeneous Bell pair, showing residual logical error rate versus physical Bell pairs consumed.]
   {Other conditions and definitions are as in Figure \ref{fig:graph_CSS7_CSS7}.}
   \label{fig:graph_SCD3_SCD3}
  \end{center}
\end{figure*}
  The difference between {\it purification before encoding} and purification of physical Bell pairs
  indicates how many logical errors are introduced during the encoding process.
  {\it Purification after encoding} may show that the surface code distance 3
  is more suitable than the Steane [[7,1,3]], however, because those local gate error rates
  are greater than the threshold of the Steane [[7,1,3]] code it is hard to judge fairly.

 Figure \ref{fig:graph_resource_vs_pl} plots the number of consumed raw physical Bell pairs versus logical error rate in the output state for the heterogeneous Steane [[7,1,3]]-surface code distance 3 case. 
  This heterogeneous result falls near the average of those two baseline homogeneous simulations above.
   The
   $\operatorname{max}\left|\frac{\textrm{heterogeneous result}}{\textrm{average of homogeneous result}} - 1 \right|$
   is 0.086. The
   $\operatorname{average}\left|\frac{\textrm{heterogeneous result}}{\textrm{average of homogeneous result}}\right|$
   is 1.007. Since the depth of the circuit of the heterogeneous simulation is aligned to the longer depth of the two codes, the heterogeneous result is a bit higher than the average of the homogeneous simulations.
\begin{figure*}[t]
  \begin{center}
   \includegraphics[width=15cm]{fig/./plot2.pdf}
\caption[Results of simulation of creation of a Steane {[[7,1,3]]}-surface code distance 3 heterogeneous Bell pair, showing
   residual logical error rate versus physical Bell pairs consumed.]
   {Other conditions and definitions are as in Figure \ref{fig:graph_CSS7_CSS7}.}
   \label{fig:graph_resource_vs_pl}
  \end{center}
\end{figure*}

The numbers of raw Bell pairs consumed declines as the local gate error rate is lowered.
This is because the influence of the local gate error rate shrinks relative to the infidelity of generated raw Bell pairs.
If the system is free from local gate error, the numbers of raw Bell pairs consumed by the three schemes must converge.
At $p=10^{-5}$, the required number of raw Bell pairs of the schemes are essentially identical and
they require about 26 raw Bell pairs to achieve four rounds of purification.
Higher efficiency would require improving the initial fidelity of $F=0.85$.

At any error rate and with any number of rounds of purification from 0 to 4,
{\it purification before encoding} and {\it purification after encoding}
result in fidelity worse than simple purification of physical Bell pairs.
This suggests that errors accumulated during encoding are difficult to correct.
On the other hand, {\it purification after encoding with strict post-selection} 
gives better results than simple purification, at the expense of consuming more raw physical Bell pairs.
This difference is noticeable at $p=10^{-3}$; 49 raw physical Bell pairs are used
to create an encoded Bell pair purified four rounds.
The local gate error rate is so high that an eigenvalue of -1 is often found
at the measurement in purification and the output Bell pair is discarded.
For {\it purification after encoding with post-selection}, the residual error rate after $n$ rounds of purification
is similar at any $p$, but resource demands change. It converts local errors into ``loss'', or discarded states.
Therefore {\it purification after encoding with post-selection} is dominated by the original raw Bell pair infidelity.
At $p=10^{-3}$, {\it purification after encoding} also requires more raw physical Bell pairs than the other schemes, because the error rate after purification is so high that
the success probability of purification is poor.

Though more rounds of purification are supposed to result in smaller logical error rate, 
three rounds of purification of {\it purification after encoding} at $p=10^{-3}$ give an error rate {\it larger}
than that of two rounds. The local gate error rate is too high and
purification introduces more errors than it suppresses on odd-numbered purification rounds.

{\it Purification after encoding with strict post-selection}
gives similar results for the two local gate error rates $p=10^{-4}$ and $p=10^{-5}$.
The difference is a small number of consumed raw physical Bell pairs.
Even at $p = 10^{-3}$, we see that four rounds of purification drives
the residual error rate down almost to 0.1\%.  From this fact we
conclude that $p = 10^{-3}$ will be a good enough local gate error
rate to allow us to create heterogeneously encoded Bell pairs,
suitable for many purposes, from raw physical Bell pairs of F=0.85.


\clearpage
\chapter{Conclusion}
\label{chap:conc}
\label{chap:conclusion}

In this Chapter, the conclusion of this dissertation is stated.
First the overview of the discussion is summarized.
Next the problems,
the proposed architecture,
and the difficulties to solve the problems are stated.
Finally the future work, extendability and importance of this work is summarized.

In this dissertation, a practically scalable distributed quantum computer architecture is proposed
in which quantum imperfections including state errors, dynamic losses and static losses are tolerated by quantum error correction,
space-efficient error correcting code is employed in memories,
and fault-tolerant internal and external data transfer is designed to connect components.
Numerical analysis of overhead to tolerate imperfections is completed by analyzing overhead of static losses in this work.
It is shown that static losses are tolerable. However, the required resource increase emphasizes the importance of
division of computation chip like classical multi-CPU computer and quantum distributed computation.
To alleviate the obvious problem of expensive resource requirement, space-saving error correcting code is developed,
which should be employed in quantum memories.
To connect components such as memories, computation area and network interface,
the overhead of fault-tolerant communication of components is numerically analyzed.
The importance of parallelism and scalability of internal communication are shown.

To realize a quantum computer, tolerance against imperfections and its overhead
is the first thing we have to consider.
Fabrication technology for quantum computation is rapidly developing,
however, there still large possibility to have problematic devices on the computation chip
and it may be difficult to completely get rid of them in the future.
Analysis based on ideal assumptions was done, in which concrete methods to deal with static losses are not given.

To realize a quantum computer which is large enough to execute quantum algorithm to solve meaningful problems,
resource efficiency is important.
Quantum error correcting code requires much computational resource.
A logical qubit of the typical surface code, the planar code, consists of $(2d-1)^2$ physical qubits where $d$ is the code distance.
Factoring $N$ bits requires $2N+2$ logical qubits, excluding ancilla qubits for gates and resources for communication.
Integration of quantum devices depends on the physical system employed and is limited yet.
It is actually infeasible to implement everything on a computation chip, maybe in a machine.
Quantum error correcting code which has high error threshold and has small resource requirement has not been found yet.

To realize a quantum computer which has multiple quantum components,
fault-tolerant internal networking between any components supporting code conversion is required.
Employing multiple error correcting codes had been proposed
Scalable internal routing by optical crossbar had been proposed.
Concrete means to bridge components which employ heterogeneous error correcting codes and
analysis of its overhead are required.

We propose a scalable distributed quantum computer architecture,
which tolerates quantum imperfections,
has several areas for several purposes in which
each area consists of an array of same components of the feasible size
to adapt to the overheads caused by tolerating imperfections,
and has a core routing module of an optical crossbar
which routes photons from components to create entanglement between arbitrary pair of components
where imperfections are managed by purification with post-selection
after encoding entangled qubits to arbitrary codes.
The purposes are computation, memory, ancilla state generation for complex gates
and computer networking.

\section{Difficulties in solving problems}
In the analysis of an adaptation of the surface code for static losses,
which are manifested as faulty devices on quantum computation chips occurring during fabrication.
With this fundamental analysis of static loss and its influence,
independent analysis has now been conducted for the three major
imperfections of quantum computation for the surface code:
state error, dynamic loss, and static loss.
The ultimate goal of investigating faulty devices is
to support collection of a large pool of sufficiently
fault-tolerant quantum computation chips which will be the arrays of components in the computation/memory areas
of distributed quantum computers.
The method we employ to work around static losses is the code deformation; merging stabilizer units
by removing the lost qubits.
The surface code on the perfect lattice has complete regularity, but in our case,
there come many size of stabilizers in which the order of error syndrome collection is not unified
and whose duration differ.
To achieve a practical method, we need new two algorithms which is adapted to the quantum error occurrence nature.
We determined the order of error syndrome collection by solving traveling salesman problem.
There would be a difficulty that solving for big stabilizers may not finish in realistic time, however,
later by simulation it is shown that the size of the biggest stabilizer has the strongest correlation
with the logical error rate of the defective lattice
so that lattices holding such big stabilizers should be discarded anyway; hence this difficulty can be ignored.
Another difficulty is how to schedule stabilizers of heterogeneous duration.
We solved this difficulty by focusing on the fact that stabilizers of lower frequency may accumulate more physical errors
and are easy to lengthen error chains.
Therefore we give priority to larger stabilizers hence schedule larger stabilizers first and then
schedule small stabilizers repeatedly when member qubits are free.

To develop the space-saving surface code,
we have shown an extension of the deformation-based surface code
which is capable of close placement by measuring superstabilizers which produce deformation-based qubits.
The universality of gates largely influences the flexibility of an error correcting code,
which is required for quantum CPU and is useful for quantum memory to realize read and store operations.
The space-saving characteristics is the most important factor of this new code,
we have to have both characteristics in the same time.
We solve this difficulty by a lattice surgery-like CNOT gate for the deformation-based qubits
which requires fewer physical qubits than the braiding CNOT gate.
Meanwhile we achieve direct conversion from the defect-based surface code to
the deformation-based surface code, which can be used as state injection
for the deformation-based surface code,
then we achieve arbitrary single qubit gate by combining CNOT gate and the arbitrary state injection,
hence a universal gate set is completed.
The acceptability of close placement and the space-saving CNOT gate
allow deformation-based qubits
to be packed more tightly than planar code qubits and defect-based qubits.

We have proposed and analyzed a generalized method for creating heterogeneously encoded Bell pairs that 
can be used for interoperability between encoded quantum components.
This is the first step in examining the full design of 
interconnection of quantum computers/routers/repeaters utilizing different error mitigation techniques.
Acquired encoded Bell pairs must have reasonably high fidelity.
The fidelity of raw physical Bell pairs are determined by the implementation of components,
optical fibers and BSA hence it is physically limited.
Hence the method how to increase the fidelity is important
and to find the means in which the achievement and resource consumption meets is the difficult thing.
To solve this problem, we considered three ways of entanglement purification,
{\it purification before encoding},
{\it purification after encoding} and
{\it purification after encoding with strict post-selection}
and we investigated the relationship of the achieved fidelity and the resource consumption of those three
by numerical simulation .
	
\section{Quantitative results and overheads}
We analyzed our approach against faulty losses by simulation to investigate the relationship
between the logical error rates and lattice characteristics of simulated defective lattices.
Our approach is to merge stabilizers broken by faulty data qubits to a superstabilizer
and to work around faulty ancilla qubits using SWAP gates,
without changing the original role of the qubits.
Our simulation with single faulty device revealed that faulty qubits at the periphery reduce the logical error rate less than those in the center.
Even a single fault has a large impact on the residual error rate.
Our simulation with randomly placed faulty devices showed that
at $95\%$ yield, the impact on net error rate is significant but many of the chips still achieve break-even by $p=10^{-3}$,
and therefore could be used in a real-world setting.
At $90\%$ yield, very few chips achieve break-even.
At $80\%$ yield, almost no chips are usable.
Those facts establish the goals for experimental research to build the surface code quantum computer.
The simulation of randomly placed faulty devices
also showed that discarding bad lattices makes the ensemble better,
showing the trade-off between the cost by discarding and the strength of fault-tolerance of an ensemble.
Discarding makes the effective code distance of the defective lattice ensemble result in two more longer effective code distance at distance 9 and 13 at $95\%$ yield.
$90\%$ yield shows error suppression at obviously practical physical error rates, at around $p=0.1\%$,
and discarding works for $90\%$ yield.
With a low physical error rate, $90\%$ yield may be sufficient to build a quantum computer.
At $80\%$ yield, only very weak error suppression is observed even at $p=0.1\%$.
Even 90\%-discarding does not show enough error suppression and distance 17 has the highest error rate.
We conclude that $80\%$ yield is not suitable for building quantum computer,
using the surface code without addtional architectural support.
The randomly faulty lattice simulation also revealed that
the average of the $CDQ$ and the average of the $CQ$ of $Z$ stabilizers 
show the strongest correlations to simulated residual error rate among
a set of proposed metrics for chip evaluation.
The $CDQ$ and $CQ$ is the product of the ``cycle'' which is the average steps the stabilizer is measured by
and the number of data qubits/the number of qubits involved in the stabilizer, respectively.
Therefore the accumulated error possibilities in a stabilizer may be the most correlated factor
to the logical error rate.

We have shown theoretical basic concepts of the deformation-based surface code
but have not calculated the error suppression ability
since that of the surface code has been investigated well.
The superstabilizers which compose deformation-based qubits
require $4d_e + 9$ steps for stabilizer measurements
where $d_e$ is the effective code distance.
Our placement design preserves logical qubits 
as any logical operator passes through a chain of normal stabilizers that compose the effective code distance, $d_e$.
Hence, by adding $3$ to the original code distance,
the long stabilizer measurement does not degrade the error suppression efficiency.
The deformation-based surface code should have residual error rate similar to the conventional surface code of code distance three shorter,
and hence conventional error analysis for the surface code can be applied to the deformation-based surface code.
Our design requires
$\frac{25d_e^2+170d_e+289}{4}$
physical qubits for a logical qubit,
compared to the $16d^2 -16d + 4$ physical qubits required in the conventional design.
Our design would halve the resource required to build a large scale quantum computer.

Our purification simulations have shown that {\it purification after encoding with strict post-selection} is
a better preparation method than our other two candidates.
Strict post-selection of two rounds of purification results in better fidelity
than error correction of four rounds of purification at all error rates,
and better physical Bell pair efficiency.
Since the threshold of the error rate of the Steane [[7,1,3]] code is around $10^{-4}$,
our simulations of {\it purification before encoding} and
{\it purification after encoding} of $\sim 10^{-4}$ do not show an advantage compared to simple physical purification;
however, strict post-selection does.
{\it Purification after encoding with strict post-selection} has a higher threshold than the normal encoding and purification do.
With initial $F = 0.85$, we can almost reach a residual error rate of
$10^{-3}$ 
using 4 rounds of purification, for physical Bell pairs at $p =
10^{-5}$ or post-selected heterogeneous pairs at $p = 10^{-4}$.
QEC generally corrects up to $e \le \lfloor \frac{d-1}{2}\rfloor$
errors per block, and detects but mis-corrects $\frac{d-1}{2} < e < d$
errors. Post-selection eliminates this mis-correction possibility,
leaving only groups of $d$ errors or errors that occur after syndrome extraction in the state.
The structure of CSS codes is so self-similar that
we expect that the analysis will be useful for evaluating
other hardware models and CSS codes.

\section{Future work}
We have shown solutions to three problems and analyzed the overhead of them separately.
Hence the biggest future work should be the all-in-one analysis of the proposed architecture.

\subsection{Better way to tolerant against static losses}
Faulty data qubits result in merging plaquettes and deepen the stabilizer circuit
hence lengthen the ``cycle''.
Faulty ancilla qubits result in requiring more SWAP gates to walk through data qubits and ancilla qubits
surrounding the faulty ancilla qubits.
However, our data also shows that the number of faulty ancilla qubits has weak correlation to the residual error rate.
Therefore, utilizing an ancilla qubit
to substitute for neighboring faulty data qubit and keeping stabilizer sizes at four qubits (three qubits at boundaries)
may be an effective solution against static losses.

There is a range in the degree of problems a quantum device has.
Some of problematic devices do not work at all.
Some of them work but their error rate is higher than regular ones.
Some of them execute most gates properly but have problem for some gate.
The definition of the problems for a quantum device depends on its purpose.
If the definition is fulfilled, it should be a problem.
To consider such difference would help designing high performance quantum computer.

\subsection{More denser placement of the deformation-based code}
We have shown the placement in which four logical qubits are grouped to be a unit for the global placement.
This design is to enable lattice surgery both on the smooth boundaries and on the rough boundaries.
By restricting this flexibility, more space-saving placement may be found.

\subsection{Combination of purification methods}
We employed and simulated
{\it purification before encoding},
{\it purification after encoding} and
{\it purification after encoding with strict post-selection}
separately.
{\it Purification before encoding} actually works as increase the fidelity of source Bell pairs for encoding,
hence there is a possibility that the combination of
{\it Purification before encoding} and {\it purification after encoding with strict post-selection}
would result in better balance between the achievement and the resource consumption.

\section{Diversion of the architecture}
\begin{figure}[t]
 \begin{center}
  \includegraphics[width=7cm]{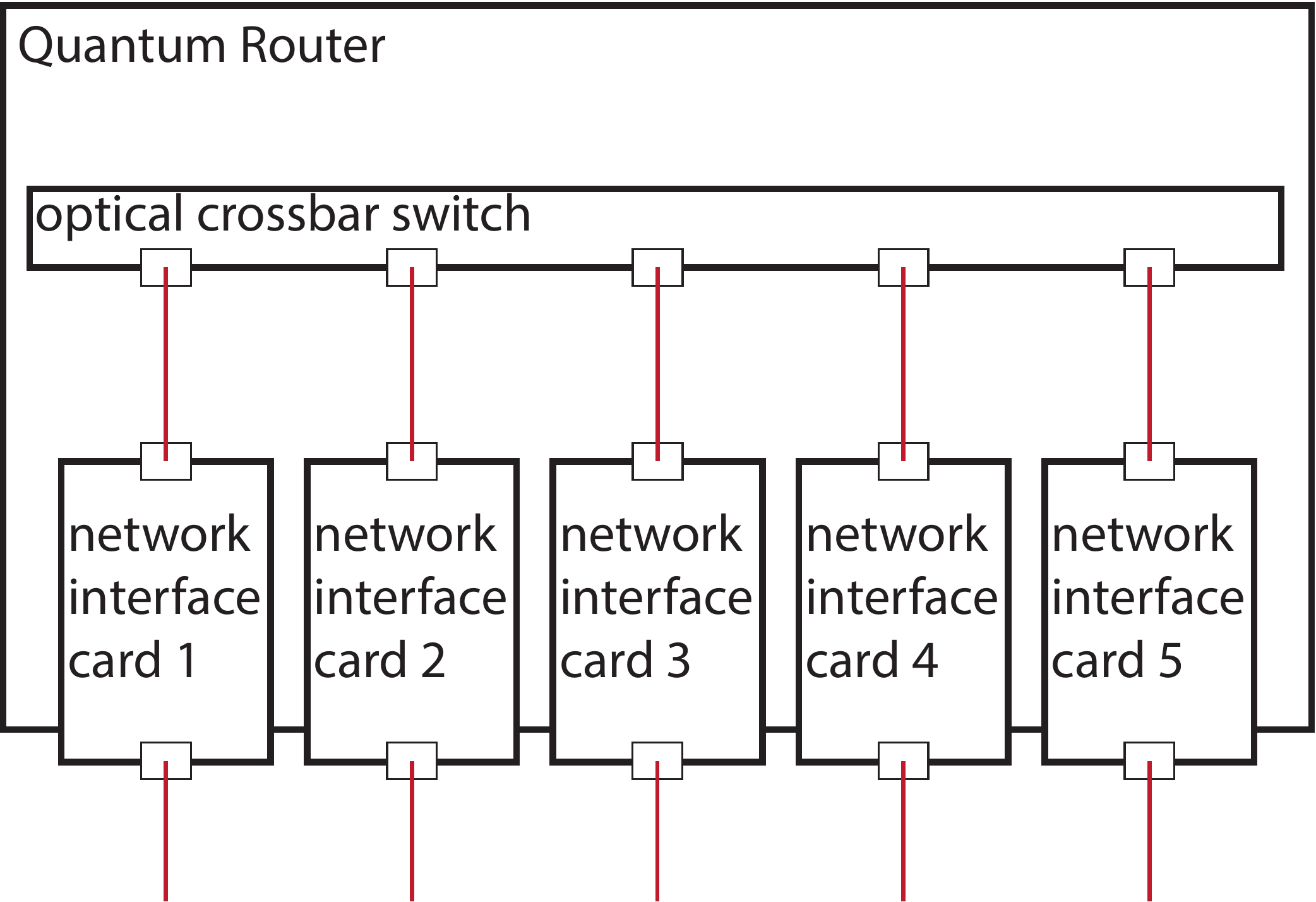}
  \caption[Architecture of a scalable quantum repeater supporting routing.]
  {Each red line describes a photonic connection.
  The crossbar switch 
  switches interconnections of network interface cards.
  Each network interface card is connected to another quantum
  repeater or computer at the open ends of the red lines.
  By using a crossbar switch, interconnections do not
  interrupt each other so that this architecture is scalable.
  The first step to create a heterogeneously encoded Bell pair is the creation
  of a physical Bell pair via the optical crossbar switch.
  Next step is encoding each half of the physical Bell pair via local gates within the respective NICs.
  }
  \label{fig:router_arch}
 \end{center}
\end{figure}

Meanwhile, the crossbar switch can be the core module of a quantum router architecture
as shown in Figure \ref{fig:router_arch}~\cite{PhysRevA.93.042338}.
This router architecture is derived from the architecture of conventional classical router
and it has several network interface cards connected to the crossbar switch.
Combining the quantum computer architecture and the router architecture
and exchanging the logical qubit units with quantum computation chips of the surface code
and of other complemental codes,
a quantum computer architecture which supports networking and utilizes the benefit of the multiple code model is achieved.

The analysis presented here is useful not only in the abstract, but
also serves as a first step toward a hardware design for a
multi-protocol quantum router (the boxes in Figure~\ref{fig:router_arch}).
Such a router may be built on a quantum multicomputer architecture,
with several small quantum computers coupled internally via a local optical
network~\cite{van-meter10:dist_arch_ijqi,jiang07:PhysRevA.76.062323,kim09:_integ_optic_ion_trap,oi06:_dist-ion-trap-qec}.
This allows hardware architects to build separate,
small devices to connect to each type of network, then to create Bell
pairs between these devices using the method described in this paper.
In addition, this method can be used within large-scale quantum
computers that wish to use different quantum error correcting codes
for different purposes, such as long-term memory or ancilla state
preparation.

This scheme is internal to a single repeater at the border of two networks, and will allow
effective end-to-end communication where errors across links are more important than errors within a repeater node.
It therefore can serve as a building block for a quantum Internet.

\section{The importance of this work}
The feasibility of the quantum computer strongly depends on the error management scheme
and on how to get the required computation power from small components available with technologies at that time.
This work shows how to get the computation power by connecting such components with limited power,
with tolerating quantum imperfections.
My architecture and the way to design will be referred from any future work to design quantum computer architectures dedicated
to specific fundamental physical technologies.

\clearpage
\phantomsection
\addcontentsline{toc}{chapter}{Bibliography}
\bibliographystyle{plain}

\bibliography{paper-reviews/quantum/algorithm,paper-reviews/quantum/atom,paper-reviews/quantum/blind-quantum-computation,paper-reviews/quantum/error,paper-reviews/quantum/mechanical,paper-reviews/quantum/nv-center,paper-reviews/quantum/optical,paper-reviews/quantum/qkd,paper-reviews/quantum/quantum-annealing,paper-reviews/quantum/quantum-computer-architecture,paper-reviews/quantum/quantum-computer-general,paper-reviews/quantum/quantum-dot,paper-reviews/quantum/quantum-error-correcting-code,paper-reviews/quantum/quantum-language,paper-reviews/quantum/quantum-logic,paper-reviews/quantum/quantum-network,paper-reviews/quantum/quantum-satellite,paper-reviews/quantum/silicon-donor,paper-reviews/quantum/superconducting,paper-reviews/quantum/classical}


\clearpage
\appendix
\chapter{Supplemental graphs for the defective lattice analysis}
This appendix shows supplemental graphs to visualize the effect of culling and raw data of the surface code on the defective lattice.
\section{Graphs to compare culled pools}
\label{subsec:overlapped_graphs}
\begin{landscape}
\begin{figure}[t]
 \begin{center}
  \includegraphics[width=6cm]{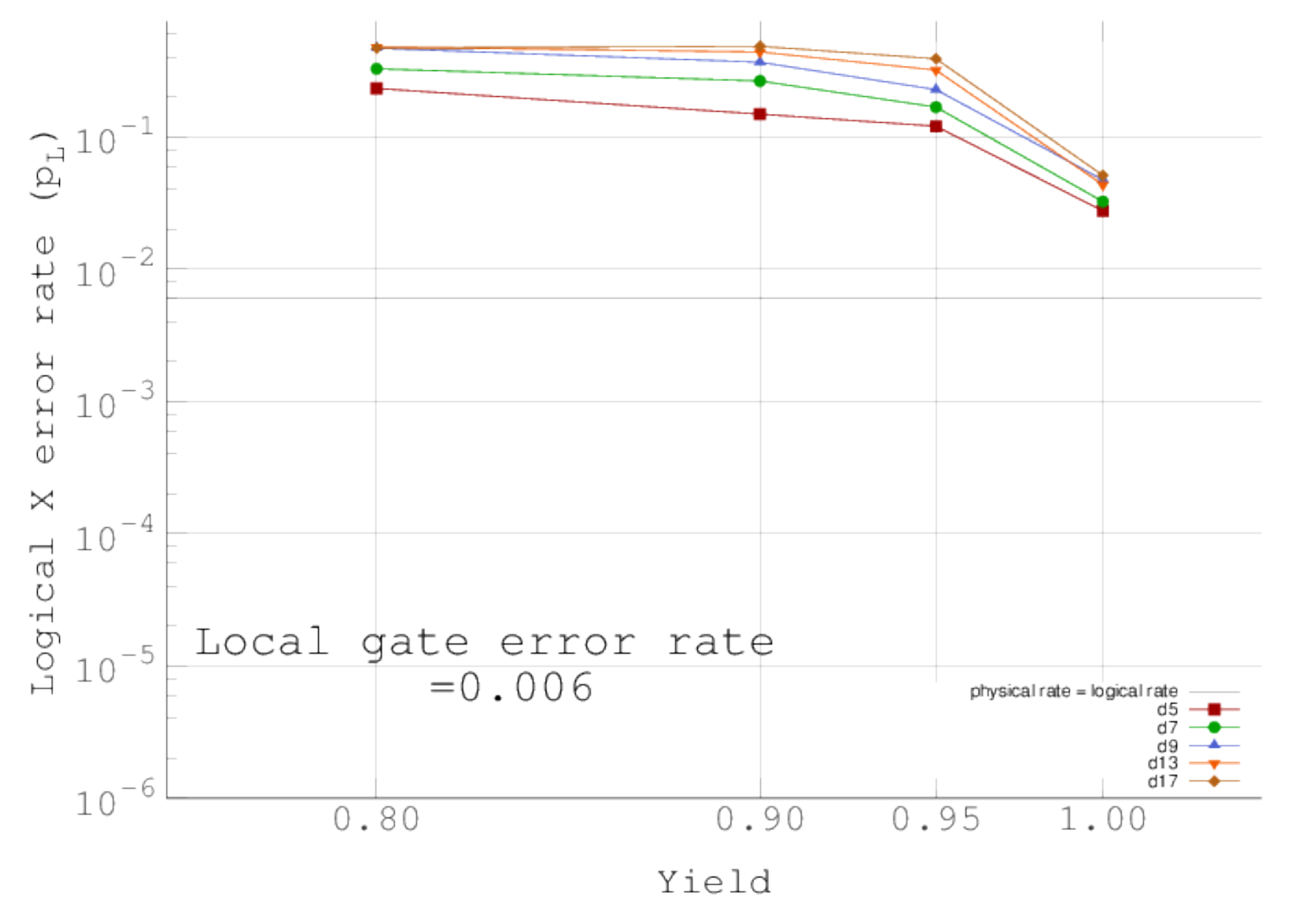}
  \includegraphics[width=6cm]{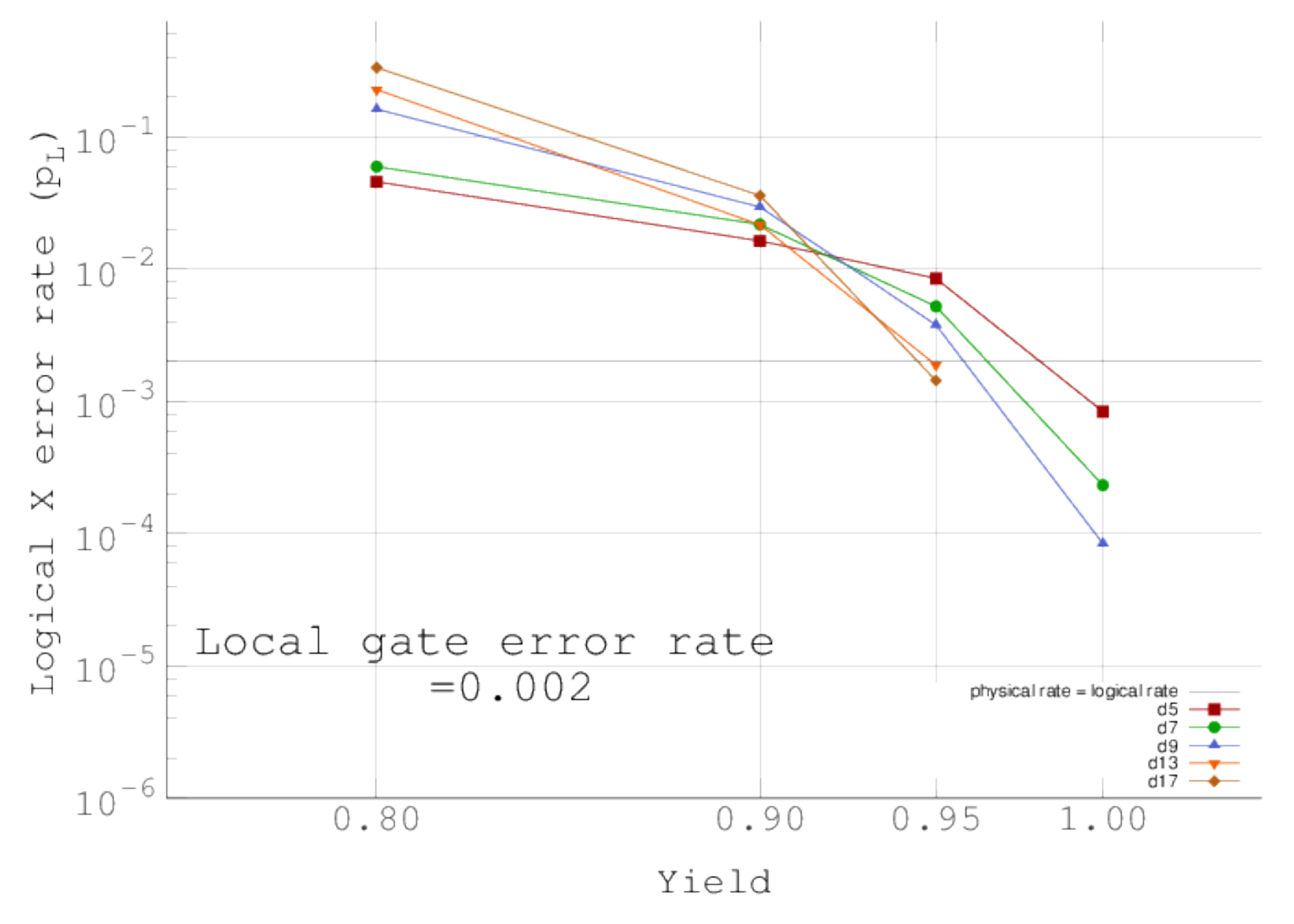}
  \includegraphics[width=6cm]{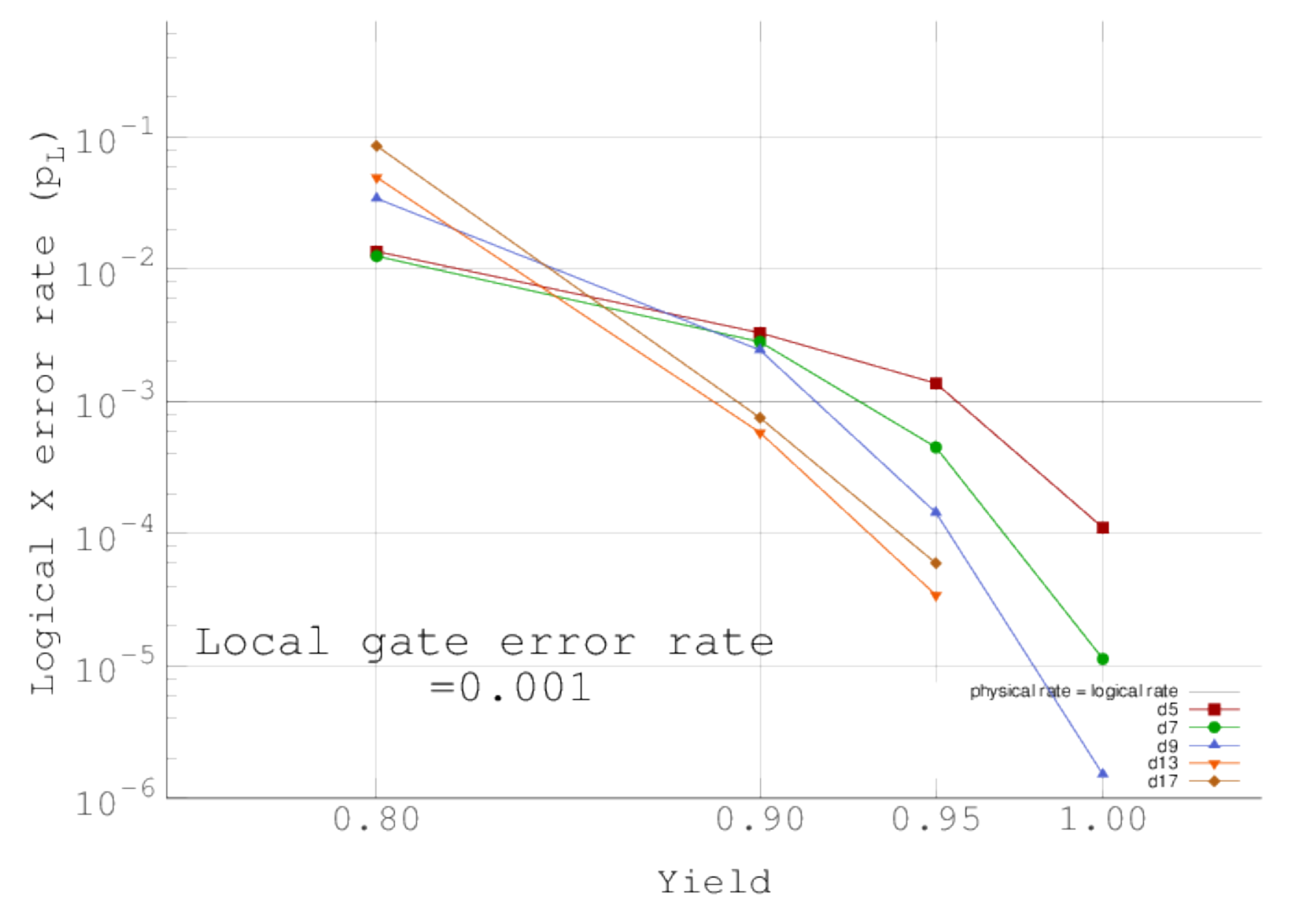}\\
  \includegraphics[width=6cm]{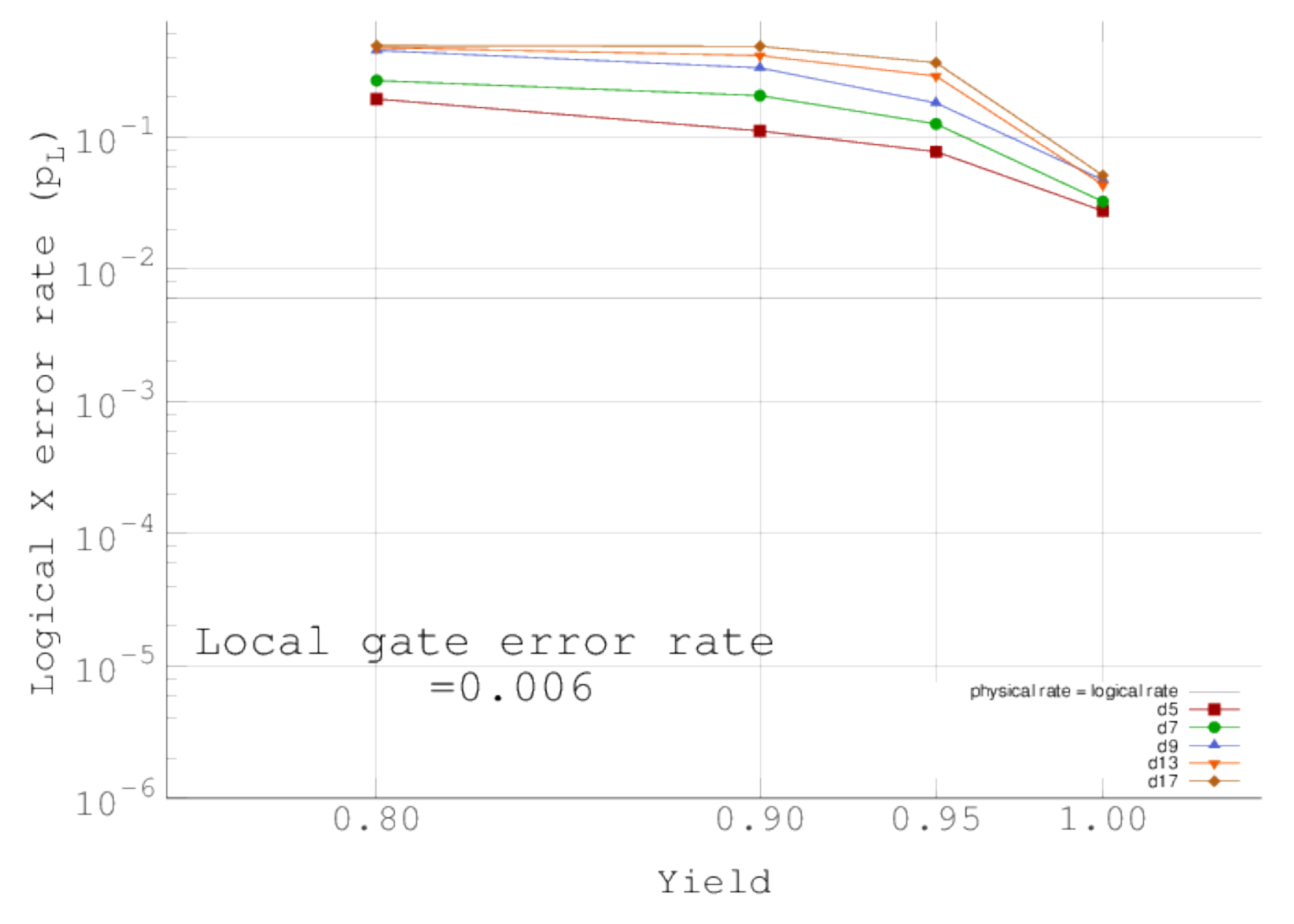}
  \includegraphics[width=6cm]{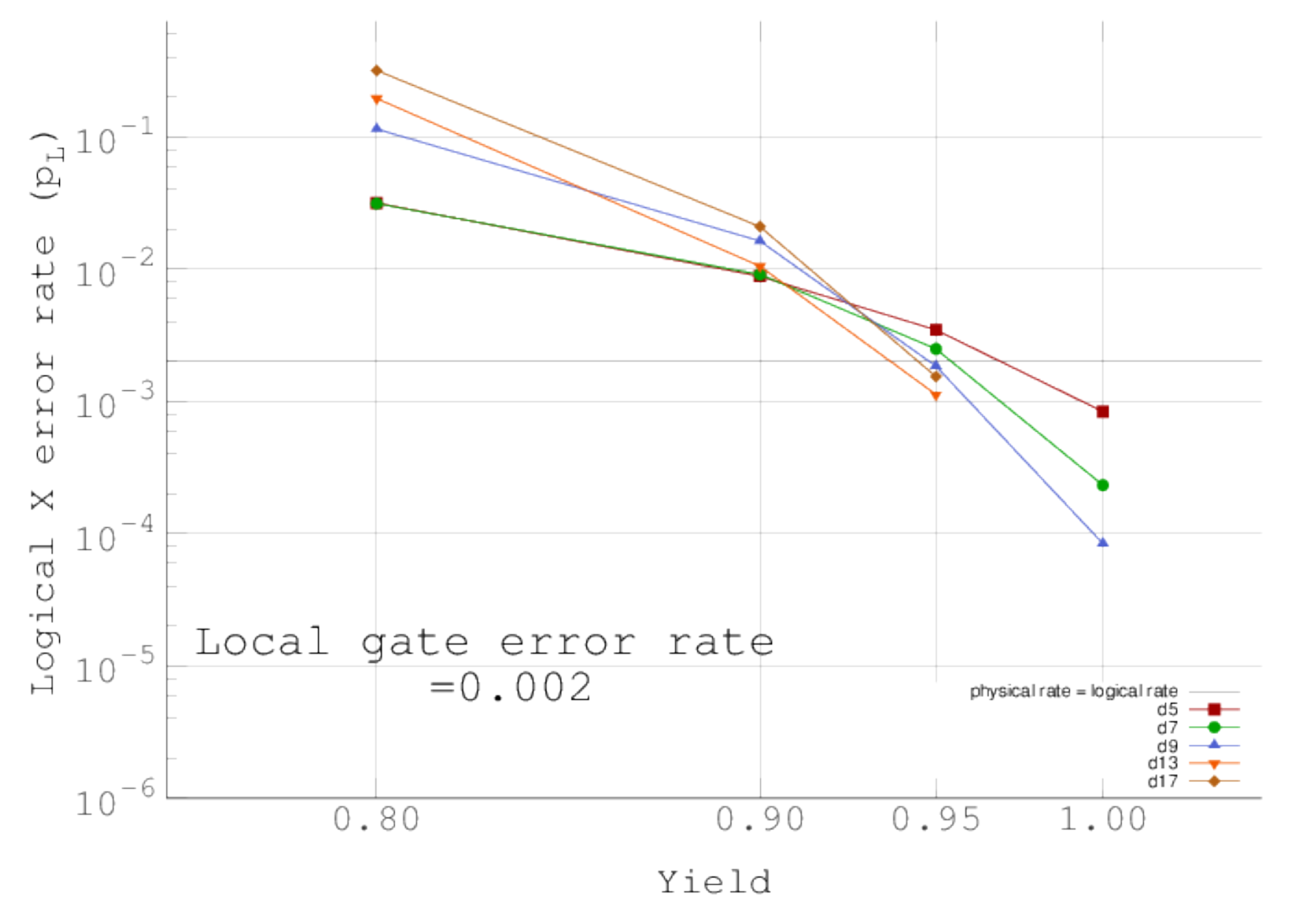}
  \includegraphics[width=6cm]{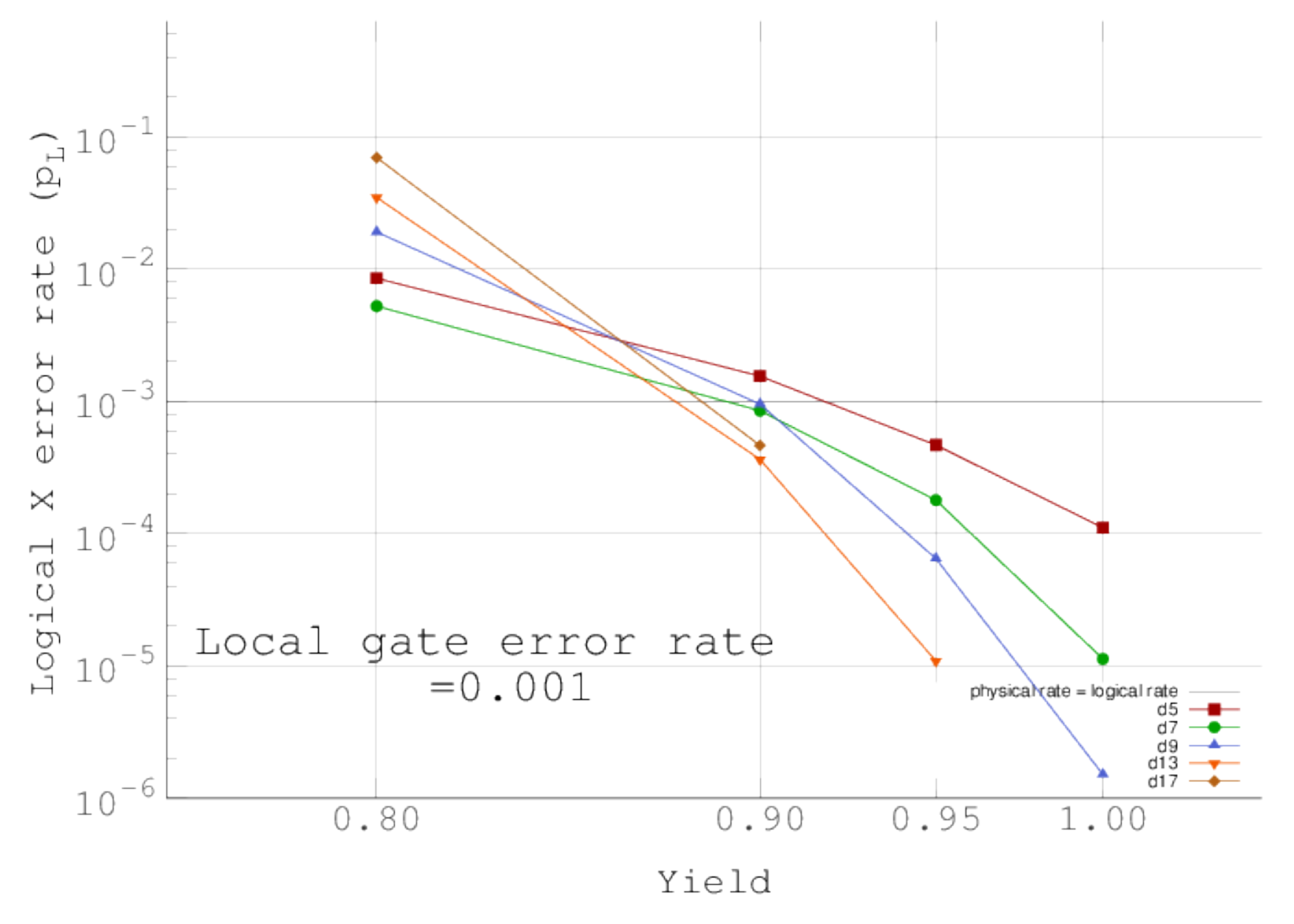}\\
  \includegraphics[width=6cm]{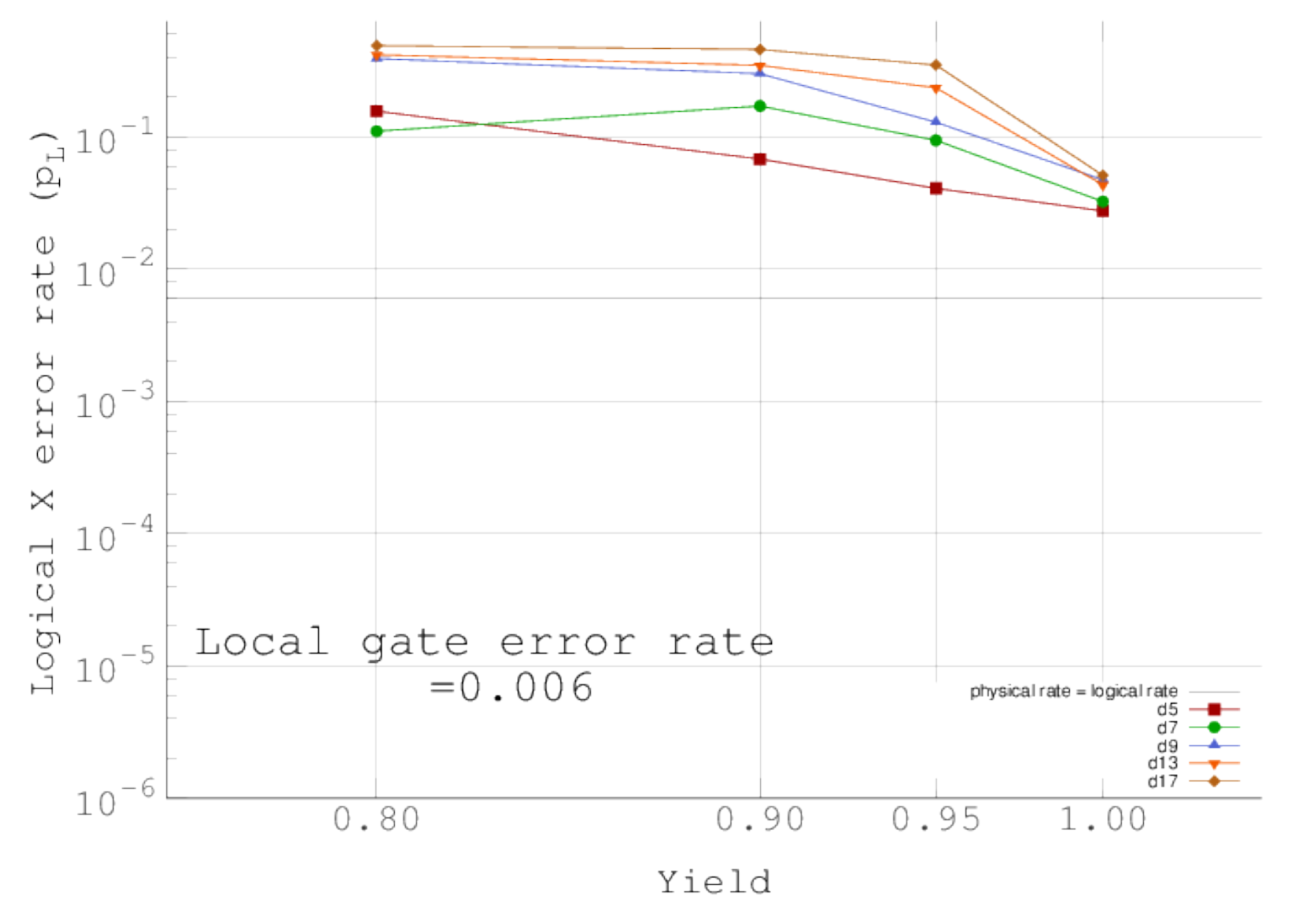}
  \includegraphics[width=6cm]{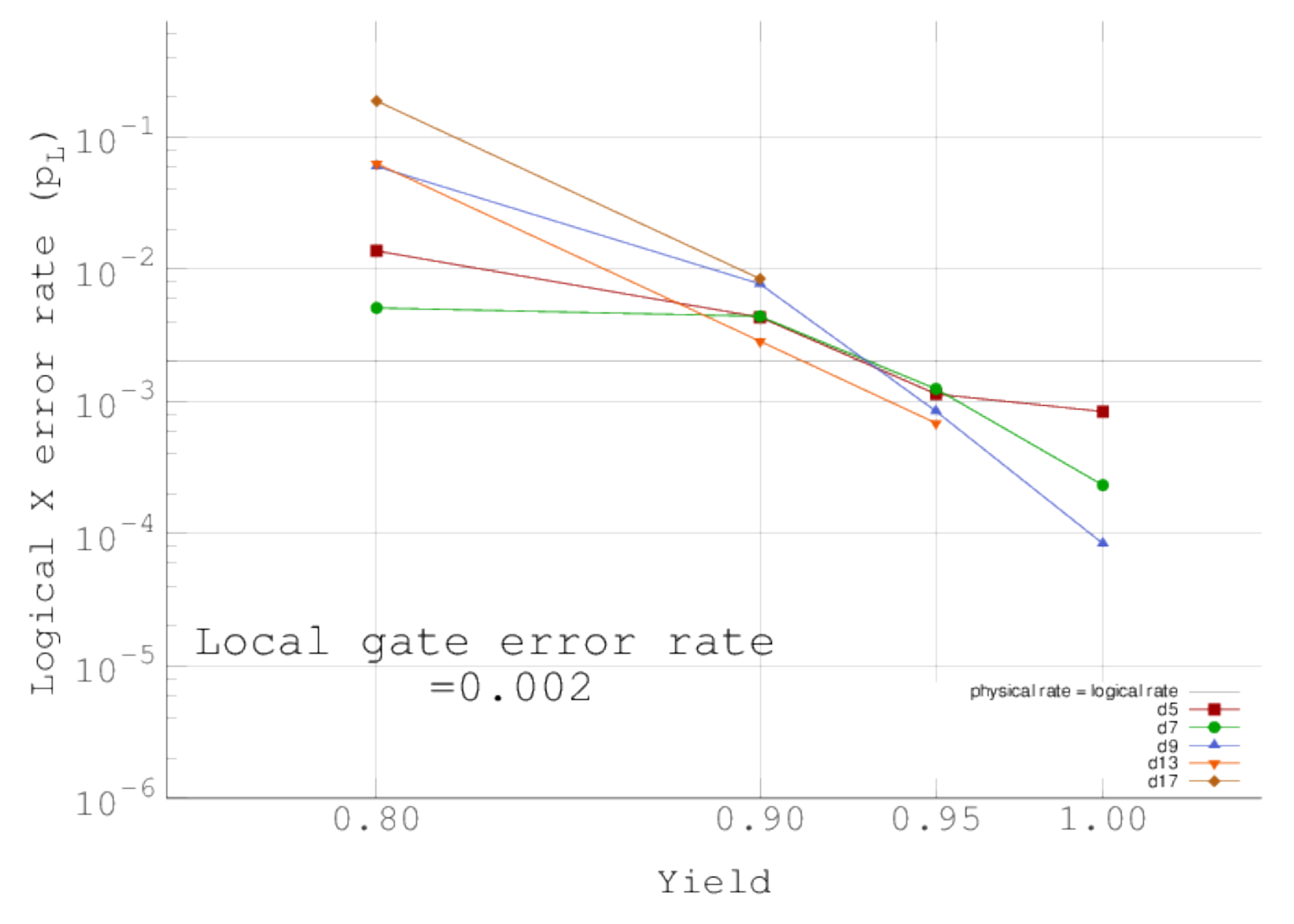}
  \includegraphics[width=6cm]{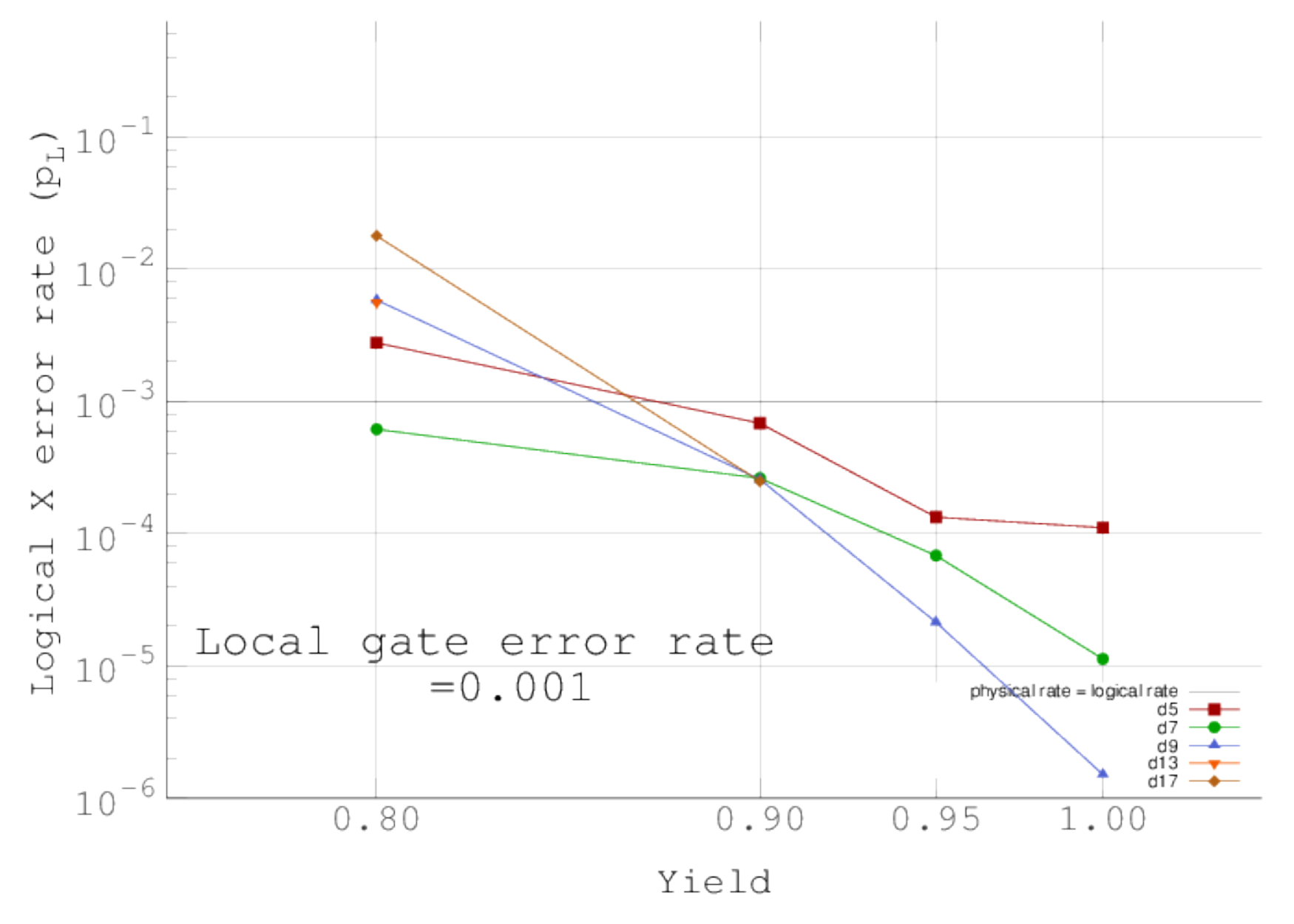}
  \caption[The graphs between yields and logical error rates at specific physical error rates.]
{The top row is of no-culled graphs,
the middle row is of 50\%-culled graphs
and the bottom row is of 90\%-culled graphs.
The left column is for $p=0.6\%$, 
the middle column is for $p=0.2\%$ 
and the right column is for $p=0.1\%$.
}
  \label{fig:cull-comp}
 \end{center}
\end{figure}
\end{landscape}

Figure \ref{fig:cull-comp}
shows the graphs between yields and logical error rates at specific physical error rates.

\section{Scatterplots of randomly defective lattices}
\label{subsec:scatterplots}
\begin{figure}[t]
 \begin{center}
  \includegraphics[width=15cm]{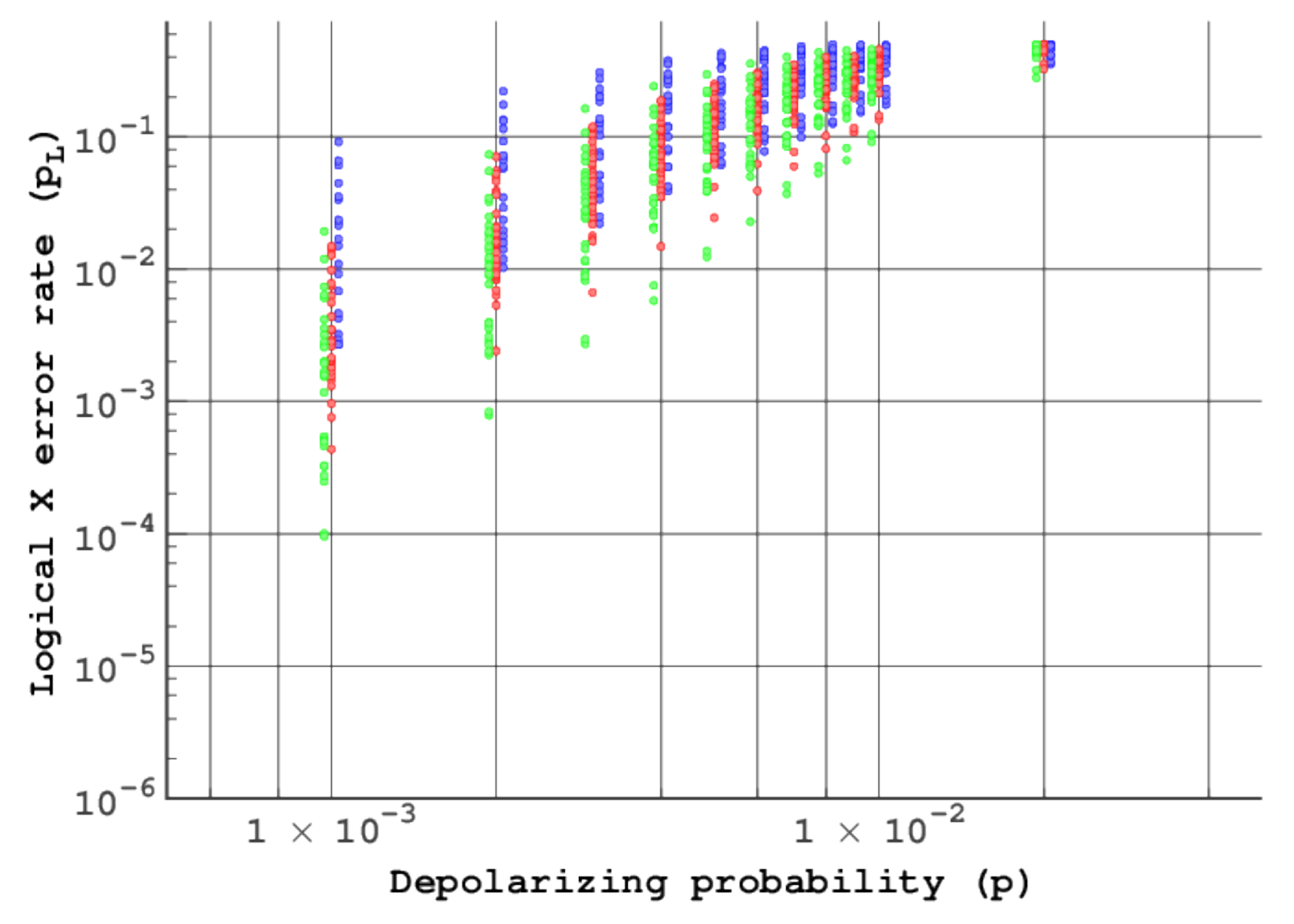}
  \caption[Scatterplot of $d=5$ with one dot per chip.]
  {Green dots are of $y=95\%$, red dots are of $y=90\%$ and blue dots are of $y=80\%$.
  Blue and green data are offset from the vertical line for visibility.
  }
  \label{fig:scatter_d5}
 \end{center}
\end{figure}

\begin{figure}[t]
 \begin{center}
  \includegraphics[width=15cm]{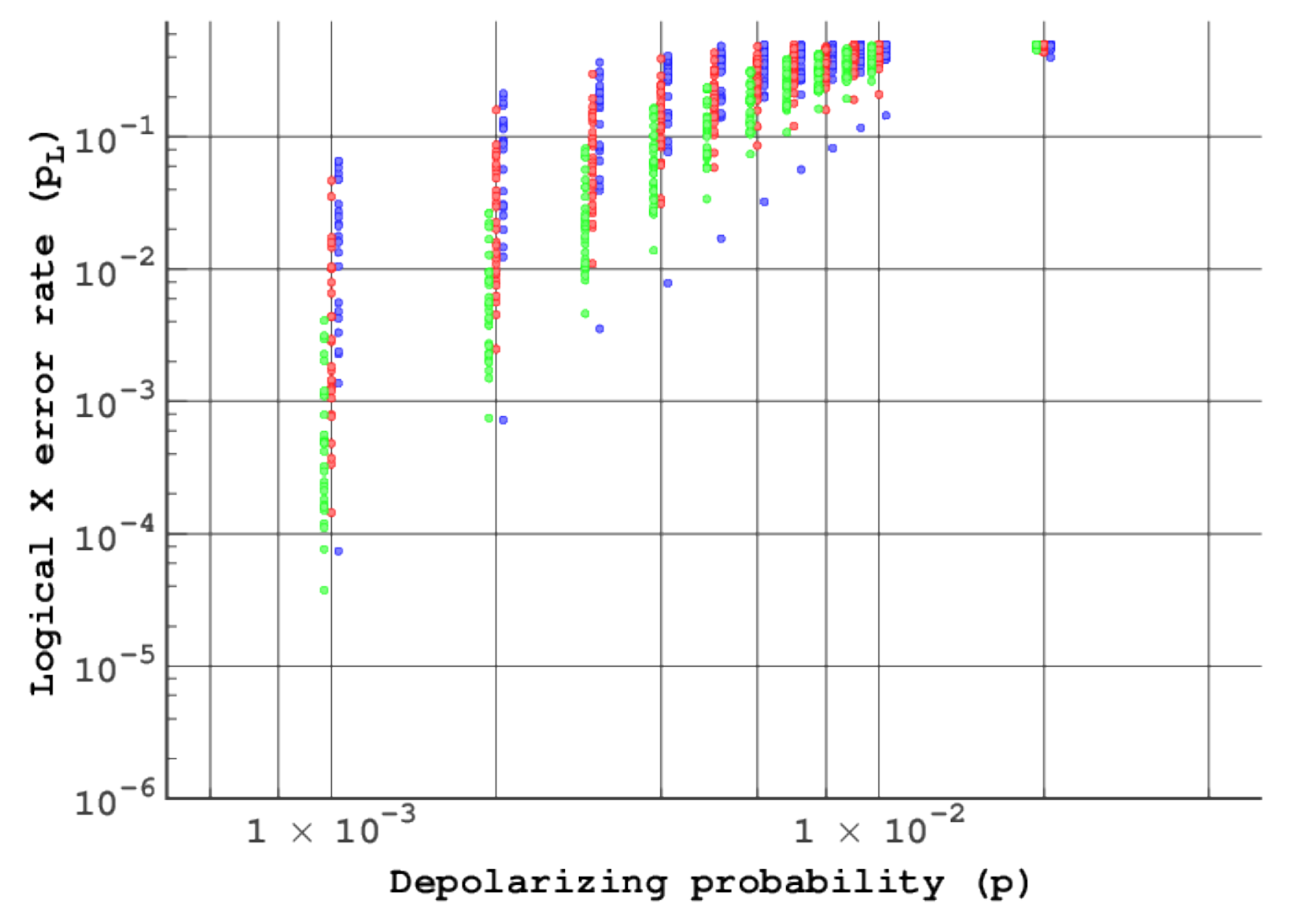}
  \caption[Scatterplot of $d=7$ with one dot per chip.]
  {Green dots are of $y=95\%$, red dots are of $y=90\%$ and blue dots are of $y=80\%$.
  Blue and green data are offset from the vertical line for visibility.
  }
  \label{fig:scatter_d7}
 \end{center}
\end{figure}

\begin{figure}[t]
 \begin{center}
  \includegraphics[width=15cm]{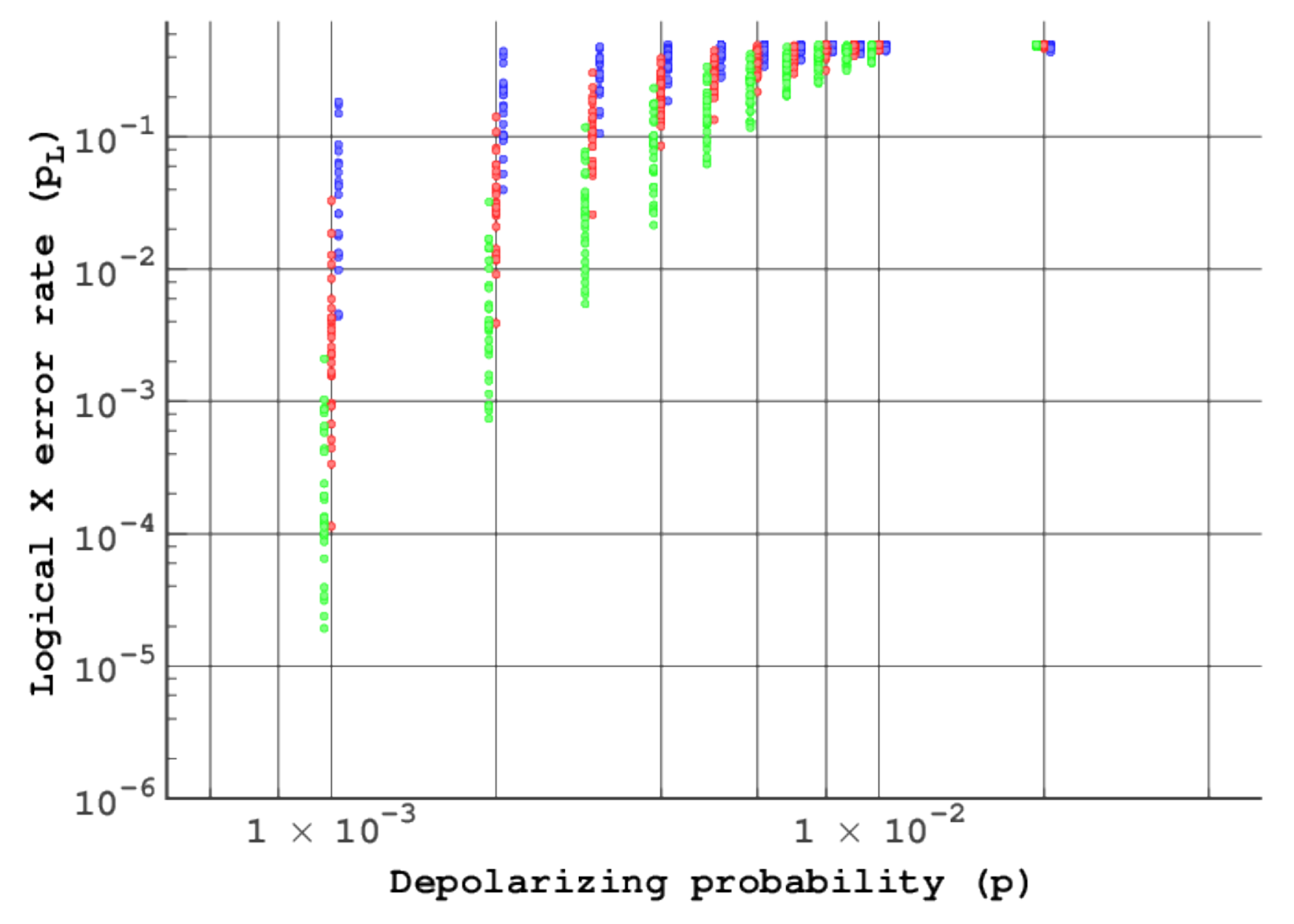}
  \caption[Scatterplot of $d=9$ with one dot per chip.]
  {Green dots are of $y=95\%$, red dots are of $y=90\%$ and blue dots are of $y=80\%$.
  Blue and green data are offset from the vertical line for visibility.
  }
  \label{fig:scatter_d9}
 \end{center}
\end{figure}

\begin{figure}[t]
 \begin{center}
  \includegraphics[width=15cm]{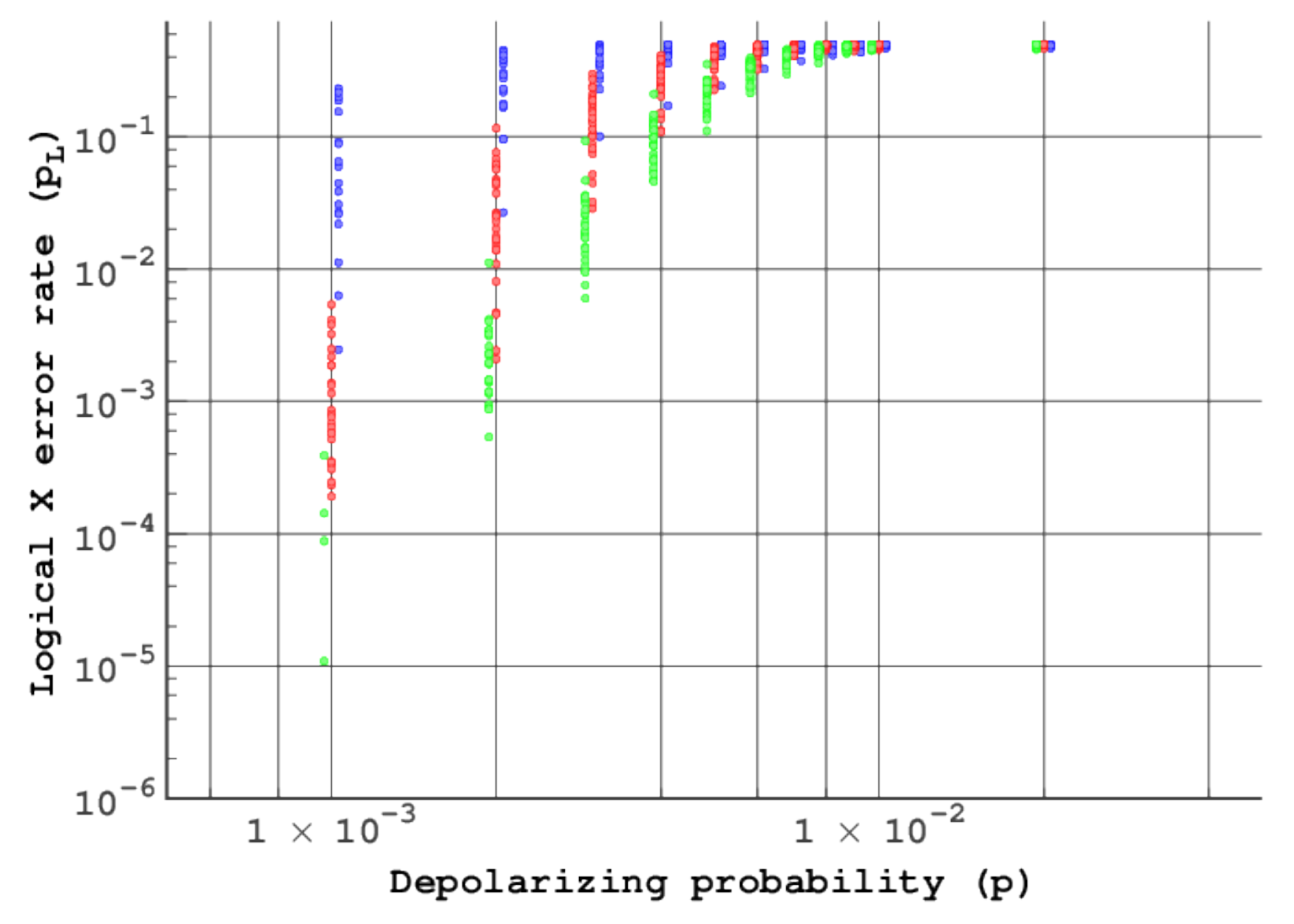}
  \caption[Scatterplot of $d=13$ with one dot per chip.]
  {Green dots are of $y=95\%$, red dots are of $y=90\%$ and blue dots are of $y=80\%$.
  Blue and green data are offset from the vertical line for visibility.
  }
  \label{fig:scatter_d13}
 \end{center}
\end{figure}

\begin{figure}[t]
 \begin{center}
  \includegraphics[width=15cm]{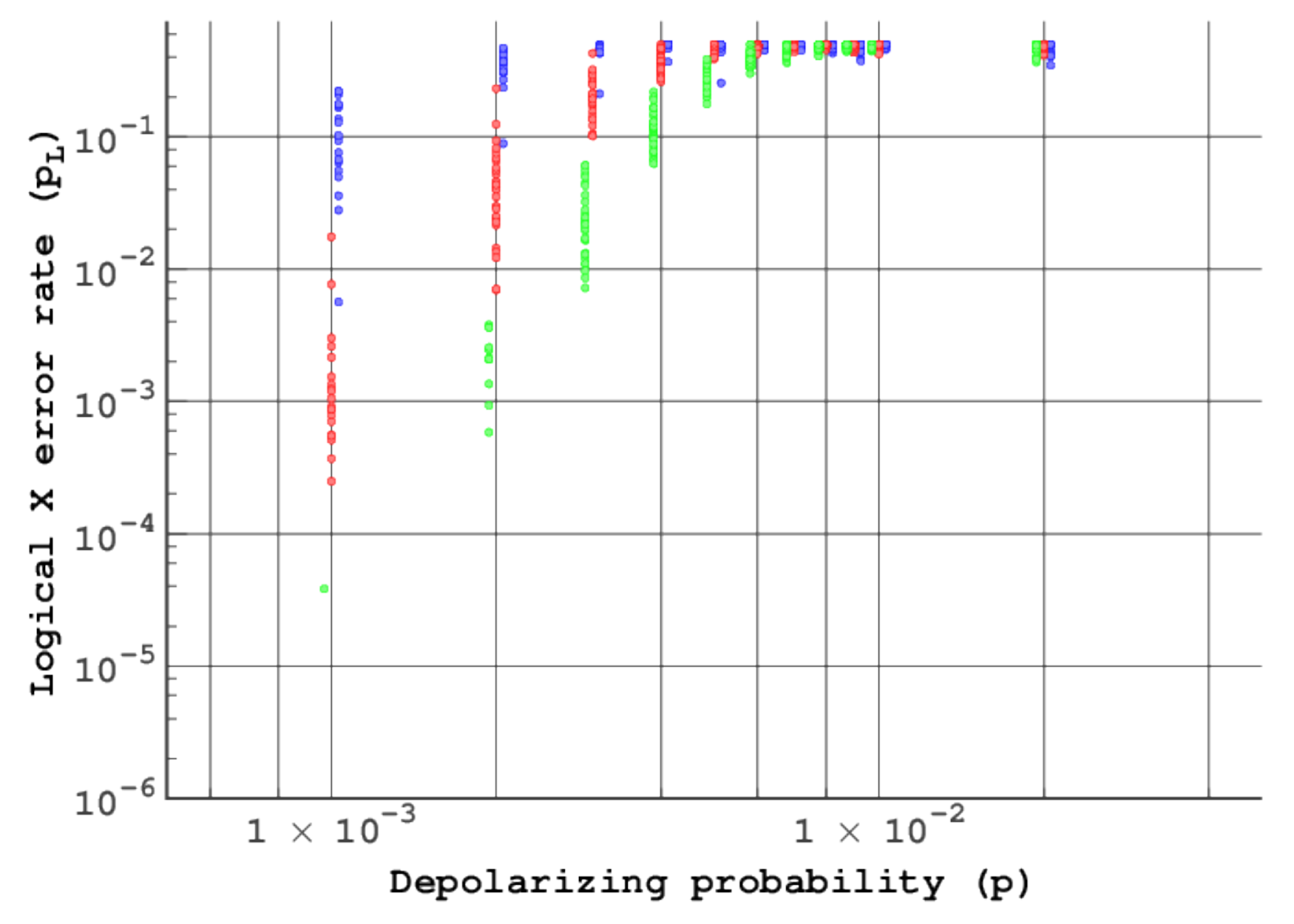}
  \caption[Scatterplot of $d=17$ with one dot per chip.]
  {Green dots are of $y=95\%$, red dots are of $y=90\%$ and blue dots are of $y=80\%$.
  Blue and green data are offset from the vertical line for visibility.
  }
  \label{fig:scatter_d17}
 \end{center}
\end{figure}

Figures \ref{fig:scatter_d5}, \ref{fig:scatter_d7}, \ref{fig:scatter_d9}, \ref{fig:scatter_d13} and \ref{fig:scatter_d17}
show the scatter plots of raw data of randomly defective lattices.
  Figure \ref{fig:scatter_d7} shows an outlier chip.
  Actually the lattice on the chip has $40 \times$ worse logical $Z$ error rate as logical $X$ error rate.
  The lattice by chance has faulty devices 
  which deform the left and the right boundaries to be close
  with preserving the top and the bottom boundaries apart.
  Because of the largely deformed shape of the lattice,
  the usable area of the lattice is narrow and there are only few faulty devices on the usable area
  which increase logical error rates.
  Hence the chip exhibits stronger tolerance against logical $X$ error than others.


\chapter{Supplemental data in encoding Bell pairs heterogeneously}
\section{Simulation data for $F=0.85$ raw Bell pairs on multi-NIC router architecture}
\label{data}
Table \ref{tab:purificationresult} shows our baseline simulation results using physical entanglement only with no encoding.
Table \ref{tab:purify_before_encode} shows the simulated results of {\it purification before encoding} for a Bell pair of a single layer of the Steane [[7,1,3]] code and a distance 3 surface code.
Table \ref{tab:purify_after_encode} shows the simulated results of the scheme {\it purification after encoding} of the same codes.
Table \ref{tab:purify_after_encode_strict} shows the simulated results of the scheme {\it purification after encoding with strict post-selection}.
Since purification at the level of encoded qubits consists of logical gates, {\it purification before encoding} has a much smaller KQ than the other two schemes.
{\it Purification after encoding with strict post-selection} discards more qubits than {\it purification after encoding} does to create a purified encoded Bell pair, so that {\it purification after encoding with strict post-selection} also results in a larger KQ.
Table \ref{tab:purify_after_encode_strict_css_phys} shows the simulated results of the scheme {\it purification after encoding with strict post-selection} between the Steane [[7,1,3]] code and the non-encoded physical half.
Table \ref{tab:purify_after_encode_strict_surface_phys} shows the simulated results of the scheme {\it purification after encoding with strict post-selection} between the distance three surface code and the non-encoded physical half.
   \begin{center}
 \begin{table*}
   \begin{center}
    \caption[Our baseline case, discrete simulation using physical entanglement purification only.]
    {The merged error rate is the probability that either X error or Z error occurs.
    The physical Bell pair inefficiency is $(\#\ created\ raw\ Bell\ pairs)/(\#\ purified\ Bell\ pairs)$.
    KQ is $\#qubit \times \#steps$. In this simulation, KQ is the number of chances that errors may occur.
    }
    \label{tab:purificationresult}
    (a)The local gate error rate is $10^{-3}$.\\
    \begin{tabular}[t]{|l||l|l|l|r|r|r|r|}
     \hline
     \#purifi- &X error &Z error & Merged     & Phys.      & KQ & \#single  & \#two\\
	 cation  & rate   & rate   & error      & Bell Pair  &    & qubit     & qubit\\
               &        &        & rate       & Ineff.     &    & gate      & gate \\ 
     \hline
0 & 0.113 & 0.1 & 0.156 & 1.0 & 88 & 86 & 1 \\
1 & 0.096 & 0.0197 & 0.106 & 2.5 & 98 & 91 & 5 \\
2 & 0.0247 & 0.0154 & 0.036 & 6.0 & 122 & 103 & 14 \\
3 & 0.0248 & 0.00498 & 0.0275 & 12.6 & 167 & 125 & 32 \\
4 & 0.00753 & 0.00523 & 0.0103 & 26.4 & 262 & 173 & 70 \\
     \hline
    \end{tabular}
    
    (b)The local gate error rate is $10^{-4}$.\\
    \begin{tabular}[t]{|l||l|l|l|r|r|r|r|}
     \hline
     \#purifi- &X error &Z error & Merged     & Phys.      & KQ & \#single  & \#two\\
	 cation  & rate   & rate   & error      & Bell Pair  &    & qubit     & qubit\\
               &        &        & rate       & Ineff.     &    & gate      & gate \\ 
     \hline
0 & 0.109 & 0.0968 & 0.151 & 1.0 & 88 & 86 & 1 \\
1 & 0.0915 & 0.0151 & 0.0988 & 2.5 & 98 & 91 & 5 \\
2 & 0.0183 & 0.0104 & 0.0271 & 6.0 & 122 & 103 & 14 \\
3 & 0.0183 & 0.000796 & 0.0189 & 12.4 & 166 & 125 & 32 \\
4 & 0.00125 & 0.000796 & 0.00182 & 25.7 & 258 & 171 & 68 \\
     \hline
    \end{tabular}
    
     (c)The local gate error rate is $10^{-5}$.\\
    \begin{tabular}[t]{|l||l|l|l|r|r|r|r|}
     \hline
     \#purifi- &X error &Z error & Merged     & Phys.      & KQ & \#single  & \#two\\
	 cation  & rate   & rate   & error      & Bell Pair  &    & qubit     & qubit\\
               &        &        & rate       & Ineff.     &    & gate      & gate \\ 
     \hline
0 & 0.112 & 0.0963 & 0.152 & 1.0 & 88 & 86 & 1 \\
1 & 0.0928 & 0.0152 & 0.101 & 2.5 & 98 & 91 & 5 \\
2 & 0.0177 & 0.0102 & 0.0262 & 6.0 & 121 & 103 & 14 \\
3 & 0.0176 & 0.000381 & 0.0179 & 12.4 & 166 & 125 & 32 \\
4 & 0.000633 & 0.000371 & 0.000975 & 25.6 & 257 & 171 & 68 \\
     \hline
    \end{tabular} 
   \end{center}
 \end{table*}
   \end{center}
 
   \begin{center}
 \begin{table*}
   \begin{center}
    \caption[Simulation results of {\it purification before encoding} for a Bell pair of a single layer of the
    Steane {[[7,1,3]]} code and a distance 3 surface code.]
    {Other conditions and definitions are as in Table \ref{tab:purificationresult}.
    }
    \label{tab:purify_before_encode}
    (a)The local gate error rate is $10^{-3}$.\\
    \begin{tabular}[t]{|l||l|l|l|r|r|r|r|}
     \hline
     \#purifi- &X error &Z error & Merged     & Phys.      & KQ & \#single  & \#two\\
	 cation  & rate   & rate   & error      & Bell Pair  &    & qubit     & qubit\\
               &        &        & rate       & Ineff.     &    & gate      & gate \\ 
     \hline
0 & 0.128 & 0.12 & 0.188 & 1.0 & 5402 & 4130 & 636 \\
1 & 0.111 & 0.0379 & 0.135 & 2.5 & 5412 & 4135 & 640 \\
2 & 0.0402 & 0.0355 & 0.0674 & 6.0 & 5436 & 4147 & 649 \\
3 & 0.0421 & 0.0258 & 0.061 & 12.6 & 5481 & 4170 & 667 \\
4 & 0.0231 & 0.0251 & 0.0424 & 26.4 & 5576 & 4217 & 705 \\
\hline
    \end{tabular}
    
    (b)The local gate error rate is $10^{-4}$.\\
    \begin{tabular}[t]{|l||l|l|l|r|r|r|r|}
     \hline
     \#purifi- &X error &Z error & Merged     & Phys.      & KQ & \#single  & \#two\\
	 cation  & rate   & rate   & error      & Bell Pair  &    & qubit     & qubit\\
               &        &        & rate       & Ineff.     &    & gate      & gate \\ 
     \hline
0 & 0.114 & 0.0976 & 0.155 & 1.0 & 5402 & 4130 & 636 \\
1 & 0.0927 & 0.0173 & 0.102 & 2.5 & 5412 & 4135 & 640 \\
2 & 0.02 & 0.0136 & 0.0315 & 6.0 & 5436 & 4147 & 649 \\
3 & 0.0201 & 0.00293 & 0.0224 & 12.4 & 5480 & 4169 & 667 \\
4 & 0.00298 & 0.0029 & 0.00529 & 25.7 & 5572 & 4215 & 703 \\
     \hline
    \end{tabular}
    
     (c)The local gate error rate is $10^{-5}$.\\
    \begin{tabular}[t]{|l||l|l|l|r|r|r|r|}
     \hline
     \#purifi- &X error &Z error & Merged     & Phys.      & KQ & \#single  & \#two\\
	 cation  & rate   & rate   & error      & Bell Pair  &    & qubit     & qubit\\
               &        &        & rate       & Ineff.     &    & gate      & gate \\ 
     \hline
0 & 0.11 & 0.0953 & 0.152 & 1.0 & 5402 & 4130 & 636 \\
1 & 0.0922 & 0.015 & 0.0999 & 2.5 & 5412 & 4135 & 640 \\
2 & 0.0183 & 0.0106 & 0.0273 & 6.0 & 5436 & 4147 & 649 \\
3 & 0.0177 & 0.000598 & 0.0182 & 12.4 & 5480 & 4169 & 667 \\
4 & 0.000797 & 0.000583 & 0.00132 & 25.6 & 5571 & 4215 & 703 \\
     \hline
    \end{tabular} 
   \end{center}
 \end{table*}
  \end{center}
  
   \begin{center}
 \begin{table*}
   \begin{center}
    \caption[Simulation results of the scheme {\it purification after encoding}
    between the Steane {[[7,1,3]]} code and the distance three surface code.]
    {Other conditions and definitions are as in Table \ref{tab:purificationresult}.
    }
    \label{tab:purify_after_encode}
    (a)The local gate error rate is $10^{-3}$.\\
    \begin{tabular}[t]{|l||l|l|l|r|r|r|r|}
     \hline
     \#purifi- &X error &Z error & Merged     & Phys.      & KQ & \#single  & \#two\\
	 cation  & rate   & rate   & error      & Bell Pair  &    & qubit     & qubit\\
               &        &        & rate       & Ineff.     &    & gate      & gate \\ 
    \hline
0 & 0.126 & 0.115 & 0.181 & 1.0 & 5402 & 4130 & 636 \\
1 & 0.131 & 0.0311 & 0.146 & 2.6 & 6892 & 5477 & 722 \\
2 & 0.0412 & 0.039 & 0.0726 & 7.2 & 10967 & 9159 & 956 \\
3 & 0.0693 & 0.0137 & 0.079 & 16.1 & 19068 & 16480 & 1425 \\
4 & 0.0278 & 0.031 & 0.0559 & 37.6 & 38536 & 34071 & 2550 \\
     \hline
    \end{tabular} 

    (b)The local gate error rate is $10^{-4}$.\\
    \begin{tabular}[t]{|l||l|l|l|r|r|r|r|}
     \hline
     \#purifi- &X error &Z error & Merged     & Phys.      & KQ & \#single  & \#two\\
	 cation  & rate   & rate   & error      & Bell Pair  &    & qubit     & qubit\\
               &        &        & rate       & Ineff.     &    & gate      & gate \\ 
    \hline
0 & 0.11 & 0.0983 & 0.154 & 1.0 & 5402 & 4130 & 636 \\
1 & 0.0958 & 0.0159 & 0.104 & 2.5 & 6776 & 5370 & 716 \\
2 & 0.0198 & 0.0125 & 0.0303 & 6.1 & 10061 & 8335 & 907 \\
3 & 0.0213 & 0.0012 & 0.0222 & 12.7 & 16134 & 13814 & 1262 \\
4 & 0.00222 & 0.00237 & 0.00442 & 26.5 & 28864 & 25300 & 2005 \\
     \hline
    \end{tabular} 

    (c)The local gate error rate is $10^{-5}$.\\
    \begin{tabular}[t]{|l||l|l|l|r|r|r|r|}
     \hline
     \#purifi- &X error &Z error & Merged     & Phys.      & KQ & \#single  & \#two\\
	 cation  & rate   & rate   & error      & Bell Pair  &    & qubit     & qubit\\
               &        &        & rate       & Ineff.     &    & gate      & gate \\ 
    \hline
0 & 0.108 & 0.0957 & 0.148 & 1.0 & 5402 & 4130 & 636 \\
1 & 0.0931 & 0.015 & 0.101 & 2.5 & 6759 & 5355 & 715 \\
2 & 0.018 & 0.01 & 0.0264 & 6.0 & 9961 & 8244 & 901 \\
3 & 0.0178 & 0.000395 & 0.0182 & 12.4 & 15880 & 13584 & 1247 \\
4 & 0.000729 & 0.000515 & 0.00123 & 25.7 & 28149 & 24652 & 1965 \\
     \hline
    \end{tabular} 
   \end{center}
 \end{table*}
  \end{center}

   \begin{table*}
   \begin{center}
    \caption[Simulation results of the scheme {\it purification after encoding with strict post-selection}
    between the Steane {[[7,1,3]]} code and the distance three surface code.]
    {Other conditions and definitions are as in Table \ref{tab:purificationresult}.
    The values at $p=10^{-4}$ and $10^{-5}$ demonstrate that the residual error rate saturates after more than one round of purification.
    }
    \label{tab:purify_after_encode_strict}
    (a)The local gate error rate is $10^{-3}$.\\
    \begin{tabular}[t]{|l||l|l|l|r|r|r|r|}
     \hline
     \#purifi- &X error &Z error & Merged     & Phys.      & KQ & \#single  & \#two\\
	 cation  & rate   & rate   & error      & Bell Pair  &    & qubit     & qubit\\
               &        &        & rate       & Ineff.     &    & gate      & gate \\ 
    \hline
0 & 0.13 & 0.118 & 0.184 & 1.0 & 5402 & 4130 & 636 \\
1 & 0.114 & 0.0146 & 0.121 & 3.0 & 7173 & 5732 & 737 \\
2 & 0.018 & 0.0113 & 0.0276 & 9.3 & 12748 & 10774 & 1053 \\
3 & 0.0193 & 0.000387 & 0.0196 & 21.2 & 23398 & 20403 & 1661 \\
4 & 0.000776 & 0.000414 & 0.00118 & 48.8 & 47963 & 42611 & 3063 \\
     \hline
    \end{tabular} 

    (b)The local gate error rate is $10^{-4}$.\\
    \begin{tabular}[t]{|l||l|l|l|r|r|r|r|}
     \hline
     \#purifi- &X error &Z error & Merged     & Phys.      & KQ & \#single  & \#two\\
	 cation  & rate   & rate   & error      & Bell Pair  &    & qubit     & qubit\\
               &        &        & rate       & Ineff.     &    & gate      & gate \\ 
    \hline
0 & 0.108 & 0.0981 & 0.152 & 1.0 & 5402 & 4130 & 636 \\
1 & 0.0948 & 0.0149 & 0.102 & 2.5 & 6797 & 5389 & 717 \\
2 & 0.0177 & 0.0104 & 0.0267 & 6.2 & 10173 & 8437 & 913 \\
3 & 0.0176 & 0.000343 & 0.0179 & 13.1 & 16442 & 14094 & 1278 \\
4 & 0.000545 & 0.000326 & 0.00087 & 27.3 & 29540 & 25912 & 2042 \\
     \hline
    \end{tabular} 

    (c)The local gate error rate is $10^{-5}$.\\
    \begin{tabular}[t]{|l||l|l|l|r|r|r|r|}
     \hline
     \#purifi- &X error &Z error & Merged     & Phys.      & KQ & \#single  & \#two\\
	 cation  & rate   & rate   & error      & Bell Pair  &    & qubit     & qubit\\
               &        &        & rate       & Ineff.     &    & gate      & gate \\ 
    \hline
0 & 0.109 & 0.0917 & 0.146 & 1.0 & 5402 & 4130 & 636 \\
1 & 0.0921 & 0.0153 & 0.0996 & 2.5 & 6766 & 5361 & 715 \\
2 & 0.0181 & 0.0101 & 0.0268 & 6.0 & 9968 & 8251 & 902 \\
3 & 0.0176 & 0.000333 & 0.0179 & 12.4 & 15913 & 13614 & 1249 \\
4 & 0.000572 & 0.000317 & 0.000887 & 25.8 & 28213 & 24710 & 1968 \\
     \hline
    \end{tabular} 
\end{center}
 \end{table*}

   \begin{table*}
   \begin{center}
    \caption[Simulation results of the scheme {\it purification after encoding with strict post-selection}
    between the Steane {[[7,1,3]]} code and non-encoded physical half.]
    {Other conditions and definitions are as in Table \ref{tab:purificationresult}.
    }
    \label{tab:purify_after_encode_strict_css_phys}
    (a)The local gate error rate is $10^{-3}$.\\
    \begin{tabular}[t]{|l||l|l|l|r|r|r|r|}
     \hline
     \#purifi- &X error &Z error & Merged     & Phys.      & KQ & \#single  & \#two\\
	 cation  & rate   & rate   & error      & Bell Pair  &    & qubit     & qubit\\
               &        &        & rate       & Ineff.     &    & gate      & gate \\ 
    \hline
0 & 0.116 & 0.107 & 0.166 & 1.0 & 4260 & 3660 & 300 \\
1 & 0.103 & 0.0171 & 0.111 & 2.6 & 5367 & 4709 & 330 \\
2 & 0.022 & 0.013 & 0.0325 & 6.9 & 8234 & 7425 & 409 \\
3 & 0.0227 & 0.00142 & 0.0236 & 15.0 & 13692 & 12595 & 559 \\
4 & 0.0028 & 0.00155 & 0.00386 & 32.5 & 25548 & 23825 & 884 \\
     \hline
    \end{tabular}
    
    (b)The local gate error rate is $10^{-4}$.\\
    \begin{tabular}[t]{|l||l|l|l|r|r|r|r|}
     \hline
     \#purifi- &X error &Z error & Merged     & Phys.      & KQ & \#single  & \#two\\
	 cation  & rate   & rate   & error      & Bell Pair  &    & qubit     & qubit\\
               &        &        & rate       & Ineff.     &    & gate      & gate \\ 
    \hline
0 & 0.11 & 0.0951 & 0.151 & 1.0 & 4260 & 3660 & 300 \\
1 & 0.0948 & 0.0154 & 0.103 & 2.5 & 5281 & 4627 & 328 \\
2 & 0.0179 & 0.0103 & 0.0266 & 6.1 & 7713 & 6929 & 395 \\
3 & 0.0181 & 0.000438 & 0.0184 & 12.6 & 12201 & 11179 & 518 \\
4 & 0.00077 & 0.000429 & 0.00115 & 26.2 & 21554 & 20033 & 775 \\
     \hline
    \end{tabular} 

    (c)The local gate error rate is $10^{-5}$.\\
    \begin{tabular}[t]{|l||l|l|l|r|r|r|r|}
     \hline
     \#purifi- &X error &Z error & Merged     & Phys.      & KQ & \#single  & \#two\\
	 cation  & rate   & rate   & error      & Bell Pair  &    & qubit     & qubit\\
               &        &        & rate       & Ineff.     &    & gate      & gate \\ 
    \hline
0 & 0.11 & 0.097 & 0.152 & 1.0 & 4260 & 3660 & 300 \\
1 & 0.0929 & 0.0155 & 0.101 & 2.5 & 5274 & 4620 & 328 \\
2 & 0.0182 & 0.0101 & 0.0267 & 6.0 & 7659 & 6878 & 393 \\
3 & 0.0176 & 0.00035 & 0.0179 & 12.4 & 12069 & 11053 & 515 \\
4 & 0.000577 & 0.000334 & 0.000904 & 25.7 & 21200 & 19697 & 766 \\
     \hline
    \end{tabular} 
\end{center}
 \end{table*}

    \begin{table*}
   \begin{center}
    \caption[Simulation results of the scheme {\it purification after encoding with strict post-selection}
    between the distance three surface code and non-encoded physical half.]
    {Other conditions and definitions are as in Table \ref{tab:purificationresult}.
    }
    \label{tab:purify_after_encode_strict_surface_phys}
    (a)The local gate error rate is $10^{-3}$.\\
    \begin{tabular}[t]{|l||l|l|l|r|r|r|r|}
     \hline
     \#purifi- &X error &Z error & Merged     & Phys.      & KQ & \#single  & \#two\\
	 cation  & rate   & rate   & error      & Bell Pair  &    & qubit     & qubit\\
               &        &        & rate       & Ineff.     &    & gate      & gate \\ 
    \hline
0 & 0.13 & 0.113 & 0.181 & 1.0 & 1188 & 914 & 137 \\
1 & 0.116 & 0.0172 & 0.125 & 2.8 & 1985 & 1611 & 202 \\
2 & 0.0224 & 0.0135 & 0.0334 & 8.2 & 4321 & 3652 & 393 \\
3 & 0.0227 & 0.00145 & 0.0236 & 18.3 & 8739 & 7511 & 755 \\
4 & 0.0028 & 0.00154 & 0.00386 & 40.9 & 18625 & 16148 & 1566 \\
     \hline
    \end{tabular}
    
    (b)The local gate error rate is $10^{-4}$.\\
    \begin{tabular}[t]{|l||l|l|l|r|r|r|r|}
     \hline
     \#purifi- &X error &Z error & Merged     & Phys.      & KQ & \#single  & \#two\\
	 cation  & rate   & rate   & error      & Bell Pair  &    & qubit     & qubit\\
               &        &        & rate       & Ineff.     &    & gate      & gate \\ 
    \hline
0 & 0.114 & 0.0959 & 0.154 & 1.0 & 1188 & 914 & 137 \\
1 & 0.0943 & 0.0157 & 0.102 & 2.5 & 1864 & 1504 & 193 \\
2 & 0.0179 & 0.0105 & 0.0267 & 6.2 & 3481 & 2915 & 328 \\
3 & 0.018 & 0.000445 & 0.0184 & 12.9 & 6482 & 5533 & 580 \\
4 & 0.000808 & 0.000438 & 0.0012 & 26.9 & 12728 & 10982 & 1104 \\
     \hline
    \end{tabular} 

    (c)The local gate error rate is $10^{-5}$.\\
    \begin{tabular}[t]{|l||l|l|l|r|r|r|r|}
     \hline
     \#purifi- &X error &Z error & Merged     & Phys.      & KQ & \#single  & \#two\\
	 cation  & rate   & rate   & error      & Bell Pair  &    & qubit     & qubit\\
               &        &        & rate       & Ineff.     &    & gate      & gate \\ 
    \hline
0 & 0.108 & 0.0934 & 0.147 & 1.0 & 1188 & 914 & 137 \\
1 & 0.091 & 0.0152 & 0.0988 & 2.5 & 1848 & 1490 & 192 \\
2 & 0.0179 & 0.0105 & 0.0267 & 6.0 & 3410 & 2853 & 323 \\
3 & 0.0176 & 0.000336 & 0.0179 & 12.4 & 6294 & 5368 & 565 \\
4 & 0.000589 & 0.00033 & 0.000911 & 25.8 & 12266 & 10578 & 1067 \\
     \hline
    \end{tabular} 
\end{center}
 \end{table*}

\section{Simulation data for Bell pair creation via local gates}
\label{sec:appendixB}
Data in this appendix is of simulations in which raw Bell pairs are
created by local gates, two initializations, an Hadamard gate, an identity gate
and a CNOT gate.
Table~\ref{tab:lg_phys_phys} shows the simulated results using physical entanglement only with no encoding.
Table~\ref{tab:lg_purify_after_encode} shows the simulated results of the scheme {\it purification after encoding} for a Bell pair of a single layer of the Steane [[7,1,3]] code and a distance 3 surface code.
Table~\ref{tab:lg_purify_after_encode_strict} shows the simulated results of the scheme {\it purification after encoding with strict post-selection} for a Bell pair of a single layer of the Steane [[7,1,3]] code and a distance 3 surface code.

The fidelity of raw Bell pairs created via local gates is much better
than 0.85, lowering the need for purification.   However,
architectures that can use this method are more limited in
scalability.  Thus, this method may be used for standalone code
converters, but will not be the preferred method when building
scalable quantum internetworking repeaters.
\begin{table*}
  \begin{center}
    \caption[Results of simulation in which raw Bell pairs are created using local gates,
   and using physical entanglement purification only.]
   {Other conditions and definitions are as in Table \ref{tab:purificationresult}.}
    \label{tab:lg_phys_phys}
    (a)The local gate error rate is $10^{-3}$.\\
    \begin{tabular}[t]{|l||l|l|l|r|r|r|r|}
      \hline
      \#purifi- &X error &Z error & Merged     & Phys.      & KQ & \#single  & \#two\\
	 cation  & rate   & rate   & error      & Bell Pair  &    & qubit     & qubit\\
               &        &        & rate       & Ineff.     &    & gate      & gate \\ 
      \hline
      0 & 0.00444 & 0.00447 & 0.00736 & 1.0 & 88 & 90 & 1 \\
      1 & 0.00777 & 0.0047 & 0.0101 & 2.0 & 96 & 98 & 4 \\
      2 & 0.00629 & 0.00452 & 0.00858 & 4.1 & 112 & 115 & 10 \\
      \hline
    \end{tabular} 
    
    (b)The local gate error rate is $10^{-4}$.\\
    \begin{tabular}[t]{|l||l|l|l|r|r|r|r|}
      \hline
      \#purifi- &X error &Z error & Merged     & Phys.      & KQ & \#single  & \#two\\
	 cation  & rate   & rate   & error      & Bell Pair  &    & qubit     & qubit\\
               &        &        & rate       & Ineff.     &    & gate      & gate \\ 
      \hline
      0 & 0.000445 & 0.00045 & 0.000731 & 1.0 & 88 & 90 & 1 \\
      1 & 0.000768 & 0.000456 & 0.000995 & 2.0 & 96 & 98 & 4 \\
      2 & 0.000644 & 0.000468 & 0.000875 & 4.0 & 112 & 114 & 10 \\      
      \hline
    \end{tabular} 
    
    (c)The local gate error rate is $10^{-5}$.\\
    \begin{tabular}[t]{|l||l|l|l|r|r|r|r|}
      \hline
      \#purifi- &X error &Z error & Merged     & Phys.      & KQ & \#single  & \#two\\
	 cation  & rate   & rate   & error      & Bell Pair  &    & qubit     & qubit\\
               &        &        & rate       & Ineff.     &    & gate      & gate \\ 
      \hline
      0 & 4.46e-05 & 4.64e-05 & 7.5e-05 & 1.0 & 88 & 90 & 1 \\
      1 & 7.79e-05 & 4.57e-05 & 0.000101 & 2.0 & 96 & 98 & 4 \\
      2 & 6.35e-05 & 4.38e-05 & 8.51e-05 & 4.0 & 112 & 114 & 10 \\
      \hline
    \end{tabular} 
  \end{center}
\end{table*}

\begin{table*}
  \begin{center}
    \caption[Simulation results of {\it purification after encoding} 
      for a Bell pair of a single layer of the
      Steane {[[7,1,3]]} code and a distance 3 surface code.
      Raw Bell pairs are created using local gates.]{
      Other conditions and definitions are as in Table \ref{tab:purificationresult}.}
    \label{tab:lg_purify_after_encode}
    (a)The local gate error rate is $10^{-3}$.\\
    \begin{tabular}[t]{|l||l|l|l|r|r|r|r|}
      \hline
      \#purifi- &X error &Z error & Merged     & Phys.      & KQ & \#single  & \#two\\
	 cation  & rate   & rate   & error      & Bell Pair  &    & qubit     & qubit\\
               &        &        & rate       & Ineff.     &    & gate      & gate \\ 
      \hline
      0 & 0.0221 & 0.0255 & 0.0419 & 1.0 & 5402 & 4134 & 636 \\
      1 & 0.0371 & 0.00451 & 0.0399 & 2.2 & 6521 & 5148 & 702 \\
      2 & 0.00697 & 0.0102 & 0.0162 & 5.0 & 9147 & 7526 & 857 \\
      \hline
    \end{tabular} 
    
    (b)The local gate error rate is $10^{-4}$.\\
    \begin{tabular}[t]{|l||l|l|l|r|r|r|r|}
      \hline
      \#purifi- &X error &Z error & Merged     & Phys.      & KQ & \#single  & \#two\\
	 cation  & rate   & rate   & error      & Bell Pair  &    & qubit     & qubit\\
               &        &        & rate       & Ineff.     &    & gate      & gate \\ 
      \hline
      0 & 0.00217 & 0.00263 & 0.00426 & 1.0 & 5402 & 4134 & 636 \\
      1 & 0.00354 & 0.000246 & 0.00369 & 2.0 & 6380 & 5018 & 695 \\
      2 & 0.000249 & 0.00041 & 0.000617 & 4.1 & 8369 & 6818 & 814 \\
      \hline
    \end{tabular} 

    (c)The local gate error rate is $10^{-5}$.\\
    \begin{tabular}[t]{|l||l|l|l|r|r|r|r|}
      \hline
      \#purifi- &X error &Z error & Merged     & Phys.      & KQ & \#single  & \#two\\
	 cation  & rate   & rate   & error      & Bell Pair  &    & qubit     & qubit\\
               &        &        & rate       & Ineff.     &    & gate      & gate \\ 
      \hline
      0 & 0.000214 & 0.000251 & 0.000413 & 1.0 & 5402 & 4134 & 636 \\
      1 & 0.000351 & 2.2e-05 & 0.000364 & 2.0 & 6366 & 5005 & 694 \\
      2 & 2.24e-05 & 3.56e-05 & 5.44e-05 & 4.0 & 8296 & 6751 & 810 \\
      \hline
    \end{tabular} 
  \end{center}
\end{table*}

\begin{table*}
  \begin{center}
    \caption[Simulation results of {\it purification after encoding with post-selection} 
      for a Bell pair of a single layer of the
      Steane {[[7,1,3]]} code and a distance 3 surface code.
   Raw Bell pairs are created using local gates.]
   {Other conditions and definitions are as in Table \ref{tab:purificationresult}.
      Note that one round of purification at $p=10^{-5}$ finds only 4 residual Z errors in 100 million output Bell pairs
      and that two rounds of purification at $p=10^{-5}$ find only 1 residual X error and only 2 residual Z errors in 100 million output Bell pairs.
    }
    \label{tab:lg_purify_after_encode_strict}
    (a)The local gate error rate is $10^{-3}$.\\
    \begin{tabular}[t]{|l||l|l|l|r|r|r|r|}
      \hline
      \#purifi- &X error &Z error & Merged     & Phys.      & KQ & \#single  & \#two\\
	 cation  & rate   & rate   & error      & Bell Pair  &    & qubit     & qubit\\
               &        &        & rate       & Ineff.     &    & gate      & gate \\ 
      \hline
      0 & 0.0203 & 0.0251 & 0.0402 & 1.0 & 5402 & 4134 & 636 \\
      1 & 0.0303 & 0.000142 & 0.0304 & 2.4 & 6705 & 5316 & 712 \\
      2 & 0.000323 & 0.000246 & 0.000556 & 6.3 & 10230 & 8511 & 917 \\
      \hline
    \end{tabular} 
    
    (b)The local gate error rate is $10^{-4}$.\\
    \begin{tabular}[t]{|l||l|l|l|r|r|r|r|}
      \hline
      \#purifi- &X error &Z error & Merged     & Phys.      & KQ & \#single  & \#two\\
	 cation  & rate   & rate   & error      & Bell Pair  &    & qubit     & qubit\\
               &        &        & rate       & Ineff.     &    & gate      & gate \\ 
      \hline
      0 & 0.0021 & 0.00261 & 0.00418 & 1.0 & 5402 & 4134 & 636 \\
      1 & 0.00301 & 1.41e-06 & 0.00301 & 2.0 & 6395 & 5033 & 696 \\
      2 & 2.69e-06 & 2.18e-06 & 4.81e-06 & 4.2 & 8446 & 6888 & 819 \\
      \hline
    \end{tabular} 
    
    (c)The local gate error rate is $10^{-5}$.\\
    \begin{tabular}[t]{|l||l|l|l|r|r|r|r|}
      \hline
      \#purifi- &X error &Z error & Merged     & Phys.      & KQ & \#single  & \#two\\
	 cation  & rate   & rate   & error      & Bell Pair  &    & qubit     & qubit\\
               &        &        & rate       & Ineff.     &    & gate      & gate \\ 
      \hline
      0 & 0.000203 & 0.000251 & 0.000405 & 1.0 & 5402 & 4134 & 636 \\
      1 & 0.000298 & 4e-08 & 0.000298 & 2.0 & 6367 & 5007 & 694 \\
      2 & 1e-08 & 2e-08 & 3e-08 & 4.0 & 8304 & 6758 & 811 \\      
      \hline
    \end{tabular} 
  \end{center}
\end{table*}

\chapter{List of Papers and Presentations}
\section{First Author Papers and Presentations}
\subsection{Peer-Reviewed Journals}
\begin{enumerate}
 \item S. Nagayama, B.-S. Choi, S. Devitt, S. Suzuki, and R. Van Meter. Interoperability
in encoded quantum repeater networks. \textit{Phys. Rev. A}, 93:042338, Apr 2016.
 \item S. Nagayama, A. G. Fowler, D. Horsman, S. J. Devitt, and R. Van Meter. Surface
code error correction on a defective lattice, \textit{New Journal of Physics}, 19(2):023050, 2017.
 \item S. Nagayama, T. Satoh, and R. Van Meter. State injection, lattice surgery, and dense packing of the deformation-based surface code. \textit{Phys. Rev. A}, 95:012321, Jan 2017.
\end{enumerate}
\subsection{International Conferences}
\subsubsection{Oral Presentation}
\begin{enumerate}
 \item Shota Nagayama, Clare Horsman, Austin Fowler and Rodney Van Meter. Surface Code Quantum Computation on a Defective Physical Lattice. \textit{JFLI meeting on quantum information and computation}, http://jfli.nii.ac.jp/medias/wordpress/?p=832, Paris, French, March 5th, 2013.
\end{enumerate}
\subsubsection{Poster Presentations}
\begin{enumerate}
 \item Shota Nagayama and Rodney Van Meter. RaQoon2: Extension of Internet Key Exchange to Use Quantum Key Distribution. \textit{Updating Quantum Cryptography and Communications 2010} Japan, Nov 2010.
 \item Shota Nagayama and Rodney Van Meter. Defective Qubits in Surface Code Quantum Computation on a Fixed Lattice. \textit{Asian Conference on Quantum Information Science 2010}, Tokyo Univ. Japan, Aug 2010.
 \item Shota Nagayama, Clare Horsman, Austin Fowler and Rodney Van Meter. Surface Code Quantum Computation on a Defective Physical Lattice. \textit{Quantum Error Correction 2011}, Los Angeles, USA, Dec 2011.
 \item Shota Nagayama, Byung-Soo Choi, Simon Devitt, Shigeya Suzuki, Rodney Van Meter. Error and resource estimation of generalized heterogeneous coded Bell pairs. \textit{Workshop for Quantum Repeaters and Networks}, http://wqrn.pratt.duke.edu/ Monterey, USA, 2015.
 \item Shota Nagayama, Shigeya Suzuki, Takahiko Satoh, Takaaki Matsuo and Rodney Van Meter. Scalable quantum router architecture with code interoperability. \textit{QCMC2016}, https://qcmc.quantumlah.org/, Singapore, Jul. 2016.
\end{enumerate}
\subsubsection{Internet Drafts}
\begin{enumerate}
 \item S. Nagayama and R. Van Meter. draft-nagayama-ipsecme-ipsec-with-qkd-00. Oct. 2009. expires April 22, 2010.
 \item S. Nagayama and R. Van Meter. draft-nagayama-ipsecme-ipsec-with-qkd-01. Oct. 2014. expires April 30, 2015.
\end{enumerate}

\subsection{National Conferences and Workshops}
\begin{enumerate}
 \item 
\begin{CJK}{UTF8}{min}
量子鍵配送を使ったIPsecのためのIKE拡張
\end{CJK}
. \textit{Internet Conference 2009}, Kyoto, Japan. Oct. 2009.
 \item Shota Nagayama and Rodney Van Meter. Defective Qubits in Surface Code Quantum Computation on a Fixed Lattice. \textit{FIRST Project Summer School}, Okinawa, Japan. Aug. 2010.
 \item Shota Nagayama and Rodney Van Meter. Defective Qubits in Surface Code Quantum Computation on a Fixed Lattice. \textit{FIRST All-hands meeting}, Kyoto, Japan. Dec. 2010.
 \item Shota Nagayama, Clare Horsman, Austin Fowler and Rodney Van Meter. Surface code quantum computation on a defective lattice. \textit{FIRST Project Summer School}, Okinawa, Japan. Jul. 2012.
\end{enumerate}

\subsection{Teaching}
\begin{enumerate}
 \item FIRST/Quantum Cybernetics/CREST Joint 1.5-day Surface Code Quantum Error Correction Tutorial/Workshop.
 Videos (Japanese) available at http://www.soi.wide.ad.jp/class/cgi/class\_top.cgi?20110030, Feb. 2011.
\end{enumerate}

\subsection{Other Presentations}
\begin{enumerate}
 \item IPsec with QKD, WIDE Camp Spring 2009. Mar. 2009.
 \item Surface code on defective lattice, The University of Melbourne. Apr. 2011.
 \item Surface Code Quantum Computation on a Defective Physical Lattice. Microsoft Research, Seattle. Jan. 2013.
 \item Surface Code Quantum Computation on a Defective Physical Lattice. Georgia Institute of Technology. Feb. 2013.
 \item Surface Code Quantum Computation on a Defective Physical Lattice. Kyoto. Jan. 2013.
 \item Heterogeneously encoded Bell pairs. Ritsumeikan University, Shiga, Japan. Jul. 2015.
\end{enumerate}

\section{Non-First Author Papers and Presentations}
\subsection{Peer-Reviewed Journals}
\begin{enumerate}
 \item Takahiko Satoh, Kaori Ishizaki, Shota Nagayama and Rodney Van Meter. Analysis of quantum network coding for realistic repeater networks. \textit{Physical Review A}, 93:032302. Mar. 2016.
 \item Takahiko Satoh, Shota Nagayama, and Rodney Van Meter. The Network Impact of Hijacking a Quantum Repeater. arXiv:1701.04587, Jan. 2017.
 \item Rodney Van Meter, Takahiko Satoh, Shota Nagayama, Takaaki Matsuo and Shigeya Suzuki. Optimizing Timing of High-Success-Probability Quantum Repeaters. arXiv:1701.04586, Jan. 2017.
\end{enumerate}

\subsection{International Conferences}
\subsubsection{Poster Presentations}
\begin{enumerate}
 \item Takahiko Satoh, Shota Nagayama and Rodney Van Meter.  A reversible ternary adder for quantum computation. \textit{Asian Conference on Quantum Information Science 2010}, Tokyo Univ. Japan, Aug 2007.
 \item Takaaki Matsuo, Takahiko Satoh, Shota Nagayama, Shigeya Suzuki and Rodney Van Meter. Wide-area topology of a Quantum Internet. \textit{QCMC2016}, https://qcmc.quantumlah.org/, Singapore, Jul. 2016.
 \item Rodney Van Meter, Shigeya Suzuki, Shota Nagayama, Takahiko Satoh, Takaaki Matsuo, Amin Taherkhani, Simon Devitt and Joe Touch. Large-Scale Simulation of the Quantum Internet. \textit{QCMC2016}, https://qcmc.quantumlah.org/, Singapore, Jul. 2016.
 \item Takahiko Satoh, Shigeya Suzuki, Shota Nagayama, Takaaki Matsuo and Rodney Van Meter. Routing on a Quantum Internet. \textit{QCMC2016}, https://qcmc.quantumlah.org/, Singapore, Jul. 2016.
 \item Shigeya Suzuki, Rodney Van Meter, Shota Nagayama, Takahiko Satoh and Takaaki Matsuo. Architecture of software simulation of a Quantum Internet. \textit{QCMC2016}, https://qcmc.quantumlah.org/, Singapore, Jul. 2016.
\end{enumerate}

\end{document}